\documentclass[superscriptaddress,amsfonts,amssymb,amsmath,floats,twocolumn,footinbib,aps,prb,longbibliography,shownopacs]{revtex4-2}
\usepackage[pdftex]{graphicx}
\usepackage{subfigure}
\usepackage{dsfont}
\usepackage{lipsum}
\usepackage{dcolumn}
\usepackage{bm}
\usepackage{color}
\usepackage{physics}
\usepackage{multirow}
\usepackage[pdftex,colorlinks=true, linkcolor=blue,citecolor=blue,filecolor=blue]{hyperref}
\usepackage{natbib}
\usepackage{url}

\usepackage{systeme}

\usepackage{comment}
\usepackage{bbold}

\usepackage{array}
\newcommand{\PreserveBackslash}[1]{\let\temp=\\#1\let\\=\temp}
\newcolumntype{C}[1]{>{\PreserveBackslash\centering}p{#1}}
\newcolumntype{R}[1]{>{\PreserveBackslash\raggedleft}p{#1}}
\newcolumntype{L}[1]{>{\PreserveBackslash\raggedright}p{#1}}
\usepackage{pifont}

\begin{document}
\title{Theories for charge-driven nematicity in kagome metals}

\author{Francesco Grandi} 
\affiliation{Institut f\"ur Theorie der Statistischen Physik, RWTH Aachen University, 52056 Aachen, Germany}
\author{Michael A. Sentef}
\affiliation{Institute for Theoretical Physics and Bremen Center for Computational Materials Science, University of Bremen, 28359 Bremen, Germany}
\affiliation{Max Planck Institute for the Structure and Dynamics of Matter, Center for Free-Electron Laser Science (CFEL), Luruper Chaussee 149, 22761 Hamburg, Germany}
\author{Dante M. Kennes}
\affiliation{Institut f\"ur Theorie der Statistischen Physik, RWTH Aachen University, 52056 Aachen, Germany}
\affiliation{JARA-Fundamentals of Future Information Technology, 52056 Aachen, Germany}
\affiliation{Max Planck Institute for the Structure and Dynamics of Matter, Center for Free-Electron Laser Science (CFEL), Luruper Chaussee 149, 22761 Hamburg, Germany}
\author{Ronny Thomale}
\affiliation{Institut f\"ur Theoretische Physik und Astrophysik and W\"urzburg-Dresden Cluster of Excellence ct.qmat, Universit\"at W\"urzburg, 97074 W\"urzburg, Germany}

\begin{abstract}
Starting from a low-energy continuum model for the band dispersion of the $2 \times 2$ charge-ordered phase of the kagome metals $A$V$_3$Sb$_5$ ($A=$ K, Rb, Cs), we show that nematicity can develop in this state driven either by three inequivalent $1 \times 4$ charge fluctuations preemptive of a $1 \times 4$ charge order (CO), or by an actual zero momentum $d$-wave charge Pomeranchuk instability (PI). We perform an analysis that starts from a Kohn-Luttinger theory in the particle-hole sector, which allows us to establish a criterion for the development of an attractive nematic channel near the onset of the $1 \times 4$ CO and near the $d$-wave charge PI, respectively. We derive an effective charge-fermion model for the $d$-wave PI with a nematic susceptibility given via a random phase approximation (RPA) summation. By contrast, for the finite momentum CO, the RPA scheme breaks down and needs to be improved upon by including Aslamazov-Larkin contributions to the nematic pairing vertex. We then move to the derivation of the Ginzburg-Landau potentials for the $1 \times 4$ CO and for the $d$-wave PI, and we obtain the corresponding analytical expression for the nematic susceptibility at the nematic transition temperature T $ \sim \text{T}_\text{nem}$ in both cases. The nematic response functions obtained in this way are interpreted starting from the two charge-fermion models, and we underline under which assumptions one recovers the Ginzburg-Landau result. Finally, we show an enhancement of the nematic character that is rooted in the coupling of the order parameters to elastic deformations. Our work establishes a relation between the nematicity observed in some of the iron-based superconductors, where the nematic phase might be driven by spin fluctuations, and the vanadium-based kagome metals, where charge fluctuations likely induce nematicity. The two microscopic mechanisms we propose for the stabilization of the nematic state in $A$V$_3$Sb$_5$, i.e., the zero-momentum $d$-wave PI and the fluctuations of the finite momentum CO, are distinguishable by diffusive scattering experiments, meaning that it is possible to gauge which of the two theories, if any, is the most likely to describe this phase. Both mechanisms might also be relevant for the recently discovered titanium-based family $A$Ti$_3$Sb$_5$, where nematicity has also been observed.
\end{abstract} 
\maketitle

\section{Introduction} \label{sec:intro}
Nematicity is a property originally observed in classical liquid crystals, where the microscopic structure of the constituent molecules and their anisotropic interactions lead to a breaking, on average, of the rotational symmetry in the system but preserving the spatial isotropy of the liquid. When nematicity occurs together with stratification so that the molecules are arranged in layers and show correlations in their positions, the liquid crystal is in a smectic phase \cite{Stephen1974_RMP}. In the smectic phase, no long-range crystalline order is present within each layer, and, as a general rule, nematicity occurs at higher temperatures than smecticity. The first quantum analog of the nematic state was described in the context of interacting spins on a lattice \cite{Blume1969_JAP,Andreev1984_JETP}, where the rotational symmetry that is broken is the internal SU$(2)$ spin symmetry \cite{Kohama2019_PNAS,Caci2023_PRB}. Another quantum counterpart of liquid crystals has been found in correlated Fermi liquids \cite{Abanov1995_PRB}. On a lattice, the properties of the Fermi liquid are only invariant under the action of the discrete rotations C$_n$ ($n \in \{ 1,2,3,4,6 \}$) of the point group symmetry of the host crystal, and the continuous translational symmetry is always broken down to a discrete one. Nevertheless, when the discrete rotational symmetry of the lattice is broken, the system is in a nematic phase, while if also the discrete translational symmetry breaks along one direction, the state is called smectic.

\noindent
As given, the definition of nematicity is purely phenomenological, meaning that it provides room for several microscopic mechanisms for the stabilization of a (generic) nematic state. In some cases, the nematic character might be driven by a structural transition of the system, while in some other cases the transition might be electronically driven. The existence of a finite coupling between the electronic and the structural degrees of freedom makes the problem of understanding the leading instability very challenging both from the experimental and the theoretical points of view. In experiments, since the electronic and the structural transitions take place together, one might hardly separate among the two conditions, even if time-resolved measurements might provide indications on the most important contribution. In the theoretical works, if a model takes into account both the degrees of freedom on the same footing, one might get a different result concerning the most relevant contribution to the instability depending by the values of the parameters employed for the description of the problem, and, in several cases, it is hard to gauge their values against experiments, making most of the theoretical predictions not conclusive.

\noindent
Even when it is clear that the nematic character comes from the electrons, still there are several possibilities for the microscopic mechanism driving the transition. Nevertheless, one can identify two macro-types of nematic states \cite{Yamase2021_JPSJ}: 1) A symmetry breaking which leads to some bilinear fermionic operator with non-zero expectation value such that the translational symmetry of the problem is preserved while the rotational one is broken; 2) A symmetry breaking with quadrilinear fermionic operators order parameter, i.e., the nematic character, in this second case, is driven by the anisotropy of the fluctuations in some degrees of freedom, a topic connected to intertwined vestigial orders \cite{Fradkin2015_RMP,Svistunov2015_SSM,Fernandes2019_ARCMP} and to the original observation of nematicity in liquid crystals, which sees this property as arising from a melted or ``unsuccessful'' smectic phase. In the first case, the nematic character of the phase naturally emerges from the real- or momentum-space structure of the order parameter, which, below the critical temperature, becomes different from zero. For instance, by considering the distortions of the Fermi surface of a system projected on harmonics with angular momentum $l$ and on the charge or the spin sectors, one might get a charge (spin) $l$-wave Pomeranchuk instability (PI) \cite{Pomeranchuk1958_JETP} where the order parameter has the form of a bilinear in the fermionic operators \cite{Quintanilla2006_PRB,Chubukov2018_JETP}. This instability might take place in several systems, however, it does not always correspond to a nematic state, as it happens for $l=0$ ($s$-wave), where one gets phase separation (Stoner instability \cite{Stoner1938_PRSL}) in the charge (spin) channel. For $l=1$ ($p$-wave) PI, conservation laws for the total charge and spin prevent the onset of charge- and spin-current order parameters \cite{Wu2018_PRB}, however the corresponding state would be nematic \cite{Hirsch1990_PRB}. Instead, no restrictions are found for the order parameter of an $l=2$ ($d$-wave) PI, which also corresponds to a nematic state \cite{Halboth2000_PRL,Yamase2000_JPSJ,Zacharias2009_PRB,Oganesyan2001_PRB,Valenzuela2008_NJP,Maslov2010_PRB}. In case 2), the fluctuations of some degree of freedom are assumed to become large enough, and this might naturally occur nearby a second order phase transition. The lower temperature phase \footnote{Here, we speak about ``low-temperature phase'' to keep the analogy with the phenomenology of the liquid crystals described before, but, in general condensed matter systems, the smectic state might be stabilized at higher or lower pressure or doping.} might even break the translational symmetry of the problem, leading to a smectic state \cite{Fernandes2012_PRB}. Another possibility consists in the presence of a quantum critical point (QCP) at zero temperature \cite{Hertz1976_PRB,Millis1993_PRB,Loehneysen2007_RMP,Sachdev2011_CUP}, which might also lead to strong fluctuations nearby it. These fluctuations might give rise to non-Fermi liquid behaviors \cite{Efetov2013_NatPhys,Licciardello2019_Nat} and they might provide a pairing glue between electrons leading to superconductivity, and this becomes particularly relevant in the proximity of a nematic QCP \cite{Lederer2015_PRL,Lederer2017_PNAS,Ishida2022_PNAS,Sur2023_NatComm}. The poorly-understood relation between nematicity and unconventional superconductivity explains the huge interest about the former phase: It might explain the origins of the latter \cite{Chu2012_Sci,Hosoi2016_PNAS,Auvray2019_NatComm,Boehmer2022_NatPhys}.

\noindent
Coming to a more concrete example, in several of the iron-based superconductors \cite{Fernandes2022_Nat}, one observes the following phenomenology: Below a critical temperature T$_\text{nem}$, an Ising $\mathbb{Z}_2$ symmetry is broken and the nematic character seems to have electronic origin \cite{Chu2012_Sci}. Only by lowering the temperature further, at T$_\text{N} < \text{T}_\text{nem}$, the system develops an orbital order together with a spin-density wave that breaks the SU$(2)$ spin symmetry and the discrete translational symmetry of the lattice along one direction and, for this reason, this phase might be referred to as smectic \cite{Ming2011_PNAS,Fang2008_PRB,Xu2008_PRB}. There are two main theories, based on the above-mentioned scenario 2), which explain the onset of nematicity in this class of compounds. The first one relates the nematic properties of the system to orbital fluctuations \cite{Kontani2011_PRB,Stanev2013_PRB,Yamase2013_PRB,Yamase2013_PRB}, while the second one invokes a major role of anisotropic spin fluctuations which should be large right above the critical temperature T$_\text{N}$ \cite{Golubovic1988_PRB,Fernandes2012_PRB,Fernandes2014_NatPhys,Boemer2016_CRP}. The breaking of the Ising $\mathbb{Z}_2$ symmetry corresponds to two ordering vectors for the broken rotational symmetry of the lattice (C$_4 \rightarrow$ C$_2$), i.e., $\mathbf{\bar{Q}}_1 \propto (0, 1)$ or $\mathbf{\bar{Q}}_2 \propto (1,0)$ (for 1-Fe unit cell)  \cite{Vojta2009_AdvPhys,Fradkin2010_AnnRev,Fernandes2014_NatPhys}. The spin density wave fluctuations might also determine the symmetry of the superconducting gap at lower temperatures \cite{Mazin2008_PRL}. Understanding the most prominent character of the fluctuations that drive nematicity is particularly challenging because the two orders (orbital and spin density wave) are coupled and one can induce the other \cite{Fanfarillo2015_PRB,Fanfarillo2018_PRB}. This leads to a typical \textit{the chicken or the egg} problem, the solution of which might rely on the presence of different regimes where one mechanism dominates over the other depending by the magnitude of the Fermi energy of the system \cite{Chubukov2016_PRX}. Also the development of a zero-momentum orbital order, i.e., a $d$-wave PI, can explain the onset of nematicity in the iron-based superconductor, and this mechanism would rely on scenario 1) rather than on scenario 2) \cite{Zhai2009_PRB,Thorsmoelle2016_PRB}.

\begin{figure}
    \centerline{\includegraphics[width=0.5\textwidth]{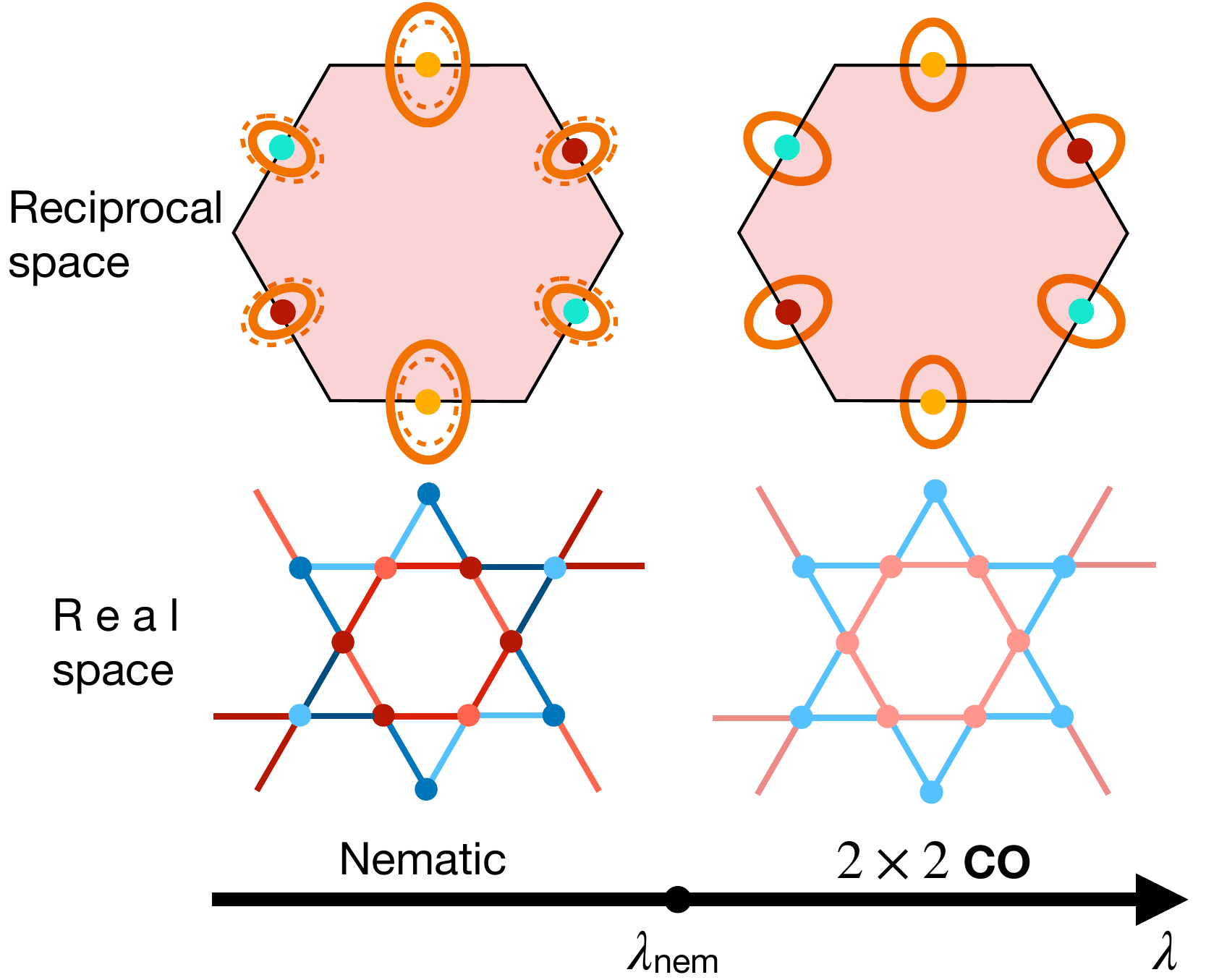}}
	\caption{ \textbf{Nematicity from the d-wave charge Pomeranchuk instability -} Scheme of the d-wave charge Pomeranchuk instability in reciprocal (upper row) and real (lower row) space. By tuning an external control parameter $\lambda$, the system goes through a nematic phase transition at $\lambda = \lambda_\text{nem}$. In reciprocal space, the instability corresponds to a breaking of the rotational symmetry by the Fermi surface of the $2 \times 2$ charge ordered phase. In real space, the instability consists of an intra-unit cell charge order (for a more extensive discussion of the possible shapes of the intra-unit cell charge order, see \cite{Grandi2023_PRB}). For $\lambda > \lambda_\text{nem}$, the Fermi surface of the system and the $2 \times 2$ charge order are sixfold symmetric. When $\lambda < \lambda_\text{nem}$, both the Fermi surface and the charge order break the C$_6$ symmetry of the kagome lattice.} \label{fig:PI_nem}
\end{figure}

\noindent
In the vanadium-based kagome metals $A$V$_3$Sb$_5$ ($A =$ Cs, K, Rb), signatures of nematicity have been reported in several experiments \cite{Xiang2021_NatComm,Wu2022_PRB2,Xu2022_NatPhys,Nie2022_Nat,Zheng2022_Nat,Sur2023_NatComm}. However, the underlying phenomenology is different with respect to the one observed in most of the iron-based superconductors. At a critical temperature T$_\text{co} \sim 90$K, these systems develop an unconventional charge order (CO) with $2 \times 2$ ($\times 1$ \cite{Li2022_NatComm}, $\times 2$ \cite{Jiang2021_Nat_Mat,Li2021_PRX,Liang2021_PRX} or $\times 4$ \cite{Ortiz2021_PRX}) in-plane (out-of-plane) lattice reconstruction which \textit{might} (or not \cite{Saykin2023_PRL}) break time reversal symmetry without any signature of magnetism \cite{Yang2020_SciAdv,Yu2021_PRB,Neupert2022_NatPhys,Khasanov2022_PRR,Mielke2022_Nat}. At T$_\text{sc} \sim 1$K, a superconducting phase is observed \cite{Ortiz2021_PRM,Yin2021_ChinPhysLett}, which might inherit the unconventional properties of the higher temperature state \cite{Ortiz2020_PRL,Tazai2022_SciAdv,Guguchia2023_NatComm}. Particularly, the superconducting state has a two-dome structure, as observed by applying an external pressure \cite{Chen2021_PRL,Wang2021_PRR,Du2021_PRB,Zheng2022_Nat} or by doping (CsV$_3$Sb$_{5-x}$Sn$_x$ \cite{CapaSalinas2023_FEM} and CsV$_{3-x}$Ti$_x$Sb$_5$ \cite{Yang2022_SciBull,Wu2024_ResSq}) which further complicates the description of this phase. Interestingly, the double dome structure is not observed by hole doping the potassium and the rubidium compounds (K,Rb)V$_3$Sb$_{5-x}$Sn$_x$ \cite{Oey2022_PRM2}. The different origin of the superconductivity in the two domes is suggested by the different shape of the superconductive gap, which has U- and V-shape, respectively \cite{Yang2022_SciBull}. Immediately below T$_\text{co}$, a ``weak'' nematicity is observed, probably due to a $\pi$-shift of the CO between consecutive kagome layers, while each vanadium plane preserves the p$6$mm wallpaper group symmetry \cite{Ratcliff2021_PRM,Christensen2021_PRB,Jin2024_PRL}. At T$_\text{nem} \sim 30-50$K, several experiments performed on CsV$_3$Sb$_5$, the family-member which has the most controversial properties, report the transition to a ``strong'' nematic state, where each kagome layer has a lower rotational symmetry (C$_6 \rightarrow$ C$_2$) \cite{Nie2022_Nat,Li2022_NatComm}. Since the kagome lattice has three independent directions, the onset of nematicity in this context breaks a Potts $\mathbb{Z}_3$ symmetry \cite{Hecker2018_npj,Cho2020_NatComm,Tazai2023_NatComm}, corresponding to the ordering vectors $\mathbf{Q}_1 \propto (-\sqrt{3}, -1) / 2$, $\mathbf{Q}_2 \propto (0, 1)$ and $\mathbf{Q}_3 \propto (\sqrt{3}, -1)/2$. Other experiments find, in the same temperature range, the stabilization of an additional $1 \times 4$ CO that simultaneously breaks the rotational and the discrete translational symmetry of the $2 \times 2$ CO \cite{Zhao2021_Nat,Guo2022_Nat,Li2022_NatComm,Li2022_PRB,Li2023_PRX}. For this reason, this state might be regarded as smectic, considering the $2 \times 2$ CO as the ``pristine'' or ``parent'' state of the system \cite{Li2023_NatPhys}. Other experiments do not find, instead, any transition at T$_\text{nem}$ \cite{Liu2024_PRX,Frachet2024_PRL}, suggesting nevertheless that both the nematic and the $1 \times 4$ COs might be stabilized by applying a tiny perturbation to the system such as a small magnetic field or a small strain \cite{Guo2024_NatPhys}. The relation between superconductivity and nematicity in this class of compounds is suggested by the presence of strong nematic fluctuations corresponding to the maximum of the lower pressure/doping dome \cite{Sur2023_NatComm}, i.e., the ``superconducting glue'' might be provided by the nematic fluctuations related to a quantum critical point in one dome \cite{Lederer2017_PNAS}.

\begin{figure}
    \centerline{\includegraphics[width=0.5\textwidth]{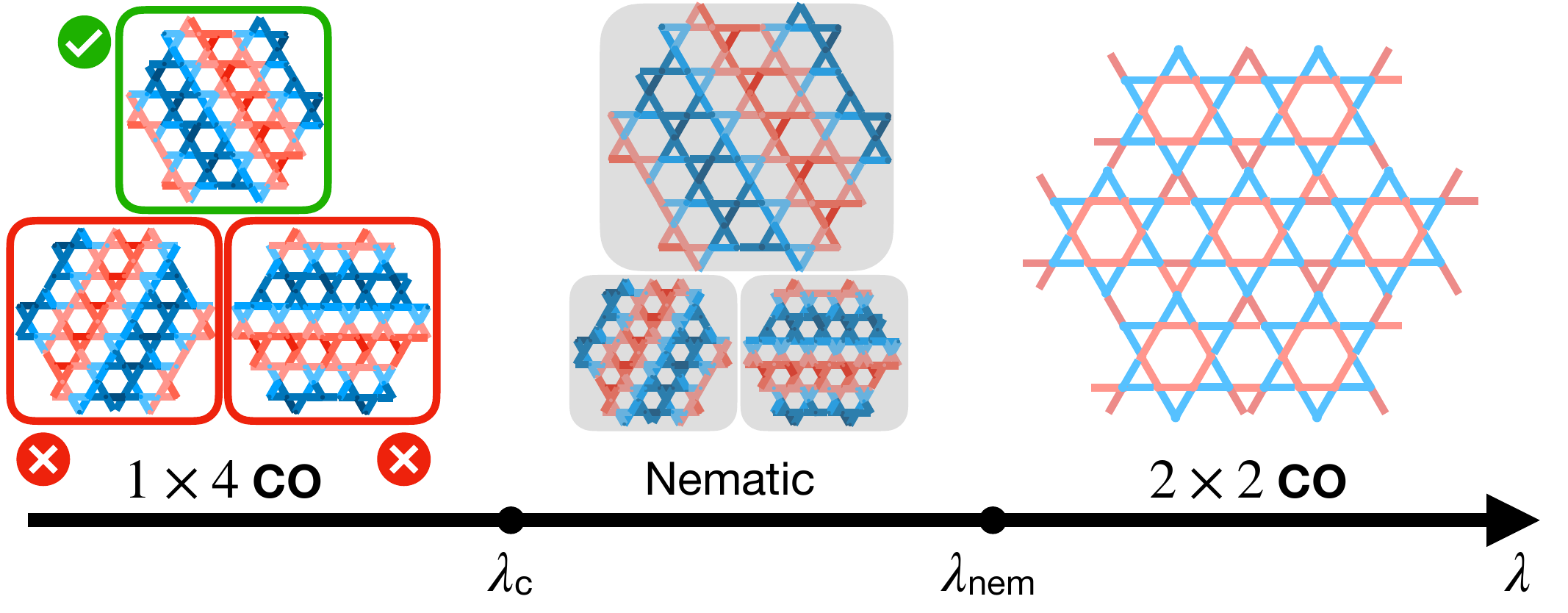}}
	\caption{ \textbf{Nematicity from the fluctuations of the $1 \times 4$ charge order -} Scheme of the nematicity driven by the fluctuations of the $1 \times 4$ charge order. By tuning an externally controllable parameter $\lambda$, the system moves from the $2 \times 2$ charge order ($\lambda > \lambda_\text{nem}$) to the $1 \times 4$ charge order ($\lambda < \lambda_\text{c}$). The $1 \times 4$ charge order breaks a $\mathbb{Z}_3$ symmetry since just one of the three equivalent states is selected by the symmetry breaking. For intermediate values of the control parameter ($\lambda_\text{c} < \lambda < \lambda_\text{nem}$), the system experiences strong fluctuations of the three one-dimensional charge orders (represented as shaded grey areas on top of the one dimensional charge orders in the figure). The fluctuations in one of the three possible directions become stronger than the other two, leading to a breaking of the rotational part of the point group symmetry of the lattice (in the figure, one of the three shaded grey areas is larger than the other two). Even in the nematic state, a $\mathbb{Z}_3$ symmetry is broken since each of the three channels might become the one where the fluctuations are the strongest.} \label{fig:1D_co_nem}
\end{figure}

\noindent
As opposed to the iron-based superconductors, the spin order seems absent in the whole phase diagram of the kagome metals, suggesting a minor contribution of the spin fluctuations to the physics of $A$V$_3$Sb$_5$ \cite{Zheng2022_Nat}. The orbital degree of freedom might play some role, even if it is not so clear, so far, what this might be \cite{Wu2021_PRL}. Indeed, most of the theoretical approaches towards the physics of kagome metals has focused on models with a single orbital per site, in some cases even neglecting the sublattice character of the states near the Fermi energy~\cite{Denner2021_PRL,Park2021_PRB,Lin2021_PRB,Dong2023_PRB,Grandi2023_PRB,Profe2024_PRR}. Instead, charge fluctuations are found to play a role already above T$_\text{co}$ \cite{Chen2022_PRL,Yang2023_PRB,Subires2023_NatComm} and the temperature-pressure phase diagram shows the presence of several CO phases \cite{Zheng2022_Nat,Kautzsch2023_npjQM}, so that, close to the phase boundaries, the fluctuations of the symmetry broken states should become strong \cite{Wu2024_ResSq}. As suggested by recent calculations \cite{Tian2024_arXiv}, even the low-temperature superconducting character might be strongly influenced, if not determined, by charge fluctuations. In this regard, instead of a spin-fermion model \cite{Abanov2003_AdvPhys,Chubukov2009_PRL}, the physics observed in the kagome metals might be described by a charge-fermion model, in analogy with the relevant role charge fluctuations might play in the cuprates \cite{Castellani1995_PRL,Kivelson1998_Nat,Caprara2005_PRL,Torchinsky2013_NatMat}.

\noindent
Here, we suggest two possibilities for the onset of nematicity in the vanadium-based kagome metals as driven by the charge degree of freedom: This might happen either through a transition to a $d$-wave charge PI (see Fig.~\ref{fig:PI_nem} for a schematic representation of this state) or by spatially anisotropic finite momentum CO fluctuations (see Fig.~\ref{fig:1D_co_nem} for a schematic representation of the mechanism guiding this transition). In support of these ideas, we perform a Kohn-Luttinger analysis \cite{Kohn1965_PRL,Luttinger1966_PR,Chubukov1993_PRB,Gonzalez2008_PRB} of a model \cite{Li2023_PRX} for the electronic states at the Fermi level of the $2 \times 2$ CO in the particle-hole sector in the presence of electronic interactions \cite{Dong2023_PRB,Dong2023_PRB2}, which allows us to obtain the renormalized nematic vertices in the proximity of the zero- and finite-momentum ordered states. These vertices allow us to define two charge-fermion models describing the interactions among the electrons at the Fermi level, driving the transition to the nematic state in the two situations. Additionally, we describe how to restore the validity of the Ward identities for spin and charge conservation for the finite momentum charge-fermion model by including the Aslamazov-Larkin diagrams, i.e., by going beyond the random phase approximation (RPA) scheme. The analysis of the symmetry broken nematic phases is then supplied by the derivation of the Ginzburg-Landau potentials for the $d$-wave charge PI and the $1 \times 4$ CO. Since from the experimental point of view, it is not clear if these states are actually stable or not, we focus on a natural probe both of nematicity and of nematic fluctuations \cite{Chu2012_Sci,Kuo2013_PRB,Massat2016_PNAS}, i.e., on the nematic susceptibility as can be defined in the two cases and we connect the expressions obtained from the Ginzburg-Landau analysis with the microscopic understanding coming from the charge fermion models.

\noindent
The article is organized as follows: In Sec.~\ref{sec:model}, we introduce the microscopic model to describe the main properties of the $2 \times 2$ charge order. In Sec.~\ref{sec:KL_analysis}, we perform a Kohn-Luttinger analysis in the nematic channel. Secs.~\ref{sec:Q4aCO}-\ref{sec:Q1aCO} are devoted to the derivation of the Ginzburg-Landau theories for the two ordering mechanisms. In Sec.~\ref{sec:rise_nem}, we discuss the nematic susceptibilities, which represent a natural probe of the nematic fluctuations. In Sec.~\ref{sec:micro_nem}, we discuss the relation between the microscopic expression of the nematic susceptibilities one can obtain in the two cases starting from the corresponding charge-fermion model and the Ginzburg-Landau expressions. Finally, Sec.~\ref{sec:discussion} is devoted to concluding remarks on the the role of the $2 \times 2$ charge fluctuations explaining a few recent experimental findings and on the coupling of the nematic order parameters to the elastic deformations of the solid.

\begin{figure}
    \centerline{\includegraphics[width=0.5\textwidth]{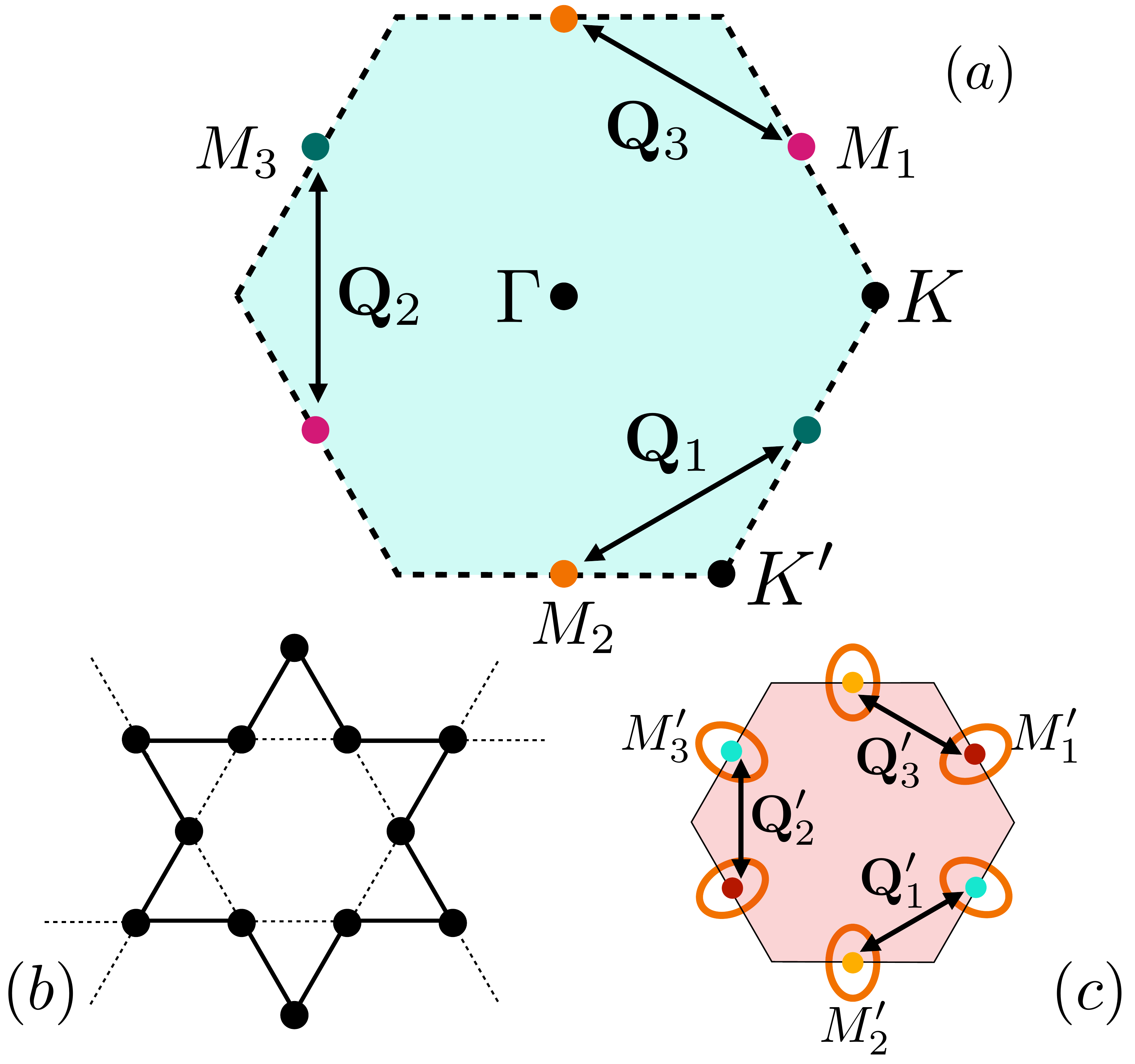}}
	\caption{ \textbf{Fermi surface of the charge-ordered state -} (a) Brillouin zone of the pristine kagome lattice with van Hove singularities \cite{Smollett1952_PPSA,VanHove1953_PR} at the three inequivalent momenta $\mathbf{M}_1 = \pi (\sqrt{3}, 1) / (2 \sqrt{3} a)$, $\mathbf{M}_2 = \pi (0, -1) / (\sqrt{3} a)$ and $\mathbf{M}_3 = \pi (- \sqrt{3}, 1) / (2 \sqrt{3} a)$, with $a$ the nearest-neighbor distance. The ordering wave vectors for the $2 \times 2$ instability are depicted as $\mathbf{Q}_{1,2,3}$ \cite{Venderbos2016_PRB}. (b) Putative $2 \times 2$ charge-bond order with trihexagonal shape. Solid (dashed) lines describe a strong (weak) hopping. (c) Reduced Brillouin zone of the $2 \times 2$ charge-ordered state (the area is $1/4$ of the one shown in panel (a)). The orange ellipses are a sketch of the Fermi surface, centered at the high-symmetry points $\mathbf{M}_i' = \mathbf{M}_i/2$. The $\mathbf{Q}'$ vectors are defined as $\mathbf{Q}'_i = \mathbf{M}'_j - \mathbf{M}'_k$, with $( i,j,k )$ an even permutation of $( 1,2,3 )$.
    } \label{fig:hole_pockets}
\end{figure}

\section{Microscopic model} \label{sec:model}
In the reduced Brillouin zone, the Fermi surface of the system has three inequivalent pockets located at the reconstructed M points (M') \cite{Zhou2022_NatComm,Lin2022_PRB,Dong2023_PRB,Varma2023_PRB,Ortiz2021_PRX,Shrestha2022_PRB} with hole-like character related to the states coming from the $d$ orbitals of the vanadium atoms \cite{Li2023_PRX}, see Fig.~\ref{fig:hole_pockets}. A similar band structure hosting only hole pockets is relevant for some iron-based compounds \cite{Wu2024_NatPhys}, where a suppression of the spin-density wave fluctuations was found in this case \cite{Thomale2011_PRL}, which might lead to $d$-wave superconductivity \cite{Agterberg1999_PRB}. The relevance of the hole pockets for the formation of the superconducting state in the kagome metals is in agreement with the experimental findings that observe an increase of T$_\text{sc}$ by hole doping and a decrease by electron doping \cite{Oey2022_PRM1,Oey2022_PRM2,Yang2022_SciBull,CapaSalinas2023_FEM}. In the following, we assume that the elliptical hole pockets are the relevant features of the Fermi surface at zero doping. Even if the system hosts a circular electronic pocket coming from the $p$ orbitals of the antimony atoms centered at $\Gamma$, this should not be relevant for the properties of kagome metals at zero doping and no externally applied pressure given the strong mismatch between the radius of the electron and hole pockets, which should not provide any relevant nesting among the two. Nevertheless, at finite tin doping \cite{Oey2022_PRM1}, the central Fermi pocket is tuned above the Fermi level of the system, i.e., at some point a nesting condition will be reached between the electron and the hole pockets, potentially opening an additional channel for the stabilization of the $1 \times 4$ CO. The analysis of this different model goes beyond the present study.

\noindent
We start from a minimal three-band model (six bands counting the spin degeneracy) for the electronic structure in the $2 \times 2$ charge-ordered state with hole-pockets centered at $M_1'$, $M_2'$ and $M_3'$. To keep the calculations simple, we consider elliptic paraboloid dispersions, see Fig.~\ref{fig:hole_pockets}c: 
\begin{align} \label{eq:dispersions}	
	& \varepsilon_{2, \mathbf{k}} = - \sum_{a=x,y} \frac{k_a^2}{2 m_a} , \ \ \ \  \varepsilon_{1/3, \mathbf{k}} =  \varepsilon_{2, R (\pm \frac{\pi}{3}) \mathbf{k}} , 
\end{align}
where $m_x = m (1 + \delta)$ and $m_y = m (1 - \delta)$ are the electronic masses along the $k_x$ and $k_y$ directions of reciprocal space, respectively, proportional to the inverse hopping $m \propto 1/t$ \cite{Eremin2010_PRB,Vorontsov2010_PRB}, while $\delta$ quantifies the ellipticity of the bands, i.e., the deviation of the pockets from being perfectly circular. $R(\phi)$ is the rotation matrix about an angle $\phi$ in the momentum plane $\mathbf{k} = (k_x, k_y)$. We take $t=1$ as our energy unit. To convert to physical units, the value of the hopping is $t \sim 0.3$eV \cite{Denner2021_PRL}. Since the dispersions Eq.~\eqref{eq:dispersions} describe the physics of the system in the $2 \times 2$ CO state, our theory applies for temperatures $T$ that fulfill T$_\text{sc} \sim 1\text{K} < \text{T} < \text{T}_\text{co} \sim 100$K, i.e., $3 \times 10^{-6} \lesssim k_B T / t \lesssim 3 \times 10^{-4}$, with $k_B$ the Boltzmann constant. The noninteracting Hamiltonian is
\begin{align} \label{eq:nonint_ham}
	& \mathcal{H}_0 = \sum_{i=1,2,3} \sum_{\mathbf{k}, \sigma} \big( \varepsilon_{i, \mathbf{k}} - \mu \big) c^\dagger_{i, \mathbf{k}, \sigma} c_{i, \mathbf{k}, \sigma} ,
\end{align}
where $c^\dagger_{i, \mathbf{k}, \sigma}$ ($c_{i, \mathbf{k}, \sigma}$) is the creation (destruction) operator of an electron in band $\varepsilon_{i, \mathbf{k}}$ with spin $\sigma = \uparrow, \downarrow$ and momentum $\mathbf{k}$, and $\mu$ is the chemical potential. 
We take $\mu/(t \Lambda^2) = - 3 \times 10^{-5}$ ($\Lambda$ is the momentum/energy cutoff) \cite{Park2021_PRB}. In Eq.~\eqref{eq:nonint_ham}, the momenta of the $i$-th band's fermions are defined relative to $\mathbf{M}_i'$.

\noindent
The structure of the interactions is described within the g-ology model \cite{Nandkishore2012_NatPhys,Park2021_PRB,Tsvelik2023_PRB}:
\begin{align} \label{eq:ham_patch_int}
	& \mathcal{H}_\text{int} = \frac{1}{2 N} \sum_{|\mathbf{k}_1|, \cdots , |\mathbf{k}_4| < \Lambda} \delta \big( \mathbf{k}_1 + \mathbf{k}_2 - \mathbf{k}_3 - \mathbf{k}_4 \big) \nonumber \\
	& \Big[ \sum_{i \neq j, \sigma, \sigma'} \big( g_1 c^\dagger_{i, \mathbf{k}_1, \sigma} c^\dagger_{j, \mathbf{k}_2, \sigma'} c_{i, \mathbf{k}_3, \sigma'} c_{j, \mathbf{k}_4, \sigma}  \nonumber \\
	& + g_2 c^\dagger_{i, \mathbf{k}_1, \sigma} c^\dagger_{j, \mathbf{k}_2, \sigma'} c_{j, \mathbf{k}_3, \sigma'} c_{i, \mathbf{k}_4, \sigma} \nonumber \\
	& + g_3 c^\dagger_{i, \mathbf{k}_1, \sigma} c^\dagger_{i, \mathbf{k}_2, \sigma'} c_{j, \mathbf{k}_3, \sigma'} c_{j, \mathbf{k}_4, \sigma} \big) \nonumber \\
	&+ g_4 \sum_{i, \sigma, \sigma'} c^\dagger_{i, \mathbf{k}_1, \sigma} c^\dagger_{i, \mathbf{k}_2, \sigma'} c_{i, \mathbf{k}_3, \sigma'} c_{i, \mathbf{k}_4, \sigma} \Big] ,
\end{align}
with $N$ the number of unit cells in the system. Here the couplings are the interpatch exchange interaction $g_1$,  interpatch density-density interaction $g_2$, umklapp scattering $g_3$, and intrapatch density-density interaction $g_4$, respectively.


\begin{figure*}	\centerline{\includegraphics[width=1.00\textwidth]{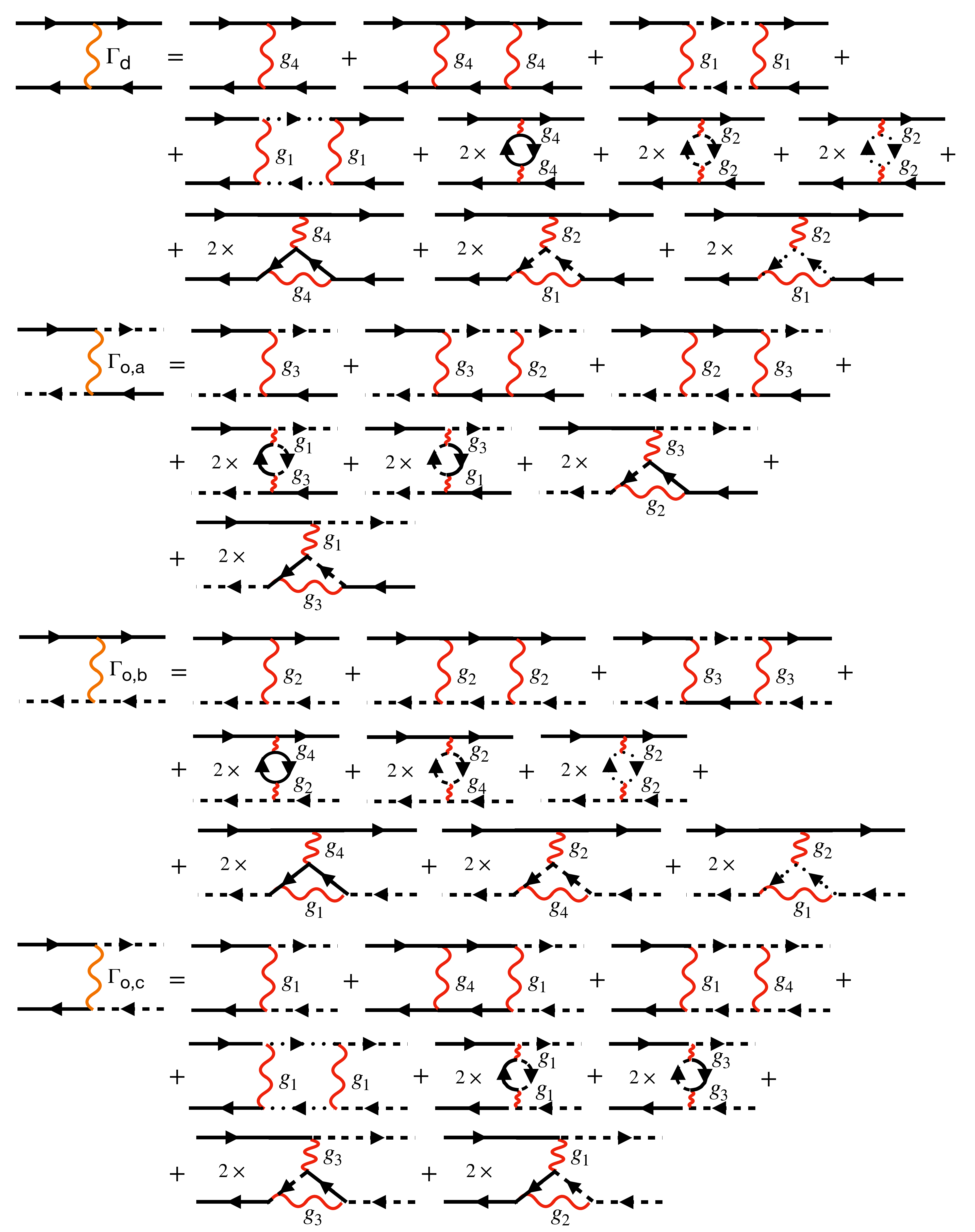}}
	\caption{ \textbf{Irreducible vertices at the second order in the interactions -} Diagrammatic representation of the irreducible vertices $\Gamma_\text{d}$ (intrapatch density-density), $\Gamma_\text{o,a}$ (umklapp), $\Gamma_\text{o,b}$ (interpatch density density) and $\Gamma_\text{o,c}$ (interpatch exchange) up to the second order in the interactions $g_1$, $g_2$, $g_3$ and $g_4$, having excluded the renormalizations coming from particle-particle response functions. The ladder, the bubble and the wine-glass diagrams appear at the second order. A few of them cancel out.} \label{fig:vertices_sec_ord}
\end{figure*}

\section{Kohn-Luttinger analysis} \label{sec:KL_analysis}
In this section, we analyze the conditions under which the model introduced above has a tendency towards either a $4 \times 4$ charge ordering or a $d$-wave charge Pomeranchuk instability. For now we will not investigate whether or not these are the leading instabilities of the model. Indeed, to properly address this point, the orbital/sublattice character of the states at the Fermi level has to be determined, but this information is lacking at present. Essentially, we \textit{assume} that the combined role of the sublattice/orbital degrees of freedom consists of reducing the tendency of the system to develop any spin ordered state \cite{Kiesel2012_PRB}. Nevertheless, we stress how recent analysis suggests a charge-fluctuations scenario to be preferred over a dominant spin-fluctuations one for a large number of lattice sites in the unit-cell of the system and for long-range repulsion \cite{Braz2024_arXiv}. We perform a Kohn-Luttinger analysis of the particle-hole vertices and we project them on the interesting channels to see if and under which conditions the system shows an instability \cite{Kohn1965_PRL,Luttinger1966_PR,Maiti2013_AIPCP}. We analyze the renormalization of the vertices to the second order in the interactions, and we neglect the renormalization coming from the particle-particle response function because we assume the main instability to take place in the particle-hole sector. We write the particle-hole response function as:
\begin{align}
	& \Pi_\text{ph}^{i j} (q) = i \int \frac{d^2 \mathbf{p} d \omega'}{(2 \pi \hbar)^3} \mathcal{G}_{0,i} (\mathbf{p}, \omega') \mathcal{G}_{0,j} (\mathbf{p} + \mathbf{q}, \omega' + \omega) < 0 , \nonumber
\end{align}
with $q = (\mathbf{q}, \omega)$ and $\mathcal{G}_{0,i} (k) = \frac{1}{-\omega + \epsilon_{i} (k) - \mu}$ the single-particle Green's function for the $i$-th band. In the following, we define the intraband and the interband particle-hole response function, respectively, as:
\begin{align}
	& \Pi_\text{ph} (\mathbf{0}) =  i \int \frac{d^2 \mathbf{p} d \omega'}{(2 \pi \hbar)^3} \mathcal{G}_{0,i} (\mathbf{p}, \omega') \mathcal{G}_{0,i} (\mathbf{p} + \mathbf{q}, \omega' + \omega) , \nonumber \\
	& \Pi_\text{ph} (\mathbf{Q}') =  i \int \frac{d^2 \mathbf{p} d \omega'}{(2 \pi \hbar)^3} \mathcal{G}_{0,i} (\mathbf{p}, \omega') \mathcal{G}_{0,j} (\mathbf{p} + \mathbf{Q}' + \mathbf{q}, \omega' + \omega) , \nonumber
\end{align}
with $i \neq j$ and small momentum ($\mathbf{q} \approx \mathbf{0}$) and energy ($\omega \approx 0$) transfer.

\noindent
The expressions for the irreducible vertices can be computed at the second order in perturbation theory by assuming a small value of the coupling interactions $g_1$, $g_2$, $g_3$ and $g_4$. The corresponding analytical expressions can be read from Fig.~\ref{fig:vertices_sec_ord}, leading to:
\begin{align}
	& \Gamma_\text{d} = g_4 + [g_4^2 + 2 g_1^2 - 4 g_2 ( g_2 - g_1)] \Pi_\text{ph} (\mathbf{0}) , \nonumber \\
	& \Gamma_\text{o,a} = g_3 + 2 g_3 (2 g_2 - g_1) \Pi_\text{ph} (\mathbf{Q}') , \nonumber \\
	& \Gamma_\text{o,b} = g_2 - 2 (g_2 - g_1) (g_2 + g_4) \Pi_\text{ph} (\mathbf{0}) \nonumber \\
	& + (g_2^2 + g_3^2) \Pi_\text{ph} (\mathbf{Q}') , \nonumber \\
	& \Gamma_\text{o,c} = g_1 + g_1 (2 g_4 + g_1) \Pi_\text{ph} (\mathbf{0}) + 2 g_1 ( g_2 - g_1 ) \Pi_\text{ph} (\mathbf{Q}') , \nonumber 
\end{align}
where the second order contribution proportional to $\Pi_\text{ph} (\mathbf{0})$ or to $\Pi_\text{ph} (\mathbf{Q}')$ is, by the perturbative construction, smaller than the first order term. In most of the physically relevant situations, $g_i > 0$, $i=1,2,3,4$, since they represent repulsive interactions. Moreover, all the interactions are generally of the same order of magnitude $g_1 \sim g_2 \sim g_3 \sim g_4$ (even if, in many cases, $g_4 > g_1, g_2, g_3$). The considerations above, together with the perturbative nature of the calculation, suggest that it is not possible for the second order contribution to overcome the magnitude of the linear terms. In the following, we analyze the tendency of the system towards the development of the $1 \times 4$ CO and the $d$-wave charge PI.

\begin{figure*}	\centerline{\includegraphics[width=1.0\textwidth]{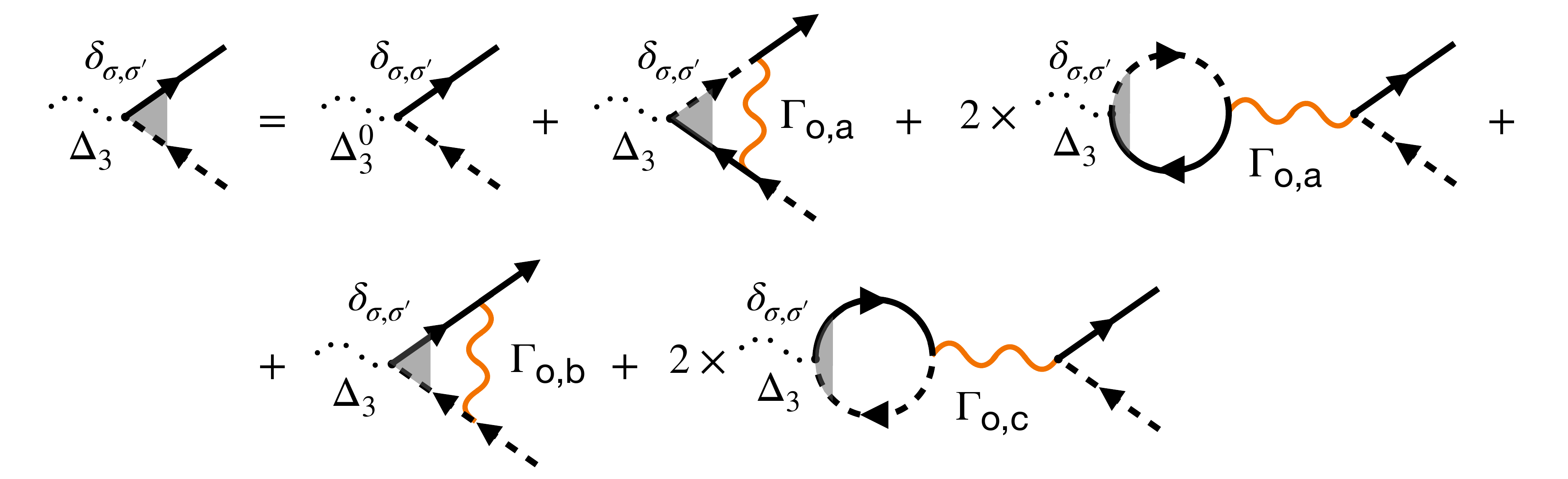}}
	\caption{ \textbf{Diagrammatic expression of the $1 \times 4$ CO instability -} Diagrammatic representation of the interacting vertex for the $Q_{4 a}$ CO instability dressed by interactions.} \label{fig:Q4CO_inst_diag}
\end{figure*}

\subsubsection{Instability in the $1 \times 4$ charge order channel}
We introduce an infinitesimal coupling in the $Q_{4a}$ CO $\Delta_j^0 | \epsilon_{jlm} | \sum_{\mathbf{k}, \sigma} c^\dagger_{l,\mathbf{k}, \sigma} c_{m,\mathbf{k}, \sigma}$, $j,l,m = 1,2,3$ and where $\epsilon_{jlm}$ is the Levi-Civita tensor. We then dress this vertex by interactions as shown in Fig.~\ref{fig:Q4CO_inst_diag} \cite{Chubukov2008_PRB,Chichinadze2020_PRB,Dong2023_PRB2}. We arrive at the relation:
\begin{align}
	\begin{pmatrix}
		& \Delta_1 \\
		& \Delta_2 \\
		& \Delta_3
	\end{pmatrix} =
	\begin{pmatrix}
		& \Delta_1^0 \\
		& \Delta_2^0 \\
		& \Delta_3^0
	\end{pmatrix} -
	\Pi_\text{ph} (\mathbf{Q}') \Gamma^{Q_{4a}} \
	\begin{pmatrix}
		1 & 0 & 0 \\
		0 & 1 & 0 \\
		0 & 0 & 1
	\end{pmatrix}
	\begin{pmatrix}
		& \Delta_1 \\
		& \Delta_2 \\
		& \Delta_3
	\end{pmatrix} , \nonumber
\end{align}
which is already diagonal and it has the same form for the three order parameters, leading to:
\begin{align}
	\Delta_j = \frac{\Delta_j^0}{1 + \Gamma^{Q_{4a}} \Pi_\text{ph} (\mathbf{Q}')} . \nonumber
\end{align}
In the previous expression, we introduced the effective vertex $\Gamma^{Q_{4a}} = \Gamma_\text{o,b} - 2 \Gamma_\text{o,c} - \Gamma_\text{o,a}$. To get an enhancement of the $Q_{4a}$ instability, we must have $\Gamma^{Q_{4a}} > 0$ since $\Pi_\text{ph} (\mathbf{Q}') < 0$. The vertex can be written as 
\begin{align}
    \Gamma^{Q_{4a}} & = g_2 - 2 g_1 - g_3  \nonumber \\
    & - 2 \big[ (g_2 - g_1) (g_2 + g_4) + g_1 (2 g_4 + g_1) \big] \Pi_\text{ph} (\mathbf{0}) \nonumber \\
    & + \big[ (g_2^2 + g_3^2) - 4 \big( (g_1 - g_3 )^2 + g_1 (g_3 - g_2) \big) \nonumber \\
    & - 2 g_3 (2 g_2 - g_1) \big] \Pi_\text{ph} (\mathbf{Q}') . \nonumber
\end{align}
At the first order in the interactions, this expression becomes $\Gamma^{Q_{4a}} = g_2 - 2 g_1 - g_3$. If we assume $g_2 = g_4 = 0$, we get $\Gamma^{Q_{4a}} = -(2 g_1 + g_3) - 2 g_1^2 \Pi_\text{ph} (\mathbf{0}) - (4 g_1^2 - 6 g_1 g_3 + 3 g_3^2)\Pi_\text{ph} (\mathbf{Q}') \sim -(2 g_1 + g_3)$ at the leading order. If we assume, instead, $g_1 = g_2 = g_3 = g_4 = g$, the leading order expression is $\Gamma^{Q_{4a}} \sim -2 g < 0$, i.e., in these cases there is no tendency towards the $Q_{4a}$ CO. Generally speaking, the first order contribution to $\Gamma^{Q_{4a}}$ is positive when $g_2 > 2 g_1 + g_3$. In the following, we assume the first order contributions to $ \Gamma^{Q_{4a}}$ to be different from zero and dominant over the second order ones, so that  $\Gamma^{Q_{4a}}=g_2 - 2 g_1 - g_3$.

\begin{figure*}	\centerline{\includegraphics[width=1.0\textwidth]{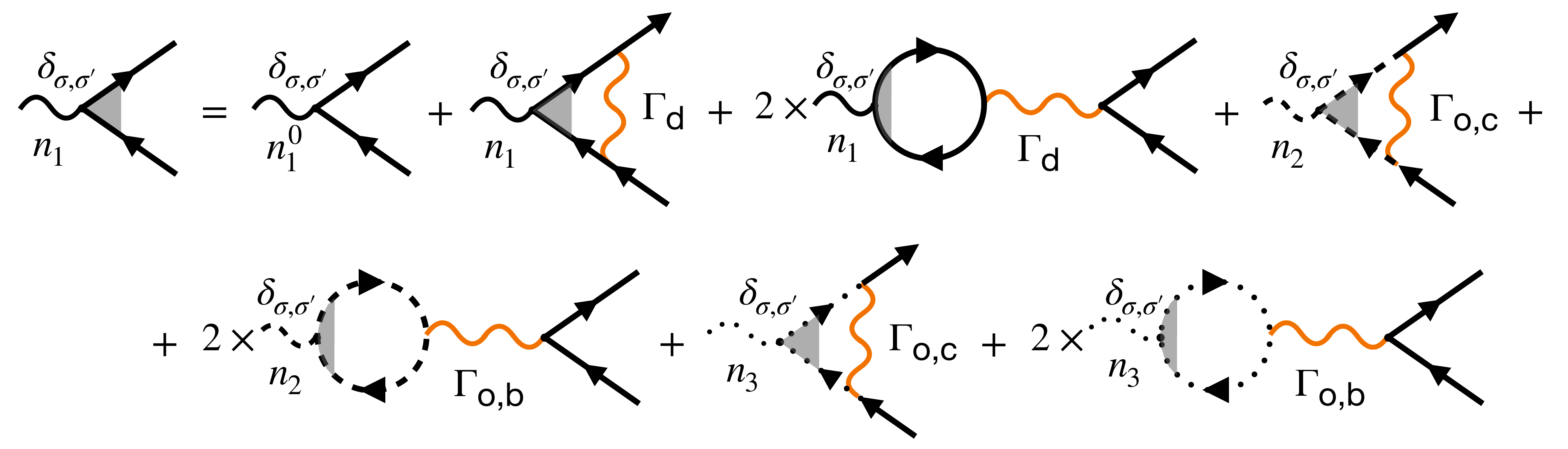}}
	\caption{ \textbf{Diagrammatic expression of the Pomeranchuk instability -} Diagrammatic representation of the interacting vertex for the charge Pomeranchuk instability dressed by interactions.} \label{fig:pom_inst_diag}
\end{figure*}

\subsubsection{Instability in the charge Pomeranchuk channel}
We introduce an infinitesimal coupling in the charge PI $n_j^0 \sum_{\mathbf{k}, \sigma} c^\dagger_{j,\mathbf{k}, \sigma} c_{j,\mathbf{k}, \sigma}$, $j = 1,2,3$. We then dress this vertex by interactions as shown in Fig.~\ref{fig:pom_inst_diag} \cite{Chubukov2008_PRB,Chichinadze2020_PRB,Dong2023_PRB2}. This way, one arrives at the relation:
\begin{align}
	\begin{pmatrix}
		& n_1 \\
		& n_2 \\
		& n_3
	\end{pmatrix} =
	\begin{pmatrix}
		& n_1^0 \\
		& n_2^0 \\
		& n_3^0
	\end{pmatrix} +
	\Pi_\text{ph} (\mathbf{0}) \ \Upsilon
	\begin{pmatrix}
		& n_1 \\
		& n_2 \\
		& n_3
	\end{pmatrix} \nonumber ,
\end{align}
having defined:
\begin{align}
	\Upsilon = 
	\begin{pmatrix}
		\Gamma_\text{d} & 2 \Gamma_{\text{o,b}} - \Gamma_{\text{o,c}} & 2 \Gamma_{\text{o,b}} - \Gamma_{\text{o,c}} \\
		2 \Gamma_{\text{o,b}} - \Gamma_{\text{o,c}} & \Gamma_\text{d} & 2 \Gamma_{\text{o,b}} - \Gamma_{\text{o,c}} \\
		2 \Gamma_{\text{o,b}} - \Gamma_{\text{o,c}} & 2 \Gamma_{\text{o,b}} - \Gamma_{\text{o,c}} & \Gamma_\text{d}
	\end{pmatrix} \nonumber .
\end{align}
In the diagonal basis, one obtains:
\begin{align}
	v = 
	v^0 
	- \Pi_\text{ph} (\mathbf{0}) \ 
	\begin{pmatrix}
		\Gamma^\text{d-PI} & 0 & 0 \\
		0 & \Gamma^\text{d-PI} & 0 \\
		0 & 0 & \Gamma^\text{s-PI} 
	\end{pmatrix} 
	v \nonumber ,
\end{align}
having introduced the vector:
\begin{align}
	v = 
	\begin{pmatrix}
		\sqrt{\frac{2}{3}} \big( n_1 - \frac{n_2 + n_3}{2} \big) \\
		\frac{n_2 -n_3}{\sqrt{2}} \\
		\frac{n_1 + n_2 + n_3}{\sqrt{3}}
	\end{pmatrix} \nonumber
\end{align}
and with $\Gamma^\text{d-PI} = 2 \Gamma_{\text{o,b}} - \Gamma_\text{d} - \Gamma_{\text{o,c}}$ and $\Gamma^\text{s-PI} = 2 \Gamma_{\text{o,c}} - \Gamma_\text{d} - 4 \Gamma_{\text{o,b}} $. $\Gamma^\text{d-PI}$ is doubly degenerate, in agreement with the two-dimensional nature of the $\text{E}_{2}$ irrep. By solving the relation above, one obtains the relations for the dressed order parameters \cite{Chubukov2008_PRB,Maiti2010_PRB}:
\begin{align}
	& n_1 - \frac{n_2 + n_3}{2} = \frac{ n_1^0 - \frac{n_2^0 + n_3^0}{2}}{1 + \Gamma^\text{d-PI} \Pi_\text{ph} (\mathbf{0})}  , \nonumber \\ 
	& n_2 -n_3 = \frac{n_2^0 - n_3^0}{1 + \Gamma^\text{d-PI} \Pi_\text{ph} (\mathbf{0})} , \nonumber \\
	& n_1 + n_2 + n_3 = \frac{n_1^0 + n_2^0 + n_3^0}{1 + \Gamma^\text{s-PI} \Pi_\text{ph} (\mathbf{0})} . \nonumber
\end{align}
We can explicitly write the expressions for the interaction vertices in the s- and d-PI channels:
\begin{align}
    \Gamma^\text{s-PI} & = 2 g_1 - g_4 - 4 g_2  \nonumber \\
    & - \big[ g_4 (g_4 - 4 g_2) + 4 (3 g_2 + g_4) (g_1 - g_2) \big] \Pi_\text{ph} (\mathbf{0}) \nonumber \\
    & - 4 \big[ (g_2 - g_1) (g_1 + g_2) + g_1 (g_2 + g_3) \big] \Pi_\text{ph} (\mathbf{Q}') \nonumber ,
\end{align}
\begin{align}
    \Gamma^\text{d-PI} & = 2 g_2 - g_4 - g_1 \nonumber \\
    & - \big( g_4^2 + 3 g_1^2 + 4 g_2 g_4 - 2 g_1 g_4 \big) \Pi_\text{ph} (\mathbf{0}) \nonumber \\
    & - 2 \big[ (g_1 - g_2) (g_1 + g_2) - g_1 (g_2 + g_3) \big] \Pi_\text{ph} (\mathbf{Q}') \nonumber .
\end{align}
At the lowest order in the interactions, one obtains $\Gamma^\text{d-PI} = 2 g_2 - g_4 - g_1$ and $\Gamma^\text{s-PI} = 2 g_1 - g_4 - 4 g_2$. If we assume the zero-momentum exchanged interactions to be dominant, i.e., $g_2, g_4 \gg g_1 = g_3$, the leading contribution to the instabilities are $\Gamma^\text{s-PI} \sim - (g_4 + 4 g_2)$ and $\Gamma^\text{d-PI} \sim (2 g_2 - g_4)$. While the s-wave channel is always negative $\Gamma^\text{s-PI} < 0$, the $d$-wave one becomes positive when $g_2 > g_4/2$, a condition that might safely be reached. If we consider, instead, the condition $g_1 = g_2 = g_3 = g_4 = g$, we obtain $\Gamma^\text{s-PI} \sim - 3 g$ and $\Gamma^\text{d-PI} \sim - 2 g^2 \big( 3 \Pi_\text{ph} (\mathbf{0}) - 2 \Pi_\text{ph} (\mathbf{Q}') \big)$. Once again, the system does not show any tendency towards phase separation (s-wave charge PI) for positive interactions \cite{Chubukov2018_JETP}, while it has a tendency towards nematicity ($d$-wave charge PI) when $3 \Pi_\text{ph} (\mathbf{0}) < 2 \Pi_\text{ph} (\mathbf{Q}')$ \cite{Gonzalez2008_PRB}. In the following, we assume the first order contributions to $\Gamma^\text{d-PI}$ to be different from zero and dominant over the second order ones, so that  $\Gamma^\text{d-PI}= 2 g_2 - g_4 - g_1$.

\begin{figure*}	\centerline{\includegraphics[width=1.0\textwidth]{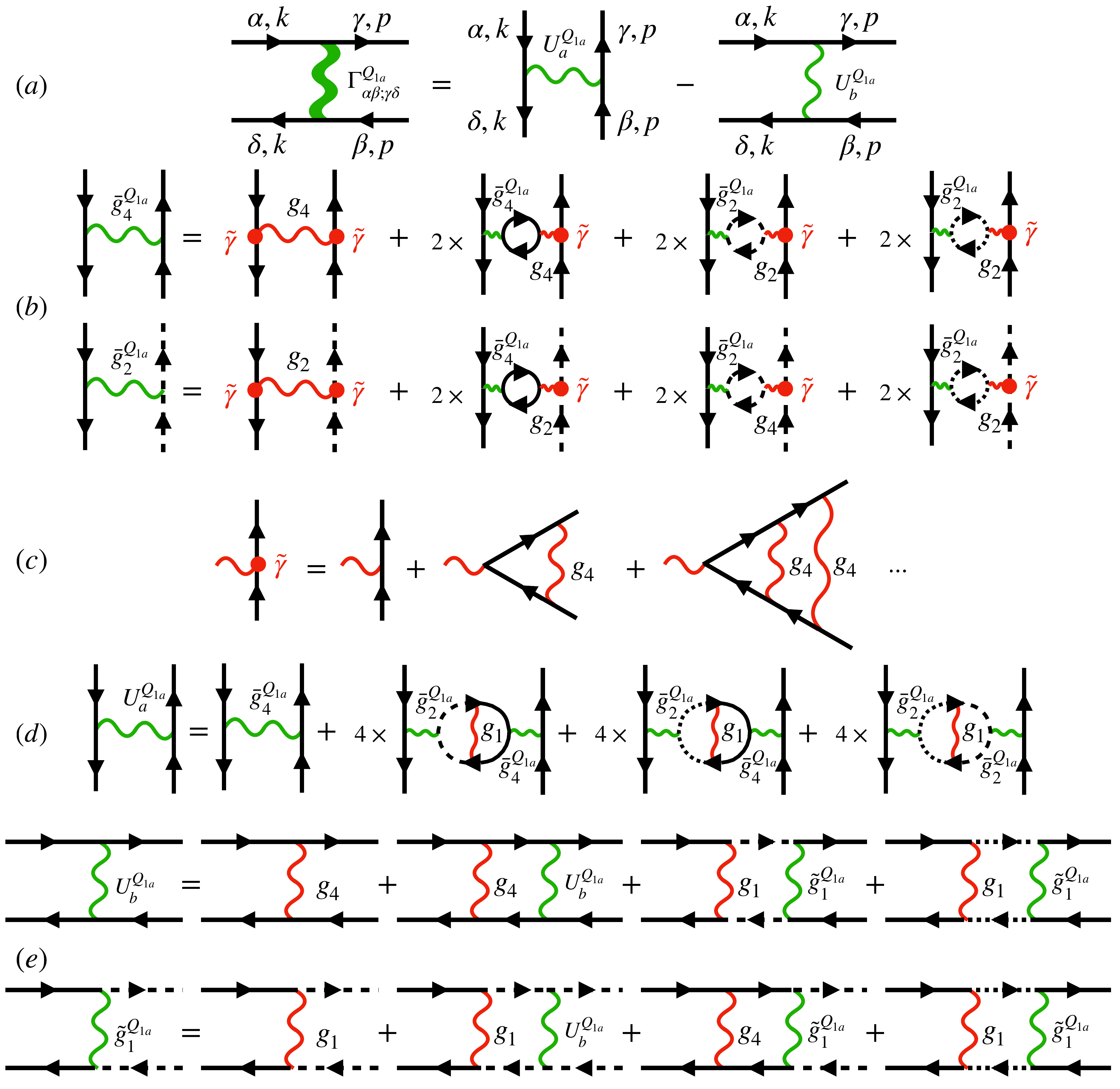}}
	\caption{\textbf{Diagrams for the renormalized interactions near the onset of the $d$-wave Pomeranchuk instability -} (a) Diagrammatic representation of Eq.~\eqref{eq:gamma_vert_pom}. (b) Relevant diagrams containing the polarization $\Pi_\text{ph} (\mathbf{0})$ to the renormalization of the interactions $g_4$ and $g_2$. (c) Diagrammatic representation of the dressed vertex (red dot). (d) Expression of the effective interaction $U_a^{Q_{1a}}$ including the first order correction due to $g_1$. (e) Coupled ladder series for the computation of $U_b^{Q_{1a}}$.} \label{fig:diagrams_Ua}
\end{figure*}

\subsubsection{Pairing interaction in the nematic channel near the onset of the $d$-wave charge Pomeranchuk instability}
In this and in the next sections we aim to analyze the pairing interaction in the nematic channel that arises in the proximity of the charge ordered states analyzed before \cite{Dong2023_PRB2}:
\begin{align} \label{eq:gamma_vert_pom}
    \Gamma_{\alpha, \beta; \gamma, \delta}^{Q_{1a}} (k,p ; p, k) = U_a^{Q_{1a}} \delta_{\alpha \delta} \delta_{\beta \gamma} - U_b^{Q_{1a}} \delta_{\alpha \gamma} \delta_{\beta \delta} ,
\end{align}
where $\alpha$, $\beta$, $\gamma$ and $\delta$ are spin indexes and $U_a^{Q_{1a}}$ ($U_b^{Q_{1a}}$) represents the fully dressed irreducible interaction with zero ($k-p$) momentum transfer. We first focus on the $d$-wave charge PI. In this case, the diagrammatic representation of Eq.~\eqref{eq:gamma_vert_pom} is provided in Fig.~\ref{fig:diagrams_Ua}(a). In the proximity of the phase transition, the most relevant diagrams are the ones that contain a zero-momentum polarization bubble $\Pi_\text{ph} (\mathbf{0})$. The diagrammatic expression of $U_a^{Q_{1a}}$ is related to the renormalized interaction $\bar{g}_4^{Q_{1a}}$, digrammatically shown in Fig.~\ref{fig:diagrams_Ua}(b), which includes the ladder series for the vertex renormalization (Fig.~\ref{fig:diagrams_Ua}(c)):
\begin{align}
    \tilde{\gamma} = 1 - g_4 \Pi_\text{ph} (\mathbf{0}) + g_4^2 \Pi_\text{ph}^2 (\mathbf{0}) + \cdots = \frac{1}{1 + g_4 \Pi_\text{ph} (\mathbf{0})} , \nonumber
\end{align}
as well as bubble diagrams and the renormalized interaction $\bar{g}_2^{Q_{1a}}$. We can write analytically the coupled equations diagrammatically shown in Fig.~\ref{fig:diagrams_Ua}(b) as:
\begin{align}
    & \bar{g}_4^{Q_{1a}} = \tilde{\gamma}^2 g_4 + 2 \tilde{\gamma} g_4 \bar{g}_4^{Q_{1a}} \Pi_\text{ph} (\mathbf{0}) + 4 \tilde{\gamma} g_2 \bar{g}_2^{Q_{1a}} \Pi_\text{ph} (\mathbf{0}) , \nonumber \\
    & \bar{g}_2^{Q_{1a}} = \tilde{\gamma}^2 g_2 + 2 \tilde{\gamma} g_2 \bar{g}_4^{Q_{1a}} \Pi_\text{ph} (\mathbf{0}) + 2 \tilde{\gamma} \bar{g}_2^{Q_{1a}} (g_2 + g_4) \Pi_\text{ph} (\mathbf{0}) . \nonumber
\end{align}
The solution of these equations leads to:
\begin{align}
    & \bar{g}_4^{Q_{1a}} = \frac{1}{1 - g_4 \Pi_\text{ph} (\mathbf{0})} \Big[ \frac{g_4}{1 + g_4 \Pi_\text{ph} (\mathbf{0})}  \nonumber \\
    & - \frac{4 g_2^2 \Pi_\text{ph} (\mathbf{0})}{ \Big[ 1 - (g_4 + 4 g_2) \Pi_\text{ph} (\mathbf{0}) \Big] \Big[ 1 + (2 g_2 - g_4) \Pi_\text{ph} (\mathbf{0}) \Big]} \Big], \nonumber  \\
    & \bar{g}_2^{Q_{1a}} = \frac{g_2}{\Big[ 1 - (g_4 + 4 g_2) \Pi_\text{ph} (\mathbf{0}) \Big] \Big[ 1 + (2 g_2 - g_4) \Pi_\text{ph} (\mathbf{0}) \Big]} . \nonumber
\end{align}
To get the expression of the effective interaction $U_a^{Q_{1a}}$, we include the effect of the interaction $g_1$ at the first order, as diagrammatically shown in Fig.~\ref{fig:diagrams_Ua}(d). The corresponding analytical expression reads:
\begin{align} \label{eq:expr_Ua}
    U_a^{Q_{1a}} = \bar{g}_4^{Q_{1a}} + 4 g_1 \Pi_\text{ph}^2 (\mathbf{0}) \bar{g}_2^{Q_{1a}} (2 \bar{g}_4^{Q_{1a}} + \bar{g}_2^{Q_{1a}}) .
\end{align}
Near the onset of the phase transition, i.e, when $(g_4 - 2 g_2) \Pi_\text{ph} (\mathbf{0}) \approx 1$, we can rewrite $\bar{g}_2^{Q_{1a}}$ and $\bar{g}_4^{Q_{1a}}$, keeping only the divergent part, as:
\begin{align}
    & \bar{g}_4^{Q_{1a}} \approx \frac{1}{3 \Pi_\text{ph} (\mathbf{0})} \frac{1}{1 + (2 g_2 - g_4) \Pi_\text{ph} (\mathbf{0})}, \nonumber \\
    & \bar{g}_2^{Q_{1a}} \approx - \frac{1}{6 \Pi_\text{ph} (\mathbf{0})} \frac{1}{1 + (2 g_2 - g_4) \Pi_\text{ph} (\mathbf{0})} , \nonumber
\end{align}
which, substituted into Eq.~\eqref{eq:expr_Ua}, lead to:
\begin{align}
    U_a^{Q_{1a}} & \approx \frac{1}{3 \Pi_\text{ph} (\mathbf{0})} \frac{1}{1 + (2 g_2 - g_4) \Pi_\text{ph} (\mathbf{0})} \nonumber \\
    & \Big[ 1 + \frac{g_1 \Pi_\text{ph} (\mathbf{0})}{1 + (2 g_2 - g_4) \Pi_\text{ph} (\mathbf{0})} \Big] \nonumber \\
    & \approx \frac{1}{3 \Pi_\text{ph} (\mathbf{0})} \frac{1}{1 + (2 g_2 - g_4 - g_1) \Pi_\text{ph} (\mathbf{0})} . \nonumber
\end{align}
By proceeding in a similar manner, one can compute $U_b^{Q_{1a}}$ considering the diagrammatic expressions shown in Fig.~\ref{fig:diagrams_Ua}(e). The corresponding analytical expressions are:
\begin{align}
    & U_b^{Q_{1a}} = g_4 - g_4 U_b^{Q_{1a}} \Pi_\text{ph} (\mathbf{0}) - 2 g_1 \tilde{g}_1^{Q_{1a}} \Pi_\text{ph} (\mathbf{0}) \nonumber \\
    & \tilde{g}_1^{Q_{1a}} = g_1 - \big[ g_1 U_b^{Q_{1a}} + (g_1 + g_4) \tilde{g}_1^{Q_{1a}} \big] \Pi_\text{ph} (\mathbf{0}) , \nonumber
\end{align}
with solutions:
\begin{align}
    & \tilde{g}_1^{Q_{1a}} = \frac{g_1}{\big[ 1 - (g_1 - g_4) \Pi_\text{ph} (\mathbf{0}) \big] \big[ 1 + (2 g_1 + g_4) \Pi_\text{ph} (\mathbf{0}) \big]} \nonumber \\
    & U_b^{Q_{1a}} = \frac{1}{1 + g_4 \Pi_\text{ph} (\mathbf{0})} \Big[ g_4 \nonumber \\
    & - \frac{2 g_1^2 \Pi_\text{ph} (\mathbf{0})}{ \big[ 1 - (g_1 - g_4) \Pi_\text{ph} (\mathbf{0}) \big] \big[ 1 + (2 g_1 + g_4) \Pi_\text{ph} (\mathbf{0}) \big]} \Big] . \nonumber
\end{align}
From the expression of $U_b^{Q_{1a}}$, it becomes clear that it is not divergent in the proximity of the phase transition, thus it can be neglected. Finally, one arrives at the effective nematic pairing interaction:
\begin{align} \label{eq:gamma_vert_pom_fin}
    \Gamma^{Q_{1a}}_{\alpha, \beta; \gamma, \delta} \approx \frac{1}{3 \Pi_\text{ph} (\mathbf{0})} \frac{\delta_{\alpha \delta} \delta_{\beta \gamma}}{1 + \Gamma^\text{d-PI} \Pi_\text{ph} (\mathbf{0})} .
\end{align}
We notice that the effective interaction Eq.~\eqref{eq:gamma_vert_pom_fin} is strongly enhanced in the proximity of the phase transition, and it is actually divergent. Moreover, for our choice of the particle-hole polarization, $\Pi_\text{ph} (\mathbf{0})$ is negative, meaning that Eq.~\eqref{eq:gamma_vert_pom_fin} is attractive. Finally, the spin structure of the vertex $\delta_{\alpha \delta} \delta_{\beta \gamma}$ suggests that the nematic interaction is mediated by both zero-momentum charge and spin fluctuations, as it becomes apparent by using the SU$(2)$ Fierz identity $\delta_{\alpha \delta} \delta_{\beta \gamma} = \frac{1}{2} \big( \delta_{\alpha \gamma} \delta_{\beta \delta} + \boldsymbol{\sigma}_{\alpha \gamma} \cdot \boldsymbol{\sigma}_{\beta \delta} \big)$. The equal contribution coming from charge and spin fluctuations suggests that the charge-fermion model arising from this attractive interaction satisfies the Ward identities for charge and spin already at the RPA level, without the need to correct the effective interaction with other diagrams \cite{Chubukov2009_PRL,Maslov2010_PRB,Chubukov2014_PRB}.

\begin{figure*}	\centerline{\includegraphics[width=1.0\textwidth]{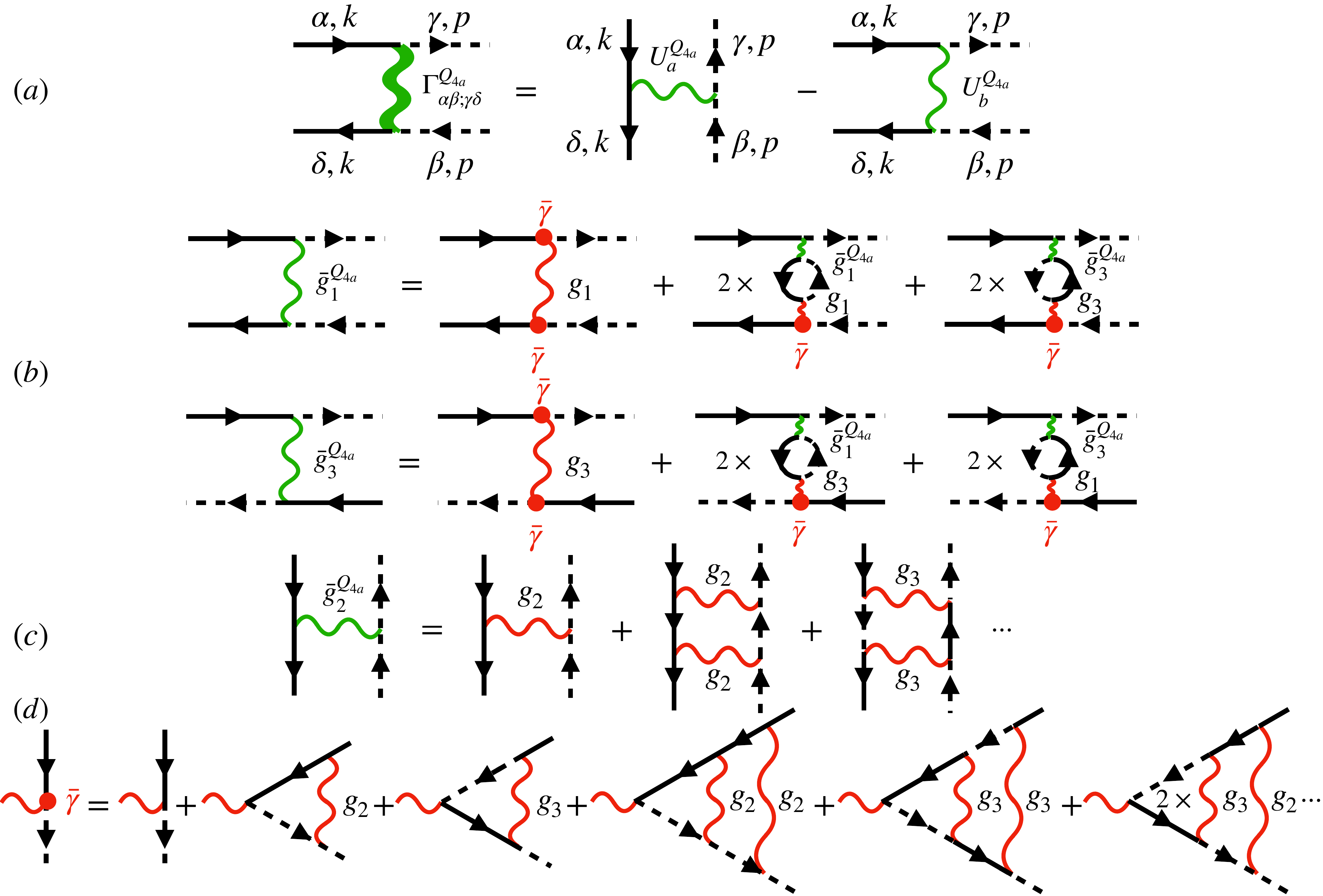}}
	\caption{ \textbf{Diagrams for the renormalized interactions near the onset of the $Q_{4a}$ CO -} (a) Diagrammatic representation of Eq.~\eqref{eq_sup:gamma_vert_co}, focusing on the interaction between band $1$ and band $2$ electrons. (b) Relevant diagrams containing the polarization $\Pi_\text{ph} (\mathbf{Q})$ to the renormalization of the interactions $g_1$ (equal to $U_b^{Q_{4a}}$) and $g_3$. (c) Perturbative series for the renormalization of the interaction $g_2$ (equal to $U_a^{Q_{4a}}$). (d) Diagrammatic representation of the dressed vertices (red dots).} \label{fig:diagrams_UaQ4a}
\end{figure*}

\subsubsection{Pairing interaction in the nematic channel near the onset of the $Q_{4a}$ CO}
In the proximity of the onset of the $Q_{4a}$ CO, the nematic pairing vertex
\begin{align} \label{eq_sup:gamma_vert_co}
    \Gamma_{\alpha, \beta; \gamma, \delta}^{Q_{4a}} (k,p ; p, k) = U_a^{Q_{4a}} \delta_{\alpha \delta} \delta_{\beta \gamma} - U_b^{Q_{4a}} \delta_{\alpha \gamma} \delta_{\beta \delta} ,
\end{align}
takes the diagrammatic form shown in Fig.~\ref{fig:diagrams_UaQ4a}(a). The expression of $U_a^{Q_{4a}}$ is equal to the renormalized interaction $\bar{g}_2^{Q_{4a}}$, and it can be obtained summing the ladder series in Fig.~\ref{fig:diagrams_UaQ4a}(c):
\begin{align}
    & U_a^{Q_{4a}} = \bar{g}_2^{Q_{4a}} = \frac{1}{1 + g_2 \Pi_\text{ph} (\mathbf{Q'})} \Big[ g_2  \nonumber \\
    & - \frac{g_3^2 \Pi_\text{ph} (\mathbf{Q'})}{(1 + g_2 \Pi_\text{ph} (\mathbf{Q'}))^2 (1 - g_3^2 \Pi_\text{ph}^2 (\mathbf{Q'}))} \Big] . \nonumber
\end{align}
The computation of $U_b^{Q_{4a}}$ is more involved, and one needs to solve a system of coupled equations for $\bar{g}_1^{Q_{4a}}$ and $\bar{g}_3^{Q_{4a}}$ as shown in Fig.~\ref{fig:diagrams_UaQ4a}(b):
\begin{align} \label{eq:coupled_eq_UaQ4a}
    & \bar{g}_1^{Q_{4a}} = \bar{\gamma}^2 g_1 + 2 \bar{\gamma} ( g_1 \bar{g}_1^{Q_{4a}} + g_3 \bar{g}_3^{Q_{4a}} ) \Pi_\text{ph} (\mathbf{Q'}) , \nonumber \\
    & \bar{g}_3^{Q_{4a}} = \bar{\gamma}^2 g_3 + 2 \bar{\gamma} ( g_1 \bar{g}_3^{Q_{4a}} + g_3 \bar{g}_1^{Q_{4a}} ) \Pi_\text{ph} (\mathbf{Q'}) .
\end{align}
The renormalized vertex $\bar{\gamma}$ is computed from the diagrams in Fig.~\ref{fig:diagrams_UaQ4a}(c), leading to:
\begin{align}
    \bar{\gamma} = \frac{1}{1 + g_2 \Pi_\text{ph} (\mathbf{Q}') + g_3 \Pi_\text{ph} (\mathbf{Q}')}
\end{align}
The solution of the system of equations Eq.~\eqref{eq:coupled_eq_UaQ4a} is:
\begin{align}
    & \bar{g}_3^{Q_{4a}} = \frac{g_3}{1 - (2 g_1 + g_3 - g_2 ) \Pi_\text{ph} (\mathbf{Q}')} \nonumber \\
    & \frac{1}{\big[ 1 + (g_2 + g_3) \Pi_\text{ph} (\mathbf{Q}') \big] \big[ 1 - (2 g_1 - 3 g_3 - g_2 ) \Pi_\text{ph} (\mathbf{Q}') \big]} , \nonumber \\
    & U_b^{Q_{4a}} = \bar{g}_1^{Q_{4a}} = \frac{1}{1 + (g_2 + g_3 - 2 g_1 ) \Pi_\text{ph} (\mathbf{Q}')}  \nonumber \\
    & \frac{1}{1 + (g_2 + g_3) \Pi_\text{ph} (\mathbf{Q}')} \Big[ g_1 + \frac{2 g_3^2 \Pi_\text{ph} (\mathbf{Q}')}{1 + (g_2 - 2 g_1 - g_3 ) \Pi_\text{ph} (\mathbf{Q}')} \nonumber \\
    &  \frac{1}{1 + (g_2 - 2 g_1 + 3 g_3 ) \Pi_\text{ph} (\mathbf{Q}')} \Big] . \nonumber
\end{align}

\begin{figure*}	\centerline{\includegraphics[width=1.0\textwidth]{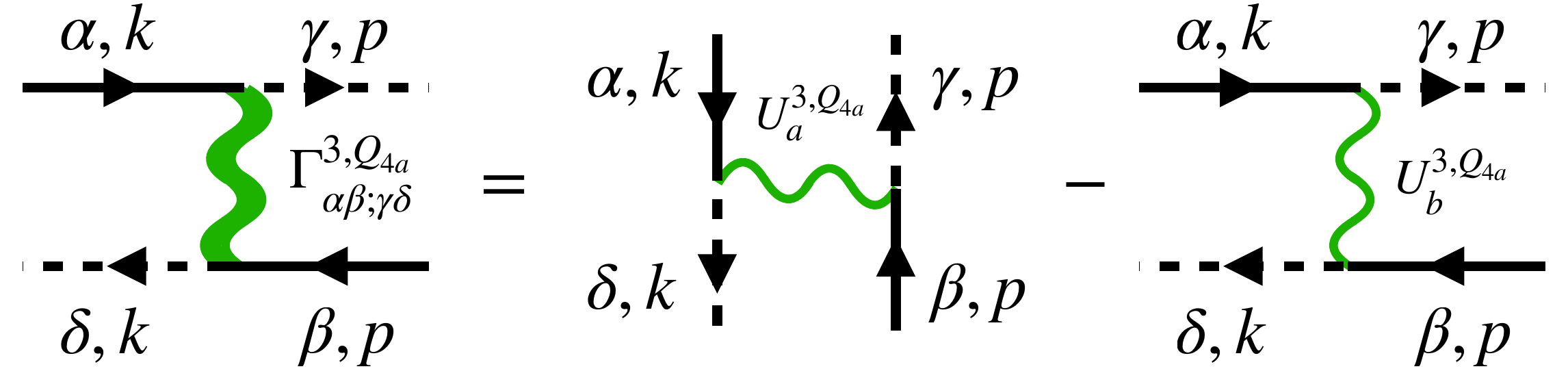}}
	\caption{\textbf{Diagrams for the renormalized umklapp interaction near the onset of the $Q_{4a}$ CO -} Diagrammatic representation of Eq.~\eqref{eq:gamma_vert_Q4a3_fin}, focusing on the interaction between band $1$ and band $2$ electrons. In this case, $U_a^{3, Q_{4a}} = U_b^{3, Q_{4a}}$.} \label{fig:diagrams_UaQ4a3}
\end{figure*}

\noindent
In the proximity of the $Q_{4a}$ CO phase transition, i.e., when $(2 g_1 + g_3 - g_2) \Pi_\text{ph} (\mathbf{Q}') \approx 1$, the effective nematic pairing interactions becomes:
\begin{align} \label{eq:gamma_vert_Q4a_fin}
    \Gamma^{Q_{4a}}_{\alpha, \beta; \gamma, \delta} \approx - \frac{1}{8 (g_1 + g_3) \Pi_\text{ph}^2 (\mathbf{Q}')} \frac{\delta_{\alpha \gamma} \delta_{\beta \delta}}{1 + \Gamma^{Q_{4a}} \Pi_\text{ph} (\mathbf{Q}')} ,
\end{align}
given that $U_a^{Q_{4a}}$ does not diverge when $(2 g_1 + g_3 - g_2) \Pi_\text{ph} (\mathbf{Q}') = 1$. Eq.~\eqref{eq:gamma_vert_Q4a_fin} is divergent at the phase transition. Moreover, it is negative, i.e., attractive. Finally, the interaction is mediated only by finite momentum charge fluctuations, i.e., as it is, the interaction Eq.~\eqref{eq:gamma_vert_Q4a_fin} violates the Ward identities. We notice that the interaction $\bar{g}_3^{Q_{4a}}$ also diverges at the phase transition, leading to an interacting vertex (retaining only the divergent part) of the form:
\begin{align} \label{eq:gamma_vert_Q4a3_fin}
    & \Gamma^{3, Q_{4a}}_{\alpha, \beta; \gamma, \delta} = \frac{\bar{g}_3^{Q_{4a}}}{2} (\boldsymbol{\sigma}_{\alpha \gamma} \cdot \boldsymbol{\sigma}_{\beta \delta} - \delta_{\alpha \gamma} \delta_{\beta \delta} ) \nonumber \\
    & \approx \frac{1}{8 (g_1 + g_3) \Pi_\text{ph}^2 (\mathbf{Q}')} \frac{ \frac{1}{2} \big( \boldsymbol{\sigma}_{\alpha \gamma} \cdot \boldsymbol{\sigma}_{\beta \delta} - \delta_{\alpha \gamma} \delta_{\beta \delta} \big)}{1 + \Gamma^{Q_{4a}} \Pi_\text{ph} (\mathbf{Q}')} ,
\end{align}
which is a positive (repulsive) divergent quantity in the spin channel ($\boldsymbol{\sigma}_{\alpha \gamma} \cdot \boldsymbol{\sigma}_{\beta \delta}$) and a negative (attractive) divergent quantity in the charge channel ($\delta_{\alpha \gamma} \delta_{\beta \delta}$). The diagrammatic representation of Eq.~\eqref{eq:gamma_vert_Q4a3_fin} is reported in Fig.~\ref{fig:diagrams_UaQ4a3}.

\begin{figure*}	\centerline{\includegraphics[width=1.0\textwidth]{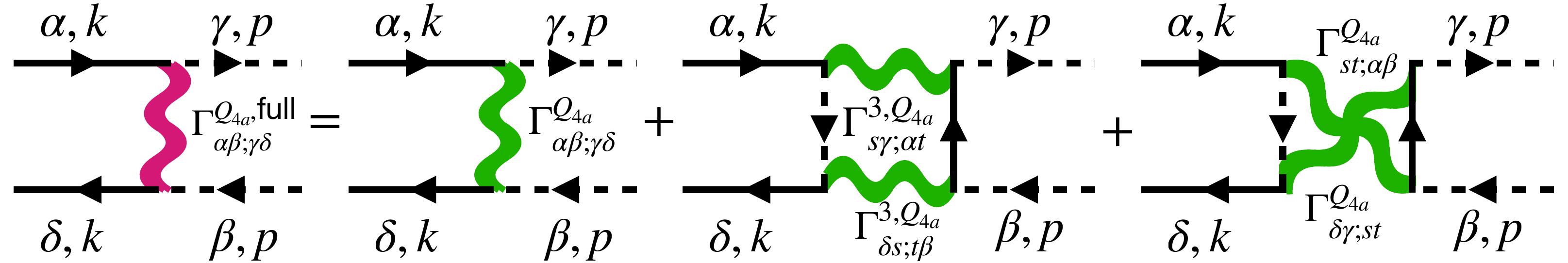}}
	\caption{ \textbf{Renormalization of the charge-fermion vertex -} The first diagram is the RPA vertex coming from $\Gamma^{Q_{4a}}$. The last two diagrams are Aslamazov-Larkin contributions that restore the Ward identities. The first of these two diagrams comes from $\Gamma^{3,Q_{4a}}$ , while the second from $\Gamma^{Q_{4a}}$.} \label{fig:diagrams_UaQ4afull}
\end{figure*}

\noindent
By combining the interaction vertices in Eqs.~\eqref{eq:gamma_vert_Q4a_fin}-\eqref{eq:gamma_vert_Q4a3_fin}, one can renormalize the charge-fermion vertex Eq.~\eqref{eq:gamma_vert_Q4a_fin} restoring charge and spin conservations. The full vertex is represented diagrammatically in Fig.~\ref{fig:diagrams_UaQ4afull}. The spin structure of the first Aslamazov-Larkin diagram that appears in Fig.~\ref{fig:diagrams_UaQ4afull} has the form:
\begin{align}
    & \frac{1}{4} \sum_{s,t} (\boldsymbol{\sigma}_{\alpha s} \cdot \boldsymbol{\sigma}_{t \gamma} - \delta_{\alpha s} \delta_{t \gamma} ) (\boldsymbol{\sigma}_{s \delta} \cdot \boldsymbol{\sigma}_{\beta t} - \delta_{s \delta} \delta_{\beta t} ) \nonumber \\
    & = \boldsymbol{\sigma}_{\alpha \gamma} \cdot \boldsymbol{\sigma}_{\beta \delta} + 3 \delta_{\alpha \gamma} \delta_{\beta \delta} , \nonumber
\end{align}
while the spin structure of the second Aslamazov-Larkin diagram in Fig.~\ref{fig:diagrams_UaQ4afull} is:
\begin{align}
    \sum_{s,t} \delta_{\alpha s} \delta_{\beta t} \delta_{\delta s} \delta_{\gamma t} = \frac{1}{2} \big( \delta_{\alpha \gamma} \delta_{\beta \delta} + \boldsymbol{\sigma}_{\alpha \gamma} \cdot \boldsymbol{\sigma}_{\beta \delta} \big) . \nonumber
\end{align}
Using the same notation of \cite{Chubukov2014_PRB}, we arrive writing the contribution to the full vertex coming from the Aslamazov-Larkin diagrams as:
\begin{align}
    \Gamma^{\text{full}, \text{AL}}_{\alpha, \beta; \gamma, \delta} = \frac{1}{4} \bar{g} \xi^2 \big( 5 \delta_{\alpha \gamma} \delta_{\beta \delta} + \boldsymbol{\sigma}_{\alpha \gamma} \cdot \boldsymbol{\sigma}_{\beta \delta} \big) , \nonumber
\end{align}
where the minus sign between the two diagrams has been discussed previously \cite{Chubukov2014_PRL}. The contribution coming from the RPA diagram is
\begin{align}
    \Gamma^{\text{full}, \text{RPA}}_{\alpha, \beta; \gamma, \delta} = - \bar{g} \xi^2 \delta_{\alpha \gamma} \delta_{\beta \delta} , \nonumber
\end{align}
where the minus sign comes from the negative value of the vertex Eq.~\eqref{eq:gamma_vert_Q4a_fin}. In the previous expressions, $\bar{g}$ represents the effective coupling between the charge fluctuations and the fermions and $\xi$ is the CO correlation length (further details can be found in \cite{Chubukov2014_PRB}). Adding up the two contributions, one obtains:
\begin{align}
    & \Gamma^\text{full}_{\alpha, \beta; \gamma, \delta} = \Gamma^{\text{full}, \text{RPA}}_{\alpha, \beta; \gamma, \delta} + \Gamma^{\text{full}, \text{AL}}_{\alpha, \beta; \gamma, \delta} \nonumber \\
    & = \frac{1}{4} \bar{g} \xi^2 \big( \delta_{\alpha \gamma} \delta_{\beta \delta} + \boldsymbol{\sigma}_{\alpha \gamma} \cdot \boldsymbol{\sigma}_{\beta \delta} \big) , \nonumber
\end{align}
indeed leading to a restoration of the Ward identities.
\noindent
While the crossed and uncrossed Aslamazov-Larkin diagrams are found to cancel out in the charge sector for $q=0$ spin \cite{Chubukov2014_PRL} and charge \cite{Mayrhofer2024_PRB} instabilities, we find them to add up near a finite momentum CO. The explanation of this difference comes from the different spin structure of the two diagrams, see Fig.~\ref{fig:diagrams_UaQ4afull}, which does not lead to a cancellation of the two contributions despite the fact that they have a different sign.

\subsubsection{Charge-fermion models}
From the analysis above, one can write the interaction between fermions at the Fermi surface with collective charge fluctuations, which represents a microscopic theory for the interaction near a finite momentum CO quantum critical point. The interaction reads 
\begin{align}
    \mathcal{H}_\text{ch-ch} = & \sum_{\mathbf{q},i,j,l} \sum_{\alpha,\beta,\gamma, \delta} V_{i j l ;\alpha \beta, \gamma \delta} (\mathbf{q}) \nonumber \\
    &\sum_{\mathbf{k},\mathbf{p}} c^\dagger_{l,\mathbf{p},\gamma} c^\dagger_{j,\mathbf{k},\delta} c_{j,\mathbf{k} - \mathbf{q},\alpha} c_{l,\mathbf{p} + \mathbf{q},\beta} , \nonumber
\end{align}
where $i,j$ and $l$ run over the band indices, and where 
\begin{align}
    V_{i j l ;\alpha \beta, \gamma \delta} (\mathbf{q}) = | \epsilon_{ijl} | V_i (\mathbf{q}) (\delta_{\alpha \gamma} \delta_{\beta \delta} + \boldsymbol{\sigma}_{\alpha \gamma} \cdot \boldsymbol{\sigma}_{\beta \delta}) . \nonumber
\end{align}
In the expression above, the interaction potential $V_i (\mathbf{q})$ is peaked at $\mathbf{Q}_i'$ such that near $\mathbf{Q}_i'$ it can be approximated as
\begin{align}
    V_i (\mathbf{q}) = \frac{\bar{g}}{\xi^{-2} + | \mathbf{q} - \mathbf{Q}_i' |^2} .\nonumber
\end{align}
From the previous analysis, it follows that the nematic susceptibility for the $d$-wave charge Pomeranchuk instability has the standard RPA expression \cite{Maslov2010_PRB}, while in the case of the $Q_{4a}$ CO it has a form similar to what has been discussed for iron-based superconductors \cite{Karahasanovic2015_PRB,Kretzschmar2016_NatPhys,Hinojosa2016_PRB}, but with charge fluctuations instead of spin fluctuations.

\noindent
Now that we have shown that a divergent attraction exists in the nematic channel due to the presence of charge-fluctuations in the proximity of the $d$-wave charge PI and of the $Q_{4a}$ CO, we analyze the precise form of these phases by performing a Ginzburg-Landau analysis of the symmetry broken states the system tends to or even stabilizes.

\section{Ginzburg-Landau theory for the $\mathbf{Q_{4 a}}$ charge order} \label{sec:Q4aCO}
For the $Q_{4 a}$ CO, the patch model interaction Eq.~\eqref{eq:ham_patch_int} can be rewritten as:
\begin{align} 
	& \mathcal{H}_\text{int}^{Q_{4 a}} \approx - N  \frac{g_{\text{co}}}{2}  \sum_{i} \rho_i \rho_i , \label{eq:cc_int} 
\end{align}
where $g_\text{co} = \Gamma^{Q_{4a}}$ and $\rho_i = \frac{|\epsilon_{i j l}|}{2 N} \sum_\sigma \sum_{\mathbf{q} < \Lambda} c^\dagger_{j, \mathbf{q}, \sigma} c_{l, \mathbf{q}, \sigma}$ is the charge operator with momentum $\mathbf{Q}_i' = \mathbf{M}_j' - \mathbf{M}_l'$, i.e., it describes an instability towards charge ordered states that might stabilize an additional $1 \times 4$ or $4 \times 4$ CO besides the $2 \times 2$ one.

\begin{figure}
    \centerline{\includegraphics[width=0.5\textwidth]{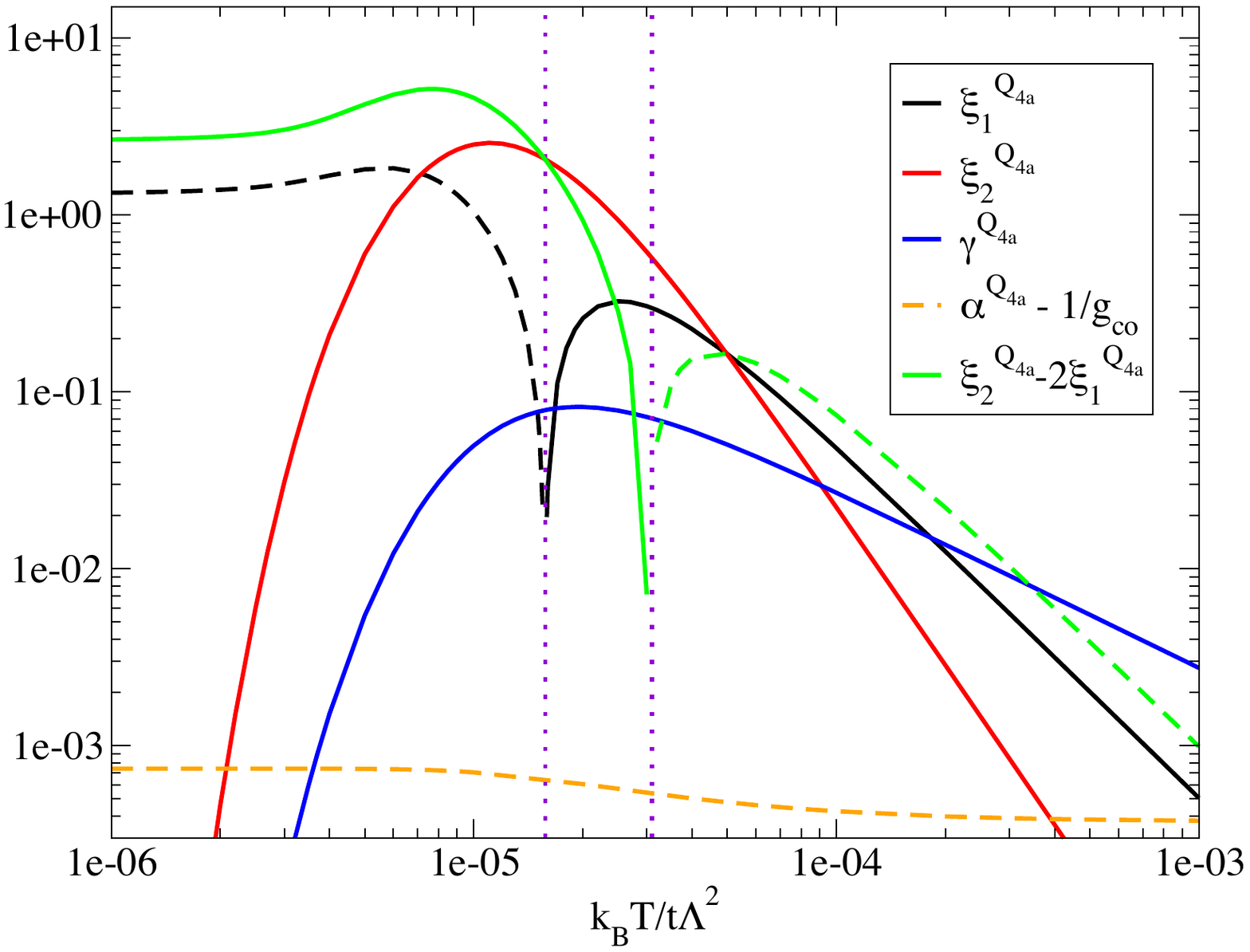}}
	\caption{ \textbf{Temperature evolution of the Ginzburg-Landau coefficients -} Log-log plot of the absolute value of the coefficients $\alpha^{Q_{4 a}}$, $\gamma^{Q_{4 a}}$, $\xi_1^{Q_{4 a}}$ and $\xi_2^{Q_{4 a}}$ evaluated asymptotically in the limit of $\mu, k_B T \ll t \Lambda^2$ (see the SM for their analytical expression \cite{Suppl_mat}) at $\mu/t \Lambda^2 = -3 \times 10^{-5}$ ($\Lambda = 1$). Solid (dashed) lines indicate positive (negative) values. The vertical dotted lines indicate the position of the zeros of $\xi_1^{Q_{4 a}}$ and $\xi_2^{Q_{4 a}} - 2 \xi_1^{Q_{4 a}}$. The band structure parameters are $m=1$, $\delta = 0.3$.} \label{fig:coeff_vs_temp}
\end{figure}

\noindent
By decoupling the interaction Eq.~\eqref{eq:cc_int} in the Hubbard-Stratonovich (HS) sense and by integrating out the fermions, we derive the effective free energy in terms of the three fields $\Delta_1$, $\Delta_2$ and $\Delta_3$ up to the fourth order \cite{McMillan1975_PRB}:
\begin{align} \label{eq:eff_free_en_fluct}
	& \mathcal{F}_\text{eff}^{Q_{4 a}} = \sum_i  \int_k \Delta_i (k)  \chi^{-1} (k) \Delta_i (-k) \nonumber \\
	& + \frac{\gamma^{Q_{4 a}}}{3} \int_x \Delta_1 (x) \Delta_2 (x) \Delta_3 (x) + \frac{\xi_1^{Q_{4 a}}}{4} \int_x \big( \sum_i \Delta_i^2 (x) \big)^2 \nonumber \\
	& + \frac{\xi_2^{Q_{4 a}} - 2 \xi_1^{Q_{4 a}}}{4} \int_x \sum_{i < j} \Delta_i^2 (x) \Delta_j^2 (x),
\end{align}
having included the contribution of the fluctuations of the order parameters, and where $2 \chi^{-1} (k) = \alpha^{Q_{4 a}} + \bar{\gamma} | \nu_n | + \nu_n^2 + \bar{f} (\mathbf{k})$ is the charge propagator in principle peaked at one of the wave vectors $\mathbf{Q}_i'$, the ordering vector corresponding to the order parameter $\Delta_i$ \cite{Caprara2007_PRB,Caprara2011_PRB,Caprara2015_PRB}. However, the momentum $\mathbf{k}$ in the expression above is taken with respect to the ordering vectors $\mathbf{Q}_i'$, so in this regard the dependence of the charge propagator on the index $i$ does not appear. $\nu_n = 2 \pi T n$ is the bosonic Matsubara frequency, due to the bosonic nature of the excitations of the charge-ordered state, $k = (i \nu_n, \mathbf{k})$, $\bar{\gamma}$ is the Landau damping coefficient, and $\bar{f} (\mathbf{k}) = k_x^2 + k_y^2$ \cite{Fernandes2012_PRB,Paul2017_PRL}. The explicit expressions for the coefficients in Eq.~\eqref{eq:eff_free_en_fluct}, which depend on temperature and chemical potential, are
\begin{align}
	& \alpha^{Q_{4 a}} = \frac{1}{g_\text{co}} + \frac{1}{2} \int_k \mathcal{G}_{0,1} (k) \mathcal{G}_{0,2} (k), \nonumber \\
	& \gamma^{Q_{4 a}} = \frac{3}{2} \int_k \mathcal{G}_{0,1} (k) \mathcal{G}_{0,2} (k)  \mathcal{G}_{0,3} (k) , \nonumber \\
	& \xi_1^{Q_{4 a}} = \frac{1}{8} \int_k \mathcal{G}_{0,1}^2 (k) \mathcal{G}_{0,2}^2 (k) , \nonumber \\
	& \xi_2^{Q_{4 a}} = \frac{1}{4} \int_k \mathcal{G}_{0,1} (k) \mathcal{G}_{0,2} (k) \mathcal{G}_{0,3}^2 (k) , \nonumber
\end{align}
with $\mathcal{G}_{0,i}^{-1} (i \omega_n, \mathbf{k}) = -i \omega_n + \varepsilon_{i, \mathbf{k}} - \mu$ the electronic Green's functions and $\omega_n = (2 n + 1) \pi T$ the fermionic Matsubara frequencies. Their temperature dependence is shown in Fig.~\ref{fig:coeff_vs_temp} \cite{Classen2020_PRB,Park2021_PRB}. In Eq.~\eqref{eq:eff_free_en_fluct}, the integrals are defined such that $\int_k = T \sum_n \int_{-\Lambda}^\Lambda \frac{d^2 k}{(2 \pi \Lambda)^2}$, $\int_x = \int_0^\beta d \tau \int d^2 r$ and $x = (\tau, \mathbf{r})$, with $\tau$ the time variable and $\mathbf{r}$ the position. The quartic part of the interaction can be decoupled by introducing three auxiliary HS fields $A (x) \rightarrow \sum_i \Delta_i^2 (x) / \sqrt{3}$, $O_1 (x) \rightarrow N_1(x) = \big( \Delta_1^2 (x) - \frac{\Delta_2^2 (x) + \Delta_3^2 (x)}{2} \big) \sqrt{2/3}$ and $O_2 (x) \rightarrow N_2 (x) = (\Delta_2^2 (x) - \Delta_3^2 (x)) / \sqrt{2}$ ($\mathbf{N} = (N_1, N_2)$), where $A (x)$ transforms as an A$_1$ irreducible representation (irrep), while $\mathbf{O}^t (x) = (O_1(x), O_2(x))$ transforms as the two dimensional E$_2$ irrep of D$_{6\text{h}}$. We use the Gaussian approximation rewriting the fields $\Delta_i (x) = \Delta_i + \delta \Delta_i (x)$. Keeping only the quadratic contributions in the fluctuations $\delta \Delta_i (x)$ and integrating them out \cite{Hecker2018_npj,Dolgirev2020_PRB,Grandi2021_PRB}, the free energy becomes
\begin{align} \label{eq:free_en_eff_fin}
	\mathcal{F}_\text{eff}^{Q_{4 a}} = & - \int_k \Big[ \frac{A^2}{\xi_1^{Q_{4 a}} + \xi_2^{Q_{4 a}}} - \frac{2 (O_1^2 + O_2^2)}{\xi_2^{Q_{4 a}} - 2 \xi_1^{Q_{4 a}}} \Big] \nonumber \\
    & + \frac{1}{2} \sum_i \int_k  \ln \big( \chi^{-1}_i (k)  \big) .
\end{align}
Here we have assumed $\Delta_i = 0$ and
\begin{align}
	& \chi^{-1}_1 (k) = \tilde{\chi}^{-1} (k) + O_1 \sqrt{2/3} , \nonumber \\
	& \chi^{-1}_2 (k) = \tilde{\chi}^{-1} (k) - O_1 / \sqrt{6} + O_2 / \sqrt{2} , \nonumber \\
	& \chi^{-1}_3 (k) = \tilde{\chi}^{-1} (k) - O_1 / \sqrt{6} - O_2 / \sqrt{2} , \nonumber
\end{align}
with $\tilde{\chi}^{-1} (k) = \chi^{-1} (k) + A / \sqrt{3}$, where the field $A$ renormalizes the bare mass term due to fluctuations ($\alpha^{Q_{4 a}} \rightarrow \alpha^{Q_{4 a}} + A / \sqrt{3}$). Additional details of the derivation can be found in the SM \cite{Suppl_mat}.

\noindent
The temperature dependence of the coefficients in Fig.~\ref{fig:coeff_vs_temp} shows that, below some temperature, $\xi_1^{Q_{4 a}} < 0$, indicating an instability of the theory, which might be cured by higher-order contributions to the Ginzburg-Landau free energy \cite{Fernandes2012_PRB,Fernandes2012_PRBE}. We focus on the temperature range where the theory is stable. At the mean-field level, a negative $\xi_2^{Q_{4 a}} - 2 \xi_1^{Q_{4 a}} < 0$ prefers to stabilize a $3$Q CO, i.e., a state with $\Delta_1$, $\Delta_2$ and $\Delta_3$ different from zero ($4 \times 4$ CO), while $\xi_2^{Q_{4 a}} - 2 \xi_1^{Q_{4 a}} > 0$ points towards a $1$Q solution, corresponding to a $1 \times 4$ CO. Since the quartic coefficient $\xi_2^{Q_{4 a}} - 2 \xi_1^{Q_{4 a}}$ changes from negative to positive values, one would expect a transition from the $4 \times 4$ to the $1 \times 4$ CO. However, the mass renormalization coming from the $A$ field might suppress the $4 \times 4$ CO in the temperature range where it appears at the mean-field level. Indeed, this phase is not observed experimentally. Nevertheless, $A$ might become small enough to stabilize the $1 \times 4$ CO close to the temperature where $\xi_2^{Q_{4 a}} - 2 \xi_1^{Q_{4 a}}$ changes sign.

\section{Ginzburg-Landau theory for the $\mathbf{Q_{1 a}}$ $d$-wave Pomeranchuk instability} \label{sec:Q1aCO}
For the $Q_{1 a}$ $d$-wave charge PI, the patch model interaction Eq.~\eqref{eq:ham_patch_int} can be rewritten as
\begin{align}
	& \mathcal{H}_\text{int}^{Q_{1 a}} \approx - N  \frac{1}{2} \sum_{i, j} g_{ij} n_i n_j , \label{eq:cc_int_zero}
\end{align}
with $g_{ij} = g_\text{d} = -g_4$ if $i=j$, $g_{ij} = g_\text{o} = -2 g_2 + g_1$ if $i \neq j$ and $n_i = \frac{1}{2 N} \sum_\sigma \sum_{\mathbf{q} < \Lambda} c^\dagger_{i, \mathbf{q}, \sigma} c_{i, \mathbf{q}, \sigma}$ is the charge operator with zero momentum, i.e., Eq.~\eqref{eq:cc_int_zero} describes an instability towards a $1 \times 1$ charge-ordered state within the $2 \times 2$ one, describing an intra-unit-cell CO.

\noindent
Considering the interaction Eq.~\eqref{eq:cc_int_zero} and the effective band structure Eq.~\eqref{eq:nonint_ham}, one might decouple the interacting part of the Hamiltonian by introducing two HS fields $O_1 \rightarrow N_1 = \big( n_1 - \frac{n_2 + n_3}{2} \big) \sqrt{2/3}$ and $O_2 \rightarrow N_2 = (n_2 - n_3) / \sqrt{2}$. By integrating out the fermions and by expanding the resulting expression for small $O_1$ and $O_2$, one arrives at the effective free energy (up to irrelevant constants)
\begin{align} \label{eq:free_en_zeromom}
	\mathcal{F}_\text{eff}^{Q_{1 a}} & \approx \frac{\alpha^{Q_{1 a}}}{2} \sum_{i=1}^2 O_i^2 - \frac{\gamma^{Q_{1 a}}}{3} O_1 (O_1^2 - 3 O_2^2) \nonumber \\
    & + \frac{\xi^{Q_{1 a}}}{4} \big( \sum_{i=1}^2 O_i^2 \big)^2 ,
\end{align}
having defined the Ginzburg-Landau coefficients
\begin{align}
	& \alpha^{Q_{1 a}} = \frac{1}{g_\text{nem}^{Q_{1a}}} + \int_k \mathcal{G}_{0,1}^2 (k) , \nonumber \\
	& \gamma^{Q_{1 a}} = \frac{3}{2} \int_k \mathcal{G}_{0,1}^3 (k) , \nonumber \\
	& \xi^{Q_{1 a}} = \frac{1}{8} \int_k \mathcal{G}_{0,1}^4 (k) , \nonumber
\end{align}
with $g_\text{nem}^{Q_{1a}} = g_\text{d} - g_\text{o} = \Gamma^{\text{d-PI}}$. The free energy Eq.~\eqref{eq:free_en_zeromom} describes a three states Potts model \cite{Straley1973_JPA,Baek2011_PRE,Hecker2018_npj,Little2020_NatMat,Cho2020_NatComm,Fernandes2020_SciAdv,Kimura2022_PRB}. Details concerning the derivation of the expressions above can be found in the SM \cite{Suppl_mat}.

\section{Rise of nematicity} \label{sec:rise_nem}
We envisage two ways in which nematicity might arise in the kagome metals: On the one hand, nematicity might manifest itself as a vestigial order of the $1$Q $1 \times 4$ charge order related to the finite momentum charge fluctuations. Within the saddle point approximation, one can write the corresponding zero-momentum nematic susceptibility as \cite{Fernandes2010_PRL,Fernandes2012_SuScTe,Khodas2015_PRB,Karahasanovic2015_PRB,Yamase2015_NJP,Fernandes2017_RPP}
\begin{align} \label{eq:chi_nem}
	& (\hat{\chi}^{Q_{4a}}_\text{nem})_{l m} = - \delta_{l m} \frac{g_\text{nem}^{Q_{4a}} \int_k \tilde{\chi}^2 (k)}{1 + g_\text{nem}^{Q_{4a}} \int_k \tilde{\chi}^2 (k)} ,
\end{align}
with $l, m = 1,2$, corresponding to the two-components of the E$_2$ irrep, and having introduced the nematic interaction $g_\text{nem}^{Q_{4a}} = - \frac{1}{4} (\xi_2^{Q_{4 a}} - 2 \xi_1^{Q_{4 a}})$. If $\int_k \tilde{\chi}^2 (k)$ becomes large, it might produce a divergence of the nematic susceptibility in the temperature range where $\xi_2^{Q_{4 a}} - 2 \xi_1^{Q_{4 a}} > 0$. This condition is surely satisfied near the transition to the $1 \times 4$ CO.

\noindent
On the other hand, nematicity might occur as a symmetry breaking in the zero momentum order parameters, which, as such, would not lead to an increase of the unit cell of the system. More explicitly, the $Q_{1a}$ nematic response function becomes \cite{Gallais2016_CRP,Gallais2016_PRL}:
\begin{align} \label{eq:chi_nem_zeromom}
	(\hat{\chi}^{Q_{1a}}_\text{nem})_{l m} = - \delta_{l m} \frac{g_\text{nem}^{Q_{1a}} \int_k \mathcal{G}_{0,i}^{2} (k)}{1 + g_\text{nem}^{Q_{1a}} \int_k \mathcal{G}_{0,i}^{2} (k)} ,
\end{align}
with $i=1,2$ or $3$ and $g_\text{nem}^{Q_{1a}} > 0$. From Eq.~\eqref{eq:chi_nem_zeromom}, one can obtain an analytic expression for the divergence of the nematic susceptibility relating the hopping strength of the effective dispersions Eq.~\eqref{eq:dispersions} to the other parameters of the problem (see the SM for the explicit relation and for additional details on the calculations sketched above \cite{Suppl_mat}). The diagonal structure of Eqs.~\eqref{eq:chi_nem}-\eqref{eq:chi_nem_zeromom} reflects the degeneracy of the E$_2$ irrep \textit{at} the instability level. However, \textit{below} the instability level, the C$_{2\text{v}}$ nematic state is actually preferred \cite{Kiesel2013_PRL,Wang2013_PRB}. Differently from the case of a system with D$_{4\text{h}}$ symmetry, where only the B$_{1\text{g}}$ symmetry channel gives a contribution different from zero, while the B$_{2\text{g}}$ one goes to zero \cite{Kretzschmar2016_NatPhys}, here we obtain a finite contribution in both the degenerate components of E$_2$.

\begin{table*}[t]
  \centering
  \begin{tabular}{ | C{4.8cm} | C{4.8cm} | C{5.8cm} |}
    \hline
    $ $ & Iron-based superconductors & Vanadium-based kagome metals \\ \hline
     Broken symmetry & $\mathbb{Z}_2$ Ising & $\mathbb{Z}_3$ Potts \\ \hline
    Fluctuation scenario & Spin-driven & Charge-driven \\ \hline
    Standard symmetry breaking & Orbital order & d-wave charge Pomeranchuk instability \\
    \hline
  \end{tabular}
    \caption{ \textbf{Different nematic character of the iron-based superconductors and of the vanadium-based kagome metals -} Summary of the main differences between the nematicity in the iron-based superconductors and in the vanadium-based kagome metals.} \label{table:nem}
\end{table*}

\section{Microscopic interpretation of the nematic response functions} \label{sec:micro_nem}
The analysis performed in Sec.~\ref{sec:KL_analysis} allows us to introduce two microscopic models that can explain the onset of nematicity in the kagome metals. These models represent a transposition of the spin-fermion \cite{Abanov2000_PRL} and of the fermion-boson model \cite{Garst2010_PRB} to the charge fluctuations cases, i.e., we introduce two charge-fermion models. The first model is described by the effective zero-momentum charge-charge interaction Eq.~\eqref{eq:gamma_vert_pom_fin}, while the second one comes from the effective finite momentum interaction depicted in Fig.~\ref{fig:diagrams_UaQ4afull}. The nematic susceptibility for the zero-momentum charge fluctuations (d-PI) can be described at the RPA level \cite{Maslov2010_PRB}, in agreement with the result of the Ginzburg-Landau analysis we performed, compare Eq.~\eqref{eq:chi_nem_zeromom} and Fig.~\ref{fig:diagrams_Ua} (b) and (d).

\noindent
The nematic susceptibility of the finite-momentum charge fluctuations is related to the Aslamazov-Larkin diagrams, i.e., its representation goes beyond the RPA scheme. Indeed, the triangular fermionic loops that appear in these diagrams (by closing the external lines in Fig.~\ref{fig:diagrams_UaQ4afull} with a zero-momentum vertex) become large in the presence of nesting \cite{Paul2014_PRB}, i.e., when the ellipticity of the hole pockets is small enough, and one can approximate the nematic susceptibility by the contribution coming from the Aslamazov-Larkin diagrams, in analogy with the relevant role played by these diagrams in determining the $d$-wave Raman response function in the cuprates \cite{Caprara2005_PRL}. If we assume the charge fluctuations to have a smaller energy scale than the electronic degree of freedom, we might replace the triangular fermionic loop by a constant. By considering the corresponding ladder series, one arrives at Eq.~\eqref{eq:chi_nem} \cite{Hinojosa2016_PRB,Fernandes2017_RPP}.

\section{Discussion} \label{sec:discussion}
There is a third possibility that we have not analyzed so far for the onset of nematicity in the kagome metals, and this is related to the fluctuations of the three charge order parameters stabilized immediately below $T_\text{co}$. While the thermal fluctuations might be important above the critical temperature $T > T_\text{co}$, as shown experimentally \cite{Chen2022_PRL}, we do not expect them to play a relevant contribution for $T < T_\text{co}$. Moreover, a recent experiment \cite{Asaba2024_NatPhys} has found an odd nematic state above $T_\text{co}$ in the cesium member of the family, which points towards a nematic order parameter belonging to the E$_{1 \text{u}}$ or E$_{2 \text{u}}$ irreps. Any theory for the charge order stabilized below $T_\text{co}$ that considers it as a purely two-dimensional ordered state fails to catch this feature. Since this phase might actually have an out-of-plane component, one would need to derive a microscopic model that leads to a $2 \times 2 \times 2$ charge order, described by an effective theory that takes into account the three-dimensional nature of the order parameters \cite{Christensen2021_PRB}. The development of this theory goes beyond the scope of the present work.

\noindent
The coupling of the elastic deformations of the E$_2$ irrep to the nematic order parameter renormalizes the nematic couplings $g_\text{nem}^{Q_{na}}$ ($n=1,4$) appearing in Eqs.~\eqref{eq:chi_nem}-\eqref{eq:chi_nem_zeromom}. Indeed, considering the nemato-elastic interaction $\mathcal{F}_\text{int} = \tilde{g} \int_x \boldsymbol{\epsilon}_{\text{E}_2} (x) \cdot \mathbf{N} (x)$ where $\boldsymbol{\epsilon}_{\text{E}_2} = ( \epsilon_{\text{E}_{2,1}}, \epsilon_{\text{E}_{2,2}} )$ is the deformation tensor and the elastic contribution $\mathcal{F}_\text{el} = c_{\text{E}_2} (\epsilon_{\text{E}_{2,1}}^2 + \epsilon_{\text{E}_{2,2}}^2)$, with $c_{\text{E}_2}$ an element of the stiffness matrix \cite{Grandi2023_PRB}, by integrating out the deformation tensor we can rewrite the nematic interactions as $g_\text{nem}^{Q_{4a}} \rightarrow  g_\text{nem}^{Q_{4a}} - 3 \tilde{g}^2/(8 c_{\text{E}_2})$ and $g_\text{nem}^{Q_{1a}} \rightarrow  g_\text{nem}^{Q_{1a}} + \tilde{g}^2/(2 c_{\text{E}_2})$. Generally speaking, the propensity to nematicity is increased by the coupling to the lattice \cite{Paul2014_PRB}. If, instead of considering the coupling to the classical elastic deformations, we would consider the coupling to quantum phonons, we would obtain a frequency and momentum dependence of the renormalized nematic interaction \cite{Karahasanovic2015_PRB,Gallais2016_PRL}.

\noindent
One can also consider the effect of finite stress fields on the system. The free energy of the problem after having integrated out the elastic deformations would correspond to a linear coupling between the stress field and the nematic operator $\mathbf{N}$. The presence of strain can induce a non-zero expectation value of $N_1$ and $N_2$, irrespective of the electronic effects previously discussed. The crucial role played by stress in stabilizing the nematic state seems to be in agreement with recent experimental findings \cite{Guo2024_NatPhys} (see the SM for additional details regarding this calculation \cite{Suppl_mat}).

\noindent
To conclude, we have derived two microscopic theories to explain the nematicity and the nematic fluctuations observed in vanadium-based kagome metals, starting from a purely electronic model of phenomenological inspiration. In one case, the nematic state is described as a zero-momentum d-wave charge Pomeranchuk instability, while in the other case, it is stabilized by the anisotropic character of the fluctuations of the $1 \times 4$ charge order observed in several experiments. The two mechanisms might be relevant in different regions of the phase diagram \cite{Zheng2022_Nat}, and diffusive scattering techniques, which are sensitive to the spatial/momentum distribution of the fluctuations \cite{Chen2022_PRL}, might rule out the mechanism of nematicity driven by the $1 \times 4$ charge fluctuations. This way, the d-wave charge Pomeranchuk instability would remain the most likely candidate to explain the nematic character of the state. The correlation between nematicity and superconductivity observed in the kagome metals \cite{Sur2023_NatComm} suggests that charge fluctuations might play a role in forming the superconducting pairs \cite{Tian2024_arXiv}. This finding not only adds to our understanding of superconductivity but also highlights the complementary nature of the kagome metals to the iron pnictides, where superconductivity is believed to be related to spin fluctuations. From the methodological point of view, the spin- (charge-) fermion model that describes the interaction of the electrons at the Fermi surface with the spin (charge) fluctuations in the iron-based superconductors (kagome metals) cannot be derived at the random phase approximation level, but have to be supplied by the Aslamazov-Larkin diagrams. Moreover, while the iron-based superconductors likely show a spin-driven nematicity that breaks a $\mathbb{Z}_2$ Ising symmetry, the kagome metals likely show a charge-driven nematicity that breaks a $\mathbb{Z}_3$ Potts symmetry, see Tab.~\ref{table:nem}. A part from the one-dimensional charge-fluctuation scenario, we have identified another mechanism for the onset of nematicity in the kagome metals related to a symmetry breaking of the d-wave charge Pomeranchuk kind, which reminds the orbital-driven nematicity of the iron pnictides.

\begin{figure}
    \centerline{\includegraphics[width=0.5\textwidth]{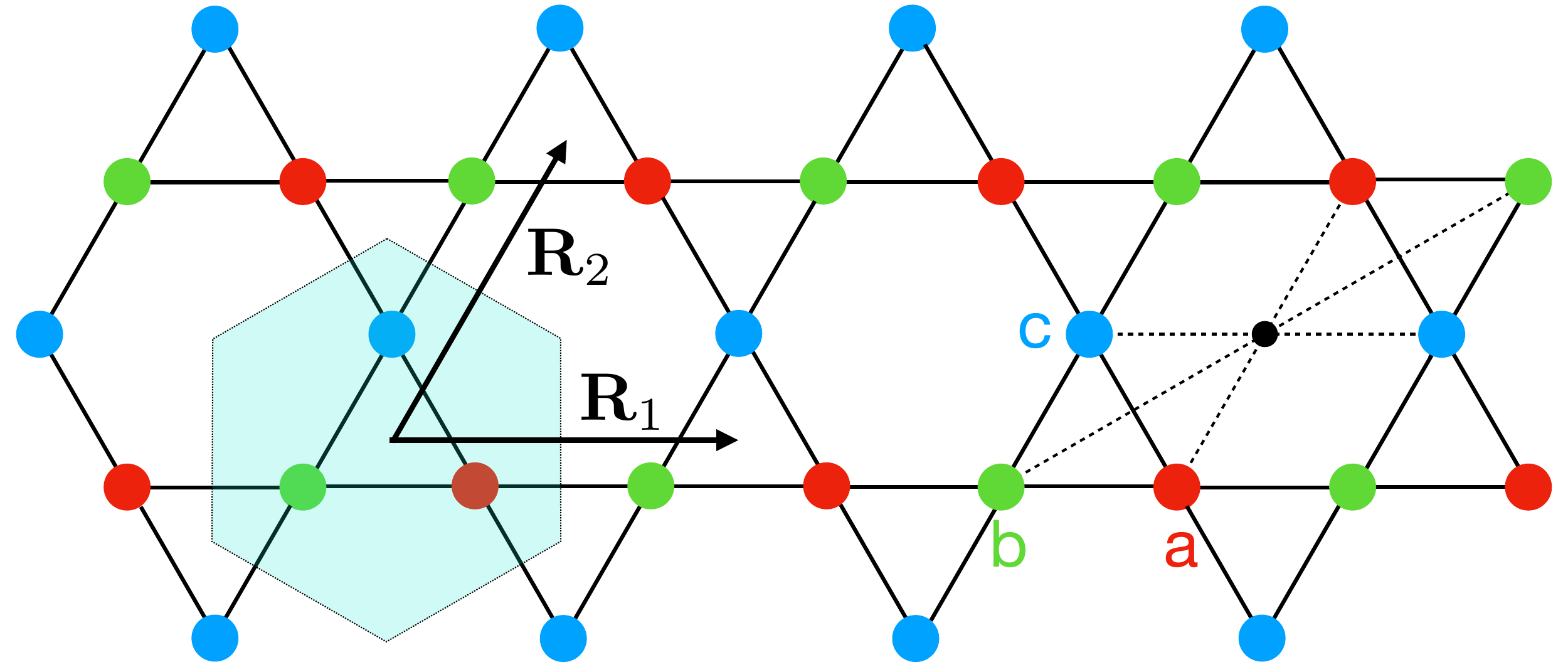}}
	\caption{ \textbf{d-wave charge Pomeranchuk instability in the titanium compounds -} In the titanium-based kagome metals, the d-wave charge Pomeranchuk instability corresponds to an intra-unit cell CO within the pristine unit cell of the kagome lattice which hosts three inequivalent atoms (named a, b and c in the figure). When the three sublattices become different one from the other, the system still preserves the inversion symmetry, as shown in the rightmost part of the image.} \label{fig:nematicity_titanium}
\end{figure}

\noindent
Despite the three family members of the vanadium-based kagome metals might be characterized by a similar effective noninteracting model once they enter the $2 \times 2$ charge-ordered state with three inequivalent hole pockets at the reconstructed M-points \cite{Li2023_PRX}, still the interactions might be different for them, leading to a different propensity towards the two instabilities we have identified.

\noindent
Finally, the proposed mechanism for the $Q_{1 a}$ nematicity suggested in this article might also be relevant for the class of compounds $A$Ti$_3$Sb$_5$ ($A=$ K, Rb, Cs) given the presence of (rhombic-like) hole-pockets at the Fermi level at the non reconstructed M points of the Brillouin zone. The corresponding real-space representation of the intra-unit cell charge order is schematically shown in Fig.~\ref{fig:nematicity_titanium}. This might provide an analogy between the nematic characters of the vanadium and titanium kagome families \cite{Hu2023_NatPhys,Yang2024_NatComm}.

\begin{acknowledgements}
F.G. acknowledges stimulating discussions with Mara Caltapanides, Armando Consiglio, Matteo D\"{u}rrnagel, Ammon Fisher, Hendrik Hohmann, Lennart Klebl, Giacomo Passetti and Jonas B. Profe on the content of this article and related subjects. Simulations were performed with computing resources granted by RWTH Aachen University under project rwth1230. 
F.G. and D.M.K. acknowledge support by the DFG via Germany’s Excellence Strategy$-$Cluster of Excellence Matter and Light for Quantum Computing (ML$4$Q, Project No. EXC $2004/1$, Grant No. $390534769$), within the RTG 1995 and within the Priority Program SPP 2244 ``2DMP''. MAS was funded by the European Union (ERC, CAVMAT, project no. 101124492). 
R.T. acknowledges support from the DFG through QUAST FOR $5249-449872909$ (Project P$3$), through Project-ID $258499086-$SFB $1170$, and from the W\"urzburg-Dresden Cluster of Excellence on Complexity and Topology in Quantum Matter$-$ct.qmat Project$-$ID $390858490-$EXC $2147$. 
\end{acknowledgements}


\begin{thebibliography}{180}%
\makeatletter
\providecommand \@ifxundefined [1]{%
 \@ifx{#1\undefined}
}%
\providecommand \@ifnum [1]{%
 \ifnum #1\expandafter \@firstoftwo
 \else \expandafter \@secondoftwo
 \fi
}%
\providecommand \@ifx [1]{%
 \ifx #1\expandafter \@firstoftwo
 \else \expandafter \@secondoftwo
 \fi
}%
\providecommand \natexlab [1]{#1}%
\providecommand \enquote  [1]{``#1''}%
\providecommand \bibnamefont  [1]{#1}%
\providecommand \bibfnamefont [1]{#1}%
\providecommand \citenamefont [1]{#1}%
\providecommand \href@noop [0]{\@secondoftwo}%
\providecommand \href [0]{\begingroup \@sanitize@url \@href}%
\providecommand \@href[1]{\@@startlink{#1}\@@href}%
\providecommand \@@href[1]{\endgroup#1\@@endlink}%
\providecommand \@sanitize@url [0]{\catcode `\\12\catcode `\$12\catcode
  `\&12\catcode `\#12\catcode `\^12\catcode `\_12\catcode `\%12\relax}%
\providecommand \@@startlink[1]{}%
\providecommand \@@endlink[0]{}%
\providecommand \url  [0]{\begingroup\@sanitize@url \@url }%
\providecommand \@url [1]{\endgroup\@href {#1}{\urlprefix }}%
\providecommand \urlprefix  [0]{URL }%
\providecommand \Eprint [0]{\href }%
\providecommand \doibase [0]{https://doi.org/}%
\providecommand \selectlanguage [0]{\@gobble}%
\providecommand \bibinfo  [0]{\@secondoftwo}%
\providecommand \bibfield  [0]{\@secondoftwo}%
\providecommand \translation [1]{[#1]}%
\providecommand \BibitemOpen [0]{}%
\providecommand \bibitemStop [0]{}%
\providecommand \bibitemNoStop [0]{.\EOS\space}%
\providecommand \EOS [0]{\spacefactor3000\relax}%
\providecommand \BibitemShut  [1]{\csname bibitem#1\endcsname}%
\let\auto@bib@innerbib\@empty
\bibitem [{\citenamefont {Stephen}\ and\ \citenamefont
  {Straley}(1974)}]{Stephen1974_RMP}%
  \BibitemOpen
  \bibfield  {author} {\bibinfo {author} {\bibfnamefont {M.~J.}\ \bibnamefont
  {Stephen}}\ and\ \bibinfo {author} {\bibfnamefont {J.~P.}\ \bibnamefont
  {Straley}},\ }\href {https://doi.org/10.1103/RevModPhys.46.617} {\bibfield
  {journal} {\bibinfo  {journal} {Rev. Mod. Phys.}\ }\textbf {\bibinfo {volume}
  {46}},\ \bibinfo {pages} {617} (\bibinfo {year} {1974})}\BibitemShut
  {NoStop}%
\bibitem [{\citenamefont {Blume}\ and\ \citenamefont
  {Hsieh}(1969)}]{Blume1969_JAP}%
  \BibitemOpen
  \bibfield  {author} {\bibinfo {author} {\bibfnamefont {M.}~\bibnamefont
  {Blume}}\ and\ \bibinfo {author} {\bibfnamefont {Y.~Y.}\ \bibnamefont
  {Hsieh}},\ }\href {https://doi.org/10.1063/1.1657616} {\bibfield  {journal}
  {\bibinfo  {journal} {Journal of Applied Physics}\ }\textbf {\bibinfo
  {volume} {40}},\ \bibinfo {pages} {1249} (\bibinfo {year}
  {1969})}\BibitemShut {NoStop}%
\bibitem [{\citenamefont {Andreev}\ and\ \citenamefont
  {Grishchuk}(1984)}]{Andreev1984_JETP}%
  \BibitemOpen
  \bibfield  {author} {\bibinfo {author} {\bibfnamefont {A.~F.}\ \bibnamefont
  {Andreev}}\ and\ \bibinfo {author} {\bibfnamefont {I.~A.}\ \bibnamefont
  {Grishchuk}},\ }\href@noop {} {\bibfield  {journal} {\bibinfo  {journal}
  {Sov. Phys. JETP}\ }\textbf {\bibinfo {volume} {60}},\ \bibinfo {pages} {267}
  (\bibinfo {year} {1984})}\BibitemShut {NoStop}%
\bibitem [{\citenamefont {Kohama}\ \emph {et~al.}(2019)\citenamefont {Kohama},
  \citenamefont {Ishikawa}, \citenamefont {Matsuo}, \citenamefont {Kindo},
  \citenamefont {Shannon},\ and\ \citenamefont {Hiroi}}]{Kohama2019_PNAS}%
  \BibitemOpen
  \bibfield  {author} {\bibinfo {author} {\bibfnamefont {Y.}~\bibnamefont
  {Kohama}}, \bibinfo {author} {\bibfnamefont {H.}~\bibnamefont {Ishikawa}},
  \bibinfo {author} {\bibfnamefont {A.}~\bibnamefont {Matsuo}}, \bibinfo
  {author} {\bibfnamefont {K.}~\bibnamefont {Kindo}}, \bibinfo {author}
  {\bibfnamefont {N.}~\bibnamefont {Shannon}},\ and\ \bibinfo {author}
  {\bibfnamefont {Z.}~\bibnamefont {Hiroi}},\ }\href
  {https://doi.org/10.1073/pnas.1821969116} {\bibfield  {journal} {\bibinfo
  {journal} {Proceedings of the National Academy of Sciences}\ }\textbf
  {\bibinfo {volume} {116}},\ \bibinfo {pages} {10686} (\bibinfo {year}
  {2019})}\BibitemShut {NoStop}%
\bibitem [{\citenamefont {Caci}\ \emph {et~al.}(2023)\citenamefont {Caci},
  \citenamefont {M\"uhlbacher}, \citenamefont {Ueltschi},\ and\ \citenamefont
  {Wessel}}]{Caci2023_PRB}%
  \BibitemOpen
  \bibfield  {author} {\bibinfo {author} {\bibfnamefont {N.}~\bibnamefont
  {Caci}}, \bibinfo {author} {\bibfnamefont {P.}~\bibnamefont {M\"uhlbacher}},
  \bibinfo {author} {\bibfnamefont {D.}~\bibnamefont {Ueltschi}},\ and\
  \bibinfo {author} {\bibfnamefont {S.}~\bibnamefont {Wessel}},\ }\href
  {https://doi.org/10.1103/PhysRevB.107.L020409} {\bibfield  {journal}
  {\bibinfo  {journal} {Phys. Rev. B}\ }\textbf {\bibinfo {volume} {107}},\
  \bibinfo {pages} {L020409} (\bibinfo {year} {2023})}\BibitemShut {NoStop}%
\bibitem [{\citenamefont {Abanov}\ \emph {et~al.}(1995)\citenamefont {Abanov},
  \citenamefont {Kalatsky}, \citenamefont {Pokrovsky},\ and\ \citenamefont
  {Saslow}}]{Abanov1995_PRB}%
  \BibitemOpen
  \bibfield  {author} {\bibinfo {author} {\bibfnamefont {A.}~\bibnamefont
  {Abanov}}, \bibinfo {author} {\bibfnamefont {V.}~\bibnamefont {Kalatsky}},
  \bibinfo {author} {\bibfnamefont {V.~L.}\ \bibnamefont {Pokrovsky}},\ and\
  \bibinfo {author} {\bibfnamefont {W.~M.}\ \bibnamefont {Saslow}},\ }\href
  {https://doi.org/10.1103/PhysRevB.51.1023} {\bibfield  {journal} {\bibinfo
  {journal} {Phys. Rev. B}\ }\textbf {\bibinfo {volume} {51}},\ \bibinfo
  {pages} {1023} (\bibinfo {year} {1995})}\BibitemShut {NoStop}%
\bibitem [{\citenamefont {Yamase}(2021)}]{Yamase2021_JPSJ}%
  \BibitemOpen
  \bibfield  {author} {\bibinfo {author} {\bibfnamefont {H.}~\bibnamefont
  {Yamase}},\ }\href {https://doi.org/10.7566/JPSJ.90.111011} {\bibfield
  {journal} {\bibinfo  {journal} {Journal of the Physical Society of Japan}\
  }\textbf {\bibinfo {volume} {90}},\ \bibinfo {pages} {111011} (\bibinfo
  {year} {2021})}\BibitemShut {NoStop}%
\bibitem [{\citenamefont {Fradkin}\ \emph {et~al.}(2015)\citenamefont
  {Fradkin}, \citenamefont {Kivelson},\ and\ \citenamefont
  {Tranquada}}]{Fradkin2015_RMP}%
  \BibitemOpen
  \bibfield  {author} {\bibinfo {author} {\bibfnamefont {E.}~\bibnamefont
  {Fradkin}}, \bibinfo {author} {\bibfnamefont {S.~A.}\ \bibnamefont
  {Kivelson}},\ and\ \bibinfo {author} {\bibfnamefont {J.~M.}\ \bibnamefont
  {Tranquada}},\ }\href {https://doi.org/10.1103/RevModPhys.87.457} {\bibfield
  {journal} {\bibinfo  {journal} {Rev. Mod. Phys.}\ }\textbf {\bibinfo {volume}
  {87}},\ \bibinfo {pages} {457} (\bibinfo {year} {2015})}\BibitemShut
  {NoStop}%
\bibitem [{\citenamefont {Svistunov}\ \emph {et~al.}(2015)\citenamefont
  {Svistunov}, \citenamefont {Babaev},\ and\ \citenamefont
  {Prokof\'ev}}]{Svistunov2015_SSM}%
  \BibitemOpen
  \bibfield  {author} {\bibinfo {author} {\bibfnamefont {B.~V.}\ \bibnamefont
  {Svistunov}}, \bibinfo {author} {\bibfnamefont {E.~S.}\ \bibnamefont
  {Babaev}},\ and\ \bibinfo {author} {\bibfnamefont {N.~V.}\ \bibnamefont
  {Prokof\'ev}},\ }\href {https://doi.org/10.1201/b18346} {\emph {\bibinfo
  {title} {{Superfluid States of Matter}}}},\ \bibinfo {edition} {1st}\ ed.\
  (\bibinfo  {publisher} {Taylor \& Francis Group},\ \bibinfo {year}
  {2015})\BibitemShut {NoStop}%
\bibitem [{\citenamefont {Fernandes}\ \emph {et~al.}(2019)\citenamefont
  {Fernandes}, \citenamefont {Orth},\ and\ \citenamefont
  {Schmalian}}]{Fernandes2019_ARCMP}%
  \BibitemOpen
  \bibfield  {author} {\bibinfo {author} {\bibfnamefont {R.~M.}\ \bibnamefont
  {Fernandes}}, \bibinfo {author} {\bibfnamefont {P.~P.}\ \bibnamefont
  {Orth}},\ and\ \bibinfo {author} {\bibfnamefont {J.}~\bibnamefont
  {Schmalian}},\ }\href
  {https://doi.org/10.1146/annurev-conmatphys-031218-013200} {\bibfield
  {journal} {\bibinfo  {journal} {Annual Review of Condensed Matter Physics}\
  }\textbf {\bibinfo {volume} {10}},\ \bibinfo {pages} {133} (\bibinfo {year}
  {2019})}\BibitemShut {NoStop}%
\bibitem [{\citenamefont {Pomeranchuk}(1958)}]{Pomeranchuk1958_JETP}%
  \BibitemOpen
  \bibfield  {author} {\bibinfo {author} {\bibfnamefont {I.}~\bibnamefont
  {Pomeranchuk}},\ }\href {http://jetp.ras.ru/cgi-bin/dn/e_008_02_0361.pdf}
  {\bibfield  {journal} {\bibinfo  {journal} {Sov. Phys. JETP}\ }\textbf
  {\bibinfo {volume} {8}} (\bibinfo {year} {1958})}\BibitemShut {NoStop}%
\bibitem [{\citenamefont {Quintanilla}\ and\ \citenamefont
  {Schofield}(2006)}]{Quintanilla2006_PRB}%
  \BibitemOpen
  \bibfield  {author} {\bibinfo {author} {\bibfnamefont {J.}~\bibnamefont
  {Quintanilla}}\ and\ \bibinfo {author} {\bibfnamefont {A.~J.}\ \bibnamefont
  {Schofield}},\ }\href {https://doi.org/10.1103/PhysRevB.74.115126} {\bibfield
   {journal} {\bibinfo  {journal} {Phys. Rev. B}\ }\textbf {\bibinfo {volume}
  {74}},\ \bibinfo {pages} {115126} (\bibinfo {year} {2006})}\BibitemShut
  {NoStop}%
\bibitem [{\citenamefont {Chubukov}\ \emph {et~al.}(2018)\citenamefont
  {Chubukov}, \citenamefont {Klein},\ and\ \citenamefont
  {Maslov}}]{Chubukov2018_JETP}%
  \BibitemOpen
  \bibfield  {author} {\bibinfo {author} {\bibfnamefont {A.~V.}\ \bibnamefont
  {Chubukov}}, \bibinfo {author} {\bibfnamefont {A.}~\bibnamefont {Klein}},\
  and\ \bibinfo {author} {\bibfnamefont {D.~L.}\ \bibnamefont {Maslov}},\
  }\href {https://doi.org/10.1134/S1063776118110122} {\bibfield  {journal}
  {\bibinfo  {journal} {Journal of Experimental and Theoretical Physics}\
  }\textbf {\bibinfo {volume} {127}},\ \bibinfo {pages} {826} (\bibinfo {year}
  {2018})}\BibitemShut {NoStop}%
\bibitem [{\citenamefont {Stoner}(1938)}]{Stoner1938_PRSL}%
  \BibitemOpen
  \bibfield  {author} {\bibinfo {author} {\bibfnamefont {E.~C.}\ \bibnamefont
  {Stoner}},\ }\href {https://doi.org/10.1098/rspa.1938.0066} {\bibfield
  {journal} {\bibinfo  {journal} {Proceedings of the Royal Society of London.
  Series A. Mathematical and Physical Sciences}\ }\textbf {\bibinfo {volume}
  {165}},\ \bibinfo {pages} {372} (\bibinfo {year} {1938})}\BibitemShut
  {NoStop}%
\bibitem [{\citenamefont {Wu}\ \emph {et~al.}(2018)\citenamefont {Wu},
  \citenamefont {Klein},\ and\ \citenamefont {Chubukov}}]{Wu2018_PRB}%
  \BibitemOpen
  \bibfield  {author} {\bibinfo {author} {\bibfnamefont {Y.-M.}\ \bibnamefont
  {Wu}}, \bibinfo {author} {\bibfnamefont {A.}~\bibnamefont {Klein}},\ and\
  \bibinfo {author} {\bibfnamefont {A.~V.}\ \bibnamefont {Chubukov}},\ }\href
  {https://doi.org/10.1103/PhysRevB.97.165101} {\bibfield  {journal} {\bibinfo
  {journal} {Phys. Rev. B}\ }\textbf {\bibinfo {volume} {97}},\ \bibinfo
  {pages} {165101} (\bibinfo {year} {2018})}\BibitemShut {NoStop}%
\bibitem [{\citenamefont {Hirsch}(1990)}]{Hirsch1990_PRB}%
  \BibitemOpen
  \bibfield  {author} {\bibinfo {author} {\bibfnamefont {J.~E.}\ \bibnamefont
  {Hirsch}},\ }\href {https://doi.org/10.1103/PhysRevB.41.6820} {\bibfield
  {journal} {\bibinfo  {journal} {Phys. Rev. B}\ }\textbf {\bibinfo {volume}
  {41}},\ \bibinfo {pages} {6820} (\bibinfo {year} {1990})}\BibitemShut
  {NoStop}%
\bibitem [{\citenamefont {Halboth}\ and\ \citenamefont
  {Metzner}(2000)}]{Halboth2000_PRL}%
  \BibitemOpen
  \bibfield  {author} {\bibinfo {author} {\bibfnamefont {C.~J.}\ \bibnamefont
  {Halboth}}\ and\ \bibinfo {author} {\bibfnamefont {W.}~\bibnamefont
  {Metzner}},\ }\href {https://doi.org/10.1103/PhysRevLett.85.5162} {\bibfield
  {journal} {\bibinfo  {journal} {Phys. Rev. Lett.}\ }\textbf {\bibinfo
  {volume} {85}},\ \bibinfo {pages} {5162} (\bibinfo {year}
  {2000})}\BibitemShut {NoStop}%
\bibitem [{\citenamefont {Yamase}\ and\ \citenamefont
  {Kohno}(2000)}]{Yamase2000_JPSJ}%
  \BibitemOpen
  \bibfield  {author} {\bibinfo {author} {\bibfnamefont {H.}~\bibnamefont
  {Yamase}}\ and\ \bibinfo {author} {\bibfnamefont {H.}~\bibnamefont {Kohno}},\
  }\href {https://doi.org/10.1143/JPSJ.69.332} {\bibfield  {journal} {\bibinfo
  {journal} {Journal of the Physical Society of Japan}\ }\textbf {\bibinfo
  {volume} {69}},\ \bibinfo {pages} {332} (\bibinfo {year} {2000})}\BibitemShut
  {NoStop}%
\bibitem [{\citenamefont {Zacharias}\ \emph {et~al.}(2009)\citenamefont
  {Zacharias}, \citenamefont {W\"olfle},\ and\ \citenamefont
  {Garst}}]{Zacharias2009_PRB}%
  \BibitemOpen
  \bibfield  {author} {\bibinfo {author} {\bibfnamefont {M.}~\bibnamefont
  {Zacharias}}, \bibinfo {author} {\bibfnamefont {P.}~\bibnamefont
  {W\"olfle}},\ and\ \bibinfo {author} {\bibfnamefont {M.}~\bibnamefont
  {Garst}},\ }\href {https://doi.org/10.1103/PhysRevB.80.165116} {\bibfield
  {journal} {\bibinfo  {journal} {Phys. Rev. B}\ }\textbf {\bibinfo {volume}
  {80}},\ \bibinfo {pages} {165116} (\bibinfo {year} {2009})}\BibitemShut
  {NoStop}%
\bibitem [{\citenamefont {Oganesyan}\ \emph {et~al.}(2001)\citenamefont
  {Oganesyan}, \citenamefont {Kivelson},\ and\ \citenamefont
  {Fradkin}}]{Oganesyan2001_PRB}%
  \BibitemOpen
  \bibfield  {author} {\bibinfo {author} {\bibfnamefont {V.}~\bibnamefont
  {Oganesyan}}, \bibinfo {author} {\bibfnamefont {S.~A.}\ \bibnamefont
  {Kivelson}},\ and\ \bibinfo {author} {\bibfnamefont {E.}~\bibnamefont
  {Fradkin}},\ }\href {https://doi.org/10.1103/PhysRevB.64.195109} {\bibfield
  {journal} {\bibinfo  {journal} {Phys. Rev. B}\ }\textbf {\bibinfo {volume}
  {64}},\ \bibinfo {pages} {195109} (\bibinfo {year} {2001})}\BibitemShut
  {NoStop}%
\bibitem [{\citenamefont {Valenzuela}\ and\ \citenamefont
  {Vozmediano}(2008)}]{Valenzuela2008_NJP}%
  \BibitemOpen
  \bibfield  {author} {\bibinfo {author} {\bibfnamefont {B.}~\bibnamefont
  {Valenzuela}}\ and\ \bibinfo {author} {\bibfnamefont {M.~A.~H.}\ \bibnamefont
  {Vozmediano}},\ }\href {https://doi.org/10.1088/1367-2630/10/11/113009}
  {\bibfield  {journal} {\bibinfo  {journal} {New Journal of Physics}\ }\textbf
  {\bibinfo {volume} {10}},\ \bibinfo {pages} {113009} (\bibinfo {year}
  {2008})}\BibitemShut {NoStop}%
\bibitem [{\citenamefont {Maslov}\ and\ \citenamefont
  {Chubukov}(2010)}]{Maslov2010_PRB}%
  \BibitemOpen
  \bibfield  {author} {\bibinfo {author} {\bibfnamefont {D.~L.}\ \bibnamefont
  {Maslov}}\ and\ \bibinfo {author} {\bibfnamefont {A.~V.}\ \bibnamefont
  {Chubukov}},\ }\href {https://doi.org/10.1103/PhysRevB.81.045110} {\bibfield
  {journal} {\bibinfo  {journal} {Phys. Rev. B}\ }\textbf {\bibinfo {volume}
  {81}},\ \bibinfo {pages} {045110} (\bibinfo {year} {2010})}\BibitemShut
  {NoStop}%
\bibitem [{Note1()}]{Note1}%
  \BibitemOpen
  \bibinfo {note} {Here, we speak about ``low-temperature phase'' to keep the
  analogy with the phenomenology of the liquid crystals described before, but,
  in general condensed matter systems, the smectic state might be stabilized at
  higher or lower pressure or doping.}\BibitemShut {Stop}%
\bibitem [{\citenamefont {Fernandes}\ \emph
  {et~al.}(2012{\natexlab{a}})\citenamefont {Fernandes}, \citenamefont
  {Chubukov}, \citenamefont {Knolle}, \citenamefont {Eremin},\ and\
  \citenamefont {Schmalian}}]{Fernandes2012_PRB}%
  \BibitemOpen
  \bibfield  {author} {\bibinfo {author} {\bibfnamefont {R.~M.}\ \bibnamefont
  {Fernandes}}, \bibinfo {author} {\bibfnamefont {A.~V.}\ \bibnamefont
  {Chubukov}}, \bibinfo {author} {\bibfnamefont {J.}~\bibnamefont {Knolle}},
  \bibinfo {author} {\bibfnamefont {I.}~\bibnamefont {Eremin}},\ and\ \bibinfo
  {author} {\bibfnamefont {J.}~\bibnamefont {Schmalian}},\ }\href
  {https://doi.org/10.1103/PhysRevB.85.024534} {\bibfield  {journal} {\bibinfo
  {journal} {Phys. Rev. B}\ }\textbf {\bibinfo {volume} {85}},\ \bibinfo
  {pages} {024534} (\bibinfo {year} {2012}{\natexlab{a}})}\BibitemShut
  {NoStop}%
\bibitem [{\citenamefont {Hertz}(1976)}]{Hertz1976_PRB}%
  \BibitemOpen
  \bibfield  {author} {\bibinfo {author} {\bibfnamefont {J.~A.}\ \bibnamefont
  {Hertz}},\ }\href {https://doi.org/10.1103/PhysRevB.14.1165} {\bibfield
  {journal} {\bibinfo  {journal} {Phys. Rev. B}\ }\textbf {\bibinfo {volume}
  {14}},\ \bibinfo {pages} {1165} (\bibinfo {year} {1976})}\BibitemShut
  {NoStop}%
\bibitem [{\citenamefont {Millis}(1993)}]{Millis1993_PRB}%
  \BibitemOpen
  \bibfield  {author} {\bibinfo {author} {\bibfnamefont {A.~J.}\ \bibnamefont
  {Millis}},\ }\href {https://doi.org/10.1103/PhysRevB.48.7183} {\bibfield
  {journal} {\bibinfo  {journal} {Phys. Rev. B}\ }\textbf {\bibinfo {volume}
  {48}},\ \bibinfo {pages} {7183} (\bibinfo {year} {1993})}\BibitemShut
  {NoStop}%
\bibitem [{\citenamefont {L\"ohneysen}\ \emph {et~al.}(2007)\citenamefont
  {L\"ohneysen}, \citenamefont {Rosch}, \citenamefont {Vojta},\ and\
  \citenamefont {W\"olfle}}]{Loehneysen2007_RMP}%
  \BibitemOpen
  \bibfield  {author} {\bibinfo {author} {\bibfnamefont {H.~v.}\ \bibnamefont
  {L\"ohneysen}}, \bibinfo {author} {\bibfnamefont {A.}~\bibnamefont {Rosch}},
  \bibinfo {author} {\bibfnamefont {M.}~\bibnamefont {Vojta}},\ and\ \bibinfo
  {author} {\bibfnamefont {P.}~\bibnamefont {W\"olfle}},\ }\href
  {https://doi.org/10.1103/RevModPhys.79.1015} {\bibfield  {journal} {\bibinfo
  {journal} {Rev. Mod. Phys.}\ }\textbf {\bibinfo {volume} {79}},\ \bibinfo
  {pages} {1015} (\bibinfo {year} {2007})}\BibitemShut {NoStop}%
\bibitem [{\citenamefont {Sachdev}(2011)}]{Sachdev2011_CUP}%
  \BibitemOpen
  \bibfield  {author} {\bibinfo {author} {\bibfnamefont {S.}~\bibnamefont
  {Sachdev}},\ }\href@noop {} {\emph {\bibinfo {title} {{Quantum Phase
  Transitions}}}},\ \bibinfo {edition} {2nd}\ ed.\ (\bibinfo  {publisher}
  {Cambridge University Press},\ \bibinfo {year} {2011})\BibitemShut {NoStop}%
\bibitem [{\citenamefont {Efetov}\ \emph {et~al.}(2013)\citenamefont {Efetov},
  \citenamefont {Meier},\ and\ \citenamefont {P\'{e}pin}}]{Efetov2013_NatPhys}%
  \BibitemOpen
  \bibfield  {author} {\bibinfo {author} {\bibfnamefont {K.~B.}\ \bibnamefont
  {Efetov}}, \bibinfo {author} {\bibfnamefont {H.}~\bibnamefont {Meier}},\ and\
  \bibinfo {author} {\bibfnamefont {C.}~\bibnamefont {P\'{e}pin}},\ }\href
  {https://doi.org/10.1038/nphys2641} {\bibfield  {journal} {\bibinfo
  {journal} {Nature Physics}\ }\textbf {\bibinfo {volume} {9}},\ \bibinfo
  {pages} {442} (\bibinfo {year} {2013})}\BibitemShut {NoStop}%
\bibitem [{\citenamefont {Licciardello}\ \emph {et~al.}(2019)\citenamefont
  {Licciardello}, \citenamefont {Buhot}, \citenamefont {Lu}, \citenamefont
  {Ayres}, \citenamefont {Kasahara}, \citenamefont {Matsuda}, \citenamefont
  {Shibauchi},\ and\ \citenamefont {Hussey}}]{Licciardello2019_Nat}%
  \BibitemOpen
  \bibfield  {author} {\bibinfo {author} {\bibfnamefont {S.}~\bibnamefont
  {Licciardello}}, \bibinfo {author} {\bibfnamefont {J.}~\bibnamefont {Buhot}},
  \bibinfo {author} {\bibfnamefont {J.}~\bibnamefont {Lu}}, \bibinfo {author}
  {\bibfnamefont {J.}~\bibnamefont {Ayres}}, \bibinfo {author} {\bibfnamefont
  {S.}~\bibnamefont {Kasahara}}, \bibinfo {author} {\bibfnamefont
  {Y.}~\bibnamefont {Matsuda}}, \bibinfo {author} {\bibfnamefont
  {T.}~\bibnamefont {Shibauchi}},\ and\ \bibinfo {author} {\bibfnamefont
  {N.~E.}\ \bibnamefont {Hussey}},\ }\href
  {https://doi.org/10.1038/s41586-019-0923-y} {\bibfield  {journal} {\bibinfo
  {journal} {Nature}\ }\textbf {\bibinfo {volume} {567}},\ \bibinfo {pages}
  {213} (\bibinfo {year} {2019})}\BibitemShut {NoStop}%
\bibitem [{\citenamefont {Lederer}\ \emph {et~al.}(2015)\citenamefont
  {Lederer}, \citenamefont {Schattner}, \citenamefont {Berg},\ and\
  \citenamefont {Kivelson}}]{Lederer2015_PRL}%
  \BibitemOpen
  \bibfield  {author} {\bibinfo {author} {\bibfnamefont {S.}~\bibnamefont
  {Lederer}}, \bibinfo {author} {\bibfnamefont {Y.}~\bibnamefont {Schattner}},
  \bibinfo {author} {\bibfnamefont {E.}~\bibnamefont {Berg}},\ and\ \bibinfo
  {author} {\bibfnamefont {S.~A.}\ \bibnamefont {Kivelson}},\ }\href
  {https://doi.org/10.1103/PhysRevLett.114.097001} {\bibfield  {journal}
  {\bibinfo  {journal} {Phys. Rev. Lett.}\ }\textbf {\bibinfo {volume} {114}},\
  \bibinfo {pages} {097001} (\bibinfo {year} {2015})}\BibitemShut {NoStop}%
\bibitem [{\citenamefont {Lederer}\ \emph {et~al.}(2017)\citenamefont
  {Lederer}, \citenamefont {Schattner}, \citenamefont {Berg},\ and\
  \citenamefont {Kivelson}}]{Lederer2017_PNAS}%
  \BibitemOpen
  \bibfield  {author} {\bibinfo {author} {\bibfnamefont {S.}~\bibnamefont
  {Lederer}}, \bibinfo {author} {\bibfnamefont {Y.}~\bibnamefont {Schattner}},
  \bibinfo {author} {\bibfnamefont {E.}~\bibnamefont {Berg}},\ and\ \bibinfo
  {author} {\bibfnamefont {S.~A.}\ \bibnamefont {Kivelson}},\ }\href
  {https://doi.org/10.1073/pnas.1620651114} {\bibfield  {journal} {\bibinfo
  {journal} {Proceedings of the National Academy of Sciences}\ }\textbf
  {\bibinfo {volume} {114}},\ \bibinfo {pages} {4905} (\bibinfo {year}
  {2017})}\BibitemShut {NoStop}%
\bibitem [{\citenamefont {Ishida}\ \emph {et~al.}(2022)\citenamefont {Ishida},
  \citenamefont {Onishi}, \citenamefont {Tsujii}, \citenamefont {Mukasa},
  \citenamefont {Qiu}, \citenamefont {Saito}, \citenamefont {Sugimura},
  \citenamefont {Matsuura}, \citenamefont {Mizukami}, \citenamefont
  {Hashimoto},\ and\ \citenamefont {Shibauchi}}]{Ishida2022_PNAS}%
  \BibitemOpen
  \bibfield  {author} {\bibinfo {author} {\bibfnamefont {K.}~\bibnamefont
  {Ishida}}, \bibinfo {author} {\bibfnamefont {Y.}~\bibnamefont {Onishi}},
  \bibinfo {author} {\bibfnamefont {M.}~\bibnamefont {Tsujii}}, \bibinfo
  {author} {\bibfnamefont {K.}~\bibnamefont {Mukasa}}, \bibinfo {author}
  {\bibfnamefont {M.}~\bibnamefont {Qiu}}, \bibinfo {author} {\bibfnamefont
  {M.}~\bibnamefont {Saito}}, \bibinfo {author} {\bibfnamefont
  {Y.}~\bibnamefont {Sugimura}}, \bibinfo {author} {\bibfnamefont
  {K.}~\bibnamefont {Matsuura}}, \bibinfo {author} {\bibfnamefont
  {Y.}~\bibnamefont {Mizukami}}, \bibinfo {author} {\bibfnamefont
  {K.}~\bibnamefont {Hashimoto}},\ and\ \bibinfo {author} {\bibfnamefont
  {T.}~\bibnamefont {Shibauchi}},\ }\href
  {https://doi.org/10.1073/pnas.2110501119} {\bibfield  {journal} {\bibinfo
  {journal} {Proceedings of the National Academy of Sciences}\ }\textbf
  {\bibinfo {volume} {119}},\ \bibinfo {pages} {e2110501119} (\bibinfo {year}
  {2022})}\BibitemShut {NoStop}%
\bibitem [{\citenamefont {Sur}\ \emph {et~al.}(2023)\citenamefont {Sur},
  \citenamefont {Kim}, \citenamefont {Kim},\ and\ \citenamefont
  {Kim}}]{Sur2023_NatComm}%
  \BibitemOpen
  \bibfield  {author} {\bibinfo {author} {\bibfnamefont {Y.}~\bibnamefont
  {Sur}}, \bibinfo {author} {\bibfnamefont {K.-T.}\ \bibnamefont {Kim}},
  \bibinfo {author} {\bibfnamefont {S.}~\bibnamefont {Kim}},\ and\ \bibinfo
  {author} {\bibfnamefont {K.~H.}\ \bibnamefont {Kim}},\ }\href
  {https://doi.org/10.1038/s41467-023-39495-1} {\bibfield  {journal} {\bibinfo
  {journal} {Nature Communications}\ }\textbf {\bibinfo {volume} {14}},\
  \bibinfo {pages} {3899} (\bibinfo {year} {2023})}\BibitemShut {NoStop}%
\bibitem [{\citenamefont {Chu}\ \emph {et~al.}(2012)\citenamefont {Chu},
  \citenamefont {Kuo}, \citenamefont {Analytis},\ and\ \citenamefont
  {Fisher}}]{Chu2012_Sci}%
  \BibitemOpen
  \bibfield  {author} {\bibinfo {author} {\bibfnamefont {J.-H.}\ \bibnamefont
  {Chu}}, \bibinfo {author} {\bibfnamefont {H.-H.}\ \bibnamefont {Kuo}},
  \bibinfo {author} {\bibfnamefont {J.~G.}\ \bibnamefont {Analytis}},\ and\
  \bibinfo {author} {\bibfnamefont {I.~R.}\ \bibnamefont {Fisher}},\ }\href
  {https://doi.org/10.1126/science.1221713} {\bibfield  {journal} {\bibinfo
  {journal} {Science}\ }\textbf {\bibinfo {volume} {337}},\ \bibinfo {pages}
  {710} (\bibinfo {year} {2012})}\BibitemShut {NoStop}%
\bibitem [{\citenamefont {Hosoi}\ \emph {et~al.}(2016)\citenamefont {Hosoi},
  \citenamefont {Matsuura}, \citenamefont {Ishida}, \citenamefont {Wang},
  \citenamefont {Mizukami}, \citenamefont {Watashige}, \citenamefont
  {Kasahara}, \citenamefont {Matsuda},\ and\ \citenamefont
  {Shibauchi}}]{Hosoi2016_PNAS}%
  \BibitemOpen
  \bibfield  {author} {\bibinfo {author} {\bibfnamefont {S.}~\bibnamefont
  {Hosoi}}, \bibinfo {author} {\bibfnamefont {K.}~\bibnamefont {Matsuura}},
  \bibinfo {author} {\bibfnamefont {K.}~\bibnamefont {Ishida}}, \bibinfo
  {author} {\bibfnamefont {H.}~\bibnamefont {Wang}}, \bibinfo {author}
  {\bibfnamefont {Y.}~\bibnamefont {Mizukami}}, \bibinfo {author}
  {\bibfnamefont {T.}~\bibnamefont {Watashige}}, \bibinfo {author}
  {\bibfnamefont {S.}~\bibnamefont {Kasahara}}, \bibinfo {author}
  {\bibfnamefont {Y.}~\bibnamefont {Matsuda}},\ and\ \bibinfo {author}
  {\bibfnamefont {T.}~\bibnamefont {Shibauchi}},\ }\href
  {https://doi.org/10.1073/pnas.1605806113} {\bibfield  {journal} {\bibinfo
  {journal} {Proceedings of the National Academy of Sciences}\ }\textbf
  {\bibinfo {volume} {113}},\ \bibinfo {pages} {8139} (\bibinfo {year}
  {2016})}\BibitemShut {NoStop}%
\bibitem [{\citenamefont {Auvray}\ \emph {et~al.}(2019)\citenamefont {Auvray},
  \citenamefont {Loret}, \citenamefont {Benhabib}, \citenamefont {Cazayous},
  \citenamefont {Zhong}, \citenamefont {Schneeloch}, \citenamefont {Gu},
  \citenamefont {Forget}, \citenamefont {Colson}, \citenamefont {Paul},
  \citenamefont {Sacuto},\ and\ \citenamefont {Gallais}}]{Auvray2019_NatComm}%
  \BibitemOpen
  \bibfield  {author} {\bibinfo {author} {\bibfnamefont {N.}~\bibnamefont
  {Auvray}}, \bibinfo {author} {\bibfnamefont {B.}~\bibnamefont {Loret}},
  \bibinfo {author} {\bibfnamefont {S.}~\bibnamefont {Benhabib}}, \bibinfo
  {author} {\bibfnamefont {M.}~\bibnamefont {Cazayous}}, \bibinfo {author}
  {\bibfnamefont {R.~D.}\ \bibnamefont {Zhong}}, \bibinfo {author}
  {\bibfnamefont {J.}~\bibnamefont {Schneeloch}}, \bibinfo {author}
  {\bibfnamefont {G.~D.}\ \bibnamefont {Gu}}, \bibinfo {author} {\bibfnamefont
  {A.}~\bibnamefont {Forget}}, \bibinfo {author} {\bibfnamefont
  {D.}~\bibnamefont {Colson}}, \bibinfo {author} {\bibfnamefont
  {I.}~\bibnamefont {Paul}}, \bibinfo {author} {\bibfnamefont {A.}~\bibnamefont
  {Sacuto}},\ and\ \bibinfo {author} {\bibfnamefont {Y.}~\bibnamefont
  {Gallais}},\ }\href {https://doi.org/10.1038/s41467-019-12940-w} {\bibfield
  {journal} {\bibinfo  {journal} {Nature Communications}\ }\textbf {\bibinfo
  {volume} {10}},\ \bibinfo {pages} {5209} (\bibinfo {year}
  {2019})}\BibitemShut {NoStop}%
\bibitem [{\citenamefont {B\"{o}hmer}\ \emph {et~al.}(2022)\citenamefont
  {B\"{o}hmer}, \citenamefont {Chu}, \citenamefont {Lederer},\ and\
  \citenamefont {Yi}}]{Boehmer2022_NatPhys}%
  \BibitemOpen
  \bibfield  {author} {\bibinfo {author} {\bibfnamefont {A.~E.}\ \bibnamefont
  {B\"{o}hmer}}, \bibinfo {author} {\bibfnamefont {J.-H.}\ \bibnamefont {Chu}},
  \bibinfo {author} {\bibfnamefont {S.}~\bibnamefont {Lederer}},\ and\ \bibinfo
  {author} {\bibfnamefont {M.}~\bibnamefont {Yi}},\ }\href
  {https://doi.org/10.1038/s41567-022-01833-3} {\bibfield  {journal} {\bibinfo
  {journal} {Nature Physics}\ }\textbf {\bibinfo {volume} {18}},\ \bibinfo
  {pages} {1412} (\bibinfo {year} {2022})}\BibitemShut {NoStop}%
\bibitem [{\citenamefont {Fernandes}\ \emph {et~al.}(2022)\citenamefont
  {Fernandes}, \citenamefont {Coldea}, \citenamefont {Ding}, \citenamefont
  {Fisher}, \citenamefont {Hirschfeld},\ and\ \citenamefont
  {Kotliar}}]{Fernandes2022_Nat}%
  \BibitemOpen
  \bibfield  {author} {\bibinfo {author} {\bibfnamefont {R.~M.}\ \bibnamefont
  {Fernandes}}, \bibinfo {author} {\bibfnamefont {A.~I.}\ \bibnamefont
  {Coldea}}, \bibinfo {author} {\bibfnamefont {H.}~\bibnamefont {Ding}},
  \bibinfo {author} {\bibfnamefont {I.~R.}\ \bibnamefont {Fisher}}, \bibinfo
  {author} {\bibfnamefont {P.~J.}\ \bibnamefont {Hirschfeld}},\ and\ \bibinfo
  {author} {\bibfnamefont {G.}~\bibnamefont {Kotliar}},\ }\href
  {https://doi.org/10.1038/s41586-021-04073-2} {\bibfield  {journal} {\bibinfo
  {journal} {Nature}\ }\textbf {\bibinfo {volume} {601}},\ \bibinfo {pages}
  {35} (\bibinfo {year} {2022})}\BibitemShut {NoStop}%
\bibitem [{\citenamefont {Yi}\ \emph {et~al.}(2011)\citenamefont {Yi},
  \citenamefont {Lu}, \citenamefont {Chu}, \citenamefont {Analytis},
  \citenamefont {Sorini}, \citenamefont {Kemper}, \citenamefont {Moritz},
  \citenamefont {Mo}, \citenamefont {Moore}, \citenamefont {Hashimoto},
  \citenamefont {Lee}, \citenamefont {Hussain}, \citenamefont {Devereaux},
  \citenamefont {Fisher},\ and\ \citenamefont {Shen}}]{Ming2011_PNAS}%
  \BibitemOpen
  \bibfield  {author} {\bibinfo {author} {\bibfnamefont {M.}~\bibnamefont
  {Yi}}, \bibinfo {author} {\bibfnamefont {D.}~\bibnamefont {Lu}}, \bibinfo
  {author} {\bibfnamefont {J.-H.}\ \bibnamefont {Chu}}, \bibinfo {author}
  {\bibfnamefont {J.~G.}\ \bibnamefont {Analytis}}, \bibinfo {author}
  {\bibfnamefont {A.~P.}\ \bibnamefont {Sorini}}, \bibinfo {author}
  {\bibfnamefont {A.~F.}\ \bibnamefont {Kemper}}, \bibinfo {author}
  {\bibfnamefont {B.}~\bibnamefont {Moritz}}, \bibinfo {author} {\bibfnamefont
  {S.-K.}\ \bibnamefont {Mo}}, \bibinfo {author} {\bibfnamefont {R.~G.}\
  \bibnamefont {Moore}}, \bibinfo {author} {\bibfnamefont {M.}~\bibnamefont
  {Hashimoto}}, \bibinfo {author} {\bibfnamefont {W.-S.}\ \bibnamefont {Lee}},
  \bibinfo {author} {\bibfnamefont {Z.}~\bibnamefont {Hussain}}, \bibinfo
  {author} {\bibfnamefont {T.~P.}\ \bibnamefont {Devereaux}}, \bibinfo {author}
  {\bibfnamefont {I.~R.}\ \bibnamefont {Fisher}},\ and\ \bibinfo {author}
  {\bibfnamefont {Z.-X.}\ \bibnamefont {Shen}},\ }\href
  {https://doi.org/10.1073/pnas.1015572108} {\bibfield  {journal} {\bibinfo
  {journal} {Proceedings of the National Academy of Sciences}\ }\textbf
  {\bibinfo {volume} {108}},\ \bibinfo {pages} {6878} (\bibinfo {year}
  {2011})}\BibitemShut {NoStop}%
\bibitem [{\citenamefont {Fang}\ \emph {et~al.}(2008)\citenamefont {Fang},
  \citenamefont {Yao}, \citenamefont {Tsai}, \citenamefont {Hu},\ and\
  \citenamefont {Kivelson}}]{Fang2008_PRB}%
  \BibitemOpen
  \bibfield  {author} {\bibinfo {author} {\bibfnamefont {C.}~\bibnamefont
  {Fang}}, \bibinfo {author} {\bibfnamefont {H.}~\bibnamefont {Yao}}, \bibinfo
  {author} {\bibfnamefont {W.-F.}\ \bibnamefont {Tsai}}, \bibinfo {author}
  {\bibfnamefont {J.}~\bibnamefont {Hu}},\ and\ \bibinfo {author}
  {\bibfnamefont {S.~A.}\ \bibnamefont {Kivelson}},\ }\href
  {https://doi.org/10.1103/PhysRevB.77.224509} {\bibfield  {journal} {\bibinfo
  {journal} {Phys. Rev. B}\ }\textbf {\bibinfo {volume} {77}},\ \bibinfo
  {pages} {224509} (\bibinfo {year} {2008})}\BibitemShut {NoStop}%
\bibitem [{\citenamefont {Xu}\ \emph {et~al.}(2008)\citenamefont {Xu},
  \citenamefont {M\"uller},\ and\ \citenamefont {Sachdev}}]{Xu2008_PRB}%
  \BibitemOpen
  \bibfield  {author} {\bibinfo {author} {\bibfnamefont {C.}~\bibnamefont
  {Xu}}, \bibinfo {author} {\bibfnamefont {M.}~\bibnamefont {M\"uller}},\ and\
  \bibinfo {author} {\bibfnamefont {S.}~\bibnamefont {Sachdev}},\ }\href
  {https://doi.org/10.1103/PhysRevB.78.020501} {\bibfield  {journal} {\bibinfo
  {journal} {Phys. Rev. B}\ }\textbf {\bibinfo {volume} {78}},\ \bibinfo
  {pages} {020501} (\bibinfo {year} {2008})}\BibitemShut {NoStop}%
\bibitem [{\citenamefont {Kontani}\ \emph {et~al.}(2011)\citenamefont
  {Kontani}, \citenamefont {Saito},\ and\ \citenamefont
  {Onari}}]{Kontani2011_PRB}%
  \BibitemOpen
  \bibfield  {author} {\bibinfo {author} {\bibfnamefont {H.}~\bibnamefont
  {Kontani}}, \bibinfo {author} {\bibfnamefont {T.}~\bibnamefont {Saito}},\
  and\ \bibinfo {author} {\bibfnamefont {S.}~\bibnamefont {Onari}},\ }\href
  {https://doi.org/10.1103/PhysRevB.84.024528} {\bibfield  {journal} {\bibinfo
  {journal} {Phys. Rev. B}\ }\textbf {\bibinfo {volume} {84}},\ \bibinfo
  {pages} {024528} (\bibinfo {year} {2011})}\BibitemShut {NoStop}%
\bibitem [{\citenamefont {Stanev}\ and\ \citenamefont
  {Littlewood}(2013)}]{Stanev2013_PRB}%
  \BibitemOpen
  \bibfield  {author} {\bibinfo {author} {\bibfnamefont {V.}~\bibnamefont
  {Stanev}}\ and\ \bibinfo {author} {\bibfnamefont {P.~B.}\ \bibnamefont
  {Littlewood}},\ }\href {https://doi.org/10.1103/PhysRevB.87.161122}
  {\bibfield  {journal} {\bibinfo  {journal} {Phys. Rev. B}\ }\textbf {\bibinfo
  {volume} {87}},\ \bibinfo {pages} {161122} (\bibinfo {year}
  {2013})}\BibitemShut {NoStop}%
\bibitem [{\citenamefont {Yamase}\ and\ \citenamefont
  {Zeyher}(2013)}]{Yamase2013_PRB}%
  \BibitemOpen
  \bibfield  {author} {\bibinfo {author} {\bibfnamefont {H.}~\bibnamefont
  {Yamase}}\ and\ \bibinfo {author} {\bibfnamefont {R.}~\bibnamefont
  {Zeyher}},\ }\href {https://doi.org/10.1103/PhysRevB.88.180502} {\bibfield
  {journal} {\bibinfo  {journal} {Phys. Rev. B}\ }\textbf {\bibinfo {volume}
  {88}},\ \bibinfo {pages} {180502} (\bibinfo {year} {2013})}\BibitemShut
  {NoStop}%
\bibitem [{\citenamefont {Golubovi\ifmmode~\acute{c}\else \'{c}\fi{}}\ and\
  \citenamefont {Kosti\ifmmode~\acute{c}\else
  \'{c}\fi{}}(1988)}]{Golubovic1988_PRB}%
  \BibitemOpen
  \bibfield  {author} {\bibinfo {author} {\bibfnamefont {L.}~\bibnamefont
  {Golubovi\ifmmode~\acute{c}\else \'{c}\fi{}}}\ and\ \bibinfo {author}
  {\bibfnamefont {D.}~\bibnamefont {Kosti\ifmmode~\acute{c}\else \'{c}\fi{}}},\
  }\href {https://doi.org/10.1103/PhysRevB.38.2622} {\bibfield  {journal}
  {\bibinfo  {journal} {Phys. Rev. B}\ }\textbf {\bibinfo {volume} {38}},\
  \bibinfo {pages} {2622} (\bibinfo {year} {1988})}\BibitemShut {NoStop}%
\bibitem [{\citenamefont {Fernandes}\ \emph {et~al.}(2014)\citenamefont
  {Fernandes}, \citenamefont {Chubukov},\ and\ \citenamefont
  {Schmalian}}]{Fernandes2014_NatPhys}%
  \BibitemOpen
  \bibfield  {author} {\bibinfo {author} {\bibfnamefont {R.~M.}\ \bibnamefont
  {Fernandes}}, \bibinfo {author} {\bibfnamefont {A.~V.}\ \bibnamefont
  {Chubukov}},\ and\ \bibinfo {author} {\bibfnamefont {J.}~\bibnamefont
  {Schmalian}},\ }\href {https://doi.org/10.1038/nphys2877} {\bibfield
  {journal} {\bibinfo  {journal} {Nature Physics}\ }\textbf {\bibinfo {volume}
  {10}},\ \bibinfo {pages} {97} (\bibinfo {year} {2014})}\BibitemShut {NoStop}%
\bibitem [{\citenamefont {B\"{o}hmer}\ and\ \citenamefont
  {Meingast}(2016)}]{Boemer2016_CRP}%
  \BibitemOpen
  \bibfield  {author} {\bibinfo {author} {\bibfnamefont {A.~E.}\ \bibnamefont
  {B\"{o}hmer}}\ and\ \bibinfo {author} {\bibfnamefont {C.}~\bibnamefont
  {Meingast}},\ }\href
  {https://doi.org/https://doi.org/10.1016/j.crhy.2015.07.001} {\bibfield
  {journal} {\bibinfo  {journal} {Comptes Rendus Physique}\ }\textbf {\bibinfo
  {volume} {17}},\ \bibinfo {pages} {90} (\bibinfo {year} {2016})}\BibitemShut
  {NoStop}%
\bibitem [{\citenamefont {Vojta}(2009)}]{Vojta2009_AdvPhys}%
  \BibitemOpen
  \bibfield  {author} {\bibinfo {author} {\bibfnamefont {M.}~\bibnamefont
  {Vojta}},\ }\href {https://doi.org/10.1080/00018730903122242} {\bibfield
  {journal} {\bibinfo  {journal} {Advances in Physics}\ }\textbf {\bibinfo
  {volume} {58}},\ \bibinfo {pages} {699} (\bibinfo {year} {2009})}\BibitemShut
  {NoStop}%
\bibitem [{\citenamefont {Fradkin}\ \emph {et~al.}(2010)\citenamefont
  {Fradkin}, \citenamefont {Kivelson}, \citenamefont {Lawler}, \citenamefont
  {Eisenstein},\ and\ \citenamefont {Mackenzie}}]{Fradkin2010_AnnRev}%
  \BibitemOpen
  \bibfield  {author} {\bibinfo {author} {\bibfnamefont {E.}~\bibnamefont
  {Fradkin}}, \bibinfo {author} {\bibfnamefont {S.~A.}\ \bibnamefont
  {Kivelson}}, \bibinfo {author} {\bibfnamefont {M.~J.}\ \bibnamefont
  {Lawler}}, \bibinfo {author} {\bibfnamefont {J.~P.}\ \bibnamefont
  {Eisenstein}},\ and\ \bibinfo {author} {\bibfnamefont {A.~P.}\ \bibnamefont
  {Mackenzie}},\ }\href
  {https://doi.org/10.1146/annurev-conmatphys-070909-103925} {\bibfield
  {journal} {\bibinfo  {journal} {Annual Review of Condensed Matter Physics}\
  }\textbf {\bibinfo {volume} {1}},\ \bibinfo {pages} {153} (\bibinfo {year}
  {2010})}\BibitemShut {NoStop}%
\bibitem [{\citenamefont {Mazin}\ \emph {et~al.}(2008)\citenamefont {Mazin},
  \citenamefont {Singh}, \citenamefont {Johannes},\ and\ \citenamefont
  {Du}}]{Mazin2008_PRL}%
  \BibitemOpen
  \bibfield  {author} {\bibinfo {author} {\bibfnamefont {I.~I.}\ \bibnamefont
  {Mazin}}, \bibinfo {author} {\bibfnamefont {D.~J.}\ \bibnamefont {Singh}},
  \bibinfo {author} {\bibfnamefont {M.~D.}\ \bibnamefont {Johannes}},\ and\
  \bibinfo {author} {\bibfnamefont {M.~H.}\ \bibnamefont {Du}},\ }\href
  {https://doi.org/10.1103/PhysRevLett.101.057003} {\bibfield  {journal}
  {\bibinfo  {journal} {Phys. Rev. Lett.}\ }\textbf {\bibinfo {volume} {101}},\
  \bibinfo {pages} {057003} (\bibinfo {year} {2008})}\BibitemShut {NoStop}%
\bibitem [{\citenamefont {Fanfarillo}\ \emph {et~al.}(2015)\citenamefont
  {Fanfarillo}, \citenamefont {Cortijo},\ and\ \citenamefont
  {Valenzuela}}]{Fanfarillo2015_PRB}%
  \BibitemOpen
  \bibfield  {author} {\bibinfo {author} {\bibfnamefont {L.}~\bibnamefont
  {Fanfarillo}}, \bibinfo {author} {\bibfnamefont {A.}~\bibnamefont
  {Cortijo}},\ and\ \bibinfo {author} {\bibfnamefont {B.}~\bibnamefont
  {Valenzuela}},\ }\href {https://doi.org/10.1103/PhysRevB.91.214515}
  {\bibfield  {journal} {\bibinfo  {journal} {Phys. Rev. B}\ }\textbf {\bibinfo
  {volume} {91}},\ \bibinfo {pages} {214515} (\bibinfo {year}
  {2015})}\BibitemShut {NoStop}%
\bibitem [{\citenamefont {Fanfarillo}\ \emph {et~al.}(2018)\citenamefont
  {Fanfarillo}, \citenamefont {Benfatto},\ and\ \citenamefont
  {Valenzuela}}]{Fanfarillo2018_PRB}%
  \BibitemOpen
  \bibfield  {author} {\bibinfo {author} {\bibfnamefont {L.}~\bibnamefont
  {Fanfarillo}}, \bibinfo {author} {\bibfnamefont {L.}~\bibnamefont
  {Benfatto}},\ and\ \bibinfo {author} {\bibfnamefont {B.}~\bibnamefont
  {Valenzuela}},\ }\href {https://doi.org/10.1103/PhysRevB.97.121109}
  {\bibfield  {journal} {\bibinfo  {journal} {Phys. Rev. B}\ }\textbf {\bibinfo
  {volume} {97}},\ \bibinfo {pages} {121109} (\bibinfo {year}
  {2018})}\BibitemShut {NoStop}%
\bibitem [{\citenamefont {Chubukov}\ \emph {et~al.}(2016)\citenamefont
  {Chubukov}, \citenamefont {Khodas},\ and\ \citenamefont
  {Fernandes}}]{Chubukov2016_PRX}%
  \BibitemOpen
  \bibfield  {author} {\bibinfo {author} {\bibfnamefont {A.~V.}\ \bibnamefont
  {Chubukov}}, \bibinfo {author} {\bibfnamefont {M.}~\bibnamefont {Khodas}},\
  and\ \bibinfo {author} {\bibfnamefont {R.~M.}\ \bibnamefont {Fernandes}},\
  }\href {https://doi.org/10.1103/PhysRevX.6.041045} {\bibfield  {journal}
  {\bibinfo  {journal} {Phys. Rev. X}\ }\textbf {\bibinfo {volume} {6}},\
  \bibinfo {pages} {041045} (\bibinfo {year} {2016})}\BibitemShut {NoStop}%
\bibitem [{\citenamefont {Zhai}\ \emph {et~al.}(2009)\citenamefont {Zhai},
  \citenamefont {Wang},\ and\ \citenamefont {Lee}}]{Zhai2009_PRB}%
  \BibitemOpen
  \bibfield  {author} {\bibinfo {author} {\bibfnamefont {H.}~\bibnamefont
  {Zhai}}, \bibinfo {author} {\bibfnamefont {F.}~\bibnamefont {Wang}},\ and\
  \bibinfo {author} {\bibfnamefont {D.-H.}\ \bibnamefont {Lee}},\ }\href
  {https://doi.org/10.1103/PhysRevB.80.064517} {\bibfield  {journal} {\bibinfo
  {journal} {Phys. Rev. B}\ }\textbf {\bibinfo {volume} {80}},\ \bibinfo
  {pages} {064517} (\bibinfo {year} {2009})}\BibitemShut {NoStop}%
\bibitem [{\citenamefont {Thorsm\o{}lle}\ \emph {et~al.}(2016)\citenamefont
  {Thorsm\o{}lle}, \citenamefont {Khodas}, \citenamefont {Yin}, \citenamefont
  {Zhang}, \citenamefont {Carr}, \citenamefont {Dai},\ and\ \citenamefont
  {Blumberg}}]{Thorsmoelle2016_PRB}%
  \BibitemOpen
  \bibfield  {author} {\bibinfo {author} {\bibfnamefont {V.~K.}\ \bibnamefont
  {Thorsm\o{}lle}}, \bibinfo {author} {\bibfnamefont {M.}~\bibnamefont
  {Khodas}}, \bibinfo {author} {\bibfnamefont {Z.~P.}\ \bibnamefont {Yin}},
  \bibinfo {author} {\bibfnamefont {C.}~\bibnamefont {Zhang}}, \bibinfo
  {author} {\bibfnamefont {S.~V.}\ \bibnamefont {Carr}}, \bibinfo {author}
  {\bibfnamefont {P.}~\bibnamefont {Dai}},\ and\ \bibinfo {author}
  {\bibfnamefont {G.}~\bibnamefont {Blumberg}},\ }\href
  {https://doi.org/10.1103/PhysRevB.93.054515} {\bibfield  {journal} {\bibinfo
  {journal} {Phys. Rev. B}\ }\textbf {\bibinfo {volume} {93}},\ \bibinfo
  {pages} {054515} (\bibinfo {year} {2016})}\BibitemShut {NoStop}%
\bibitem [{\citenamefont {Grandi}\ \emph {et~al.}(2023)\citenamefont {Grandi},
  \citenamefont {Consiglio}, \citenamefont {Sentef}, \citenamefont {Thomale},\
  and\ \citenamefont {Kennes}}]{Grandi2023_PRB}%
  \BibitemOpen
  \bibfield  {author} {\bibinfo {author} {\bibfnamefont {F.}~\bibnamefont
  {Grandi}}, \bibinfo {author} {\bibfnamefont {A.}~\bibnamefont {Consiglio}},
  \bibinfo {author} {\bibfnamefont {M.~A.}\ \bibnamefont {Sentef}}, \bibinfo
  {author} {\bibfnamefont {R.}~\bibnamefont {Thomale}},\ and\ \bibinfo {author}
  {\bibfnamefont {D.~M.}\ \bibnamefont {Kennes}},\ }\href
  {https://doi.org/10.1103/PhysRevB.107.155131} {\bibfield  {journal} {\bibinfo
   {journal} {Phys. Rev. B}\ }\textbf {\bibinfo {volume} {107}},\ \bibinfo
  {pages} {155131} (\bibinfo {year} {2023})}\BibitemShut {NoStop}%
\bibitem [{\citenamefont {Xiang}\ \emph {et~al.}(2021)\citenamefont {Xiang},
  \citenamefont {Li}, \citenamefont {Li}, \citenamefont {Xie}, \citenamefont
  {Yang}, \citenamefont {Wang}, \citenamefont {Yao},\ and\ \citenamefont
  {Wen}}]{Xiang2021_NatComm}%
  \BibitemOpen
  \bibfield  {author} {\bibinfo {author} {\bibfnamefont {Y.}~\bibnamefont
  {Xiang}}, \bibinfo {author} {\bibfnamefont {Q.}~\bibnamefont {Li}}, \bibinfo
  {author} {\bibfnamefont {Y.}~\bibnamefont {Li}}, \bibinfo {author}
  {\bibfnamefont {W.}~\bibnamefont {Xie}}, \bibinfo {author} {\bibfnamefont
  {H.}~\bibnamefont {Yang}}, \bibinfo {author} {\bibfnamefont {Z.}~\bibnamefont
  {Wang}}, \bibinfo {author} {\bibfnamefont {Y.}~\bibnamefont {Yao}},\ and\
  \bibinfo {author} {\bibfnamefont {H.-H.}\ \bibnamefont {Wen}},\ }\href
  {https://doi.org/10.1038/s41467-021-27084-z} {\bibfield  {journal} {\bibinfo
  {journal} {Nature Communications}\ }\textbf {\bibinfo {volume} {12}},\
  \bibinfo {pages} {6727} (\bibinfo {year} {2021})}\BibitemShut {NoStop}%
\bibitem [{\citenamefont {Wu}\ \emph {et~al.}(2022)\citenamefont {Wu},
  \citenamefont {Wang}, \citenamefont {Liu}, \citenamefont {Li}, \citenamefont
  {Xu}, \citenamefont {Yin}, \citenamefont {Gong}, \citenamefont {Tu},
  \citenamefont {Lei}, \citenamefont {Dong},\ and\ \citenamefont
  {Wang}}]{Wu2022_PRB2}%
  \BibitemOpen
  \bibfield  {author} {\bibinfo {author} {\bibfnamefont {Q.}~\bibnamefont
  {Wu}}, \bibinfo {author} {\bibfnamefont {Z.~X.}\ \bibnamefont {Wang}},
  \bibinfo {author} {\bibfnamefont {Q.~M.}\ \bibnamefont {Liu}}, \bibinfo
  {author} {\bibfnamefont {R.~S.}\ \bibnamefont {Li}}, \bibinfo {author}
  {\bibfnamefont {S.~X.}\ \bibnamefont {Xu}}, \bibinfo {author} {\bibfnamefont
  {Q.~W.}\ \bibnamefont {Yin}}, \bibinfo {author} {\bibfnamefont {C.~S.}\
  \bibnamefont {Gong}}, \bibinfo {author} {\bibfnamefont {Z.~J.}\ \bibnamefont
  {Tu}}, \bibinfo {author} {\bibfnamefont {H.~C.}\ \bibnamefont {Lei}},
  \bibinfo {author} {\bibfnamefont {T.}~\bibnamefont {Dong}},\ and\ \bibinfo
  {author} {\bibfnamefont {N.~L.}\ \bibnamefont {Wang}},\ }\href
  {https://doi.org/10.1103/PhysRevB.106.205109} {\bibfield  {journal} {\bibinfo
   {journal} {Phys. Rev. B}\ }\textbf {\bibinfo {volume} {106}},\ \bibinfo
  {pages} {205109} (\bibinfo {year} {2022})}\BibitemShut {NoStop}%
\bibitem [{\citenamefont {Xu}\ \emph {et~al.}(2022)\citenamefont {Xu},
  \citenamefont {Ni}, \citenamefont {Liu}, \citenamefont {Ortiz}, \citenamefont
  {Deng}, \citenamefont {Wilson}, \citenamefont {Yan}, \citenamefont
  {Balents},\ and\ \citenamefont {Wu}}]{Xu2022_NatPhys}%
  \BibitemOpen
  \bibfield  {author} {\bibinfo {author} {\bibfnamefont {Y.}~\bibnamefont
  {Xu}}, \bibinfo {author} {\bibfnamefont {Z.}~\bibnamefont {Ni}}, \bibinfo
  {author} {\bibfnamefont {Y.}~\bibnamefont {Liu}}, \bibinfo {author}
  {\bibfnamefont {B.~R.}\ \bibnamefont {Ortiz}}, \bibinfo {author}
  {\bibfnamefont {Q.}~\bibnamefont {Deng}}, \bibinfo {author} {\bibfnamefont
  {S.~D.}\ \bibnamefont {Wilson}}, \bibinfo {author} {\bibfnamefont
  {B.}~\bibnamefont {Yan}}, \bibinfo {author} {\bibfnamefont {L.}~\bibnamefont
  {Balents}},\ and\ \bibinfo {author} {\bibfnamefont {L.}~\bibnamefont {Wu}},\
  }\href {https://doi.org/10.1038/s41567-022-01805-7} {\bibfield  {journal}
  {\bibinfo  {journal} {Nature Physics}\ }\textbf {\bibinfo {volume} {18}},\
  \bibinfo {pages} {1470} (\bibinfo {year} {2022})}\BibitemShut {NoStop}%
\bibitem [{\citenamefont {Nie}\ \emph {et~al.}(2022)\citenamefont {Nie},
  \citenamefont {Sun}, \citenamefont {Ma}, \citenamefont {Song}, \citenamefont
  {Zheng}, \citenamefont {Liang}, \citenamefont {Wu}, \citenamefont {Yu},
  \citenamefont {Li}, \citenamefont {Shan}, \citenamefont {Zhao}, \citenamefont
  {Li}, \citenamefont {Kang}, \citenamefont {Wu}, \citenamefont {Zhou},
  \citenamefont {Liu}, \citenamefont {Xiang}, \citenamefont {Ying},
  \citenamefont {Wang}, \citenamefont {Wu},\ and\ \citenamefont
  {Chen}}]{Nie2022_Nat}%
  \BibitemOpen
  \bibfield  {author} {\bibinfo {author} {\bibfnamefont {L.}~\bibnamefont
  {Nie}}, \bibinfo {author} {\bibfnamefont {K.}~\bibnamefont {Sun}}, \bibinfo
  {author} {\bibfnamefont {W.}~\bibnamefont {Ma}}, \bibinfo {author}
  {\bibfnamefont {D.}~\bibnamefont {Song}}, \bibinfo {author} {\bibfnamefont
  {L.}~\bibnamefont {Zheng}}, \bibinfo {author} {\bibfnamefont
  {Z.}~\bibnamefont {Liang}}, \bibinfo {author} {\bibfnamefont
  {P.}~\bibnamefont {Wu}}, \bibinfo {author} {\bibfnamefont {F.}~\bibnamefont
  {Yu}}, \bibinfo {author} {\bibfnamefont {J.}~\bibnamefont {Li}}, \bibinfo
  {author} {\bibfnamefont {M.}~\bibnamefont {Shan}}, \bibinfo {author}
  {\bibfnamefont {D.}~\bibnamefont {Zhao}}, \bibinfo {author} {\bibfnamefont
  {S.}~\bibnamefont {Li}}, \bibinfo {author} {\bibfnamefont {B.}~\bibnamefont
  {Kang}}, \bibinfo {author} {\bibfnamefont {Z.}~\bibnamefont {Wu}}, \bibinfo
  {author} {\bibfnamefont {Y.}~\bibnamefont {Zhou}}, \bibinfo {author}
  {\bibfnamefont {K.}~\bibnamefont {Liu}}, \bibinfo {author} {\bibfnamefont
  {Z.}~\bibnamefont {Xiang}}, \bibinfo {author} {\bibfnamefont
  {J.}~\bibnamefont {Ying}}, \bibinfo {author} {\bibfnamefont {Z.}~\bibnamefont
  {Wang}}, \bibinfo {author} {\bibfnamefont {T.}~\bibnamefont {Wu}},\ and\
  \bibinfo {author} {\bibfnamefont {X.}~\bibnamefont {Chen}},\ }\href
  {https://doi.org/10.1038/s41586-022-04493-8} {\bibfield  {journal} {\bibinfo
  {journal} {Nature}\ }\textbf {\bibinfo {volume} {604}},\ \bibinfo {pages}
  {59} (\bibinfo {year} {2022})}\BibitemShut {NoStop}%
\bibitem [{\citenamefont {Zheng}\ \emph {et~al.}(2022)\citenamefont {Zheng},
  \citenamefont {Wu}, \citenamefont {Yang}, \citenamefont {Nie}, \citenamefont
  {Shan}, \citenamefont {Sun}, \citenamefont {Song}, \citenamefont {Yu},
  \citenamefont {Li}, \citenamefont {Zhao}, \citenamefont {Li}, \citenamefont
  {Kang}, \citenamefont {Zhou}, \citenamefont {Liu}, \citenamefont {Xiang},
  \citenamefont {Ying}, \citenamefont {Wang}, \citenamefont {Wu},\ and\
  \citenamefont {Chen}}]{Zheng2022_Nat}%
  \BibitemOpen
  \bibfield  {author} {\bibinfo {author} {\bibfnamefont {L.}~\bibnamefont
  {Zheng}}, \bibinfo {author} {\bibfnamefont {Z.}~\bibnamefont {Wu}}, \bibinfo
  {author} {\bibfnamefont {Y.}~\bibnamefont {Yang}}, \bibinfo {author}
  {\bibfnamefont {L.}~\bibnamefont {Nie}}, \bibinfo {author} {\bibfnamefont
  {M.}~\bibnamefont {Shan}}, \bibinfo {author} {\bibfnamefont {K.}~\bibnamefont
  {Sun}}, \bibinfo {author} {\bibfnamefont {D.}~\bibnamefont {Song}}, \bibinfo
  {author} {\bibfnamefont {F.}~\bibnamefont {Yu}}, \bibinfo {author}
  {\bibfnamefont {J.}~\bibnamefont {Li}}, \bibinfo {author} {\bibfnamefont
  {D.}~\bibnamefont {Zhao}}, \bibinfo {author} {\bibfnamefont {S.}~\bibnamefont
  {Li}}, \bibinfo {author} {\bibfnamefont {B.}~\bibnamefont {Kang}}, \bibinfo
  {author} {\bibfnamefont {Y.}~\bibnamefont {Zhou}}, \bibinfo {author}
  {\bibfnamefont {K.}~\bibnamefont {Liu}}, \bibinfo {author} {\bibfnamefont
  {Z.}~\bibnamefont {Xiang}}, \bibinfo {author} {\bibfnamefont
  {J.}~\bibnamefont {Ying}}, \bibinfo {author} {\bibfnamefont {Z.}~\bibnamefont
  {Wang}}, \bibinfo {author} {\bibfnamefont {T.}~\bibnamefont {Wu}},\ and\
  \bibinfo {author} {\bibfnamefont {X.}~\bibnamefont {Chen}},\ }\href
  {https://doi.org/10.1038/s41586-022-05351-3} {\bibfield  {journal} {\bibinfo
  {journal} {Nature}\ }\textbf {\bibinfo {volume} {611}},\ \bibinfo {pages}
  {682} (\bibinfo {year} {2022})}\BibitemShut {NoStop}%
\bibitem [{\citenamefont {Li}\ \emph {et~al.}(2022{\natexlab{a}})\citenamefont
  {Li}, \citenamefont {Fabbris}, \citenamefont {Said}, \citenamefont {Sun},
  \citenamefont {Jiang}, \citenamefont {Yin}, \citenamefont {Pai},
  \citenamefont {Yoon}, \citenamefont {Lupini}, \citenamefont {Nelson},
  \citenamefont {Yin}, \citenamefont {Gong}, \citenamefont {Tu}, \citenamefont
  {Lei}, \citenamefont {Cheng}, \citenamefont {Hasan}, \citenamefont {Wang},
  \citenamefont {Yan}, \citenamefont {Thomale}, \citenamefont {Lee},\ and\
  \citenamefont {Miao}}]{Li2022_NatComm}%
  \BibitemOpen
  \bibfield  {author} {\bibinfo {author} {\bibfnamefont {H.}~\bibnamefont
  {Li}}, \bibinfo {author} {\bibfnamefont {G.}~\bibnamefont {Fabbris}},
  \bibinfo {author} {\bibfnamefont {A.~H.}\ \bibnamefont {Said}}, \bibinfo
  {author} {\bibfnamefont {J.~P.}\ \bibnamefont {Sun}}, \bibinfo {author}
  {\bibfnamefont {Y.-X.}\ \bibnamefont {Jiang}}, \bibinfo {author}
  {\bibfnamefont {J.-X.}\ \bibnamefont {Yin}}, \bibinfo {author} {\bibfnamefont
  {Y.-Y.}\ \bibnamefont {Pai}}, \bibinfo {author} {\bibfnamefont
  {S.}~\bibnamefont {Yoon}}, \bibinfo {author} {\bibfnamefont {A.~R.}\
  \bibnamefont {Lupini}}, \bibinfo {author} {\bibfnamefont {C.~S.}\
  \bibnamefont {Nelson}}, \bibinfo {author} {\bibfnamefont {Q.~W.}\
  \bibnamefont {Yin}}, \bibinfo {author} {\bibfnamefont {C.~S.}\ \bibnamefont
  {Gong}}, \bibinfo {author} {\bibfnamefont {Z.~J.}\ \bibnamefont {Tu}},
  \bibinfo {author} {\bibfnamefont {H.~C.}\ \bibnamefont {Lei}}, \bibinfo
  {author} {\bibfnamefont {J.-G.}\ \bibnamefont {Cheng}}, \bibinfo {author}
  {\bibfnamefont {M.~Z.}\ \bibnamefont {Hasan}}, \bibinfo {author}
  {\bibfnamefont {Z.}~\bibnamefont {Wang}}, \bibinfo {author} {\bibfnamefont
  {B.}~\bibnamefont {Yan}}, \bibinfo {author} {\bibfnamefont {R.}~\bibnamefont
  {Thomale}}, \bibinfo {author} {\bibfnamefont {H.~N.}\ \bibnamefont {Lee}},\
  and\ \bibinfo {author} {\bibfnamefont {H.}~\bibnamefont {Miao}},\ }\href
  {https://doi.org/10.1038/s41467-022-33995-2} {\bibfield  {journal} {\bibinfo
  {journal} {Nature Communications}\ }\textbf {\bibinfo {volume} {13}},\
  \bibinfo {pages} {6348} (\bibinfo {year} {2022}{\natexlab{a}})}\BibitemShut
  {NoStop}%
\bibitem [{\citenamefont {Jiang}\ \emph {et~al.}(2021)\citenamefont {Jiang},
  \citenamefont {Yin}, \citenamefont {Denner}, \citenamefont {Shumiya},
  \citenamefont {Ortiz}, \citenamefont {Xu}, \citenamefont {Guguchia},
  \citenamefont {He}, \citenamefont {Hossain}, \citenamefont {Liu},
  \citenamefont {Ruff}, \citenamefont {Kautzsch}, \citenamefont {Zhang},
  \citenamefont {Chang}, \citenamefont {Belopolski}, \citenamefont {Zhang},
  \citenamefont {Cochran}, \citenamefont {Multer}, \citenamefont {Litskevich},
  \citenamefont {Cheng}, \citenamefont {Yang}, \citenamefont {Wang},
  \citenamefont {Thomale}, \citenamefont {Neupert}, \citenamefont {Wilson},\
  and\ \citenamefont {Hasan}}]{Jiang2021_Nat_Mat}%
  \BibitemOpen
  \bibfield  {author} {\bibinfo {author} {\bibfnamefont {Y.-X.}\ \bibnamefont
  {Jiang}}, \bibinfo {author} {\bibfnamefont {J.-X.}\ \bibnamefont {Yin}},
  \bibinfo {author} {\bibfnamefont {M.~M.}\ \bibnamefont {Denner}}, \bibinfo
  {author} {\bibfnamefont {N.}~\bibnamefont {Shumiya}}, \bibinfo {author}
  {\bibfnamefont {B.~R.}\ \bibnamefont {Ortiz}}, \bibinfo {author}
  {\bibfnamefont {G.}~\bibnamefont {Xu}}, \bibinfo {author} {\bibfnamefont
  {Z.}~\bibnamefont {Guguchia}}, \bibinfo {author} {\bibfnamefont
  {J.}~\bibnamefont {He}}, \bibinfo {author} {\bibfnamefont {M.~S.}\
  \bibnamefont {Hossain}}, \bibinfo {author} {\bibfnamefont {X.}~\bibnamefont
  {Liu}}, \bibinfo {author} {\bibfnamefont {J.}~\bibnamefont {Ruff}}, \bibinfo
  {author} {\bibfnamefont {L.}~\bibnamefont {Kautzsch}}, \bibinfo {author}
  {\bibfnamefont {S.~S.}\ \bibnamefont {Zhang}}, \bibinfo {author}
  {\bibfnamefont {G.}~\bibnamefont {Chang}}, \bibinfo {author} {\bibfnamefont
  {I.}~\bibnamefont {Belopolski}}, \bibinfo {author} {\bibfnamefont
  {Q.}~\bibnamefont {Zhang}}, \bibinfo {author} {\bibfnamefont {T.~A.}\
  \bibnamefont {Cochran}}, \bibinfo {author} {\bibfnamefont {D.}~\bibnamefont
  {Multer}}, \bibinfo {author} {\bibfnamefont {M.}~\bibnamefont {Litskevich}},
  \bibinfo {author} {\bibfnamefont {Z.-J.}\ \bibnamefont {Cheng}}, \bibinfo
  {author} {\bibfnamefont {X.~P.}\ \bibnamefont {Yang}}, \bibinfo {author}
  {\bibfnamefont {Z.}~\bibnamefont {Wang}}, \bibinfo {author} {\bibfnamefont
  {R.}~\bibnamefont {Thomale}}, \bibinfo {author} {\bibfnamefont
  {T.}~\bibnamefont {Neupert}}, \bibinfo {author} {\bibfnamefont {S.~D.}\
  \bibnamefont {Wilson}},\ and\ \bibinfo {author} {\bibfnamefont {M.~Z.}\
  \bibnamefont {Hasan}},\ }\href {https://doi.org/10.1038/s41563-021-01034-y}
  {\bibfield  {journal} {\bibinfo  {journal} {Nature Materials}\ }\textbf
  {\bibinfo {volume} {20}},\ \bibinfo {pages} {1353} (\bibinfo {year}
  {2021})}\BibitemShut {NoStop}%
\bibitem [{\citenamefont {Li}\ \emph {et~al.}(2021)\citenamefont {Li},
  \citenamefont {Zhang}, \citenamefont {Yilmaz}, \citenamefont {Pai},
  \citenamefont {Marvinney}, \citenamefont {Said}, \citenamefont {Yin},
  \citenamefont {Gong}, \citenamefont {Tu}, \citenamefont {Vescovo},
  \citenamefont {Nelson}, \citenamefont {Moore}, \citenamefont {Murakami},
  \citenamefont {Lei}, \citenamefont {Lee}, \citenamefont {Lawrie},\ and\
  \citenamefont {Miao}}]{Li2021_PRX}%
  \BibitemOpen
  \bibfield  {author} {\bibinfo {author} {\bibfnamefont {H.}~\bibnamefont
  {Li}}, \bibinfo {author} {\bibfnamefont {T.~T.}\ \bibnamefont {Zhang}},
  \bibinfo {author} {\bibfnamefont {T.}~\bibnamefont {Yilmaz}}, \bibinfo
  {author} {\bibfnamefont {Y.~Y.}\ \bibnamefont {Pai}}, \bibinfo {author}
  {\bibfnamefont {C.~E.}\ \bibnamefont {Marvinney}}, \bibinfo {author}
  {\bibfnamefont {A.}~\bibnamefont {Said}}, \bibinfo {author} {\bibfnamefont
  {Q.~W.}\ \bibnamefont {Yin}}, \bibinfo {author} {\bibfnamefont {C.~S.}\
  \bibnamefont {Gong}}, \bibinfo {author} {\bibfnamefont {Z.~J.}\ \bibnamefont
  {Tu}}, \bibinfo {author} {\bibfnamefont {E.}~\bibnamefont {Vescovo}},
  \bibinfo {author} {\bibfnamefont {C.~S.}\ \bibnamefont {Nelson}}, \bibinfo
  {author} {\bibfnamefont {R.~G.}\ \bibnamefont {Moore}}, \bibinfo {author}
  {\bibfnamefont {S.}~\bibnamefont {Murakami}}, \bibinfo {author}
  {\bibfnamefont {H.~C.}\ \bibnamefont {Lei}}, \bibinfo {author} {\bibfnamefont
  {H.~N.}\ \bibnamefont {Lee}}, \bibinfo {author} {\bibfnamefont {B.~J.}\
  \bibnamefont {Lawrie}},\ and\ \bibinfo {author} {\bibfnamefont
  {H.}~\bibnamefont {Miao}},\ }\href
  {https://doi.org/10.1103/PhysRevX.11.031050} {\bibfield  {journal} {\bibinfo
  {journal} {Phys. Rev. X}\ }\textbf {\bibinfo {volume} {11}},\ \bibinfo
  {pages} {031050} (\bibinfo {year} {2021})}\BibitemShut {NoStop}%
\bibitem [{\citenamefont {Liang}\ \emph {et~al.}(2021)\citenamefont {Liang},
  \citenamefont {Hou}, \citenamefont {Zhang}, \citenamefont {Ma}, \citenamefont
  {Wu}, \citenamefont {Zhang}, \citenamefont {Yu}, \citenamefont {Ying},
  \citenamefont {Jiang}, \citenamefont {Shan}, \citenamefont {Wang},\ and\
  \citenamefont {Chen}}]{Liang2021_PRX}%
  \BibitemOpen
  \bibfield  {author} {\bibinfo {author} {\bibfnamefont {Z.}~\bibnamefont
  {Liang}}, \bibinfo {author} {\bibfnamefont {X.}~\bibnamefont {Hou}}, \bibinfo
  {author} {\bibfnamefont {F.}~\bibnamefont {Zhang}}, \bibinfo {author}
  {\bibfnamefont {W.}~\bibnamefont {Ma}}, \bibinfo {author} {\bibfnamefont
  {P.}~\bibnamefont {Wu}}, \bibinfo {author} {\bibfnamefont {Z.}~\bibnamefont
  {Zhang}}, \bibinfo {author} {\bibfnamefont {F.}~\bibnamefont {Yu}}, \bibinfo
  {author} {\bibfnamefont {J.-J.}\ \bibnamefont {Ying}}, \bibinfo {author}
  {\bibfnamefont {K.}~\bibnamefont {Jiang}}, \bibinfo {author} {\bibfnamefont
  {L.}~\bibnamefont {Shan}}, \bibinfo {author} {\bibfnamefont {Z.}~\bibnamefont
  {Wang}},\ and\ \bibinfo {author} {\bibfnamefont {X.-H.}\ \bibnamefont
  {Chen}},\ }\href {https://doi.org/10.1103/PhysRevX.11.031026} {\bibfield
  {journal} {\bibinfo  {journal} {Phys. Rev. X}\ }\textbf {\bibinfo {volume}
  {11}},\ \bibinfo {pages} {031026} (\bibinfo {year} {2021})}\BibitemShut
  {NoStop}%
\bibitem [{\citenamefont {Ortiz}\ \emph
  {et~al.}(2021{\natexlab{a}})\citenamefont {Ortiz}, \citenamefont {Teicher},
  \citenamefont {Kautzsch}, \citenamefont {Sarte}, \citenamefont {Ratcliff},
  \citenamefont {Harter}, \citenamefont {Ruff}, \citenamefont {Seshadri},\ and\
  \citenamefont {Wilson}}]{Ortiz2021_PRX}%
  \BibitemOpen
  \bibfield  {author} {\bibinfo {author} {\bibfnamefont {B.~R.}\ \bibnamefont
  {Ortiz}}, \bibinfo {author} {\bibfnamefont {S.~M.~L.}\ \bibnamefont
  {Teicher}}, \bibinfo {author} {\bibfnamefont {L.}~\bibnamefont {Kautzsch}},
  \bibinfo {author} {\bibfnamefont {P.~M.}\ \bibnamefont {Sarte}}, \bibinfo
  {author} {\bibfnamefont {N.}~\bibnamefont {Ratcliff}}, \bibinfo {author}
  {\bibfnamefont {J.}~\bibnamefont {Harter}}, \bibinfo {author} {\bibfnamefont
  {J.~P.~C.}\ \bibnamefont {Ruff}}, \bibinfo {author} {\bibfnamefont
  {R.}~\bibnamefont {Seshadri}},\ and\ \bibinfo {author} {\bibfnamefont
  {S.~D.}\ \bibnamefont {Wilson}},\ }\href
  {https://doi.org/10.1103/PhysRevX.11.041030} {\bibfield  {journal} {\bibinfo
  {journal} {Phys. Rev. X}\ }\textbf {\bibinfo {volume} {11}},\ \bibinfo
  {pages} {041030} (\bibinfo {year} {2021}{\natexlab{a}})}\BibitemShut
  {NoStop}%
\bibitem [{\citenamefont {Saykin}\ \emph {et~al.}(2023)\citenamefont {Saykin},
  \citenamefont {Farhang}, \citenamefont {Kountz}, \citenamefont {Chen},
  \citenamefont {Ortiz}, \citenamefont {Shekhar}, \citenamefont {Felser},
  \citenamefont {Wilson}, \citenamefont {Thomale}, \citenamefont {Xia},\ and\
  \citenamefont {Kapitulnik}}]{Saykin2023_PRL}%
  \BibitemOpen
  \bibfield  {author} {\bibinfo {author} {\bibfnamefont {D.~R.}\ \bibnamefont
  {Saykin}}, \bibinfo {author} {\bibfnamefont {C.}~\bibnamefont {Farhang}},
  \bibinfo {author} {\bibfnamefont {E.~D.}\ \bibnamefont {Kountz}}, \bibinfo
  {author} {\bibfnamefont {D.}~\bibnamefont {Chen}}, \bibinfo {author}
  {\bibfnamefont {B.~R.}\ \bibnamefont {Ortiz}}, \bibinfo {author}
  {\bibfnamefont {C.}~\bibnamefont {Shekhar}}, \bibinfo {author} {\bibfnamefont
  {C.}~\bibnamefont {Felser}}, \bibinfo {author} {\bibfnamefont {S.~D.}\
  \bibnamefont {Wilson}}, \bibinfo {author} {\bibfnamefont {R.}~\bibnamefont
  {Thomale}}, \bibinfo {author} {\bibfnamefont {J.}~\bibnamefont {Xia}},\ and\
  \bibinfo {author} {\bibfnamefont {A.}~\bibnamefont {Kapitulnik}},\ }\href
  {https://doi.org/10.1103/PhysRevLett.131.016901} {\bibfield  {journal}
  {\bibinfo  {journal} {Phys. Rev. Lett.}\ }\textbf {\bibinfo {volume} {131}},\
  \bibinfo {pages} {016901} (\bibinfo {year} {2023})}\BibitemShut {NoStop}%
\bibitem [{\citenamefont {Yang}\ \emph {et~al.}(2020)\citenamefont {Yang},
  \citenamefont {Wang}, \citenamefont {Ortiz}, \citenamefont {Liu},
  \citenamefont {Gayles}, \citenamefont {Derunova}, \citenamefont
  {Gonzalez-Hernandez}, \citenamefont {\v{S}mejkal}, \citenamefont {Chen},
  \citenamefont {Parkin}, \citenamefont {Wilson}, \citenamefont {Toberer},
  \citenamefont {McQueen},\ and\ \citenamefont {Ali}}]{Yang2020_SciAdv}%
  \BibitemOpen
  \bibfield  {author} {\bibinfo {author} {\bibfnamefont {S.-Y.}\ \bibnamefont
  {Yang}}, \bibinfo {author} {\bibfnamefont {Y.}~\bibnamefont {Wang}}, \bibinfo
  {author} {\bibfnamefont {B.~R.}\ \bibnamefont {Ortiz}}, \bibinfo {author}
  {\bibfnamefont {D.}~\bibnamefont {Liu}}, \bibinfo {author} {\bibfnamefont
  {J.}~\bibnamefont {Gayles}}, \bibinfo {author} {\bibfnamefont
  {E.}~\bibnamefont {Derunova}}, \bibinfo {author} {\bibfnamefont
  {R.}~\bibnamefont {Gonzalez-Hernandez}}, \bibinfo {author} {\bibfnamefont
  {L.}~\bibnamefont {\v{S}mejkal}}, \bibinfo {author} {\bibfnamefont
  {Y.}~\bibnamefont {Chen}}, \bibinfo {author} {\bibfnamefont {S.~S.~P.}\
  \bibnamefont {Parkin}}, \bibinfo {author} {\bibfnamefont {S.~D.}\
  \bibnamefont {Wilson}}, \bibinfo {author} {\bibfnamefont {E.~S.}\
  \bibnamefont {Toberer}}, \bibinfo {author} {\bibfnamefont {T.}~\bibnamefont
  {McQueen}},\ and\ \bibinfo {author} {\bibfnamefont {M.~N.}\ \bibnamefont
  {Ali}},\ }\href {https://doi.org/10.1126/sciadv.abb6003} {\bibfield
  {journal} {\bibinfo  {journal} {Science Advances}\ }\textbf {\bibinfo
  {volume} {6}},\ \bibinfo {pages} {eabb6003} (\bibinfo {year}
  {2020})}\BibitemShut {NoStop}%
\bibitem [{\citenamefont {Yu}\ \emph {et~al.}(2021)\citenamefont {Yu},
  \citenamefont {Wu}, \citenamefont {Wang}, \citenamefont {Lei}, \citenamefont
  {Zhuo}, \citenamefont {Ying},\ and\ \citenamefont {Chen}}]{Yu2021_PRB}%
  \BibitemOpen
  \bibfield  {author} {\bibinfo {author} {\bibfnamefont {F.~H.}\ \bibnamefont
  {Yu}}, \bibinfo {author} {\bibfnamefont {T.}~\bibnamefont {Wu}}, \bibinfo
  {author} {\bibfnamefont {Z.~Y.}\ \bibnamefont {Wang}}, \bibinfo {author}
  {\bibfnamefont {B.}~\bibnamefont {Lei}}, \bibinfo {author} {\bibfnamefont
  {W.~Z.}\ \bibnamefont {Zhuo}}, \bibinfo {author} {\bibfnamefont {J.~J.}\
  \bibnamefont {Ying}},\ and\ \bibinfo {author} {\bibfnamefont {X.~H.}\
  \bibnamefont {Chen}},\ }\href {https://doi.org/10.1103/PhysRevB.104.L041103}
  {\bibfield  {journal} {\bibinfo  {journal} {Phys. Rev. B}\ }\textbf {\bibinfo
  {volume} {104}},\ \bibinfo {pages} {L041103} (\bibinfo {year}
  {2021})}\BibitemShut {NoStop}%
\bibitem [{\citenamefont {Neupert}\ \emph {et~al.}(2022)\citenamefont
  {Neupert}, \citenamefont {Denner}, \citenamefont {Yin}, \citenamefont
  {Thomale},\ and\ \citenamefont {Hasan}}]{Neupert2022_NatPhys}%
  \BibitemOpen
  \bibfield  {author} {\bibinfo {author} {\bibfnamefont {T.}~\bibnamefont
  {Neupert}}, \bibinfo {author} {\bibfnamefont {M.~M.}\ \bibnamefont {Denner}},
  \bibinfo {author} {\bibfnamefont {J.-X.}\ \bibnamefont {Yin}}, \bibinfo
  {author} {\bibfnamefont {R.}~\bibnamefont {Thomale}},\ and\ \bibinfo {author}
  {\bibfnamefont {M.~Z.}\ \bibnamefont {Hasan}},\ }\href
  {https://doi.org/10.1038/s41567-021-01404-y} {\bibfield  {journal} {\bibinfo
  {journal} {Nature Physics}\ }\textbf {\bibinfo {volume} {18}},\ \bibinfo
  {pages} {137} (\bibinfo {year} {2022})}\BibitemShut {NoStop}%
\bibitem [{\citenamefont {Khasanov}\ \emph {et~al.}(2022)\citenamefont
  {Khasanov}, \citenamefont {Das}, \citenamefont {Gupta}, \citenamefont
  {Mielke}, \citenamefont {Elender}, \citenamefont {Yin}, \citenamefont {Tu},
  \citenamefont {Gong}, \citenamefont {Lei}, \citenamefont {Ritz},
  \citenamefont {Fernandes}, \citenamefont {Birol}, \citenamefont {Guguchia},\
  and\ \citenamefont {Luetkens}}]{Khasanov2022_PRR}%
  \BibitemOpen
  \bibfield  {author} {\bibinfo {author} {\bibfnamefont {R.}~\bibnamefont
  {Khasanov}}, \bibinfo {author} {\bibfnamefont {D.}~\bibnamefont {Das}},
  \bibinfo {author} {\bibfnamefont {R.}~\bibnamefont {Gupta}}, \bibinfo
  {author} {\bibfnamefont {C.}~\bibnamefont {Mielke}}, \bibinfo {author}
  {\bibfnamefont {M.}~\bibnamefont {Elender}}, \bibinfo {author} {\bibfnamefont
  {Q.}~\bibnamefont {Yin}}, \bibinfo {author} {\bibfnamefont {Z.}~\bibnamefont
  {Tu}}, \bibinfo {author} {\bibfnamefont {C.}~\bibnamefont {Gong}}, \bibinfo
  {author} {\bibfnamefont {H.}~\bibnamefont {Lei}}, \bibinfo {author}
  {\bibfnamefont {E.~T.}\ \bibnamefont {Ritz}}, \bibinfo {author}
  {\bibfnamefont {R.~M.}\ \bibnamefont {Fernandes}}, \bibinfo {author}
  {\bibfnamefont {T.}~\bibnamefont {Birol}}, \bibinfo {author} {\bibfnamefont
  {Z.}~\bibnamefont {Guguchia}},\ and\ \bibinfo {author} {\bibfnamefont
  {H.}~\bibnamefont {Luetkens}},\ }\href
  {https://doi.org/10.1103/PhysRevResearch.4.023244} {\bibfield  {journal}
  {\bibinfo  {journal} {Phys. Rev. Research}\ }\textbf {\bibinfo {volume}
  {4}},\ \bibinfo {pages} {023244} (\bibinfo {year} {2022})}\BibitemShut
  {NoStop}%
\bibitem [{\citenamefont {Mielke}\ \emph {et~al.}(2022)\citenamefont {Mielke},
  \citenamefont {Das}, \citenamefont {Yin}, \citenamefont {Liu}, \citenamefont
  {Gupta}, \citenamefont {Jiang}, \citenamefont {Medarde}, \citenamefont {Wu},
  \citenamefont {Lei}, \citenamefont {Chang}, \citenamefont {Dai},
  \citenamefont {Si}, \citenamefont {Miao}, \citenamefont {Thomale},
  \citenamefont {Neupert}, \citenamefont {Shi}, \citenamefont {Khasanov},
  \citenamefont {Hasan}, \citenamefont {Luetkens},\ and\ \citenamefont
  {Guguchia}}]{Mielke2022_Nat}%
  \BibitemOpen
  \bibfield  {author} {\bibinfo {author} {\bibfnamefont {C.}~\bibnamefont
  {Mielke}}, \bibinfo {author} {\bibfnamefont {D.}~\bibnamefont {Das}},
  \bibinfo {author} {\bibfnamefont {J.-X.}\ \bibnamefont {Yin}}, \bibinfo
  {author} {\bibfnamefont {H.}~\bibnamefont {Liu}}, \bibinfo {author}
  {\bibfnamefont {R.}~\bibnamefont {Gupta}}, \bibinfo {author} {\bibfnamefont
  {Y.-X.}\ \bibnamefont {Jiang}}, \bibinfo {author} {\bibfnamefont
  {M.}~\bibnamefont {Medarde}}, \bibinfo {author} {\bibfnamefont
  {X.}~\bibnamefont {Wu}}, \bibinfo {author} {\bibfnamefont {H.~C.}\
  \bibnamefont {Lei}}, \bibinfo {author} {\bibfnamefont {J.}~\bibnamefont
  {Chang}}, \bibinfo {author} {\bibfnamefont {P.}~\bibnamefont {Dai}}, \bibinfo
  {author} {\bibfnamefont {Q.}~\bibnamefont {Si}}, \bibinfo {author}
  {\bibfnamefont {H.}~\bibnamefont {Miao}}, \bibinfo {author} {\bibfnamefont
  {R.}~\bibnamefont {Thomale}}, \bibinfo {author} {\bibfnamefont
  {T.}~\bibnamefont {Neupert}}, \bibinfo {author} {\bibfnamefont
  {Y.}~\bibnamefont {Shi}}, \bibinfo {author} {\bibfnamefont {R.}~\bibnamefont
  {Khasanov}}, \bibinfo {author} {\bibfnamefont {M.~Z.}\ \bibnamefont {Hasan}},
  \bibinfo {author} {\bibfnamefont {H.}~\bibnamefont {Luetkens}},\ and\
  \bibinfo {author} {\bibfnamefont {Z.}~\bibnamefont {Guguchia}},\ }\href
  {https://doi.org/10.1038/s41586-021-04327-z} {\bibfield  {journal} {\bibinfo
  {journal} {Nature}\ }\textbf {\bibinfo {volume} {602}},\ \bibinfo {pages}
  {245} (\bibinfo {year} {2022})}\BibitemShut {NoStop}%
\bibitem [{\citenamefont {Ortiz}\ \emph
  {et~al.}(2021{\natexlab{b}})\citenamefont {Ortiz}, \citenamefont {Sarte},
  \citenamefont {Kenney}, \citenamefont {Graf}, \citenamefont {Teicher},
  \citenamefont {Seshadri},\ and\ \citenamefont {Wilson}}]{Ortiz2021_PRM}%
  \BibitemOpen
  \bibfield  {author} {\bibinfo {author} {\bibfnamefont {B.~R.}\ \bibnamefont
  {Ortiz}}, \bibinfo {author} {\bibfnamefont {P.~M.}\ \bibnamefont {Sarte}},
  \bibinfo {author} {\bibfnamefont {E.~M.}\ \bibnamefont {Kenney}}, \bibinfo
  {author} {\bibfnamefont {M.~J.}\ \bibnamefont {Graf}}, \bibinfo {author}
  {\bibfnamefont {S.~M.~L.}\ \bibnamefont {Teicher}}, \bibinfo {author}
  {\bibfnamefont {R.}~\bibnamefont {Seshadri}},\ and\ \bibinfo {author}
  {\bibfnamefont {S.~D.}\ \bibnamefont {Wilson}},\ }\href
  {https://doi.org/10.1103/PhysRevMaterials.5.034801} {\bibfield  {journal}
  {\bibinfo  {journal} {Phys. Rev. Materials}\ }\textbf {\bibinfo {volume}
  {5}},\ \bibinfo {pages} {034801} (\bibinfo {year}
  {2021}{\natexlab{b}})}\BibitemShut {NoStop}%
\bibitem [{\citenamefont {Yin}\ \emph {et~al.}(2021)\citenamefont {Yin},
  \citenamefont {Tu}, \citenamefont {Gong}, \citenamefont {Fu}, \citenamefont
  {Yan},\ and\ \citenamefont {Lei}}]{Yin2021_ChinPhysLett}%
  \BibitemOpen
  \bibfield  {author} {\bibinfo {author} {\bibfnamefont {Q.}~\bibnamefont
  {Yin}}, \bibinfo {author} {\bibfnamefont {Z.}~\bibnamefont {Tu}}, \bibinfo
  {author} {\bibfnamefont {C.}~\bibnamefont {Gong}}, \bibinfo {author}
  {\bibfnamefont {Y.}~\bibnamefont {Fu}}, \bibinfo {author} {\bibfnamefont
  {S.}~\bibnamefont {Yan}},\ and\ \bibinfo {author} {\bibfnamefont
  {H.}~\bibnamefont {Lei}},\ }\href
  {https://doi.org/10.1088/0256-307X/38/3/037403} {\bibfield  {journal}
  {\bibinfo  {journal} {Chinese Physics Letters}\ }\textbf {\bibinfo {volume}
  {38}},\ \bibinfo {pages} {037403} (\bibinfo {year} {2021})}\BibitemShut
  {NoStop}%
\bibitem [{\citenamefont {Ortiz}\ \emph {et~al.}(2020)\citenamefont {Ortiz},
  \citenamefont {Teicher}, \citenamefont {Hu}, \citenamefont {Zuo},
  \citenamefont {Sarte}, \citenamefont {Schueller}, \citenamefont {Abeykoon},
  \citenamefont {Krogstad}, \citenamefont {Rosenkranz}, \citenamefont {Osborn},
  \citenamefont {Seshadri}, \citenamefont {Balents}, \citenamefont {He},\ and\
  \citenamefont {Wilson}}]{Ortiz2020_PRL}%
  \BibitemOpen
  \bibfield  {author} {\bibinfo {author} {\bibfnamefont {B.~R.}\ \bibnamefont
  {Ortiz}}, \bibinfo {author} {\bibfnamefont {S.~M.~L.}\ \bibnamefont
  {Teicher}}, \bibinfo {author} {\bibfnamefont {Y.}~\bibnamefont {Hu}},
  \bibinfo {author} {\bibfnamefont {J.~L.}\ \bibnamefont {Zuo}}, \bibinfo
  {author} {\bibfnamefont {P.~M.}\ \bibnamefont {Sarte}}, \bibinfo {author}
  {\bibfnamefont {E.~C.}\ \bibnamefont {Schueller}}, \bibinfo {author}
  {\bibfnamefont {A.~M.~M.}\ \bibnamefont {Abeykoon}}, \bibinfo {author}
  {\bibfnamefont {M.~J.}\ \bibnamefont {Krogstad}}, \bibinfo {author}
  {\bibfnamefont {S.}~\bibnamefont {Rosenkranz}}, \bibinfo {author}
  {\bibfnamefont {R.}~\bibnamefont {Osborn}}, \bibinfo {author} {\bibfnamefont
  {R.}~\bibnamefont {Seshadri}}, \bibinfo {author} {\bibfnamefont
  {L.}~\bibnamefont {Balents}}, \bibinfo {author} {\bibfnamefont
  {J.}~\bibnamefont {He}},\ and\ \bibinfo {author} {\bibfnamefont {S.~D.}\
  \bibnamefont {Wilson}},\ }\href
  {https://doi.org/10.1103/PhysRevLett.125.247002} {\bibfield  {journal}
  {\bibinfo  {journal} {Phys. Rev. Lett.}\ }\textbf {\bibinfo {volume} {125}},\
  \bibinfo {pages} {247002} (\bibinfo {year} {2020})}\BibitemShut {NoStop}%
\bibitem [{\citenamefont {Tazai}\ \emph {et~al.}(2022)\citenamefont {Tazai},
  \citenamefont {Yamakawa}, \citenamefont {Onari},\ and\ \citenamefont
  {Kontani}}]{Tazai2022_SciAdv}%
  \BibitemOpen
  \bibfield  {author} {\bibinfo {author} {\bibfnamefont {R.}~\bibnamefont
  {Tazai}}, \bibinfo {author} {\bibfnamefont {Y.}~\bibnamefont {Yamakawa}},
  \bibinfo {author} {\bibfnamefont {S.}~\bibnamefont {Onari}},\ and\ \bibinfo
  {author} {\bibfnamefont {H.}~\bibnamefont {Kontani}},\ }\href
  {https://doi.org/10.1126/sciadv.abl4108} {\bibfield  {journal} {\bibinfo
  {journal} {Science Advances}\ }\textbf {\bibinfo {volume} {8}},\ \bibinfo
  {pages} {eabl4108} (\bibinfo {year} {2022})}\BibitemShut {NoStop}%
\bibitem [{\citenamefont {Guguchia}\ \emph {et~al.}(2023)\citenamefont
  {Guguchia}, \citenamefont {Mielke}, \citenamefont {Das}, \citenamefont
  {Gupta}, \citenamefont {Yin}, \citenamefont {Liu}, \citenamefont {Yin},
  \citenamefont {Christensen}, \citenamefont {Tu}, \citenamefont {Gong},
  \citenamefont {Shumiya}, \citenamefont {Shafayat}, \citenamefont
  {Gamsakhurdashvili}, \citenamefont {Elender}, \citenamefont {Dai},
  \citenamefont {Amato}, \citenamefont {Shi}, \citenamefont {Lei},
  \citenamefont {Fernandes}, \citenamefont {Hasan}, \citenamefont {Luetkens},\
  and\ \citenamefont {Khasanov}}]{Guguchia2023_NatComm}%
  \BibitemOpen
  \bibfield  {author} {\bibinfo {author} {\bibfnamefont {Z.}~\bibnamefont
  {Guguchia}}, \bibinfo {author} {\bibfnamefont {C.}~\bibnamefont {Mielke}},
  \bibinfo {author} {\bibfnamefont {D.}~\bibnamefont {Das}}, \bibinfo {author}
  {\bibfnamefont {R.}~\bibnamefont {Gupta}}, \bibinfo {author} {\bibfnamefont
  {J.~X.}\ \bibnamefont {Yin}}, \bibinfo {author} {\bibfnamefont
  {H.}~\bibnamefont {Liu}}, \bibinfo {author} {\bibfnamefont {Q.}~\bibnamefont
  {Yin}}, \bibinfo {author} {\bibfnamefont {M.~H.}\ \bibnamefont
  {Christensen}}, \bibinfo {author} {\bibfnamefont {Z.}~\bibnamefont {Tu}},
  \bibinfo {author} {\bibfnamefont {C.}~\bibnamefont {Gong}}, \bibinfo {author}
  {\bibfnamefont {N.}~\bibnamefont {Shumiya}}, \bibinfo {author} {\bibfnamefont
  {M.~S.}\ \bibnamefont {Shafayat}}, \bibinfo {author} {\bibfnamefont
  {T.}~\bibnamefont {Gamsakhurdashvili}}, \bibinfo {author} {\bibfnamefont
  {M.}~\bibnamefont {Elender}}, \bibinfo {author} {\bibfnamefont
  {P.}~\bibnamefont {Dai}}, \bibinfo {author} {\bibfnamefont {A.}~\bibnamefont
  {Amato}}, \bibinfo {author} {\bibfnamefont {Y.}~\bibnamefont {Shi}}, \bibinfo
  {author} {\bibfnamefont {H.~C.}\ \bibnamefont {Lei}}, \bibinfo {author}
  {\bibfnamefont {R.~M.}\ \bibnamefont {Fernandes}}, \bibinfo {author}
  {\bibfnamefont {M.~Z.}\ \bibnamefont {Hasan}}, \bibinfo {author}
  {\bibfnamefont {H.}~\bibnamefont {Luetkens}},\ and\ \bibinfo {author}
  {\bibfnamefont {R.}~\bibnamefont {Khasanov}},\ }\href
  {https://doi.org/10.1038/s41467-022-35718-z} {\bibfield  {journal} {\bibinfo
  {journal} {Nature Communications}\ }\textbf {\bibinfo {volume} {14}},\
  \bibinfo {pages} {153} (\bibinfo {year} {2023})}\BibitemShut {NoStop}%
\bibitem [{\citenamefont {Chen}\ \emph {et~al.}(2021)\citenamefont {Chen},
  \citenamefont {Wang}, \citenamefont {Yin}, \citenamefont {Gu}, \citenamefont
  {Jiang}, \citenamefont {Tu}, \citenamefont {Gong}, \citenamefont {Uwatoko},
  \citenamefont {Sun}, \citenamefont {Lei}, \citenamefont {Hu},\ and\
  \citenamefont {Cheng}}]{Chen2021_PRL}%
  \BibitemOpen
  \bibfield  {author} {\bibinfo {author} {\bibfnamefont {K.~Y.}\ \bibnamefont
  {Chen}}, \bibinfo {author} {\bibfnamefont {N.~N.}\ \bibnamefont {Wang}},
  \bibinfo {author} {\bibfnamefont {Q.~W.}\ \bibnamefont {Yin}}, \bibinfo
  {author} {\bibfnamefont {Y.~H.}\ \bibnamefont {Gu}}, \bibinfo {author}
  {\bibfnamefont {K.}~\bibnamefont {Jiang}}, \bibinfo {author} {\bibfnamefont
  {Z.~J.}\ \bibnamefont {Tu}}, \bibinfo {author} {\bibfnamefont {C.~S.}\
  \bibnamefont {Gong}}, \bibinfo {author} {\bibfnamefont {Y.}~\bibnamefont
  {Uwatoko}}, \bibinfo {author} {\bibfnamefont {J.~P.}\ \bibnamefont {Sun}},
  \bibinfo {author} {\bibfnamefont {H.~C.}\ \bibnamefont {Lei}}, \bibinfo
  {author} {\bibfnamefont {J.~P.}\ \bibnamefont {Hu}},\ and\ \bibinfo {author}
  {\bibfnamefont {J.-G.}\ \bibnamefont {Cheng}},\ }\href
  {https://doi.org/10.1103/PhysRevLett.126.247001} {\bibfield  {journal}
  {\bibinfo  {journal} {Phys. Rev. Lett.}\ }\textbf {\bibinfo {volume} {126}},\
  \bibinfo {pages} {247001} (\bibinfo {year} {2021})}\BibitemShut {NoStop}%
\bibitem [{\citenamefont {Wang}\ \emph {et~al.}(2021)\citenamefont {Wang},
  \citenamefont {Chen}, \citenamefont {Yin}, \citenamefont {Ma}, \citenamefont
  {Pan}, \citenamefont {Yang}, \citenamefont {Ji}, \citenamefont {Wu},
  \citenamefont {Shan}, \citenamefont {Xu}, \citenamefont {Tu}, \citenamefont
  {Gong}, \citenamefont {Liu}, \citenamefont {Li}, \citenamefont {Uwatoko},
  \citenamefont {Dong}, \citenamefont {Lei}, \citenamefont {Sun},\ and\
  \citenamefont {Cheng}}]{Wang2021_PRR}%
  \BibitemOpen
  \bibfield  {author} {\bibinfo {author} {\bibfnamefont {N.~N.}\ \bibnamefont
  {Wang}}, \bibinfo {author} {\bibfnamefont {K.~Y.}\ \bibnamefont {Chen}},
  \bibinfo {author} {\bibfnamefont {Q.~W.}\ \bibnamefont {Yin}}, \bibinfo
  {author} {\bibfnamefont {Y.~N.~N.}\ \bibnamefont {Ma}}, \bibinfo {author}
  {\bibfnamefont {B.~Y.}\ \bibnamefont {Pan}}, \bibinfo {author} {\bibfnamefont
  {X.}~\bibnamefont {Yang}}, \bibinfo {author} {\bibfnamefont {X.~Y.}\
  \bibnamefont {Ji}}, \bibinfo {author} {\bibfnamefont {S.~L.}\ \bibnamefont
  {Wu}}, \bibinfo {author} {\bibfnamefont {P.~F.}\ \bibnamefont {Shan}},
  \bibinfo {author} {\bibfnamefont {S.~X.}\ \bibnamefont {Xu}}, \bibinfo
  {author} {\bibfnamefont {Z.~J.}\ \bibnamefont {Tu}}, \bibinfo {author}
  {\bibfnamefont {C.~S.}\ \bibnamefont {Gong}}, \bibinfo {author}
  {\bibfnamefont {G.~T.}\ \bibnamefont {Liu}}, \bibinfo {author} {\bibfnamefont
  {G.}~\bibnamefont {Li}}, \bibinfo {author} {\bibfnamefont {Y.}~\bibnamefont
  {Uwatoko}}, \bibinfo {author} {\bibfnamefont {X.~L.}\ \bibnamefont {Dong}},
  \bibinfo {author} {\bibfnamefont {H.~C.}\ \bibnamefont {Lei}}, \bibinfo
  {author} {\bibfnamefont {J.~P.}\ \bibnamefont {Sun}},\ and\ \bibinfo {author}
  {\bibfnamefont {J.-G.}\ \bibnamefont {Cheng}},\ }\href
  {https://doi.org/10.1103/PhysRevResearch.3.043018} {\bibfield  {journal}
  {\bibinfo  {journal} {Phys. Rev. Res.}\ }\textbf {\bibinfo {volume} {3}},\
  \bibinfo {pages} {043018} (\bibinfo {year} {2021})}\BibitemShut {NoStop}%
\bibitem [{\citenamefont {Du}\ \emph {et~al.}(2021)\citenamefont {Du},
  \citenamefont {Luo}, \citenamefont {Ortiz}, \citenamefont {Chen},
  \citenamefont {Duan}, \citenamefont {Zhang}, \citenamefont {Lu},
  \citenamefont {Wilson}, \citenamefont {Song},\ and\ \citenamefont
  {Yuan}}]{Du2021_PRB}%
  \BibitemOpen
  \bibfield  {author} {\bibinfo {author} {\bibfnamefont {F.}~\bibnamefont
  {Du}}, \bibinfo {author} {\bibfnamefont {S.}~\bibnamefont {Luo}}, \bibinfo
  {author} {\bibfnamefont {B.~R.}\ \bibnamefont {Ortiz}}, \bibinfo {author}
  {\bibfnamefont {Y.}~\bibnamefont {Chen}}, \bibinfo {author} {\bibfnamefont
  {W.}~\bibnamefont {Duan}}, \bibinfo {author} {\bibfnamefont {D.}~\bibnamefont
  {Zhang}}, \bibinfo {author} {\bibfnamefont {X.}~\bibnamefont {Lu}}, \bibinfo
  {author} {\bibfnamefont {S.~D.}\ \bibnamefont {Wilson}}, \bibinfo {author}
  {\bibfnamefont {Y.}~\bibnamefont {Song}},\ and\ \bibinfo {author}
  {\bibfnamefont {H.}~\bibnamefont {Yuan}},\ }\href
  {https://doi.org/10.1103/PhysRevB.103.L220504} {\bibfield  {journal}
  {\bibinfo  {journal} {Phys. Rev. B}\ }\textbf {\bibinfo {volume} {103}},\
  \bibinfo {pages} {L220504} (\bibinfo {year} {2021})}\BibitemShut {NoStop}%
\bibitem [{\citenamefont {Capa~Salinas}\ \emph {et~al.}(2023)\citenamefont
  {Capa~Salinas}, \citenamefont {Ortiz}, \citenamefont {Bales}, \citenamefont
  {Frassineti}, \citenamefont {Mitrovi\'{c}},\ and\ \citenamefont
  {Wilson}}]{CapaSalinas2023_FEM}%
  \BibitemOpen
  \bibfield  {author} {\bibinfo {author} {\bibfnamefont {A.~N.}\ \bibnamefont
  {Capa~Salinas}}, \bibinfo {author} {\bibfnamefont {B.~R.}\ \bibnamefont
  {Ortiz}}, \bibinfo {author} {\bibfnamefont {C.}~\bibnamefont {Bales}},
  \bibinfo {author} {\bibfnamefont {J.}~\bibnamefont {Frassineti}}, \bibinfo
  {author} {\bibfnamefont {V.~F.}\ \bibnamefont {Mitrovi\'{c}}},\ and\ \bibinfo
  {author} {\bibfnamefont {S.~D.}\ \bibnamefont {Wilson}},\ }\bibfield
  {journal} {\bibinfo  {journal} {Frontiers in Electronic Materials}\ }\textbf
  {\bibinfo {volume} {3}},\ \href {https://doi.org/10.3389/femat.2023.1257490}
  {10.3389/femat.2023.1257490} (\bibinfo {year} {2023})\BibitemShut {NoStop}%
\bibitem [{\citenamefont {Yang}\ \emph {et~al.}(2022)\citenamefont {Yang},
  \citenamefont {Huang}, \citenamefont {Zhang}, \citenamefont {Zhao},
  \citenamefont {Shi}, \citenamefont {Luo}, \citenamefont {Zhao}, \citenamefont
  {Qian}, \citenamefont {Tan}, \citenamefont {Hu}, \citenamefont {Zhu},
  \citenamefont {Lu}, \citenamefont {Zhang}, \citenamefont {Sun}, \citenamefont
  {Cheng}, \citenamefont {Shen}, \citenamefont {Lin}, \citenamefont {Yan},
  \citenamefont {Zhou}, \citenamefont {Wang}, \citenamefont {Pennycook},
  \citenamefont {Chen}, \citenamefont {Dong}, \citenamefont {Zhou},\ and\
  \citenamefont {Gao}}]{Yang2022_SciBull}%
  \BibitemOpen
  \bibfield  {author} {\bibinfo {author} {\bibfnamefont {H.}~\bibnamefont
  {Yang}}, \bibinfo {author} {\bibfnamefont {Z.}~\bibnamefont {Huang}},
  \bibinfo {author} {\bibfnamefont {Y.}~\bibnamefont {Zhang}}, \bibinfo
  {author} {\bibfnamefont {Z.}~\bibnamefont {Zhao}}, \bibinfo {author}
  {\bibfnamefont {J.}~\bibnamefont {Shi}}, \bibinfo {author} {\bibfnamefont
  {H.}~\bibnamefont {Luo}}, \bibinfo {author} {\bibfnamefont {L.}~\bibnamefont
  {Zhao}}, \bibinfo {author} {\bibfnamefont {G.}~\bibnamefont {Qian}}, \bibinfo
  {author} {\bibfnamefont {H.}~\bibnamefont {Tan}}, \bibinfo {author}
  {\bibfnamefont {B.}~\bibnamefont {Hu}}, \bibinfo {author} {\bibfnamefont
  {K.}~\bibnamefont {Zhu}}, \bibinfo {author} {\bibfnamefont {Z.}~\bibnamefont
  {Lu}}, \bibinfo {author} {\bibfnamefont {H.}~\bibnamefont {Zhang}}, \bibinfo
  {author} {\bibfnamefont {J.}~\bibnamefont {Sun}}, \bibinfo {author}
  {\bibfnamefont {J.}~\bibnamefont {Cheng}}, \bibinfo {author} {\bibfnamefont
  {C.}~\bibnamefont {Shen}}, \bibinfo {author} {\bibfnamefont {X.}~\bibnamefont
  {Lin}}, \bibinfo {author} {\bibfnamefont {B.}~\bibnamefont {Yan}}, \bibinfo
  {author} {\bibfnamefont {X.}~\bibnamefont {Zhou}}, \bibinfo {author}
  {\bibfnamefont {Z.}~\bibnamefont {Wang}}, \bibinfo {author} {\bibfnamefont
  {S.~J.}\ \bibnamefont {Pennycook}}, \bibinfo {author} {\bibfnamefont
  {H.}~\bibnamefont {Chen}}, \bibinfo {author} {\bibfnamefont {X.}~\bibnamefont
  {Dong}}, \bibinfo {author} {\bibfnamefont {W.}~\bibnamefont {Zhou}},\ and\
  \bibinfo {author} {\bibfnamefont {H.-J.}\ \bibnamefont {Gao}},\ }\href
  {https://doi.org/https://doi.org/10.1016/j.scib.2022.10.015} {\bibfield
  {journal} {\bibinfo  {journal} {Science Bulletin}\ }\textbf {\bibinfo
  {volume} {67}},\ \bibinfo {pages} {2176} (\bibinfo {year}
  {2022})}\BibitemShut {NoStop}%
\bibitem [{\citenamefont {Wu}\ \emph {et~al.}(2024{\natexlab{a}})\citenamefont
  {Wu}, \citenamefont {Sun}, \citenamefont {Nie}, \citenamefont {Li},
  \citenamefont {Zhao}, \citenamefont {Rao}, \citenamefont {Yu}, \citenamefont
  {Shi}, \citenamefont {Xiang}, \citenamefont {Ying}, \citenamefont {Wang},\
  and\ \citenamefont {Chen}}]{Wu2024_ResSq}%
  \BibitemOpen
  \bibfield  {author} {\bibinfo {author} {\bibfnamefont {T.}~\bibnamefont
  {Wu}}, \bibinfo {author} {\bibfnamefont {K.}~\bibnamefont {Sun}}, \bibinfo
  {author} {\bibfnamefont {L.}~\bibnamefont {Nie}}, \bibinfo {author}
  {\bibfnamefont {H.}~\bibnamefont {Li}}, \bibinfo {author} {\bibfnamefont
  {J.}~\bibnamefont {Zhao}}, \bibinfo {author} {\bibfnamefont {H.}~\bibnamefont
  {Rao}}, \bibinfo {author} {\bibfnamefont {F.}~\bibnamefont {Yu}}, \bibinfo
  {author} {\bibfnamefont {M.}~\bibnamefont {Shi}}, \bibinfo {author}
  {\bibfnamefont {Z.}~\bibnamefont {Xiang}}, \bibinfo {author} {\bibfnamefont
  {J.}~\bibnamefont {Ying}}, \bibinfo {author} {\bibfnamefont {Z.}~\bibnamefont
  {Wang}},\ and\ \bibinfo {author} {\bibfnamefont {X.}~\bibnamefont {Chen}},\
  }\href {https://doi.org/10.21203/rs.3.rs-3757459/v1} {\bibfield  {journal}
  {\bibinfo  {journal} {Research Square}\ } (\bibinfo {year}
  {2024}{\natexlab{a}})}\BibitemShut {NoStop}%
\bibitem [{\citenamefont {Oey}\ \emph {et~al.}(2022{\natexlab{a}})\citenamefont
  {Oey}, \citenamefont {Kaboudvand}, \citenamefont {Ortiz}, \citenamefont
  {Seshadri},\ and\ \citenamefont {Wilson}}]{Oey2022_PRM2}%
  \BibitemOpen
  \bibfield  {author} {\bibinfo {author} {\bibfnamefont {Y.~M.}\ \bibnamefont
  {Oey}}, \bibinfo {author} {\bibfnamefont {F.}~\bibnamefont {Kaboudvand}},
  \bibinfo {author} {\bibfnamefont {B.~R.}\ \bibnamefont {Ortiz}}, \bibinfo
  {author} {\bibfnamefont {R.}~\bibnamefont {Seshadri}},\ and\ \bibinfo
  {author} {\bibfnamefont {S.~D.}\ \bibnamefont {Wilson}},\ }\href
  {https://doi.org/10.1103/PhysRevMaterials.6.074802} {\bibfield  {journal}
  {\bibinfo  {journal} {Phys. Rev. Mater.}\ }\textbf {\bibinfo {volume} {6}},\
  \bibinfo {pages} {074802} (\bibinfo {year} {2022}{\natexlab{a}})}\BibitemShut
  {NoStop}%
\bibitem [{\citenamefont {Ratcliff}\ \emph {et~al.}(2021)\citenamefont
  {Ratcliff}, \citenamefont {Hallett}, \citenamefont {Ortiz}, \citenamefont
  {Wilson},\ and\ \citenamefont {Harter}}]{Ratcliff2021_PRM}%
  \BibitemOpen
  \bibfield  {author} {\bibinfo {author} {\bibfnamefont {N.}~\bibnamefont
  {Ratcliff}}, \bibinfo {author} {\bibfnamefont {L.}~\bibnamefont {Hallett}},
  \bibinfo {author} {\bibfnamefont {B.~R.}\ \bibnamefont {Ortiz}}, \bibinfo
  {author} {\bibfnamefont {S.~D.}\ \bibnamefont {Wilson}},\ and\ \bibinfo
  {author} {\bibfnamefont {J.~W.}\ \bibnamefont {Harter}},\ }\href
  {https://doi.org/10.1103/PhysRevMaterials.5.L111801} {\bibfield  {journal}
  {\bibinfo  {journal} {Phys. Rev. Materials}\ }\textbf {\bibinfo {volume}
  {5}},\ \bibinfo {pages} {L111801} (\bibinfo {year} {2021})}\BibitemShut
  {NoStop}%
\bibitem [{\citenamefont {Christensen}\ \emph {et~al.}(2021)\citenamefont
  {Christensen}, \citenamefont {Birol}, \citenamefont {Andersen},\ and\
  \citenamefont {Fernandes}}]{Christensen2021_PRB}%
  \BibitemOpen
  \bibfield  {author} {\bibinfo {author} {\bibfnamefont {M.~H.}\ \bibnamefont
  {Christensen}}, \bibinfo {author} {\bibfnamefont {T.}~\bibnamefont {Birol}},
  \bibinfo {author} {\bibfnamefont {B.~M.}\ \bibnamefont {Andersen}},\ and\
  \bibinfo {author} {\bibfnamefont {R.~M.}\ \bibnamefont {Fernandes}},\ }\href
  {https://doi.org/10.1103/PhysRevB.104.214513} {\bibfield  {journal} {\bibinfo
   {journal} {Phys. Rev. B}\ }\textbf {\bibinfo {volume} {104}},\ \bibinfo
  {pages} {214513} (\bibinfo {year} {2021})}\BibitemShut {NoStop}%
\bibitem [{\citenamefont {Jin}\ \emph {et~al.}(2024)\citenamefont {Jin},
  \citenamefont {Ren}, \citenamefont {Tan}, \citenamefont {Xie}, \citenamefont
  {Lu}, \citenamefont {Zhang}, \citenamefont {Ji},\ and\ \citenamefont
  {Zhang}}]{Jin2024_PRL}%
  \BibitemOpen
  \bibfield  {author} {\bibinfo {author} {\bibfnamefont {F.}~\bibnamefont
  {Jin}}, \bibinfo {author} {\bibfnamefont {W.}~\bibnamefont {Ren}}, \bibinfo
  {author} {\bibfnamefont {M.}~\bibnamefont {Tan}}, \bibinfo {author}
  {\bibfnamefont {M.}~\bibnamefont {Xie}}, \bibinfo {author} {\bibfnamefont
  {B.}~\bibnamefont {Lu}}, \bibinfo {author} {\bibfnamefont {Z.}~\bibnamefont
  {Zhang}}, \bibinfo {author} {\bibfnamefont {J.}~\bibnamefont {Ji}},\ and\
  \bibinfo {author} {\bibfnamefont {Q.}~\bibnamefont {Zhang}},\ }\href
  {https://doi.org/10.1103/PhysRevLett.132.066501} {\bibfield  {journal}
  {\bibinfo  {journal} {Phys. Rev. Lett.}\ }\textbf {\bibinfo {volume} {132}},\
  \bibinfo {pages} {066501} (\bibinfo {year} {2024})}\BibitemShut {NoStop}%
\bibitem [{\citenamefont {Hecker}\ and\ \citenamefont
  {Schmalian}(2018)}]{Hecker2018_npj}%
  \BibitemOpen
  \bibfield  {author} {\bibinfo {author} {\bibfnamefont {M.}~\bibnamefont
  {Hecker}}\ and\ \bibinfo {author} {\bibfnamefont {J.}~\bibnamefont
  {Schmalian}},\ }\href {https://doi.org/10.1038/s41535-018-0098-z} {\bibfield
  {journal} {\bibinfo  {journal} {npj Quantum Materials}\ }\textbf {\bibinfo
  {volume} {3}},\ \bibinfo {pages} {26} (\bibinfo {year} {2018})}\BibitemShut
  {NoStop}%
\bibitem [{\citenamefont {Cho}\ \emph {et~al.}(2020)\citenamefont {Cho},
  \citenamefont {Shen}, \citenamefont {Lyu}, \citenamefont {Atanov},
  \citenamefont {Chen}, \citenamefont {Lee}, \citenamefont {Hor}, \citenamefont
  {Gawryluk}, \citenamefont {Pomjakushina}, \citenamefont {Bartkowiak},
  \citenamefont {Hecker}, \citenamefont {Schmalian},\ and\ \citenamefont
  {Lortz}}]{Cho2020_NatComm}%
  \BibitemOpen
  \bibfield  {author} {\bibinfo {author} {\bibfnamefont {C.-w.}\ \bibnamefont
  {Cho}}, \bibinfo {author} {\bibfnamefont {J.}~\bibnamefont {Shen}}, \bibinfo
  {author} {\bibfnamefont {J.}~\bibnamefont {Lyu}}, \bibinfo {author}
  {\bibfnamefont {O.}~\bibnamefont {Atanov}}, \bibinfo {author} {\bibfnamefont
  {Q.}~\bibnamefont {Chen}}, \bibinfo {author} {\bibfnamefont {S.~H.}\
  \bibnamefont {Lee}}, \bibinfo {author} {\bibfnamefont {Y.~S.}\ \bibnamefont
  {Hor}}, \bibinfo {author} {\bibfnamefont {D.~J.}\ \bibnamefont {Gawryluk}},
  \bibinfo {author} {\bibfnamefont {E.}~\bibnamefont {Pomjakushina}}, \bibinfo
  {author} {\bibfnamefont {M.}~\bibnamefont {Bartkowiak}}, \bibinfo {author}
  {\bibfnamefont {M.}~\bibnamefont {Hecker}}, \bibinfo {author} {\bibfnamefont
  {J.}~\bibnamefont {Schmalian}},\ and\ \bibinfo {author} {\bibfnamefont
  {R.}~\bibnamefont {Lortz}},\ }\href
  {https://doi.org/10.1038/s41467-020-16871-9} {\bibfield  {journal} {\bibinfo
  {journal} {Nature Communications}\ }\textbf {\bibinfo {volume} {11}},\
  \bibinfo {pages} {3056} (\bibinfo {year} {2020})}\BibitemShut {NoStop}%
\bibitem [{\citenamefont {Tazai}\ \emph {et~al.}(2023)\citenamefont {Tazai},
  \citenamefont {Yamakawa},\ and\ \citenamefont {Kontani}}]{Tazai2023_NatComm}%
  \BibitemOpen
  \bibfield  {author} {\bibinfo {author} {\bibfnamefont {R.}~\bibnamefont
  {Tazai}}, \bibinfo {author} {\bibfnamefont {Y.}~\bibnamefont {Yamakawa}},\
  and\ \bibinfo {author} {\bibfnamefont {H.}~\bibnamefont {Kontani}},\ }\href
  {https://doi.org/10.1038/s41467-023-42952-6} {\bibfield  {journal} {\bibinfo
  {journal} {Nature Communications}\ }\textbf {\bibinfo {volume} {14}},\
  \bibinfo {pages} {7845} (\bibinfo {year} {2023})}\BibitemShut {NoStop}%
\bibitem [{\citenamefont {Zhao}\ \emph {et~al.}(2021)\citenamefont {Zhao},
  \citenamefont {Li}, \citenamefont {Ortiz}, \citenamefont {Teicher},
  \citenamefont {Park}, \citenamefont {Ye}, \citenamefont {Wang}, \citenamefont
  {Balents}, \citenamefont {Wilson},\ and\ \citenamefont
  {Zeljkovic}}]{Zhao2021_Nat}%
  \BibitemOpen
  \bibfield  {author} {\bibinfo {author} {\bibfnamefont {H.}~\bibnamefont
  {Zhao}}, \bibinfo {author} {\bibfnamefont {H.}~\bibnamefont {Li}}, \bibinfo
  {author} {\bibfnamefont {B.~R.}\ \bibnamefont {Ortiz}}, \bibinfo {author}
  {\bibfnamefont {S.~M.~L.}\ \bibnamefont {Teicher}}, \bibinfo {author}
  {\bibfnamefont {T.}~\bibnamefont {Park}}, \bibinfo {author} {\bibfnamefont
  {M.}~\bibnamefont {Ye}}, \bibinfo {author} {\bibfnamefont {Z.}~\bibnamefont
  {Wang}}, \bibinfo {author} {\bibfnamefont {L.}~\bibnamefont {Balents}},
  \bibinfo {author} {\bibfnamefont {S.~D.}\ \bibnamefont {Wilson}},\ and\
  \bibinfo {author} {\bibfnamefont {I.}~\bibnamefont {Zeljkovic}},\ }\href
  {https://doi.org/10.1038/s41586-021-03946-w} {\bibfield  {journal} {\bibinfo
  {journal} {Nature}\ }\textbf {\bibinfo {volume} {599}},\ \bibinfo {pages}
  {216} (\bibinfo {year} {2021})}\BibitemShut {NoStop}%
\bibitem [{\citenamefont {Guo}\ \emph {et~al.}(2022)\citenamefont {Guo},
  \citenamefont {Putzke}, \citenamefont {Konyzheva}, \citenamefont {Huang},
  \citenamefont {Gutierrez-Amigo}, \citenamefont {Errea}, \citenamefont {Chen},
  \citenamefont {Vergniory}, \citenamefont {Felser}, \citenamefont {Fischer},
  \citenamefont {Neupert},\ and\ \citenamefont {Moll}}]{Guo2022_Nat}%
  \BibitemOpen
  \bibfield  {author} {\bibinfo {author} {\bibfnamefont {C.}~\bibnamefont
  {Guo}}, \bibinfo {author} {\bibfnamefont {C.}~\bibnamefont {Putzke}},
  \bibinfo {author} {\bibfnamefont {S.}~\bibnamefont {Konyzheva}}, \bibinfo
  {author} {\bibfnamefont {X.}~\bibnamefont {Huang}}, \bibinfo {author}
  {\bibfnamefont {M.}~\bibnamefont {Gutierrez-Amigo}}, \bibinfo {author}
  {\bibfnamefont {I.}~\bibnamefont {Errea}}, \bibinfo {author} {\bibfnamefont
  {D.}~\bibnamefont {Chen}}, \bibinfo {author} {\bibfnamefont {M.~G.}\
  \bibnamefont {Vergniory}}, \bibinfo {author} {\bibfnamefont {C.}~\bibnamefont
  {Felser}}, \bibinfo {author} {\bibfnamefont {M.~H.}\ \bibnamefont {Fischer}},
  \bibinfo {author} {\bibfnamefont {T.}~\bibnamefont {Neupert}},\ and\ \bibinfo
  {author} {\bibfnamefont {P.~J.~W.}\ \bibnamefont {Moll}},\ }\bibfield
  {journal} {\bibinfo  {journal} {Nature}\ }\href
  {https://doi.org/10.1038/s41586-022-05127-9} {10.1038/s41586-022-05127-9}
  (\bibinfo {year} {2022})\BibitemShut {NoStop}%
\bibitem [{\citenamefont {Li}\ \emph {et~al.}(2022{\natexlab{b}})\citenamefont
  {Li}, \citenamefont {Wan}, \citenamefont {Li}, \citenamefont {Li},
  \citenamefont {Gu}, \citenamefont {Yang}, \citenamefont {Li}, \citenamefont
  {Wang}, \citenamefont {Yao},\ and\ \citenamefont {Wen}}]{Li2022_PRB}%
  \BibitemOpen
  \bibfield  {author} {\bibinfo {author} {\bibfnamefont {H.}~\bibnamefont
  {Li}}, \bibinfo {author} {\bibfnamefont {S.}~\bibnamefont {Wan}}, \bibinfo
  {author} {\bibfnamefont {H.}~\bibnamefont {Li}}, \bibinfo {author}
  {\bibfnamefont {Q.}~\bibnamefont {Li}}, \bibinfo {author} {\bibfnamefont
  {Q.}~\bibnamefont {Gu}}, \bibinfo {author} {\bibfnamefont {H.}~\bibnamefont
  {Yang}}, \bibinfo {author} {\bibfnamefont {Y.}~\bibnamefont {Li}}, \bibinfo
  {author} {\bibfnamefont {Z.}~\bibnamefont {Wang}}, \bibinfo {author}
  {\bibfnamefont {Y.}~\bibnamefont {Yao}},\ and\ \bibinfo {author}
  {\bibfnamefont {H.-H.}\ \bibnamefont {Wen}},\ }\href
  {https://doi.org/10.1103/PhysRevB.105.045102} {\bibfield  {journal} {\bibinfo
   {journal} {Phys. Rev. B}\ }\textbf {\bibinfo {volume} {105}},\ \bibinfo
  {pages} {045102} (\bibinfo {year} {2022}{\natexlab{b}})}\BibitemShut
  {NoStop}%
\bibitem [{\citenamefont {Li}\ \emph {et~al.}(2023{\natexlab{a}})\citenamefont
  {Li}, \citenamefont {Oh}, \citenamefont {Kang}, \citenamefont {Zhao},
  \citenamefont {Ortiz}, \citenamefont {Oey}, \citenamefont {Fang},
  \citenamefont {Ren}, \citenamefont {Jozwiak}, \citenamefont {Bostwick},
  \citenamefont {Rotenberg}, \citenamefont {Checkelsky}, \citenamefont {Wang},
  \citenamefont {Wilson}, \citenamefont {Comin},\ and\ \citenamefont
  {Zeljkovic}}]{Li2023_PRX}%
  \BibitemOpen
  \bibfield  {author} {\bibinfo {author} {\bibfnamefont {H.}~\bibnamefont
  {Li}}, \bibinfo {author} {\bibfnamefont {D.}~\bibnamefont {Oh}}, \bibinfo
  {author} {\bibfnamefont {M.}~\bibnamefont {Kang}}, \bibinfo {author}
  {\bibfnamefont {H.}~\bibnamefont {Zhao}}, \bibinfo {author} {\bibfnamefont
  {B.~R.}\ \bibnamefont {Ortiz}}, \bibinfo {author} {\bibfnamefont
  {Y.}~\bibnamefont {Oey}}, \bibinfo {author} {\bibfnamefont {S.}~\bibnamefont
  {Fang}}, \bibinfo {author} {\bibfnamefont {Z.}~\bibnamefont {Ren}}, \bibinfo
  {author} {\bibfnamefont {C.}~\bibnamefont {Jozwiak}}, \bibinfo {author}
  {\bibfnamefont {A.}~\bibnamefont {Bostwick}}, \bibinfo {author}
  {\bibfnamefont {E.}~\bibnamefont {Rotenberg}}, \bibinfo {author}
  {\bibfnamefont {J.~G.}\ \bibnamefont {Checkelsky}}, \bibinfo {author}
  {\bibfnamefont {Z.}~\bibnamefont {Wang}}, \bibinfo {author} {\bibfnamefont
  {S.~D.}\ \bibnamefont {Wilson}}, \bibinfo {author} {\bibfnamefont
  {R.}~\bibnamefont {Comin}},\ and\ \bibinfo {author} {\bibfnamefont
  {I.}~\bibnamefont {Zeljkovic}},\ }\href
  {https://doi.org/10.1103/PhysRevX.13.031030} {\bibfield  {journal} {\bibinfo
  {journal} {Phys. Rev. X}\ }\textbf {\bibinfo {volume} {13}},\ \bibinfo
  {pages} {031030} (\bibinfo {year} {2023}{\natexlab{a}})}\BibitemShut
  {NoStop}%
\bibitem [{\citenamefont {Li}\ \emph {et~al.}(2023{\natexlab{b}})\citenamefont
  {Li}, \citenamefont {Cheng}, \citenamefont {Ortiz}, \citenamefont {Tan},
  \citenamefont {Werhahn}, \citenamefont {Zeng}, \citenamefont {Johrendt},
  \citenamefont {Yan}, \citenamefont {Wang}, \citenamefont {Wilson},\ and\
  \citenamefont {Zeljkovic}}]{Li2023_NatPhys}%
  \BibitemOpen
  \bibfield  {author} {\bibinfo {author} {\bibfnamefont {H.}~\bibnamefont
  {Li}}, \bibinfo {author} {\bibfnamefont {S.}~\bibnamefont {Cheng}}, \bibinfo
  {author} {\bibfnamefont {B.~R.}\ \bibnamefont {Ortiz}}, \bibinfo {author}
  {\bibfnamefont {H.}~\bibnamefont {Tan}}, \bibinfo {author} {\bibfnamefont
  {D.}~\bibnamefont {Werhahn}}, \bibinfo {author} {\bibfnamefont
  {K.}~\bibnamefont {Zeng}}, \bibinfo {author} {\bibfnamefont {D.}~\bibnamefont
  {Johrendt}}, \bibinfo {author} {\bibfnamefont {B.}~\bibnamefont {Yan}},
  \bibinfo {author} {\bibfnamefont {Z.}~\bibnamefont {Wang}}, \bibinfo {author}
  {\bibfnamefont {S.~D.}\ \bibnamefont {Wilson}},\ and\ \bibinfo {author}
  {\bibfnamefont {I.}~\bibnamefont {Zeljkovic}},\ }\href
  {https://doi.org/10.1038/s41567-023-02176-3} {\bibfield  {journal} {\bibinfo
  {journal} {Nature Physics}\ } (\bibinfo {year}
  {2023}{\natexlab{b}})}\BibitemShut {NoStop}%
\bibitem [{\citenamefont {Liu}\ \emph {et~al.}(2024)\citenamefont {Liu},
  \citenamefont {Shi}, \citenamefont {Jiang}, \citenamefont {Rosenberg},
  \citenamefont {DeStefano}, \citenamefont {Liu}, \citenamefont {Hu},
  \citenamefont {Zhao}, \citenamefont {Wang}, \citenamefont {Yao},
  \citenamefont {Graf}, \citenamefont {Dai}, \citenamefont {Yang},
  \citenamefont {Xu},\ and\ \citenamefont {Chu}}]{Liu2024_PRX}%
  \BibitemOpen
  \bibfield  {author} {\bibinfo {author} {\bibfnamefont {Z.}~\bibnamefont
  {Liu}}, \bibinfo {author} {\bibfnamefont {Y.}~\bibnamefont {Shi}}, \bibinfo
  {author} {\bibfnamefont {Q.}~\bibnamefont {Jiang}}, \bibinfo {author}
  {\bibfnamefont {E.~W.}\ \bibnamefont {Rosenberg}}, \bibinfo {author}
  {\bibfnamefont {J.~M.}\ \bibnamefont {DeStefano}}, \bibinfo {author}
  {\bibfnamefont {J.}~\bibnamefont {Liu}}, \bibinfo {author} {\bibfnamefont
  {C.}~\bibnamefont {Hu}}, \bibinfo {author} {\bibfnamefont {Y.}~\bibnamefont
  {Zhao}}, \bibinfo {author} {\bibfnamefont {Z.}~\bibnamefont {Wang}}, \bibinfo
  {author} {\bibfnamefont {Y.}~\bibnamefont {Yao}}, \bibinfo {author}
  {\bibfnamefont {D.}~\bibnamefont {Graf}}, \bibinfo {author} {\bibfnamefont
  {P.}~\bibnamefont {Dai}}, \bibinfo {author} {\bibfnamefont {J.}~\bibnamefont
  {Yang}}, \bibinfo {author} {\bibfnamefont {X.}~\bibnamefont {Xu}},\ and\
  \bibinfo {author} {\bibfnamefont {J.-H.}\ \bibnamefont {Chu}},\ }\href
  {https://doi.org/10.1103/PhysRevX.14.031015} {\bibfield  {journal} {\bibinfo
  {journal} {Phys. Rev. X}\ }\textbf {\bibinfo {volume} {14}},\ \bibinfo
  {pages} {031015} (\bibinfo {year} {2024})}\BibitemShut {NoStop}%
\bibitem [{\citenamefont {Frachet}\ \emph {et~al.}(2024)\citenamefont
  {Frachet}, \citenamefont {Wang}, \citenamefont {Xia}, \citenamefont {Guo},
  \citenamefont {He}, \citenamefont {Maraytta}, \citenamefont {Heid},
  \citenamefont {Haghighirad}, \citenamefont {Merz}, \citenamefont {Meingast},\
  and\ \citenamefont {Hardy}}]{Frachet2024_PRL}%
  \BibitemOpen
  \bibfield  {author} {\bibinfo {author} {\bibfnamefont {M.}~\bibnamefont
  {Frachet}}, \bibinfo {author} {\bibfnamefont {L.}~\bibnamefont {Wang}},
  \bibinfo {author} {\bibfnamefont {W.}~\bibnamefont {Xia}}, \bibinfo {author}
  {\bibfnamefont {Y.}~\bibnamefont {Guo}}, \bibinfo {author} {\bibfnamefont
  {M.}~\bibnamefont {He}}, \bibinfo {author} {\bibfnamefont {N.}~\bibnamefont
  {Maraytta}}, \bibinfo {author} {\bibfnamefont {R.}~\bibnamefont {Heid}},
  \bibinfo {author} {\bibfnamefont {A.-A.}\ \bibnamefont {Haghighirad}},
  \bibinfo {author} {\bibfnamefont {M.}~\bibnamefont {Merz}}, \bibinfo {author}
  {\bibfnamefont {C.}~\bibnamefont {Meingast}},\ and\ \bibinfo {author}
  {\bibfnamefont {F.}~\bibnamefont {Hardy}},\ }\href
  {https://doi.org/10.1103/PhysRevLett.132.186001} {\bibfield  {journal}
  {\bibinfo  {journal} {Phys. Rev. Lett.}\ }\textbf {\bibinfo {volume} {132}},\
  \bibinfo {pages} {186001} (\bibinfo {year} {2024})}\BibitemShut {NoStop}%
\bibitem [{\citenamefont {Guo}\ \emph {et~al.}(2024)\citenamefont {Guo},
  \citenamefont {Wagner}, \citenamefont {Putzke}, \citenamefont {Chen},
  \citenamefont {Wang}, \citenamefont {Zhang}, \citenamefont {Gutierrez-Amigo},
  \citenamefont {Errea}, \citenamefont {Vergniory}, \citenamefont {Felser},
  \citenamefont {Fischer}, \citenamefont {Neupert},\ and\ \citenamefont
  {Moll}}]{Guo2024_NatPhys}%
  \BibitemOpen
  \bibfield  {author} {\bibinfo {author} {\bibfnamefont {C.}~\bibnamefont
  {Guo}}, \bibinfo {author} {\bibfnamefont {G.}~\bibnamefont {Wagner}},
  \bibinfo {author} {\bibfnamefont {C.}~\bibnamefont {Putzke}}, \bibinfo
  {author} {\bibfnamefont {D.}~\bibnamefont {Chen}}, \bibinfo {author}
  {\bibfnamefont {K.}~\bibnamefont {Wang}}, \bibinfo {author} {\bibfnamefont
  {L.}~\bibnamefont {Zhang}}, \bibinfo {author} {\bibfnamefont
  {M.}~\bibnamefont {Gutierrez-Amigo}}, \bibinfo {author} {\bibfnamefont
  {I.}~\bibnamefont {Errea}}, \bibinfo {author} {\bibfnamefont {M.~G.}\
  \bibnamefont {Vergniory}}, \bibinfo {author} {\bibfnamefont {C.}~\bibnamefont
  {Felser}}, \bibinfo {author} {\bibfnamefont {M.~H.}\ \bibnamefont {Fischer}},
  \bibinfo {author} {\bibfnamefont {T.}~\bibnamefont {Neupert}},\ and\ \bibinfo
  {author} {\bibfnamefont {P.~J.~W.}\ \bibnamefont {Moll}},\ }\href
  {https://doi.org/10.1038/s41567-023-02374-z} {\bibfield  {journal} {\bibinfo
  {journal} {Nature Physics}\ }\textbf {\bibinfo {volume} {20}},\ \bibinfo
  {pages} {579} (\bibinfo {year} {2024})}\BibitemShut {NoStop}%
\bibitem [{\citenamefont {Wu}\ \emph {et~al.}(2021)\citenamefont {Wu},
  \citenamefont {Schwemmer}, \citenamefont {M\"uller}, \citenamefont
  {Consiglio}, \citenamefont {Sangiovanni}, \citenamefont {Di~Sante},
  \citenamefont {Iqbal}, \citenamefont {Hanke}, \citenamefont {Schnyder},
  \citenamefont {Denner}, \citenamefont {Fischer}, \citenamefont {Neupert},\
  and\ \citenamefont {Thomale}}]{Wu2021_PRL}%
  \BibitemOpen
  \bibfield  {author} {\bibinfo {author} {\bibfnamefont {X.}~\bibnamefont
  {Wu}}, \bibinfo {author} {\bibfnamefont {T.}~\bibnamefont {Schwemmer}},
  \bibinfo {author} {\bibfnamefont {T.}~\bibnamefont {M\"uller}}, \bibinfo
  {author} {\bibfnamefont {A.}~\bibnamefont {Consiglio}}, \bibinfo {author}
  {\bibfnamefont {G.}~\bibnamefont {Sangiovanni}}, \bibinfo {author}
  {\bibfnamefont {D.}~\bibnamefont {Di~Sante}}, \bibinfo {author}
  {\bibfnamefont {Y.}~\bibnamefont {Iqbal}}, \bibinfo {author} {\bibfnamefont
  {W.}~\bibnamefont {Hanke}}, \bibinfo {author} {\bibfnamefont {A.~P.}\
  \bibnamefont {Schnyder}}, \bibinfo {author} {\bibfnamefont {M.~M.}\
  \bibnamefont {Denner}}, \bibinfo {author} {\bibfnamefont {M.~H.}\
  \bibnamefont {Fischer}}, \bibinfo {author} {\bibfnamefont {T.}~\bibnamefont
  {Neupert}},\ and\ \bibinfo {author} {\bibfnamefont {R.}~\bibnamefont
  {Thomale}},\ }\href {https://doi.org/10.1103/PhysRevLett.127.177001}
  {\bibfield  {journal} {\bibinfo  {journal} {Phys. Rev. Lett.}\ }\textbf
  {\bibinfo {volume} {127}},\ \bibinfo {pages} {177001} (\bibinfo {year}
  {2021})}\BibitemShut {NoStop}%
\bibitem [{\citenamefont {Denner}\ \emph {et~al.}(2021)\citenamefont {Denner},
  \citenamefont {Thomale},\ and\ \citenamefont {Neupert}}]{Denner2021_PRL}%
  \BibitemOpen
  \bibfield  {author} {\bibinfo {author} {\bibfnamefont {M.~M.}\ \bibnamefont
  {Denner}}, \bibinfo {author} {\bibfnamefont {R.}~\bibnamefont {Thomale}},\
  and\ \bibinfo {author} {\bibfnamefont {T.}~\bibnamefont {Neupert}},\ }\href
  {https://doi.org/10.1103/PhysRevLett.127.217601} {\bibfield  {journal}
  {\bibinfo  {journal} {Phys. Rev. Lett.}\ }\textbf {\bibinfo {volume} {127}},\
  \bibinfo {pages} {217601} (\bibinfo {year} {2021})}\BibitemShut {NoStop}%
\bibitem [{\citenamefont {Park}\ \emph {et~al.}(2021)\citenamefont {Park},
  \citenamefont {Ye},\ and\ \citenamefont {Balents}}]{Park2021_PRB}%
  \BibitemOpen
  \bibfield  {author} {\bibinfo {author} {\bibfnamefont {T.}~\bibnamefont
  {Park}}, \bibinfo {author} {\bibfnamefont {M.}~\bibnamefont {Ye}},\ and\
  \bibinfo {author} {\bibfnamefont {L.}~\bibnamefont {Balents}},\ }\href
  {https://doi.org/10.1103/PhysRevB.104.035142} {\bibfield  {journal} {\bibinfo
   {journal} {Phys. Rev. B}\ }\textbf {\bibinfo {volume} {104}},\ \bibinfo
  {pages} {035142} (\bibinfo {year} {2021})}\BibitemShut {NoStop}%
\bibitem [{\citenamefont {Lin}\ and\ \citenamefont
  {Nandkishore}(2021)}]{Lin2021_PRB}%
  \BibitemOpen
  \bibfield  {author} {\bibinfo {author} {\bibfnamefont {Y.-P.}\ \bibnamefont
  {Lin}}\ and\ \bibinfo {author} {\bibfnamefont {R.~M.}\ \bibnamefont
  {Nandkishore}},\ }\href {https://doi.org/10.1103/PhysRevB.104.045122}
  {\bibfield  {journal} {\bibinfo  {journal} {Phys. Rev. B}\ }\textbf {\bibinfo
  {volume} {104}},\ \bibinfo {pages} {045122} (\bibinfo {year}
  {2021})}\BibitemShut {NoStop}%
\bibitem [{\citenamefont {Dong}\ \emph
  {et~al.}(2023{\natexlab{a}})\citenamefont {Dong}, \citenamefont {Wang},\ and\
  \citenamefont {Zhou}}]{Dong2023_PRB}%
  \BibitemOpen
  \bibfield  {author} {\bibinfo {author} {\bibfnamefont {J.-W.}\ \bibnamefont
  {Dong}}, \bibinfo {author} {\bibfnamefont {Z.}~\bibnamefont {Wang}},\ and\
  \bibinfo {author} {\bibfnamefont {S.}~\bibnamefont {Zhou}},\ }\href
  {https://doi.org/10.1103/PhysRevB.107.045127} {\bibfield  {journal} {\bibinfo
   {journal} {Phys. Rev. B}\ }\textbf {\bibinfo {volume} {107}},\ \bibinfo
  {pages} {045127} (\bibinfo {year} {2023}{\natexlab{a}})}\BibitemShut
  {NoStop}%
\bibitem [{\citenamefont {Profe}\ \emph {et~al.}(2024)\citenamefont {Profe},
  \citenamefont {Klebl}, \citenamefont {Grandi}, \citenamefont {Hohmann},
  \citenamefont {D\"{u}rrnagel}, \citenamefont {Schwemmer}, \citenamefont
  {Thomale},\ and\ \citenamefont {Kennes}}]{Profe2024_PRR}%
  \BibitemOpen
  \bibfield  {author} {\bibinfo {author} {\bibfnamefont {J.~B.}\ \bibnamefont
  {Profe}}, \bibinfo {author} {\bibfnamefont {L.}~\bibnamefont {Klebl}},
  \bibinfo {author} {\bibfnamefont {F.}~\bibnamefont {Grandi}}, \bibinfo
  {author} {\bibfnamefont {H.}~\bibnamefont {Hohmann}}, \bibinfo {author}
  {\bibfnamefont {M.}~\bibnamefont {D\"{u}rrnagel}}, \bibinfo {author}
  {\bibfnamefont {T.}~\bibnamefont {Schwemmer}}, \bibinfo {author}
  {\bibfnamefont {R.}~\bibnamefont {Thomale}},\ and\ \bibinfo {author}
  {\bibfnamefont {D.~M.}\ \bibnamefont {Kennes}},\ }\href
  {https://doi.org/10.1103/PhysRevResearch.6.043078} {\bibfield  {journal}
  {\bibinfo  {journal} {Phys. Rev. Res.}\ }\textbf {\bibinfo {volume} {6}},\
  \bibinfo {pages} {043078} (\bibinfo {year} {2024})}\BibitemShut {NoStop}%
\bibitem [{\citenamefont {Chen}\ \emph {et~al.}(2022)\citenamefont {Chen},
  \citenamefont {Chen}, \citenamefont {Schnelle}, \citenamefont {Felser},\ and\
  \citenamefont {Gaulin}}]{Chen2022_PRL}%
  \BibitemOpen
  \bibfield  {author} {\bibinfo {author} {\bibfnamefont {Q.}~\bibnamefont
  {Chen}}, \bibinfo {author} {\bibfnamefont {D.}~\bibnamefont {Chen}}, \bibinfo
  {author} {\bibfnamefont {W.}~\bibnamefont {Schnelle}}, \bibinfo {author}
  {\bibfnamefont {C.}~\bibnamefont {Felser}},\ and\ \bibinfo {author}
  {\bibfnamefont {B.~D.}\ \bibnamefont {Gaulin}},\ }\href
  {https://doi.org/10.1103/PhysRevLett.129.056401} {\bibfield  {journal}
  {\bibinfo  {journal} {Phys. Rev. Lett.}\ }\textbf {\bibinfo {volume} {129}},\
  \bibinfo {pages} {056401} (\bibinfo {year} {2022})}\BibitemShut {NoStop}%
\bibitem [{\citenamefont {Yang}\ \emph {et~al.}(2023)\citenamefont {Yang},
  \citenamefont {Xia}, \citenamefont {Mi}, \citenamefont {Zhang}, \citenamefont
  {Gan}, \citenamefont {Wang}, \citenamefont {Chai}, \citenamefont {Zhou},
  \citenamefont {Yang}, \citenamefont {Guo},\ and\ \citenamefont
  {He}}]{Yang2023_PRB}%
  \BibitemOpen
  \bibfield  {author} {\bibinfo {author} {\bibfnamefont {K.}~\bibnamefont
  {Yang}}, \bibinfo {author} {\bibfnamefont {W.}~\bibnamefont {Xia}}, \bibinfo
  {author} {\bibfnamefont {X.}~\bibnamefont {Mi}}, \bibinfo {author}
  {\bibfnamefont {L.}~\bibnamefont {Zhang}}, \bibinfo {author} {\bibfnamefont
  {Y.}~\bibnamefont {Gan}}, \bibinfo {author} {\bibfnamefont {A.}~\bibnamefont
  {Wang}}, \bibinfo {author} {\bibfnamefont {Y.}~\bibnamefont {Chai}}, \bibinfo
  {author} {\bibfnamefont {X.}~\bibnamefont {Zhou}}, \bibinfo {author}
  {\bibfnamefont {X.}~\bibnamefont {Yang}}, \bibinfo {author} {\bibfnamefont
  {Y.}~\bibnamefont {Guo}},\ and\ \bibinfo {author} {\bibfnamefont
  {M.}~\bibnamefont {He}},\ }\href
  {https://doi.org/10.1103/PhysRevB.107.184506} {\bibfield  {journal} {\bibinfo
   {journal} {Phys. Rev. B}\ }\textbf {\bibinfo {volume} {107}},\ \bibinfo
  {pages} {184506} (\bibinfo {year} {2023})}\BibitemShut {NoStop}%
\bibitem [{\citenamefont {Subires}\ \emph {et~al.}(2023)\citenamefont
  {Subires}, \citenamefont {Korshunov}, \citenamefont {Said}, \citenamefont
  {S\'{a}nchez}, \citenamefont {Ortiz}, \citenamefont {Wilson}, \citenamefont
  {Bosak},\ and\ \citenamefont {Blanco-Canosa}}]{Subires2023_NatComm}%
  \BibitemOpen
  \bibfield  {author} {\bibinfo {author} {\bibfnamefont {D.}~\bibnamefont
  {Subires}}, \bibinfo {author} {\bibfnamefont {A.}~\bibnamefont {Korshunov}},
  \bibinfo {author} {\bibfnamefont {A.~H.}\ \bibnamefont {Said}}, \bibinfo
  {author} {\bibfnamefont {L.}~\bibnamefont {S\'{a}nchez}}, \bibinfo {author}
  {\bibfnamefont {B.~R.}\ \bibnamefont {Ortiz}}, \bibinfo {author}
  {\bibfnamefont {S.~D.}\ \bibnamefont {Wilson}}, \bibinfo {author}
  {\bibfnamefont {A.}~\bibnamefont {Bosak}},\ and\ \bibinfo {author}
  {\bibfnamefont {S.}~\bibnamefont {Blanco-Canosa}},\ }\href
  {https://doi.org/10.1038/s41467-023-36668-w} {\bibfield  {journal} {\bibinfo
  {journal} {Nature Communications}\ }\textbf {\bibinfo {volume} {14}},\
  \bibinfo {pages} {1015} (\bibinfo {year} {2023})}\BibitemShut {NoStop}%
\bibitem [{\citenamefont {Kautzsch}\ \emph {et~al.}(2023)\citenamefont
  {Kautzsch}, \citenamefont {Oey}, \citenamefont {Li}, \citenamefont {Ren},
  \citenamefont {Ortiz}, \citenamefont {Pokharel}, \citenamefont {Seshadri},
  \citenamefont {Ruff}, \citenamefont {Kongruengkit}, \citenamefont {Harter},
  \citenamefont {Wang}, \citenamefont {Zeljkovic},\ and\ \citenamefont
  {Wilson}}]{Kautzsch2023_npjQM}%
  \BibitemOpen
  \bibfield  {author} {\bibinfo {author} {\bibfnamefont {L.}~\bibnamefont
  {Kautzsch}}, \bibinfo {author} {\bibfnamefont {Y.~M.}\ \bibnamefont {Oey}},
  \bibinfo {author} {\bibfnamefont {H.}~\bibnamefont {Li}}, \bibinfo {author}
  {\bibfnamefont {Z.}~\bibnamefont {Ren}}, \bibinfo {author} {\bibfnamefont
  {B.~R.}\ \bibnamefont {Ortiz}}, \bibinfo {author} {\bibfnamefont
  {G.}~\bibnamefont {Pokharel}}, \bibinfo {author} {\bibfnamefont
  {R.}~\bibnamefont {Seshadri}}, \bibinfo {author} {\bibfnamefont
  {J.}~\bibnamefont {Ruff}}, \bibinfo {author} {\bibfnamefont {T.}~\bibnamefont
  {Kongruengkit}}, \bibinfo {author} {\bibfnamefont {J.~W.}\ \bibnamefont
  {Harter}}, \bibinfo {author} {\bibfnamefont {Z.}~\bibnamefont {Wang}},
  \bibinfo {author} {\bibfnamefont {I.}~\bibnamefont {Zeljkovic}},\ and\
  \bibinfo {author} {\bibfnamefont {S.~D.}\ \bibnamefont {Wilson}},\ }\href
  {https://doi.org/10.1038/s41535-023-00570-x} {\bibfield  {journal} {\bibinfo
  {journal} {npj Quantum Materials}\ }\textbf {\bibinfo {volume} {8}},\
  \bibinfo {pages} {37} (\bibinfo {year} {2023})}\BibitemShut {NoStop}%
\bibitem [{\citenamefont {Tian}\ and\ \citenamefont
  {Savrasov}(2024)}]{Tian2024_arXiv}%
  \BibitemOpen
  \bibfield  {author} {\bibinfo {author} {\bibfnamefont {Y.}~\bibnamefont
  {Tian}}\ and\ \bibinfo {author} {\bibfnamefont {S.~Y.}\ \bibnamefont
  {Savrasov}},\ }\href@noop {} {\bibinfo {title} {Y. Tian, {Calculated
  Unconventional Superconductivity via Charge Fluctuations in Kagome Metal
  CsV$_3$Sb$_5$}}} (\bibinfo {year} {2024}),\ \Eprint
  {https://arxiv.org/abs/2405.03137} {arXiv:2405.03137} \BibitemShut {NoStop}%
\bibitem [{\citenamefont {Abanov}\ \emph {et~al.}(2003)\citenamefont {Abanov},
  \citenamefont {Chubukov},\ and\ \citenamefont
  {Schmalian}}]{Abanov2003_AdvPhys}%
  \BibitemOpen
  \bibfield  {author} {\bibinfo {author} {\bibfnamefont {A.}~\bibnamefont
  {Abanov}}, \bibinfo {author} {\bibfnamefont {A.~V.}\ \bibnamefont
  {Chubukov}},\ and\ \bibinfo {author} {\bibfnamefont {J.}~\bibnamefont
  {Schmalian}},\ }\href {https://doi.org/10.1080/0001873021000057123}
  {\bibfield  {journal} {\bibinfo  {journal} {Advances in Physics}\ }\textbf
  {\bibinfo {volume} {52}},\ \bibinfo {pages} {119} (\bibinfo {year}
  {2003})}\BibitemShut {NoStop}%
\bibitem [{\citenamefont {Chubukov}\ and\ \citenamefont
  {Maslov}(2009)}]{Chubukov2009_PRL}%
  \BibitemOpen
  \bibfield  {author} {\bibinfo {author} {\bibfnamefont {A.~V.}\ \bibnamefont
  {Chubukov}}\ and\ \bibinfo {author} {\bibfnamefont {D.~L.}\ \bibnamefont
  {Maslov}},\ }\href {https://doi.org/10.1103/PhysRevLett.103.216401}
  {\bibfield  {journal} {\bibinfo  {journal} {Phys. Rev. Lett.}\ }\textbf
  {\bibinfo {volume} {103}},\ \bibinfo {pages} {216401} (\bibinfo {year}
  {2009})}\BibitemShut {NoStop}%
\bibitem [{\citenamefont {Castellani}\ \emph {et~al.}(1995)\citenamefont
  {Castellani}, \citenamefont {Di~Castro},\ and\ \citenamefont
  {Grilli}}]{Castellani1995_PRL}%
  \BibitemOpen
  \bibfield  {author} {\bibinfo {author} {\bibfnamefont {C.}~\bibnamefont
  {Castellani}}, \bibinfo {author} {\bibfnamefont {C.}~\bibnamefont
  {Di~Castro}},\ and\ \bibinfo {author} {\bibfnamefont {M.}~\bibnamefont
  {Grilli}},\ }\href {https://doi.org/10.1103/PhysRevLett.75.4650} {\bibfield
  {journal} {\bibinfo  {journal} {Phys. Rev. Lett.}\ }\textbf {\bibinfo
  {volume} {75}},\ \bibinfo {pages} {4650} (\bibinfo {year}
  {1995})}\BibitemShut {NoStop}%
\bibitem [{\citenamefont {Kivelson}\ \emph {et~al.}(1998)\citenamefont
  {Kivelson}, \citenamefont {Fradkin},\ and\ \citenamefont
  {Emery}}]{Kivelson1998_Nat}%
  \BibitemOpen
  \bibfield  {author} {\bibinfo {author} {\bibfnamefont {S.~A.}\ \bibnamefont
  {Kivelson}}, \bibinfo {author} {\bibfnamefont {E.}~\bibnamefont {Fradkin}},\
  and\ \bibinfo {author} {\bibfnamefont {V.~J.}\ \bibnamefont {Emery}},\ }\href
  {https://doi.org/10.1038/31177} {\bibfield  {journal} {\bibinfo  {journal}
  {Nature}\ }\textbf {\bibinfo {volume} {393}},\ \bibinfo {pages} {550}
  (\bibinfo {year} {1998})}\BibitemShut {NoStop}%
\bibitem [{\citenamefont {Caprara}\ \emph {et~al.}(2005)\citenamefont
  {Caprara}, \citenamefont {Di~Castro}, \citenamefont {Grilli},\ and\
  \citenamefont {Suppa}}]{Caprara2005_PRL}%
  \BibitemOpen
  \bibfield  {author} {\bibinfo {author} {\bibfnamefont {S.}~\bibnamefont
  {Caprara}}, \bibinfo {author} {\bibfnamefont {C.}~\bibnamefont {Di~Castro}},
  \bibinfo {author} {\bibfnamefont {M.}~\bibnamefont {Grilli}},\ and\ \bibinfo
  {author} {\bibfnamefont {D.}~\bibnamefont {Suppa}},\ }\href
  {https://doi.org/10.1103/PhysRevLett.95.117004} {\bibfield  {journal}
  {\bibinfo  {journal} {Phys. Rev. Lett.}\ }\textbf {\bibinfo {volume} {95}},\
  \bibinfo {pages} {117004} (\bibinfo {year} {2005})}\BibitemShut {NoStop}%
\bibitem [{\citenamefont {Torchinsky}\ \emph {et~al.}(2013)\citenamefont
  {Torchinsky}, \citenamefont {Mahmood}, \citenamefont {Bollinger},
  \citenamefont {Bo\v{z}ovi\'{c}},\ and\ \citenamefont
  {Gedik}}]{Torchinsky2013_NatMat}%
  \BibitemOpen
  \bibfield  {author} {\bibinfo {author} {\bibfnamefont {D.~H.}\ \bibnamefont
  {Torchinsky}}, \bibinfo {author} {\bibfnamefont {F.}~\bibnamefont {Mahmood}},
  \bibinfo {author} {\bibfnamefont {A.~T.}\ \bibnamefont {Bollinger}}, \bibinfo
  {author} {\bibfnamefont {I.}~\bibnamefont {Bo\v{z}ovi\'{c}}},\ and\ \bibinfo
  {author} {\bibfnamefont {N.}~\bibnamefont {Gedik}},\ }\href
  {https://doi.org/10.1038/nmat3571} {\bibfield  {journal} {\bibinfo  {journal}
  {Nature Materials}\ }\textbf {\bibinfo {volume} {12}},\ \bibinfo {pages}
  {387} (\bibinfo {year} {2013})}\BibitemShut {NoStop}%
\bibitem [{\citenamefont {Kohn}\ and\ \citenamefont
  {Luttinger}(1965)}]{Kohn1965_PRL}%
  \BibitemOpen
  \bibfield  {author} {\bibinfo {author} {\bibfnamefont {W.}~\bibnamefont
  {Kohn}}\ and\ \bibinfo {author} {\bibfnamefont {J.~M.}\ \bibnamefont
  {Luttinger}},\ }\href {https://doi.org/10.1103/PhysRevLett.15.524} {\bibfield
   {journal} {\bibinfo  {journal} {Phys. Rev. Lett.}\ }\textbf {\bibinfo
  {volume} {15}},\ \bibinfo {pages} {524} (\bibinfo {year} {1965})}\BibitemShut
  {NoStop}%
\bibitem [{\citenamefont {Luttinger}(1966)}]{Luttinger1966_PR}%
  \BibitemOpen
  \bibfield  {author} {\bibinfo {author} {\bibfnamefont {J.~M.}\ \bibnamefont
  {Luttinger}},\ }\href {https://doi.org/10.1103/PhysRev.150.202} {\bibfield
  {journal} {\bibinfo  {journal} {Phys. Rev.}\ }\textbf {\bibinfo {volume}
  {150}},\ \bibinfo {pages} {202} (\bibinfo {year} {1966})}\BibitemShut
  {NoStop}%
\bibitem [{\citenamefont {Chubukov}(1993)}]{Chubukov1993_PRB}%
  \BibitemOpen
  \bibfield  {author} {\bibinfo {author} {\bibfnamefont {A.~V.}\ \bibnamefont
  {Chubukov}},\ }\href {https://doi.org/10.1103/PhysRevB.48.1097} {\bibfield
  {journal} {\bibinfo  {journal} {Phys. Rev. B}\ }\textbf {\bibinfo {volume}
  {48}},\ \bibinfo {pages} {1097} (\bibinfo {year} {1993})}\BibitemShut
  {NoStop}%
\bibitem [{\citenamefont {Gonz\'alez}(2008)}]{Gonzalez2008_PRB}%
  \BibitemOpen
  \bibfield  {author} {\bibinfo {author} {\bibfnamefont {J.}~\bibnamefont
  {Gonz\'alez}},\ }\href {https://doi.org/10.1103/PhysRevB.78.205431}
  {\bibfield  {journal} {\bibinfo  {journal} {Phys. Rev. B}\ }\textbf {\bibinfo
  {volume} {78}},\ \bibinfo {pages} {205431} (\bibinfo {year}
  {2008})}\BibitemShut {NoStop}%
\bibitem [{\citenamefont {Dong}\ \emph
  {et~al.}(2023{\natexlab{b}})\citenamefont {Dong}, \citenamefont {Levitov},\
  and\ \citenamefont {Chubukov}}]{Dong2023_PRB2}%
  \BibitemOpen
  \bibfield  {author} {\bibinfo {author} {\bibfnamefont {Z.}~\bibnamefont
  {Dong}}, \bibinfo {author} {\bibfnamefont {L.}~\bibnamefont {Levitov}},\ and\
  \bibinfo {author} {\bibfnamefont {A.~V.}\ \bibnamefont {Chubukov}},\ }\href
  {https://doi.org/10.1103/PhysRevB.108.134503} {\bibfield  {journal} {\bibinfo
   {journal} {Phys. Rev. B}\ }\textbf {\bibinfo {volume} {108}},\ \bibinfo
  {pages} {134503} (\bibinfo {year} {2023}{\natexlab{b}})}\BibitemShut
  {NoStop}%
\bibitem [{\citenamefont {Kuo}\ \emph {et~al.}(2013)\citenamefont {Kuo},
  \citenamefont {Shapiro}, \citenamefont {Riggs},\ and\ \citenamefont
  {Fisher}}]{Kuo2013_PRB}%
  \BibitemOpen
  \bibfield  {author} {\bibinfo {author} {\bibfnamefont {H.-H.}\ \bibnamefont
  {Kuo}}, \bibinfo {author} {\bibfnamefont {M.~C.}\ \bibnamefont {Shapiro}},
  \bibinfo {author} {\bibfnamefont {S.~C.}\ \bibnamefont {Riggs}},\ and\
  \bibinfo {author} {\bibfnamefont {I.~R.}\ \bibnamefont {Fisher}},\ }\href
  {https://doi.org/10.1103/PhysRevB.88.085113} {\bibfield  {journal} {\bibinfo
  {journal} {Phys. Rev. B}\ }\textbf {\bibinfo {volume} {88}},\ \bibinfo
  {pages} {085113} (\bibinfo {year} {2013})}\BibitemShut {NoStop}%
\bibitem [{\citenamefont {Massat}\ \emph {et~al.}(2016)\citenamefont {Massat},
  \citenamefont {Farina}, \citenamefont {Paul}, \citenamefont {Karlsson},
  \citenamefont {Strobel}, \citenamefont {Toulemonde}, \citenamefont
  {M\'{e}asson}, \citenamefont {Cazayous}, \citenamefont {Sacuto},
  \citenamefont {Kasahara}, \citenamefont {Shibauchi}, \citenamefont
  {Matsuda},\ and\ \citenamefont {Gallais}}]{Massat2016_PNAS}%
  \BibitemOpen
  \bibfield  {author} {\bibinfo {author} {\bibfnamefont {P.}~\bibnamefont
  {Massat}}, \bibinfo {author} {\bibfnamefont {D.}~\bibnamefont {Farina}},
  \bibinfo {author} {\bibfnamefont {I.}~\bibnamefont {Paul}}, \bibinfo {author}
  {\bibfnamefont {S.}~\bibnamefont {Karlsson}}, \bibinfo {author}
  {\bibfnamefont {P.}~\bibnamefont {Strobel}}, \bibinfo {author} {\bibfnamefont
  {P.}~\bibnamefont {Toulemonde}}, \bibinfo {author} {\bibfnamefont {M.-A.}\
  \bibnamefont {M\'{e}asson}}, \bibinfo {author} {\bibfnamefont
  {M.}~\bibnamefont {Cazayous}}, \bibinfo {author} {\bibfnamefont
  {A.}~\bibnamefont {Sacuto}}, \bibinfo {author} {\bibfnamefont
  {S.}~\bibnamefont {Kasahara}}, \bibinfo {author} {\bibfnamefont
  {T.}~\bibnamefont {Shibauchi}}, \bibinfo {author} {\bibfnamefont
  {Y.}~\bibnamefont {Matsuda}},\ and\ \bibinfo {author} {\bibfnamefont
  {Y.}~\bibnamefont {Gallais}},\ }\href
  {https://doi.org/10.1073/pnas.1606562113} {\bibfield  {journal} {\bibinfo
  {journal} {Proceedings of the National Academy of Sciences}\ }\textbf
  {\bibinfo {volume} {113}},\ \bibinfo {pages} {9177} (\bibinfo {year}
  {2016})}\BibitemShut {NoStop}%
\bibitem [{\citenamefont {Smollett}(1952)}]{Smollett1952_PPSA}%
  \BibitemOpen
  \bibfield  {author} {\bibinfo {author} {\bibfnamefont {M.}~\bibnamefont
  {Smollett}},\ }\href {https://doi.org/10.1088/0370-1298/65/2/305} {\bibfield
  {journal} {\bibinfo  {journal} {Proceedings of the Physical Society. Section
  A}\ }\textbf {\bibinfo {volume} {65}},\ \bibinfo {pages} {109} (\bibinfo
  {year} {1952})}\BibitemShut {NoStop}%
\bibitem [{\citenamefont {Van~Hove}(1953)}]{VanHove1953_PR}%
  \BibitemOpen
  \bibfield  {author} {\bibinfo {author} {\bibfnamefont {L.}~\bibnamefont
  {Van~Hove}},\ }\href {https://doi.org/10.1103/PhysRev.89.1189} {\bibfield
  {journal} {\bibinfo  {journal} {Phys. Rev.}\ }\textbf {\bibinfo {volume}
  {89}},\ \bibinfo {pages} {1189} (\bibinfo {year} {1953})}\BibitemShut
  {NoStop}%
\bibitem [{\citenamefont {Venderbos}(2016)}]{Venderbos2016_PRB}%
  \BibitemOpen
  \bibfield  {author} {\bibinfo {author} {\bibfnamefont {J.~W.~F.}\
  \bibnamefont {Venderbos}},\ }\href
  {https://doi.org/10.1103/PhysRevB.93.115107} {\bibfield  {journal} {\bibinfo
  {journal} {Phys. Rev. B}\ }\textbf {\bibinfo {volume} {93}},\ \bibinfo
  {pages} {115107} (\bibinfo {year} {2016})}\BibitemShut {NoStop}%
\bibitem [{\citenamefont {Zhou}\ and\ \citenamefont
  {Wang}(2022)}]{Zhou2022_NatComm}%
  \BibitemOpen
  \bibfield  {author} {\bibinfo {author} {\bibfnamefont {S.}~\bibnamefont
  {Zhou}}\ and\ \bibinfo {author} {\bibfnamefont {Z.}~\bibnamefont {Wang}},\
  }\href {https://doi.org/10.1038/s41467-022-34832-2} {\bibfield  {journal}
  {\bibinfo  {journal} {Nature Communications}\ }\textbf {\bibinfo {volume}
  {13}},\ \bibinfo {pages} {7288} (\bibinfo {year} {2022})}\BibitemShut
  {NoStop}%
\bibitem [{\citenamefont {Lin}\ and\ \citenamefont
  {Nandkishore}(2022)}]{Lin2022_PRB}%
  \BibitemOpen
  \bibfield  {author} {\bibinfo {author} {\bibfnamefont {Y.-P.}\ \bibnamefont
  {Lin}}\ and\ \bibinfo {author} {\bibfnamefont {R.~M.}\ \bibnamefont
  {Nandkishore}},\ }\href {https://doi.org/10.1103/PhysRevB.106.L060507}
  {\bibfield  {journal} {\bibinfo  {journal} {Phys. Rev. B}\ }\textbf {\bibinfo
  {volume} {106}},\ \bibinfo {pages} {L060507} (\bibinfo {year}
  {2022})}\BibitemShut {NoStop}%
\bibitem [{\citenamefont {Varma}\ and\ \citenamefont
  {Wang}(2023)}]{Varma2023_PRB}%
  \BibitemOpen
  \bibfield  {author} {\bibinfo {author} {\bibfnamefont {C.~M.}\ \bibnamefont
  {Varma}}\ and\ \bibinfo {author} {\bibfnamefont {Z.}~\bibnamefont {Wang}},\
  }\href {https://doi.org/10.1103/PhysRevB.108.214516} {\bibfield  {journal}
  {\bibinfo  {journal} {Phys. Rev. B}\ }\textbf {\bibinfo {volume} {108}},\
  \bibinfo {pages} {214516} (\bibinfo {year} {2023})}\BibitemShut {NoStop}%
\bibitem [{\citenamefont {Shrestha}\ \emph {et~al.}(2022)\citenamefont
  {Shrestha}, \citenamefont {Chapai}, \citenamefont {Pokharel}, \citenamefont
  {Miertschin}, \citenamefont {Nguyen}, \citenamefont {Zhou}, \citenamefont
  {Chung}, \citenamefont {Kanatzidis}, \citenamefont {Mitchell}, \citenamefont
  {Welp}, \citenamefont {Popovi\ifmmode~\acute{c}\else \'{c}\fi{}},
  \citenamefont {Graf}, \citenamefont {Lorenz},\ and\ \citenamefont
  {Kwok}}]{Shrestha2022_PRB}%
  \BibitemOpen
  \bibfield  {author} {\bibinfo {author} {\bibfnamefont {K.}~\bibnamefont
  {Shrestha}}, \bibinfo {author} {\bibfnamefont {R.}~\bibnamefont {Chapai}},
  \bibinfo {author} {\bibfnamefont {B.~K.}\ \bibnamefont {Pokharel}}, \bibinfo
  {author} {\bibfnamefont {D.}~\bibnamefont {Miertschin}}, \bibinfo {author}
  {\bibfnamefont {T.}~\bibnamefont {Nguyen}}, \bibinfo {author} {\bibfnamefont
  {X.}~\bibnamefont {Zhou}}, \bibinfo {author} {\bibfnamefont {D.~Y.}\
  \bibnamefont {Chung}}, \bibinfo {author} {\bibfnamefont {M.~G.}\ \bibnamefont
  {Kanatzidis}}, \bibinfo {author} {\bibfnamefont {J.~F.}\ \bibnamefont
  {Mitchell}}, \bibinfo {author} {\bibfnamefont {U.}~\bibnamefont {Welp}},
  \bibinfo {author} {\bibfnamefont {D.}~\bibnamefont
  {Popovi\ifmmode~\acute{c}\else \'{c}\fi{}}}, \bibinfo {author} {\bibfnamefont
  {D.~E.}\ \bibnamefont {Graf}}, \bibinfo {author} {\bibfnamefont
  {B.}~\bibnamefont {Lorenz}},\ and\ \bibinfo {author} {\bibfnamefont {W.~K.}\
  \bibnamefont {Kwok}},\ }\href {https://doi.org/10.1103/PhysRevB.105.024508}
  {\bibfield  {journal} {\bibinfo  {journal} {Phys. Rev. B}\ }\textbf {\bibinfo
  {volume} {105}},\ \bibinfo {pages} {024508} (\bibinfo {year}
  {2022})}\BibitemShut {NoStop}%
\bibitem [{\citenamefont {Wu}\ \emph {et~al.}(2024{\natexlab{b}})\citenamefont
  {Wu}, \citenamefont {Jia}, \citenamefont {Yang}, \citenamefont {Hong},
  \citenamefont {Shu}, \citenamefont {Miao}, \citenamefont {Yan}, \citenamefont
  {Rong}, \citenamefont {Ai}, \citenamefont {Zhang}, \citenamefont {Yin},
  \citenamefont {Liu}, \citenamefont {Chen}, \citenamefont {Yang},
  \citenamefont {Peng}, \citenamefont {Li}, \citenamefont {Zhang},
  \citenamefont {Zhang}, \citenamefont {Yang}, \citenamefont {Wang},
  \citenamefont {Zong}, \citenamefont {Liu}, \citenamefont {Li}, \citenamefont
  {Wang}, \citenamefont {Peng}, \citenamefont {Mao}, \citenamefont {Liu},
  \citenamefont {Li}, \citenamefont {Chen}, \citenamefont {Luo}, \citenamefont
  {Wu}, \citenamefont {Xu}, \citenamefont {Zhao},\ and\ \citenamefont
  {Zhou}}]{Wu2024_NatPhys}%
  \BibitemOpen
  \bibfield  {author} {\bibinfo {author} {\bibfnamefont {D.}~\bibnamefont
  {Wu}}, \bibinfo {author} {\bibfnamefont {J.}~\bibnamefont {Jia}}, \bibinfo
  {author} {\bibfnamefont {J.}~\bibnamefont {Yang}}, \bibinfo {author}
  {\bibfnamefont {W.}~\bibnamefont {Hong}}, \bibinfo {author} {\bibfnamefont
  {Y.}~\bibnamefont {Shu}}, \bibinfo {author} {\bibfnamefont {T.}~\bibnamefont
  {Miao}}, \bibinfo {author} {\bibfnamefont {H.}~\bibnamefont {Yan}}, \bibinfo
  {author} {\bibfnamefont {H.}~\bibnamefont {Rong}}, \bibinfo {author}
  {\bibfnamefont {P.}~\bibnamefont {Ai}}, \bibinfo {author} {\bibfnamefont
  {X.}~\bibnamefont {Zhang}}, \bibinfo {author} {\bibfnamefont
  {C.}~\bibnamefont {Yin}}, \bibinfo {author} {\bibfnamefont {J.}~\bibnamefont
  {Liu}}, \bibinfo {author} {\bibfnamefont {H.}~\bibnamefont {Chen}}, \bibinfo
  {author} {\bibfnamefont {Y.}~\bibnamefont {Yang}}, \bibinfo {author}
  {\bibfnamefont {C.}~\bibnamefont {Peng}}, \bibinfo {author} {\bibfnamefont
  {C.}~\bibnamefont {Li}}, \bibinfo {author} {\bibfnamefont {S.}~\bibnamefont
  {Zhang}}, \bibinfo {author} {\bibfnamefont {F.}~\bibnamefont {Zhang}},
  \bibinfo {author} {\bibfnamefont {F.}~\bibnamefont {Yang}}, \bibinfo {author}
  {\bibfnamefont {Z.}~\bibnamefont {Wang}}, \bibinfo {author} {\bibfnamefont
  {N.}~\bibnamefont {Zong}}, \bibinfo {author} {\bibfnamefont {L.}~\bibnamefont
  {Liu}}, \bibinfo {author} {\bibfnamefont {R.}~\bibnamefont {Li}}, \bibinfo
  {author} {\bibfnamefont {X.}~\bibnamefont {Wang}}, \bibinfo {author}
  {\bibfnamefont {Q.}~\bibnamefont {Peng}}, \bibinfo {author} {\bibfnamefont
  {H.}~\bibnamefont {Mao}}, \bibinfo {author} {\bibfnamefont {G.}~\bibnamefont
  {Liu}}, \bibinfo {author} {\bibfnamefont {S.}~\bibnamefont {Li}}, \bibinfo
  {author} {\bibfnamefont {Y.}~\bibnamefont {Chen}}, \bibinfo {author}
  {\bibfnamefont {H.}~\bibnamefont {Luo}}, \bibinfo {author} {\bibfnamefont
  {X.}~\bibnamefont {Wu}}, \bibinfo {author} {\bibfnamefont {Z.}~\bibnamefont
  {Xu}}, \bibinfo {author} {\bibfnamefont {L.}~\bibnamefont {Zhao}},\ and\
  \bibinfo {author} {\bibfnamefont {X.~J.}\ \bibnamefont {Zhou}},\ }\href
  {https://doi.org/10.1038/s41567-023-02348-1} {\bibfield  {journal} {\bibinfo
  {journal} {Nature Physics}\ } (\bibinfo {year}
  {2024}{\natexlab{b}})}\BibitemShut {NoStop}%
\bibitem [{\citenamefont {Thomale}\ \emph {et~al.}(2011)\citenamefont
  {Thomale}, \citenamefont {Platt}, \citenamefont {Hanke}, \citenamefont {Hu},\
  and\ \citenamefont {Bernevig}}]{Thomale2011_PRL}%
  \BibitemOpen
  \bibfield  {author} {\bibinfo {author} {\bibfnamefont {R.}~\bibnamefont
  {Thomale}}, \bibinfo {author} {\bibfnamefont {C.}~\bibnamefont {Platt}},
  \bibinfo {author} {\bibfnamefont {W.}~\bibnamefont {Hanke}}, \bibinfo
  {author} {\bibfnamefont {J.}~\bibnamefont {Hu}},\ and\ \bibinfo {author}
  {\bibfnamefont {B.~A.}\ \bibnamefont {Bernevig}},\ }\href
  {https://doi.org/10.1103/PhysRevLett.107.117001} {\bibfield  {journal}
  {\bibinfo  {journal} {Phys. Rev. Lett.}\ }\textbf {\bibinfo {volume} {107}},\
  \bibinfo {pages} {117001} (\bibinfo {year} {2011})}\BibitemShut {NoStop}%
\bibitem [{\citenamefont {Agterberg}\ \emph {et~al.}(1999)\citenamefont
  {Agterberg}, \citenamefont {Barzykin},\ and\ \citenamefont
  {Gor'kov}}]{Agterberg1999_PRB}%
  \BibitemOpen
  \bibfield  {author} {\bibinfo {author} {\bibfnamefont {D.~F.}\ \bibnamefont
  {Agterberg}}, \bibinfo {author} {\bibfnamefont {V.}~\bibnamefont
  {Barzykin}},\ and\ \bibinfo {author} {\bibfnamefont {L.~P.}\ \bibnamefont
  {Gor'kov}},\ }\href {https://doi.org/10.1103/PhysRevB.60.14868} {\bibfield
  {journal} {\bibinfo  {journal} {Phys. Rev. B}\ }\textbf {\bibinfo {volume}
  {60}},\ \bibinfo {pages} {14868} (\bibinfo {year} {1999})}\BibitemShut
  {NoStop}%
\bibitem [{\citenamefont {Oey}\ \emph {et~al.}(2022{\natexlab{b}})\citenamefont
  {Oey}, \citenamefont {Ortiz}, \citenamefont {Kaboudvand}, \citenamefont
  {Frassineti}, \citenamefont {Garcia}, \citenamefont {Cong}, \citenamefont
  {Sanna}, \citenamefont {Mitrovi\ifmmode~\acute{c}\else \'{c}\fi{}},
  \citenamefont {Seshadri},\ and\ \citenamefont {Wilson}}]{Oey2022_PRM1}%
  \BibitemOpen
  \bibfield  {author} {\bibinfo {author} {\bibfnamefont {Y.~M.}\ \bibnamefont
  {Oey}}, \bibinfo {author} {\bibfnamefont {B.~R.}\ \bibnamefont {Ortiz}},
  \bibinfo {author} {\bibfnamefont {F.}~\bibnamefont {Kaboudvand}}, \bibinfo
  {author} {\bibfnamefont {J.}~\bibnamefont {Frassineti}}, \bibinfo {author}
  {\bibfnamefont {E.}~\bibnamefont {Garcia}}, \bibinfo {author} {\bibfnamefont
  {R.}~\bibnamefont {Cong}}, \bibinfo {author} {\bibfnamefont {S.}~\bibnamefont
  {Sanna}}, \bibinfo {author} {\bibfnamefont {V.~F.}\ \bibnamefont
  {Mitrovi\ifmmode~\acute{c}\else \'{c}\fi{}}}, \bibinfo {author}
  {\bibfnamefont {R.}~\bibnamefont {Seshadri}},\ and\ \bibinfo {author}
  {\bibfnamefont {S.~D.}\ \bibnamefont {Wilson}},\ }\href
  {https://doi.org/10.1103/PhysRevMaterials.6.L041801} {\bibfield  {journal}
  {\bibinfo  {journal} {Phys. Rev. Mater.}\ }\textbf {\bibinfo {volume} {6}},\
  \bibinfo {pages} {L041801} (\bibinfo {year}
  {2022}{\natexlab{b}})}\BibitemShut {NoStop}%
\bibitem [{\citenamefont {Eremin}\ and\ \citenamefont
  {Chubukov}(2010)}]{Eremin2010_PRB}%
  \BibitemOpen
  \bibfield  {author} {\bibinfo {author} {\bibfnamefont {I.}~\bibnamefont
  {Eremin}}\ and\ \bibinfo {author} {\bibfnamefont {A.~V.}\ \bibnamefont
  {Chubukov}},\ }\href {https://doi.org/10.1103/PhysRevB.81.024511} {\bibfield
  {journal} {\bibinfo  {journal} {Phys. Rev. B}\ }\textbf {\bibinfo {volume}
  {81}},\ \bibinfo {pages} {024511} (\bibinfo {year} {2010})}\BibitemShut
  {NoStop}%
\bibitem [{\citenamefont {Vorontsov}\ \emph {et~al.}(2010)\citenamefont
  {Vorontsov}, \citenamefont {Vavilov},\ and\ \citenamefont
  {Chubukov}}]{Vorontsov2010_PRB}%
  \BibitemOpen
  \bibfield  {author} {\bibinfo {author} {\bibfnamefont {A.~B.}\ \bibnamefont
  {Vorontsov}}, \bibinfo {author} {\bibfnamefont {M.~G.}\ \bibnamefont
  {Vavilov}},\ and\ \bibinfo {author} {\bibfnamefont {A.~V.}\ \bibnamefont
  {Chubukov}},\ }\href {https://doi.org/10.1103/PhysRevB.81.174538} {\bibfield
  {journal} {\bibinfo  {journal} {Phys. Rev. B}\ }\textbf {\bibinfo {volume}
  {81}},\ \bibinfo {pages} {174538} (\bibinfo {year} {2010})}\BibitemShut
  {NoStop}%
\bibitem [{\citenamefont {Nandkishore}\ \emph {et~al.}(2012)\citenamefont
  {Nandkishore}, \citenamefont {Levitov},\ and\ \citenamefont
  {Chubukov}}]{Nandkishore2012_NatPhys}%
  \BibitemOpen
  \bibfield  {author} {\bibinfo {author} {\bibfnamefont {R.}~\bibnamefont
  {Nandkishore}}, \bibinfo {author} {\bibfnamefont {L.~S.}\ \bibnamefont
  {Levitov}},\ and\ \bibinfo {author} {\bibfnamefont {A.~V.}\ \bibnamefont
  {Chubukov}},\ }\href {https://doi.org/10.1038/nphys2208} {\bibfield
  {journal} {\bibinfo  {journal} {Nature Physics}\ }\textbf {\bibinfo {volume}
  {8}},\ \bibinfo {pages} {158} (\bibinfo {year} {2012})}\BibitemShut {NoStop}%
\bibitem [{\citenamefont {Tsvelik}\ and\ \citenamefont
  {Sarkar}(2023)}]{Tsvelik2023_PRB}%
  \BibitemOpen
  \bibfield  {author} {\bibinfo {author} {\bibfnamefont {A.~M.}\ \bibnamefont
  {Tsvelik}}\ and\ \bibinfo {author} {\bibfnamefont {S.}~\bibnamefont
  {Sarkar}},\ }\href {https://doi.org/10.1103/PhysRevB.108.045119} {\bibfield
  {journal} {\bibinfo  {journal} {Phys. Rev. B}\ }\textbf {\bibinfo {volume}
  {108}},\ \bibinfo {pages} {045119} (\bibinfo {year} {2023})}\BibitemShut
  {NoStop}%
\bibitem [{\citenamefont {Kiesel}\ and\ \citenamefont
  {Thomale}(2012)}]{Kiesel2012_PRB}%
  \BibitemOpen
  \bibfield  {author} {\bibinfo {author} {\bibfnamefont {M.~L.}\ \bibnamefont
  {Kiesel}}\ and\ \bibinfo {author} {\bibfnamefont {R.}~\bibnamefont
  {Thomale}},\ }\href {https://doi.org/10.1103/PhysRevB.86.121105} {\bibfield
  {journal} {\bibinfo  {journal} {Phys. Rev. B}\ }\textbf {\bibinfo {volume}
  {86}},\ \bibinfo {pages} {121105} (\bibinfo {year} {2012})}\BibitemShut
  {NoStop}%
\bibitem [{\citenamefont {Braz}\ \emph {et~al.}(2024)\citenamefont {Braz},
  \citenamefont {Martins},\ and\ \citenamefont {da~Silva}}]{Braz2024_arXiv}%
  \BibitemOpen
  \bibfield  {author} {\bibinfo {author} {\bibfnamefont {L.~B.}\ \bibnamefont
  {Braz}}, \bibinfo {author} {\bibfnamefont {G.~B.}\ \bibnamefont {Martins}},\
  and\ \bibinfo {author} {\bibfnamefont {L.~G. G. V.~D.}\ \bibnamefont
  {da~Silva}},\ }\href@noop {} {\bibinfo {title} {L.~B. Braz, {Charge and spin
  fluctuations in superconductors with intersublattice and interorbital
  interactions}}} (\bibinfo {year} {2024}),\ \Eprint
  {https://arxiv.org/abs/2403.02453} {arXiv:2403.02453} \BibitemShut {NoStop}%
\bibitem [{\citenamefont {Maiti}\ and\ \citenamefont
  {Chubukov}(2013)}]{Maiti2013_AIPCP}%
  \BibitemOpen
  \bibfield  {author} {\bibinfo {author} {\bibfnamefont {S.}~\bibnamefont
  {Maiti}}\ and\ \bibinfo {author} {\bibfnamefont {A.~V.}\ \bibnamefont
  {Chubukov}},\ }\href {https://doi.org/10.1063/1.4818400} {\bibfield
  {journal} {\bibinfo  {journal} {AIP Conference Proceedings}\ }\textbf
  {\bibinfo {volume} {1550}},\ \bibinfo {pages} {3} (\bibinfo {year}
  {2013})}\BibitemShut {NoStop}%
\bibitem [{\citenamefont {Chubukov}\ \emph {et~al.}(2008)\citenamefont
  {Chubukov}, \citenamefont {Efremov},\ and\ \citenamefont
  {Eremin}}]{Chubukov2008_PRB}%
  \BibitemOpen
  \bibfield  {author} {\bibinfo {author} {\bibfnamefont {A.~V.}\ \bibnamefont
  {Chubukov}}, \bibinfo {author} {\bibfnamefont {D.~V.}\ \bibnamefont
  {Efremov}},\ and\ \bibinfo {author} {\bibfnamefont {I.}~\bibnamefont
  {Eremin}},\ }\href {https://doi.org/10.1103/PhysRevB.78.134512} {\bibfield
  {journal} {\bibinfo  {journal} {Phys. Rev. B}\ }\textbf {\bibinfo {volume}
  {78}},\ \bibinfo {pages} {134512} (\bibinfo {year} {2008})}\BibitemShut
  {NoStop}%
\bibitem [{\citenamefont {Chichinadze}\ \emph {et~al.}(2020)\citenamefont
  {Chichinadze}, \citenamefont {Classen},\ and\ \citenamefont
  {Chubukov}}]{Chichinadze2020_PRB}%
  \BibitemOpen
  \bibfield  {author} {\bibinfo {author} {\bibfnamefont {D.~V.}\ \bibnamefont
  {Chichinadze}}, \bibinfo {author} {\bibfnamefont {L.}~\bibnamefont
  {Classen}},\ and\ \bibinfo {author} {\bibfnamefont {A.~V.}\ \bibnamefont
  {Chubukov}},\ }\href {https://doi.org/10.1103/PhysRevB.102.125120} {\bibfield
   {journal} {\bibinfo  {journal} {Phys. Rev. B}\ }\textbf {\bibinfo {volume}
  {102}},\ \bibinfo {pages} {125120} (\bibinfo {year} {2020})}\BibitemShut
  {NoStop}%
\bibitem [{\citenamefont {Maiti}\ and\ \citenamefont
  {Chubukov}(2010)}]{Maiti2010_PRB}%
  \BibitemOpen
  \bibfield  {author} {\bibinfo {author} {\bibfnamefont {S.}~\bibnamefont
  {Maiti}}\ and\ \bibinfo {author} {\bibfnamefont {A.~V.}\ \bibnamefont
  {Chubukov}},\ }\href {https://doi.org/10.1103/PhysRevB.82.214515} {\bibfield
  {journal} {\bibinfo  {journal} {Phys. Rev. B}\ }\textbf {\bibinfo {volume}
  {82}},\ \bibinfo {pages} {214515} (\bibinfo {year} {2010})}\BibitemShut
  {NoStop}%
\bibitem [{\citenamefont {Chubukov}\ and\ \citenamefont
  {W\"olfle}(2014)}]{Chubukov2014_PRB}%
  \BibitemOpen
  \bibfield  {author} {\bibinfo {author} {\bibfnamefont {A.~V.}\ \bibnamefont
  {Chubukov}}\ and\ \bibinfo {author} {\bibfnamefont {P.}~\bibnamefont
  {W\"olfle}},\ }\href {https://doi.org/10.1103/PhysRevB.89.045108} {\bibfield
  {journal} {\bibinfo  {journal} {Phys. Rev. B}\ }\textbf {\bibinfo {volume}
  {89}},\ \bibinfo {pages} {045108} (\bibinfo {year} {2014})}\BibitemShut
  {NoStop}%
\bibitem [{\citenamefont {Chubukov}\ \emph {et~al.}(2014)\citenamefont
  {Chubukov}, \citenamefont {Betouras},\ and\ \citenamefont
  {Efremov}}]{Chubukov2014_PRL}%
  \BibitemOpen
  \bibfield  {author} {\bibinfo {author} {\bibfnamefont {A.~V.}\ \bibnamefont
  {Chubukov}}, \bibinfo {author} {\bibfnamefont {J.~J.}\ \bibnamefont
  {Betouras}},\ and\ \bibinfo {author} {\bibfnamefont {D.~V.}\ \bibnamefont
  {Efremov}},\ }\href {https://doi.org/10.1103/PhysRevLett.112.037202}
  {\bibfield  {journal} {\bibinfo  {journal} {Phys. Rev. Lett.}\ }\textbf
  {\bibinfo {volume} {112}},\ \bibinfo {pages} {037202} (\bibinfo {year}
  {2014})}\BibitemShut {NoStop}%
\bibitem [{\citenamefont {Mayrhofer}\ \emph {et~al.}(2024)\citenamefont
  {Mayrhofer}, \citenamefont {Chubukov},\ and\ \citenamefont
  {W\"{o}lfle}}]{Mayrhofer2024_PRB}%
  \BibitemOpen
  \bibfield  {author} {\bibinfo {author} {\bibfnamefont {R.~D.}\ \bibnamefont
  {Mayrhofer}}, \bibinfo {author} {\bibfnamefont {A.~V.}\ \bibnamefont
  {Chubukov}},\ and\ \bibinfo {author} {\bibfnamefont {P.}~\bibnamefont
  {W\"{o}lfle}},\ }\href {https://doi.org/10.1103/PhysRevB.110.205112}
  {\bibfield  {journal} {\bibinfo  {journal} {Phys. Rev. B}\ }\textbf {\bibinfo
  {volume} {110}},\ \bibinfo {pages} {205112} (\bibinfo {year}
  {2024})}\BibitemShut {NoStop}%
\bibitem [{\citenamefont {Karahasanovic}\ \emph {et~al.}(2015)\citenamefont
  {Karahasanovic}, \citenamefont {Kretzschmar}, \citenamefont {B\"ohm},
  \citenamefont {Hackl}, \citenamefont {Paul}, \citenamefont {Gallais},\ and\
  \citenamefont {Schmalian}}]{Karahasanovic2015_PRB}%
  \BibitemOpen
  \bibfield  {author} {\bibinfo {author} {\bibfnamefont {U.}~\bibnamefont
  {Karahasanovic}}, \bibinfo {author} {\bibfnamefont {F.}~\bibnamefont
  {Kretzschmar}}, \bibinfo {author} {\bibfnamefont {T.}~\bibnamefont {B\"ohm}},
  \bibinfo {author} {\bibfnamefont {R.}~\bibnamefont {Hackl}}, \bibinfo
  {author} {\bibfnamefont {I.}~\bibnamefont {Paul}}, \bibinfo {author}
  {\bibfnamefont {Y.}~\bibnamefont {Gallais}},\ and\ \bibinfo {author}
  {\bibfnamefont {J.}~\bibnamefont {Schmalian}},\ }\href
  {https://doi.org/10.1103/PhysRevB.92.075134} {\bibfield  {journal} {\bibinfo
  {journal} {Phys. Rev. B}\ }\textbf {\bibinfo {volume} {92}},\ \bibinfo
  {pages} {075134} (\bibinfo {year} {2015})}\BibitemShut {NoStop}%
\bibitem [{\citenamefont {Kretzschmar}\ \emph {et~al.}(2016)\citenamefont
  {Kretzschmar}, \citenamefont {B\"{o}hm}, \citenamefont {Karahasanovi\'{c}},
  \citenamefont {Muschler}, \citenamefont {Baum}, \citenamefont {Jost},
  \citenamefont {Schmalian}, \citenamefont {Caprara}, \citenamefont {Grilli},
  \citenamefont {Di~Castro}, \citenamefont {Analytis}, \citenamefont {Chu},
  \citenamefont {Fisher},\ and\ \citenamefont
  {Hackl}}]{Kretzschmar2016_NatPhys}%
  \BibitemOpen
  \bibfield  {author} {\bibinfo {author} {\bibfnamefont {F.}~\bibnamefont
  {Kretzschmar}}, \bibinfo {author} {\bibfnamefont {T.}~\bibnamefont
  {B\"{o}hm}}, \bibinfo {author} {\bibfnamefont {U.}~\bibnamefont
  {Karahasanovi\'{c}}}, \bibinfo {author} {\bibfnamefont {B.}~\bibnamefont
  {Muschler}}, \bibinfo {author} {\bibfnamefont {A.}~\bibnamefont {Baum}},
  \bibinfo {author} {\bibfnamefont {D.}~\bibnamefont {Jost}}, \bibinfo {author}
  {\bibfnamefont {J.}~\bibnamefont {Schmalian}}, \bibinfo {author}
  {\bibfnamefont {S.}~\bibnamefont {Caprara}}, \bibinfo {author} {\bibfnamefont
  {M.}~\bibnamefont {Grilli}}, \bibinfo {author} {\bibfnamefont
  {C.}~\bibnamefont {Di~Castro}}, \bibinfo {author} {\bibfnamefont {J.~G.}\
  \bibnamefont {Analytis}}, \bibinfo {author} {\bibfnamefont {J.-H.}\
  \bibnamefont {Chu}}, \bibinfo {author} {\bibfnamefont {I.~R.}\ \bibnamefont
  {Fisher}},\ and\ \bibinfo {author} {\bibfnamefont {R.}~\bibnamefont
  {Hackl}},\ }\href {https://doi.org/10.1038/nphys3634} {\bibfield  {journal}
  {\bibinfo  {journal} {Nature Physics}\ }\textbf {\bibinfo {volume} {12}},\
  \bibinfo {pages} {560} (\bibinfo {year} {2016})}\BibitemShut {NoStop}%
\bibitem [{\citenamefont {Hinojosa}\ \emph {et~al.}(2016)\citenamefont
  {Hinojosa}, \citenamefont {Cai},\ and\ \citenamefont
  {Chubukov}}]{Hinojosa2016_PRB}%
  \BibitemOpen
  \bibfield  {author} {\bibinfo {author} {\bibfnamefont {A.}~\bibnamefont
  {Hinojosa}}, \bibinfo {author} {\bibfnamefont {J.}~\bibnamefont {Cai}},\ and\
  \bibinfo {author} {\bibfnamefont {A.~V.}\ \bibnamefont {Chubukov}},\ }\href
  {https://doi.org/10.1103/PhysRevB.93.075106} {\bibfield  {journal} {\bibinfo
  {journal} {Phys. Rev. B}\ }\textbf {\bibinfo {volume} {93}},\ \bibinfo
  {pages} {075106} (\bibinfo {year} {2016})}\BibitemShut {NoStop}%
\bibitem [{\citenamefont {McMillan}(1975)}]{McMillan1975_PRB}%
  \BibitemOpen
  \bibfield  {author} {\bibinfo {author} {\bibfnamefont {W.~L.}\ \bibnamefont
  {McMillan}},\ }\href {https://doi.org/10.1103/PhysRevB.12.1187} {\bibfield
  {journal} {\bibinfo  {journal} {Phys. Rev. B}\ }\textbf {\bibinfo {volume}
  {12}},\ \bibinfo {pages} {1187} (\bibinfo {year} {1975})}\BibitemShut
  {NoStop}%
\bibitem [{\citenamefont {Caprara}\ \emph {et~al.}(2007)\citenamefont
  {Caprara}, \citenamefont {Grilli}, \citenamefont {Di~Castro},\ and\
  \citenamefont {Enss}}]{Caprara2007_PRB}%
  \BibitemOpen
  \bibfield  {author} {\bibinfo {author} {\bibfnamefont {S.}~\bibnamefont
  {Caprara}}, \bibinfo {author} {\bibfnamefont {M.}~\bibnamefont {Grilli}},
  \bibinfo {author} {\bibfnamefont {C.}~\bibnamefont {Di~Castro}},\ and\
  \bibinfo {author} {\bibfnamefont {T.}~\bibnamefont {Enss}},\ }\href
  {https://doi.org/10.1103/PhysRevB.75.140505} {\bibfield  {journal} {\bibinfo
  {journal} {Phys. Rev. B}\ }\textbf {\bibinfo {volume} {75}},\ \bibinfo
  {pages} {140505} (\bibinfo {year} {2007})}\BibitemShut {NoStop}%
\bibitem [{\citenamefont {Caprara}\ \emph {et~al.}(2011)\citenamefont
  {Caprara}, \citenamefont {Di~Castro}, \citenamefont {Muschler}, \citenamefont
  {Prestel}, \citenamefont {Hackl}, \citenamefont {Lambacher}, \citenamefont
  {Erb}, \citenamefont {Komiya}, \citenamefont {Ando},\ and\ \citenamefont
  {Grilli}}]{Caprara2011_PRB}%
  \BibitemOpen
  \bibfield  {author} {\bibinfo {author} {\bibfnamefont {S.}~\bibnamefont
  {Caprara}}, \bibinfo {author} {\bibfnamefont {C.}~\bibnamefont {Di~Castro}},
  \bibinfo {author} {\bibfnamefont {B.}~\bibnamefont {Muschler}}, \bibinfo
  {author} {\bibfnamefont {W.}~\bibnamefont {Prestel}}, \bibinfo {author}
  {\bibfnamefont {R.}~\bibnamefont {Hackl}}, \bibinfo {author} {\bibfnamefont
  {M.}~\bibnamefont {Lambacher}}, \bibinfo {author} {\bibfnamefont
  {A.}~\bibnamefont {Erb}}, \bibinfo {author} {\bibfnamefont {S.}~\bibnamefont
  {Komiya}}, \bibinfo {author} {\bibfnamefont {Y.}~\bibnamefont {Ando}},\ and\
  \bibinfo {author} {\bibfnamefont {M.}~\bibnamefont {Grilli}},\ }\href
  {https://doi.org/10.1103/PhysRevB.84.054508} {\bibfield  {journal} {\bibinfo
  {journal} {Phys. Rev. B}\ }\textbf {\bibinfo {volume} {84}},\ \bibinfo
  {pages} {054508} (\bibinfo {year} {2011})}\BibitemShut {NoStop}%
\bibitem [{\citenamefont {Caprara}\ \emph {et~al.}(2015)\citenamefont
  {Caprara}, \citenamefont {Colonna}, \citenamefont {Di~Castro}, \citenamefont
  {Hackl}, \citenamefont {Muschler}, \citenamefont {Tassini},\ and\
  \citenamefont {Grilli}}]{Caprara2015_PRB}%
  \BibitemOpen
  \bibfield  {author} {\bibinfo {author} {\bibfnamefont {S.}~\bibnamefont
  {Caprara}}, \bibinfo {author} {\bibfnamefont {M.}~\bibnamefont {Colonna}},
  \bibinfo {author} {\bibfnamefont {C.}~\bibnamefont {Di~Castro}}, \bibinfo
  {author} {\bibfnamefont {R.}~\bibnamefont {Hackl}}, \bibinfo {author}
  {\bibfnamefont {B.}~\bibnamefont {Muschler}}, \bibinfo {author}
  {\bibfnamefont {L.}~\bibnamefont {Tassini}},\ and\ \bibinfo {author}
  {\bibfnamefont {M.}~\bibnamefont {Grilli}},\ }\href
  {https://doi.org/10.1103/PhysRevB.91.205115} {\bibfield  {journal} {\bibinfo
  {journal} {Phys. Rev. B}\ }\textbf {\bibinfo {volume} {91}},\ \bibinfo
  {pages} {205115} (\bibinfo {year} {2015})}\BibitemShut {NoStop}%
\bibitem [{\citenamefont {Paul}\ and\ \citenamefont
  {Garst}(2017)}]{Paul2017_PRL}%
  \BibitemOpen
  \bibfield  {author} {\bibinfo {author} {\bibfnamefont {I.}~\bibnamefont
  {Paul}}\ and\ \bibinfo {author} {\bibfnamefont {M.}~\bibnamefont {Garst}},\
  }\href {https://doi.org/10.1103/PhysRevLett.118.227601} {\bibfield  {journal}
  {\bibinfo  {journal} {Phys. Rev. Lett.}\ }\textbf {\bibinfo {volume} {118}},\
  \bibinfo {pages} {227601} (\bibinfo {year} {2017})}\BibitemShut {NoStop}%
\bibitem [{\citenamefont {Classen}\ \emph {et~al.}(2020)\citenamefont
  {Classen}, \citenamefont {Chubukov}, \citenamefont {Honerkamp},\ and\
  \citenamefont {Scherer}}]{Classen2020_PRB}%
  \BibitemOpen
  \bibfield  {author} {\bibinfo {author} {\bibfnamefont {L.}~\bibnamefont
  {Classen}}, \bibinfo {author} {\bibfnamefont {A.~V.}\ \bibnamefont
  {Chubukov}}, \bibinfo {author} {\bibfnamefont {C.}~\bibnamefont
  {Honerkamp}},\ and\ \bibinfo {author} {\bibfnamefont {M.~M.}\ \bibnamefont
  {Scherer}},\ }\href {https://doi.org/10.1103/PhysRevB.102.125141} {\bibfield
  {journal} {\bibinfo  {journal} {Phys. Rev. B}\ }\textbf {\bibinfo {volume}
  {102}},\ \bibinfo {pages} {125141} (\bibinfo {year} {2020})}\BibitemShut
  {NoStop}%
\bibitem [{\citenamefont {Dolgirev}\ \emph {et~al.}(2020)\citenamefont
  {Dolgirev}, \citenamefont {Michael}, \citenamefont {Zong}, \citenamefont
  {Gedik},\ and\ \citenamefont {Demler}}]{Dolgirev2020_PRB}%
  \BibitemOpen
  \bibfield  {author} {\bibinfo {author} {\bibfnamefont {P.~E.}\ \bibnamefont
  {Dolgirev}}, \bibinfo {author} {\bibfnamefont {M.~H.}\ \bibnamefont
  {Michael}}, \bibinfo {author} {\bibfnamefont {A.}~\bibnamefont {Zong}},
  \bibinfo {author} {\bibfnamefont {N.}~\bibnamefont {Gedik}},\ and\ \bibinfo
  {author} {\bibfnamefont {E.}~\bibnamefont {Demler}},\ }\href
  {https://doi.org/10.1103/PhysRevB.101.174306} {\bibfield  {journal} {\bibinfo
   {journal} {Phys. Rev. B}\ }\textbf {\bibinfo {volume} {101}},\ \bibinfo
  {pages} {174306} (\bibinfo {year} {2020})}\BibitemShut {NoStop}%
\bibitem [{\citenamefont {Grandi}\ and\ \citenamefont
  {Eckstein}(2021)}]{Grandi2021_PRB}%
  \BibitemOpen
  \bibfield  {author} {\bibinfo {author} {\bibfnamefont {F.}~\bibnamefont
  {Grandi}}\ and\ \bibinfo {author} {\bibfnamefont {M.}~\bibnamefont
  {Eckstein}},\ }\href {https://doi.org/10.1103/PhysRevB.103.245117} {\bibfield
   {journal} {\bibinfo  {journal} {Phys. Rev. B}\ }\textbf {\bibinfo {volume}
  {103}},\ \bibinfo {pages} {245117} (\bibinfo {year} {2021})}\BibitemShut
  {NoStop}%
\bibitem [{Sup()}]{Suppl_mat}%
  \BibitemOpen
  \href@noop {} {}\bibinfo {note} {See Supplemental Material at [URL will be
  inserted by publisher] for details concerning the derivation of the
  Ginzburg-Landau theories for the $4 \times 4$ and the d-wave Pomeranchuk
  instability and the derivation of the nematic response functions. Finally, we
  discuss the effect of nonzero stress on the effective free energy of the
  problem.}\BibitemShut {Stop}%
\bibitem [{\citenamefont {Fernandes}\ \emph
  {et~al.}(2012{\natexlab{b}})\citenamefont {Fernandes}, \citenamefont
  {Chubukov}, \citenamefont {Knolle}, \citenamefont {Eremin},\ and\
  \citenamefont {Schmalian}}]{Fernandes2012_PRBE}%
  \BibitemOpen
  \bibfield  {author} {\bibinfo {author} {\bibfnamefont {R.~M.}\ \bibnamefont
  {Fernandes}}, \bibinfo {author} {\bibfnamefont {A.~V.}\ \bibnamefont
  {Chubukov}}, \bibinfo {author} {\bibfnamefont {J.}~\bibnamefont {Knolle}},
  \bibinfo {author} {\bibfnamefont {I.}~\bibnamefont {Eremin}},\ and\ \bibinfo
  {author} {\bibfnamefont {J.}~\bibnamefont {Schmalian}},\ }\href
  {https://doi.org/10.1103/PhysRevB.85.109901} {\bibfield  {journal} {\bibinfo
  {journal} {Phys. Rev. B}\ }\textbf {\bibinfo {volume} {85}},\ \bibinfo
  {pages} {109901} (\bibinfo {year} {2012}{\natexlab{b}})}\BibitemShut
  {NoStop}%
\bibitem [{\citenamefont {Straley}\ and\ \citenamefont
  {Fisher}(1973)}]{Straley1973_JPA}%
  \BibitemOpen
  \bibfield  {author} {\bibinfo {author} {\bibfnamefont {J.~P.}\ \bibnamefont
  {Straley}}\ and\ \bibinfo {author} {\bibfnamefont {M.~E.}\ \bibnamefont
  {Fisher}},\ }\href {https://doi.org/10.1088/0305-4470/6/9/007} {\bibfield
  {journal} {\bibinfo  {journal} {Journal of Physics A: Mathematical, Nuclear
  and General}\ }\textbf {\bibinfo {volume} {6}},\ \bibinfo {pages} {1310}
  (\bibinfo {year} {1973})}\BibitemShut {NoStop}%
\bibitem [{\citenamefont {Baek}\ \emph {et~al.}(2011)\citenamefont {Baek},
  \citenamefont {M\"akel\"a}, \citenamefont {Minnhagen},\ and\ \citenamefont
  {Kim}}]{Baek2011_PRE}%
  \BibitemOpen
  \bibfield  {author} {\bibinfo {author} {\bibfnamefont {S.~K.}\ \bibnamefont
  {Baek}}, \bibinfo {author} {\bibfnamefont {H.}~\bibnamefont {M\"akel\"a}},
  \bibinfo {author} {\bibfnamefont {P.}~\bibnamefont {Minnhagen}},\ and\
  \bibinfo {author} {\bibfnamefont {B.~J.}\ \bibnamefont {Kim}},\ }\href
  {https://doi.org/10.1103/PhysRevE.83.061104} {\bibfield  {journal} {\bibinfo
  {journal} {Phys. Rev. E}\ }\textbf {\bibinfo {volume} {83}},\ \bibinfo
  {pages} {061104} (\bibinfo {year} {2011})}\BibitemShut {NoStop}%
\bibitem [{\citenamefont {Little}\ \emph {et~al.}(2020)\citenamefont {Little},
  \citenamefont {Lee}, \citenamefont {John}, \citenamefont {Doyle},
  \citenamefont {Maniv}, \citenamefont {Nair}, \citenamefont {Chen},
  \citenamefont {Rees}, \citenamefont {Venderbos}, \citenamefont {Fernandes},
  \citenamefont {Analytis},\ and\ \citenamefont
  {Orenstein}}]{Little2020_NatMat}%
  \BibitemOpen
  \bibfield  {author} {\bibinfo {author} {\bibfnamefont {A.}~\bibnamefont
  {Little}}, \bibinfo {author} {\bibfnamefont {C.}~\bibnamefont {Lee}},
  \bibinfo {author} {\bibfnamefont {C.}~\bibnamefont {John}}, \bibinfo {author}
  {\bibfnamefont {S.}~\bibnamefont {Doyle}}, \bibinfo {author} {\bibfnamefont
  {E.}~\bibnamefont {Maniv}}, \bibinfo {author} {\bibfnamefont {N.~L.}\
  \bibnamefont {Nair}}, \bibinfo {author} {\bibfnamefont {W.}~\bibnamefont
  {Chen}}, \bibinfo {author} {\bibfnamefont {D.}~\bibnamefont {Rees}}, \bibinfo
  {author} {\bibfnamefont {J.~W.~F.}\ \bibnamefont {Venderbos}}, \bibinfo
  {author} {\bibfnamefont {R.~M.}\ \bibnamefont {Fernandes}}, \bibinfo {author}
  {\bibfnamefont {J.~G.}\ \bibnamefont {Analytis}},\ and\ \bibinfo {author}
  {\bibfnamefont {J.}~\bibnamefont {Orenstein}},\ }\href
  {https://doi.org/10.1038/s41563-020-0681-0} {\bibfield  {journal} {\bibinfo
  {journal} {Nature Materials}\ }\textbf {\bibinfo {volume} {19}},\ \bibinfo
  {pages} {1062} (\bibinfo {year} {2020})}\BibitemShut {NoStop}%
\bibitem [{\citenamefont {Fernandes}\ and\ \citenamefont
  {Venderbos}(2020)}]{Fernandes2020_SciAdv}%
  \BibitemOpen
  \bibfield  {author} {\bibinfo {author} {\bibfnamefont {R.~M.}\ \bibnamefont
  {Fernandes}}\ and\ \bibinfo {author} {\bibfnamefont {J.~W.~F.}\ \bibnamefont
  {Venderbos}},\ }\href {https://doi.org/10.1126/sciadv.aba8834} {\bibfield
  {journal} {\bibinfo  {journal} {Science Advances}\ }\textbf {\bibinfo
  {volume} {6}},\ \bibinfo {pages} {eaba8834} (\bibinfo {year}
  {2020})}\BibitemShut {NoStop}%
\bibitem [{\citenamefont {Kimura}\ \emph {et~al.}(2022)\citenamefont {Kimura},
  \citenamefont {Sigrist},\ and\ \citenamefont {Kawakami}}]{Kimura2022_PRB}%
  \BibitemOpen
  \bibfield  {author} {\bibinfo {author} {\bibfnamefont {K.}~\bibnamefont
  {Kimura}}, \bibinfo {author} {\bibfnamefont {M.}~\bibnamefont {Sigrist}},\
  and\ \bibinfo {author} {\bibfnamefont {N.}~\bibnamefont {Kawakami}},\ }\href
  {https://doi.org/10.1103/PhysRevB.105.035130} {\bibfield  {journal} {\bibinfo
   {journal} {Phys. Rev. B}\ }\textbf {\bibinfo {volume} {105}},\ \bibinfo
  {pages} {035130} (\bibinfo {year} {2022})}\BibitemShut {NoStop}%
\bibitem [{\citenamefont {Fernandes}\ \emph {et~al.}(2010)\citenamefont
  {Fernandes}, \citenamefont {VanBebber}, \citenamefont {Bhattacharya},
  \citenamefont {Chandra}, \citenamefont {Keppens}, \citenamefont {Mandrus},
  \citenamefont {McGuire}, \citenamefont {Sales}, \citenamefont {Sefat},\ and\
  \citenamefont {Schmalian}}]{Fernandes2010_PRL}%
  \BibitemOpen
  \bibfield  {author} {\bibinfo {author} {\bibfnamefont {R.~M.}\ \bibnamefont
  {Fernandes}}, \bibinfo {author} {\bibfnamefont {L.~H.}\ \bibnamefont
  {VanBebber}}, \bibinfo {author} {\bibfnamefont {S.}~\bibnamefont
  {Bhattacharya}}, \bibinfo {author} {\bibfnamefont {P.}~\bibnamefont
  {Chandra}}, \bibinfo {author} {\bibfnamefont {V.}~\bibnamefont {Keppens}},
  \bibinfo {author} {\bibfnamefont {D.}~\bibnamefont {Mandrus}}, \bibinfo
  {author} {\bibfnamefont {M.~A.}\ \bibnamefont {McGuire}}, \bibinfo {author}
  {\bibfnamefont {B.~C.}\ \bibnamefont {Sales}}, \bibinfo {author}
  {\bibfnamefont {A.~S.}\ \bibnamefont {Sefat}},\ and\ \bibinfo {author}
  {\bibfnamefont {J.}~\bibnamefont {Schmalian}},\ }\href
  {https://doi.org/10.1103/PhysRevLett.105.157003} {\bibfield  {journal}
  {\bibinfo  {journal} {Phys. Rev. Lett.}\ }\textbf {\bibinfo {volume} {105}},\
  \bibinfo {pages} {157003} (\bibinfo {year} {2010})}\BibitemShut {NoStop}%
\bibitem [{\citenamefont {Fernandes}\ and\ \citenamefont
  {Schmalian}(2012)}]{Fernandes2012_SuScTe}%
  \BibitemOpen
  \bibfield  {author} {\bibinfo {author} {\bibfnamefont {R.~M.}\ \bibnamefont
  {Fernandes}}\ and\ \bibinfo {author} {\bibfnamefont {J.}~\bibnamefont
  {Schmalian}},\ }\href {https://doi.org/10.1088/0953-2048/25/8/084005}
  {\bibfield  {journal} {\bibinfo  {journal} {Superconductor Science and
  Technology}\ }\textbf {\bibinfo {volume} {25}},\ \bibinfo {pages} {084005}
  (\bibinfo {year} {2012})}\BibitemShut {NoStop}%
\bibitem [{\citenamefont {Khodas}\ and\ \citenamefont
  {Levchenko}(2015)}]{Khodas2015_PRB}%
  \BibitemOpen
  \bibfield  {author} {\bibinfo {author} {\bibfnamefont {M.}~\bibnamefont
  {Khodas}}\ and\ \bibinfo {author} {\bibfnamefont {A.}~\bibnamefont
  {Levchenko}},\ }\href {https://doi.org/10.1103/PhysRevB.91.235119} {\bibfield
   {journal} {\bibinfo  {journal} {Phys. Rev. B}\ }\textbf {\bibinfo {volume}
  {91}},\ \bibinfo {pages} {235119} (\bibinfo {year} {2015})}\BibitemShut
  {NoStop}%
\bibitem [{\citenamefont {Yamase}\ and\ \citenamefont
  {Zeyher}(2015)}]{Yamase2015_NJP}%
  \BibitemOpen
  \bibfield  {author} {\bibinfo {author} {\bibfnamefont {H.}~\bibnamefont
  {Yamase}}\ and\ \bibinfo {author} {\bibfnamefont {R.}~\bibnamefont
  {Zeyher}},\ }\href {https://doi.org/10.1088/1367-2630/17/7/073030} {\bibfield
   {journal} {\bibinfo  {journal} {New Journal of Physics}\ }\textbf {\bibinfo
  {volume} {17}},\ \bibinfo {pages} {073030} (\bibinfo {year}
  {2015})}\BibitemShut {NoStop}%
\bibitem [{\citenamefont {Fernandes}\ and\ \citenamefont
  {Chubukov}(2016)}]{Fernandes2017_RPP}%
  \BibitemOpen
  \bibfield  {author} {\bibinfo {author} {\bibfnamefont {R.~M.}\ \bibnamefont
  {Fernandes}}\ and\ \bibinfo {author} {\bibfnamefont {A.~V.}\ \bibnamefont
  {Chubukov}},\ }\href {https://doi.org/10.1088/1361-6633/80/1/014503}
  {\bibfield  {journal} {\bibinfo  {journal} {Reports on Progress in Physics}\
  }\textbf {\bibinfo {volume} {80}},\ \bibinfo {pages} {014503} (\bibinfo
  {year} {2016})}\BibitemShut {NoStop}%
\bibitem [{\citenamefont {Gallais}\ and\ \citenamefont
  {Paul}(2016)}]{Gallais2016_CRP}%
  \BibitemOpen
  \bibfield  {author} {\bibinfo {author} {\bibfnamefont {Y.}~\bibnamefont
  {Gallais}}\ and\ \bibinfo {author} {\bibfnamefont {I.}~\bibnamefont {Paul}},\
  }\href {https://doi.org/https://doi.org/10.1016/j.crhy.2015.10.001}
  {\bibfield  {journal} {\bibinfo  {journal} {Comptes Rendus Physique}\
  }\textbf {\bibinfo {volume} {17}},\ \bibinfo {pages} {113} (\bibinfo {year}
  {2016})}\BibitemShut {NoStop}%
\bibitem [{\citenamefont {Gallais}\ \emph {et~al.}(2016)\citenamefont
  {Gallais}, \citenamefont {Paul}, \citenamefont {Chauvi\`ere},\ and\
  \citenamefont {Schmalian}}]{Gallais2016_PRL}%
  \BibitemOpen
  \bibfield  {author} {\bibinfo {author} {\bibfnamefont {Y.}~\bibnamefont
  {Gallais}}, \bibinfo {author} {\bibfnamefont {I.}~\bibnamefont {Paul}},
  \bibinfo {author} {\bibfnamefont {L.}~\bibnamefont {Chauvi\`ere}},\ and\
  \bibinfo {author} {\bibfnamefont {J.}~\bibnamefont {Schmalian}},\ }\href
  {https://doi.org/10.1103/PhysRevLett.116.017001} {\bibfield  {journal}
  {\bibinfo  {journal} {Phys. Rev. Lett.}\ }\textbf {\bibinfo {volume} {116}},\
  \bibinfo {pages} {017001} (\bibinfo {year} {2016})}\BibitemShut {NoStop}%
\bibitem [{\citenamefont {Kiesel}\ \emph {et~al.}(2013)\citenamefont {Kiesel},
  \citenamefont {Platt},\ and\ \citenamefont {Thomale}}]{Kiesel2013_PRL}%
  \BibitemOpen
  \bibfield  {author} {\bibinfo {author} {\bibfnamefont {M.~L.}\ \bibnamefont
  {Kiesel}}, \bibinfo {author} {\bibfnamefont {C.}~\bibnamefont {Platt}},\ and\
  \bibinfo {author} {\bibfnamefont {R.}~\bibnamefont {Thomale}},\ }\href
  {https://doi.org/10.1103/PhysRevLett.110.126405} {\bibfield  {journal}
  {\bibinfo  {journal} {Phys. Rev. Lett.}\ }\textbf {\bibinfo {volume} {110}},\
  \bibinfo {pages} {126405} (\bibinfo {year} {2013})}\BibitemShut {NoStop}%
\bibitem [{\citenamefont {Wang}\ \emph {et~al.}(2013)\citenamefont {Wang},
  \citenamefont {Li}, \citenamefont {Xiang},\ and\ \citenamefont
  {Wang}}]{Wang2013_PRB}%
  \BibitemOpen
  \bibfield  {author} {\bibinfo {author} {\bibfnamefont {W.-S.}\ \bibnamefont
  {Wang}}, \bibinfo {author} {\bibfnamefont {Z.-Z.}\ \bibnamefont {Li}},
  \bibinfo {author} {\bibfnamefont {Y.-Y.}\ \bibnamefont {Xiang}},\ and\
  \bibinfo {author} {\bibfnamefont {Q.-H.}\ \bibnamefont {Wang}},\ }\href
  {https://doi.org/10.1103/PhysRevB.87.115135} {\bibfield  {journal} {\bibinfo
  {journal} {Phys. Rev. B}\ }\textbf {\bibinfo {volume} {87}},\ \bibinfo
  {pages} {115135} (\bibinfo {year} {2013})}\BibitemShut {NoStop}%
\bibitem [{\citenamefont {Abanov}\ and\ \citenamefont
  {Chubukov}(2000)}]{Abanov2000_PRL}%
  \BibitemOpen
  \bibfield  {author} {\bibinfo {author} {\bibfnamefont {A.}~\bibnamefont
  {Abanov}}\ and\ \bibinfo {author} {\bibfnamefont {A.~V.}\ \bibnamefont
  {Chubukov}},\ }\href {https://doi.org/10.1103/PhysRevLett.84.5608} {\bibfield
   {journal} {\bibinfo  {journal} {Phys. Rev. Lett.}\ }\textbf {\bibinfo
  {volume} {84}},\ \bibinfo {pages} {5608} (\bibinfo {year}
  {2000})}\BibitemShut {NoStop}%
\bibitem [{\citenamefont {Garst}\ and\ \citenamefont
  {Chubukov}(2010)}]{Garst2010_PRB}%
  \BibitemOpen
  \bibfield  {author} {\bibinfo {author} {\bibfnamefont {M.}~\bibnamefont
  {Garst}}\ and\ \bibinfo {author} {\bibfnamefont {A.~V.}\ \bibnamefont
  {Chubukov}},\ }\href {https://doi.org/10.1103/PhysRevB.81.235105} {\bibfield
  {journal} {\bibinfo  {journal} {Phys. Rev. B}\ }\textbf {\bibinfo {volume}
  {81}},\ \bibinfo {pages} {235105} (\bibinfo {year} {2010})}\BibitemShut
  {NoStop}%
\bibitem [{\citenamefont {Paul}(2014)}]{Paul2014_PRB}%
  \BibitemOpen
  \bibfield  {author} {\bibinfo {author} {\bibfnamefont {I.}~\bibnamefont
  {Paul}},\ }\href {https://doi.org/10.1103/PhysRevB.90.115102} {\bibfield
  {journal} {\bibinfo  {journal} {Phys. Rev. B}\ }\textbf {\bibinfo {volume}
  {90}},\ \bibinfo {pages} {115102} (\bibinfo {year} {2014})}\BibitemShut
  {NoStop}%
\bibitem [{\citenamefont {Asaba}\ \emph {et~al.}(2024)\citenamefont {Asaba},
  \citenamefont {Onishi}, \citenamefont {Kageyama}, \citenamefont {Kiyosue},
  \citenamefont {Ohtsuka}, \citenamefont {Suetsugu}, \citenamefont {Kohsaka},
  \citenamefont {Gaggl}, \citenamefont {Kasahara}, \citenamefont {Murayama},
  \citenamefont {Hashimoto}, \citenamefont {Tazai}, \citenamefont {Kontani},
  \citenamefont {Ortiz}, \citenamefont {Wilson}, \citenamefont {Li},
  \citenamefont {Wen}, \citenamefont {Shibauchi},\ and\ \citenamefont
  {Matsuda}}]{Asaba2024_NatPhys}%
  \BibitemOpen
  \bibfield  {author} {\bibinfo {author} {\bibfnamefont {T.}~\bibnamefont
  {Asaba}}, \bibinfo {author} {\bibfnamefont {A.}~\bibnamefont {Onishi}},
  \bibinfo {author} {\bibfnamefont {Y.}~\bibnamefont {Kageyama}}, \bibinfo
  {author} {\bibfnamefont {T.}~\bibnamefont {Kiyosue}}, \bibinfo {author}
  {\bibfnamefont {K.}~\bibnamefont {Ohtsuka}}, \bibinfo {author} {\bibfnamefont
  {S.}~\bibnamefont {Suetsugu}}, \bibinfo {author} {\bibfnamefont
  {Y.}~\bibnamefont {Kohsaka}}, \bibinfo {author} {\bibfnamefont
  {T.}~\bibnamefont {Gaggl}}, \bibinfo {author} {\bibfnamefont
  {Y.}~\bibnamefont {Kasahara}}, \bibinfo {author} {\bibfnamefont
  {H.}~\bibnamefont {Murayama}}, \bibinfo {author} {\bibfnamefont
  {K.}~\bibnamefont {Hashimoto}}, \bibinfo {author} {\bibfnamefont
  {R.}~\bibnamefont {Tazai}}, \bibinfo {author} {\bibfnamefont
  {H.}~\bibnamefont {Kontani}}, \bibinfo {author} {\bibfnamefont {B.~R.}\
  \bibnamefont {Ortiz}}, \bibinfo {author} {\bibfnamefont {S.~D.}\ \bibnamefont
  {Wilson}}, \bibinfo {author} {\bibfnamefont {Q.}~\bibnamefont {Li}}, \bibinfo
  {author} {\bibfnamefont {H.~H.}\ \bibnamefont {Wen}}, \bibinfo {author}
  {\bibfnamefont {T.}~\bibnamefont {Shibauchi}},\ and\ \bibinfo {author}
  {\bibfnamefont {Y.}~\bibnamefont {Matsuda}},\ }\href
  {https://doi.org/10.1038/s41567-023-02272-4} {\bibfield  {journal} {\bibinfo
  {journal} {Nature Physics}\ }\textbf {\bibinfo {volume} {20}},\ \bibinfo
  {pages} {40} (\bibinfo {year} {2024})}\BibitemShut {NoStop}%
\bibitem [{\citenamefont {Hu}\ \emph {et~al.}(2023)\citenamefont {Hu},
  \citenamefont {Le}, \citenamefont {Zhang}, \citenamefont {Zhao},
  \citenamefont {Liu}, \citenamefont {Ma}, \citenamefont {Plumb}, \citenamefont
  {Radovic}, \citenamefont {Chen}, \citenamefont {Schnyder}, \citenamefont
  {Wu}, \citenamefont {Dong}, \citenamefont {Hu}, \citenamefont {Yang},
  \citenamefont {Gao},\ and\ \citenamefont {Shi}}]{Hu2023_NatPhys}%
  \BibitemOpen
  \bibfield  {author} {\bibinfo {author} {\bibfnamefont {Y.}~\bibnamefont
  {Hu}}, \bibinfo {author} {\bibfnamefont {C.}~\bibnamefont {Le}}, \bibinfo
  {author} {\bibfnamefont {Y.}~\bibnamefont {Zhang}}, \bibinfo {author}
  {\bibfnamefont {Z.}~\bibnamefont {Zhao}}, \bibinfo {author} {\bibfnamefont
  {J.}~\bibnamefont {Liu}}, \bibinfo {author} {\bibfnamefont {J.}~\bibnamefont
  {Ma}}, \bibinfo {author} {\bibfnamefont {N.~C.}\ \bibnamefont {Plumb}},
  \bibinfo {author} {\bibfnamefont {M.}~\bibnamefont {Radovic}}, \bibinfo
  {author} {\bibfnamefont {H.}~\bibnamefont {Chen}}, \bibinfo {author}
  {\bibfnamefont {A.~P.}\ \bibnamefont {Schnyder}}, \bibinfo {author}
  {\bibfnamefont {X.}~\bibnamefont {Wu}}, \bibinfo {author} {\bibfnamefont
  {X.}~\bibnamefont {Dong}}, \bibinfo {author} {\bibfnamefont {J.}~\bibnamefont
  {Hu}}, \bibinfo {author} {\bibfnamefont {H.}~\bibnamefont {Yang}}, \bibinfo
  {author} {\bibfnamefont {H.-J.}\ \bibnamefont {Gao}},\ and\ \bibinfo {author}
  {\bibfnamefont {M.}~\bibnamefont {Shi}},\ }\href
  {https://doi.org/10.1038/s41567-023-02215-z} {\bibfield  {journal} {\bibinfo
  {journal} {Nature Physics}\ } (\bibinfo {year} {2023})}\BibitemShut {NoStop}%
\bibitem [{\citenamefont {Yang}\ \emph {et~al.}(2024)\citenamefont {Yang},
  \citenamefont {Ye}, \citenamefont {Zhao}, \citenamefont {Liu}, \citenamefont
  {Yi}, \citenamefont {Zhang}, \citenamefont {Xiao}, \citenamefont {Shi},
  \citenamefont {You}, \citenamefont {Huang}, \citenamefont {Wang},
  \citenamefont {Wang}, \citenamefont {Guo}, \citenamefont {Lin}, \citenamefont
  {Shen}, \citenamefont {Zhou}, \citenamefont {Chen}, \citenamefont {Dong},
  \citenamefont {Su}, \citenamefont {Wang},\ and\ \citenamefont
  {Gao}}]{Yang2024_NatComm}%
  \BibitemOpen
  \bibfield  {author} {\bibinfo {author} {\bibfnamefont {H.}~\bibnamefont
  {Yang}}, \bibinfo {author} {\bibfnamefont {Y.}~\bibnamefont {Ye}}, \bibinfo
  {author} {\bibfnamefont {Z.}~\bibnamefont {Zhao}}, \bibinfo {author}
  {\bibfnamefont {J.}~\bibnamefont {Liu}}, \bibinfo {author} {\bibfnamefont
  {X.-W.}\ \bibnamefont {Yi}}, \bibinfo {author} {\bibfnamefont
  {Y.}~\bibnamefont {Zhang}}, \bibinfo {author} {\bibfnamefont
  {H.}~\bibnamefont {Xiao}}, \bibinfo {author} {\bibfnamefont {J.}~\bibnamefont
  {Shi}}, \bibinfo {author} {\bibfnamefont {J.-Y.}\ \bibnamefont {You}},
  \bibinfo {author} {\bibfnamefont {Z.}~\bibnamefont {Huang}}, \bibinfo
  {author} {\bibfnamefont {B.}~\bibnamefont {Wang}}, \bibinfo {author}
  {\bibfnamefont {J.}~\bibnamefont {Wang}}, \bibinfo {author} {\bibfnamefont
  {H.}~\bibnamefont {Guo}}, \bibinfo {author} {\bibfnamefont {X.}~\bibnamefont
  {Lin}}, \bibinfo {author} {\bibfnamefont {C.}~\bibnamefont {Shen}}, \bibinfo
  {author} {\bibfnamefont {W.}~\bibnamefont {Zhou}}, \bibinfo {author}
  {\bibfnamefont {H.}~\bibnamefont {Chen}}, \bibinfo {author} {\bibfnamefont
  {X.}~\bibnamefont {Dong}}, \bibinfo {author} {\bibfnamefont {G.}~\bibnamefont
  {Su}}, \bibinfo {author} {\bibfnamefont {Z.}~\bibnamefont {Wang}},\ and\
  \bibinfo {author} {\bibfnamefont {H.-J.}\ \bibnamefont {Gao}},\ }\href
  {https://doi.org/10.1038/s41467-024-53870-6} {\bibfield  {journal} {\bibinfo
  {journal} {Nature Communications}\ }\textbf {\bibinfo {volume} {15}},\
  \bibinfo {pages} {9626} (\bibinfo {year} {2024})}\BibitemShut {NoStop}%
\end{thebibliography}

%


\newpage~\newpage~

\appendix 
\onecolumngrid

\section*{Supplemental material for: 'Theories for charge-driven nematicity in kagome metals'}

In this supplemental material, we present some additional information concerning the computation of the nematic susceptibilities as emerging from the three $1 \times 4$ charge fluctuations and the zero momentum charge order. Finally, we present the calculations for the presence of a finite stress in the system and the definition of a possible nematic order parameter.

\subsection{Microscopic model for the $2 \times 2$ charge order}
In the reduced Brillouin zone (once the $2 \times 2$ charge order has settled in), the Fermi surface of the system is composed of three inequivalent pockets located at the reconstructed M points (M') \cite{Zhou2022_NatComm,Dong2023_PRB,Varma2023_PRB} with hole-like character related to the states coming from the d orbitals of the vanadium atoms \cite{Li2022_NatComm}, see Fig.~1 of the main text. The relevance of the hole pockets for the formation of the superconducting state seems to be in agreement with recent experimental findings that observe an increase in the superconducting critical temperature by hole doping and a decrease by electron doping \cite{Oey2022_PRM1,Oey2022_PRM2,Yang2022_SciBull,CapaSalinas2023_FEM}. In the following, we assume the elliptical hole pockets are the relevant features of the Fermi surface.

\noindent
We start from a minimal three-band model (six bands counting the spin degeneracy) for the electronic structure in the $2 \times 2$ charge-ordered state with hole-pockets centered at $M_1'$, $M_2'$ and $M_3'$. To keep the calculations simple, we consider elliptic paraboloid dispersions: 
\begin{align} \label{eq_sup:dispersionsQ4}
	& \varepsilon_{1, \mathbf{k}} = - \frac{k_x^2}{2} \frac{3 m_x + m_y}{4 m_x m_y} - \frac{\sqrt{3} k_x k_y}{2} \frac{m_x - m_y}{2 m_x m_y} - \frac{k_y^2}{2} \frac{m_x + 3 m_y}{4 m_x m_y} , \nonumber \\
	& \varepsilon_{2, \mathbf{k}} = - \frac{k_x^2}{2 m_x} - \frac{k_y^2}{2 m_y} , \\
	& \varepsilon_{3, \mathbf{k}} = - \frac{k_x^2}{2} \frac{3 m_x + m_y}{4 m_x m_y} + \frac{\sqrt{3} k_x k_y}{2} \frac{m_x - m_y}{2 m_x m_y} - \frac{k_y^2}{2} \frac{m_x + 3 m_y}{4 m_x m_y} , \nonumber
\end{align}
where $m_x = m (1 + \delta)$ and $m_y = m (1 - \delta)$ are the electronic masses along the $k_x$ and $k_y$ directions of reciprocal space, respectively, and they are proportional to the inverse hopping $m = 1/t$, see Fig.~1c of the main text for a sketch of the Fermi surface of the problem \cite{Eremin2010_PRB,Vorontsov2010_PRB}. In the following, we take $t=1$ as our energy unit. To convert in physical units the value of the hopping is $t \sim 0.3$eV \cite{Denner2021_PRL}. Since the dispersions Eq.~\eqref{eq_sup:dispersionsQ4} describe the physics of the system in the $2 \times 2$ charge ordered state, we consider temperatures T$_\text{sc} \sim 1\text{K} < \text{T} < \text{T}_\text{co} \sim 100$K, i.e., $3 \ 10^{-6} \lesssim k_B T / t \lesssim 3 \ 10^{-4}$, with $k_B$ the Boltzmann constant. The noninteracting Hamiltonian can thus be rewritten as:
\begin{align} \label{eq_sup:nonint_hamQ4}
	& \mathcal{H}_0 = \sum_{i=1,2,3} \sum_{\mathbf{k}, \sigma} \big( \varepsilon_{i, \mathbf{k}} - \mu \big) c^\dagger_{i, \mathbf{k}, \sigma} c_{i, \mathbf{k}, \sigma} ,
\end{align}
where $c^\dagger_{i, \mathbf{k}, \sigma}$ ($c_{i, \mathbf{k}, \sigma}$) is the creation (destruction) operator of an electron in the band $\varepsilon_{i, \mathbf{k}}$ with spin $\sigma = \uparrow, \downarrow$ and momentum $\mathbf{k}$, and $\mu$ is the chemical potential. We assume $\mu/(/t \Lambda^2) = -3 \ 10^{-5}$ \cite{Park2021_PRB}. In Eq.~\eqref{eq_sup:nonint_hamQ4}, the momenta of the $i$-th fermions are defined from $\mathbf{M}_i'$.

\noindent
Since all the bands are of hole kind, there are no interactions between electrons and holes, but only between holes. The structure of the interactions is described with the g-ology model \cite{Nandkishore2012_NatPhys,Park2021_PRB}:
\begin{align} \label{eq_sup:ham_patch_intQ4}
	\mathcal{H}_\text{int} = \frac{1}{2 N} & \sum_{|\mathbf{k}_1|, \cdots , |\mathbf{k}_4| < \Lambda} \delta \big( \mathbf{k}_1 + \mathbf{k}_2 - \mathbf{k}_3 - \mathbf{k}_4 \big) \Big[ \sum_{i \neq j, \sigma, \sigma'} \big( g_1 c^\dagger_{i, \mathbf{k}_1, \sigma} c^\dagger_{j, \mathbf{k}_2, \sigma'} c_{i, \mathbf{k}_3, \sigma'} c_{j, \mathbf{k}_4, \sigma} + g_2 c^\dagger_{i, \mathbf{k}_1, \sigma} c^\dagger_{j, \mathbf{k}_2, \sigma'} c_{j, \mathbf{k}_3, \sigma'} c_{i, \mathbf{k}_4, \sigma} \nonumber \\
	& + g_3 c^\dagger_{i, \mathbf{k}_1, \sigma} c^\dagger_{i, \mathbf{k}_2, \sigma'} c_{j, \mathbf{k}_3, \sigma'} c_{j, \mathbf{k}_4, \sigma} \big) + g_4 \sum_{i, \sigma, \sigma'} c^\dagger_{i, \mathbf{k}_1, \sigma} c^\dagger_{i, \mathbf{k}_2, \sigma'} c_{i, \mathbf{k}_3, \sigma'} c_{i, \mathbf{k}_4, \sigma} \Big] ,
\end{align}
with $N$ the number of unit cells in the system. $g_1$ ($g_2$) [$g_3$] \{$g_4$\} is the interpatch exchange (interpatch density-density) [umklapp] \{intrapatch density-density\} scattering process and $\Lambda$ is a momentum and energy cutoff.

\newpage

\subsection{Three $1 \times 4$ fluctuations and $1 \times 4$ order within the $2 \times 2$ charge ordered state}
\begin{figure}	\centerline{\includegraphics[width=1.0\textwidth]{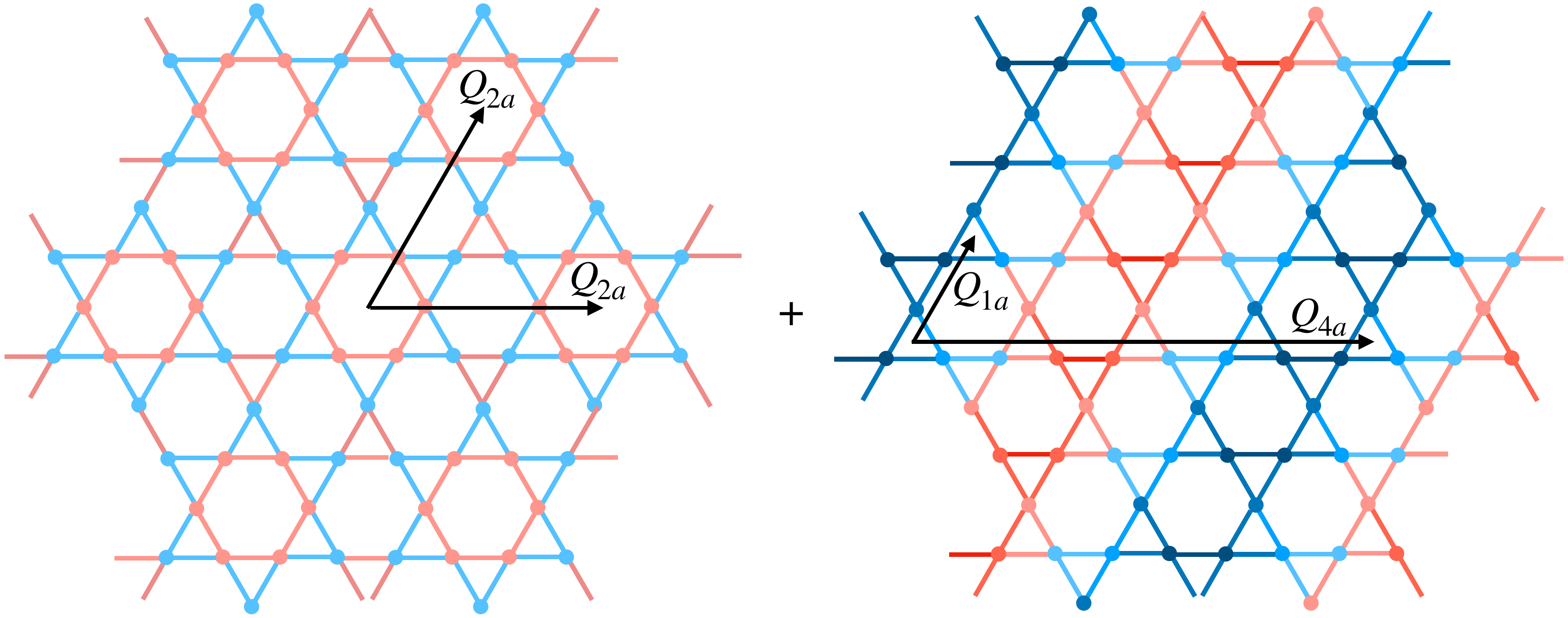}}
	\caption{ \textbf{$1 \times 4$ within the $2 \times 2$ charge order -} Putative shapes of the $2 \times 2$ charge ordered state (left) and of the $1 \times 4$ one (right) with the corresponding unit vectors. A dark blue (red) site or bond indicate low (high) occupation or weak (strong) bond strength.} \label{fig_sup:2x2_vs_1x4}
\end{figure}

The patch model interaction Eq.~\eqref{eq_sup:ham_patch_intQ4} can be rewritten as:
\begin{align} \label{eq_sup:cc_intQ4}
	\mathcal{H}_\text{int}^{Q_{4 a}} \approx - N  \frac{g_{\text{co}}}{2}  \sum_{i} \rho_i \rho_i ,
\end{align}
with $g_\text{co} = \Gamma_{Q_{4a}}$ and $\rho_i = \frac{|\epsilon_{i j l}|}{2 N} \sum_\sigma \sum_{\mathbf{q} < \Lambda} c^\dagger_{j, \mathbf{q}, \sigma} c_{l, \mathbf{q}, \sigma}$ is the charge operator with momentum $\mathbf{Q}_i' = \mathbf{M}_j' - \mathbf{M}_l'$, i.e., it describes an instability towards three one-dimensional charge ordered states that might stabilize an additional $1 \times 4$ charge order besides the $2 \times 2$ one, see Fig.~\ref{fig_sup:2x2_vs_1x4}. We assume the interaction constant $g_\text{co}$ in Eq.~\eqref{eq_sup:cc_intQ4} to be positive to ensure the stabilization of the symmetry broken state.

\subsubsection{Ginzburg-Landau theory}

By decoupling the interaction Eq.~\eqref{eq_sup:cc_intQ4} in the Hubbard-Stratonovich sense and by integrating out the fermionic variables, we arrive at the effective free energy in terms of the three fields $\Delta_1$, $\Delta_2$ and $\Delta_3$
\begin{align} \label{eq_supl:eff_free_en1}
	\mathcal{F}^{Q_{4 a}} [\Delta_1, \Delta_2, \Delta_3] = & \frac{N}{2 g_\text{co}} \sum_i \Delta_i^2 - \text{Tr} \big[ \ln \big( \hat{\mathcal{G}}_0^{-1} (k) - \hat{\mathcal{V}}_\Delta \big) \big] ,
\end{align}
where, in the absence of an interaction that involves an imaginary contribution to the charge order, the fields $\Delta_1$, $\Delta_2$ and $\Delta_3$ are real quantities, see Eq.~\eqref{eq_sup:cc_intQ4}. In Eq.~\eqref{eq_supl:eff_free_en1}, the trace ($\text{Tr}$) implies both a summation over the fermionic Matsubara frequencies $\omega_n = (2 n + 1) \pi T$, with $T$ the electronic temperature, and an integration over momentum $\mathbf{k}$, with $k = (i \omega_n, \mathbf{k})$. The Green's function reads $(\mathbf{\hat{\mathcal{G}}}_0 (k))_{i j} = \frac{\delta_{i j}}{- i \omega_n + \varepsilon_{i, \mathbf{k}} - \mu} = \delta_{i j} \mathcal{G}_{0,i} (k)$, while the interaction $\hat{\mathcal{V}}_\Delta$ has the expression:
\begin{align}
	\hat{\mathcal{V}}_\Delta = \begin{pmatrix}
	& 0 & \frac{\Delta_3}{2} & \frac{\Delta_2}{2} \\
	& \frac{\Delta_3}{2} & 0 & \frac{\Delta_1}{2} \\
	& \frac{\Delta_2}{2} & \frac{\Delta_1}{2} & 0
	\end{pmatrix} . \nonumber
\end{align}

\noindent
An expansion up to the fourth order in the Hubbard-Stratonovich fields of the last contribution to Eq.~\eqref{eq_supl:eff_free_en1} yields:
\begin{align} \label{eq_supl:expan_tr_log}
	-\text{Tr} \big[ \ln \big( \hat{\mathcal{G}}_0^{-1} (k) - \hat{\mathcal{V}}_\Delta \big) \big] & \approx - \text{Tr} \big[ \ln \big( \hat{\mathcal{G}}_0^{-1} (k) \big) \big] + \sum_{n=1}^4 \frac{\text{Tr} \big[ \big( \hat{\mathcal{G}}_0 (k) \hat{\mathcal{V}}_\Delta \big)^n \big]}{n} 
\end{align}
where $\text{Tr} \big[ \ln \big( \hat{\mathcal{G}}_0^{-1} (k) \big) \big]$ is a constant which is omitted in the following. The linear contribution ($n=1$ in the sum in Eq.~\eqref{eq_supl:expan_tr_log}) trivially goes to zero, so we remain with the quadratic, the cubic and the quartic terms in the expression above. These contributions lead to the free energy:
\begin{align} \label{eq_supl:eff_free_en}
	& \mathcal{F}^{Q_{4 a}} [\Delta_1, \Delta_2, \Delta_3] \approx \frac{\alpha}{2} \sum_i \Delta_i^2  + \frac{\gamma}{3} \Delta_1 \Delta_2 \Delta_3 + \frac{\xi_1}{4} \big( \sum_i \Delta_i^2 \big)^2 + \frac{\xi_2 - 2 \xi_1}{4} \sum_{i < j} \Delta_i^2 \Delta_j^2,
\end{align}
where $\alpha$, $\gamma$, $\xi_1$ and $\xi_2$ depend on the inverse temperature $\beta = 1/T$ and on the chemical potential $\mu$, and they have the expressions:
\begin{align} \label{eq_supl:alpha_par}
	& \alpha (\beta, \mu) = \frac{1}{g_\text{co}} + \frac{1}{2} \int_k \mathcal{G}_{0,1} (k) \mathcal{G}_{0,2} (k), \\
	& \gamma (\beta, \mu) = \frac{3}{2} \int_k \mathcal{G}_{0,1} (k) \mathcal{G}_{0,2} (k)  \mathcal{G}_{0,3} (k) , \label{eq_supl:gamma_par} \\
	& \xi_1 (\beta, \mu) = \frac{1}{8} \int_k \mathcal{G}_{0,1}^2 (k) \mathcal{G}_{0,2}^2 (k) , \label{eq_supl:eta1_par} \\
	& \xi_2 (\beta, \mu) = \frac{1}{4} \int_k \mathcal{G}_{0,1} (k) \mathcal{G}_{0,2} (k) \mathcal{G}_{0,3}^2 (k) , \label{eq_supl:eta2_par}
\end{align}
where $\int_k = T \sum_n \int_{-\Lambda}^\Lambda \frac{d^d k}{(2 \pi \Lambda)^d}$ ($d$ is the dimensionality of the problem, i.e., in the case of kagome metals, $d = 2$). The temperature dependence of $\alpha$, $\gamma$, $\xi_1$ and $\xi_2$ is shown in Fig.~6 of the main text. In Eqs.~\eqref{eq_supl:alpha_par}-\eqref{eq_supl:eta2_par}, a permutation of indexes $1$, $2$ and $3$ would lead to an equivalent expression, a property that derives from the C$_6$ symmetry of the underlying Hamiltonian. We assume the quadratic coefficient $\alpha < 0$, i.e., at the mean-field level, the system has a tendency to develop a $1 \times 4$ or a $4 \times 4$ charge order, see Fig.~6 of the main text. Moreover, also the coefficients $\gamma$ and $\xi_2$ are always positive ($\gamma > 0$ induces a preference for the so-called 3Q- charge order \cite{Park2021_PRB}), while $\xi_1$ becomes negative for $k_B T / t \Lambda^2 \lesssim 10^{-5}$, indicating a lack of stability of the Ginzburg-Landau theory in that temperature range. In that regime, we expect the fifth and the sixth order contributions to the expansion Eq.~\eqref{eq_supl:expan_tr_log} to become relevant. However, we are not interested in exploring that temperature range. When the system shows perfect nesting, i.e., $\mathcal{G}_{0,1} (k) = \mathcal{G}_{0,2} (k) = \mathcal{G}_{0,3} (k)$, $\xi_2 - 2 \xi_1 = 0$ and no nematicity can be observed.

\noindent
We generalize the expression for the free energy in Eq.~\eqref{eq_supl:eff_free_en} to include the momentum and frequency dependence of the $\Delta_i$ fields, so that $\Delta_i \rightarrow \Delta_i (k)$. This way, the prefactor of the quadratic term gets modified, at small $k$ (long wavelength and low energies), as \cite{Caprara2007_PRB,Caprara2011_PRB,Caprara2015_PRB}:
\begin{align}
	\alpha \rightarrow 2 \chi^{-1}_{i} (k) & = \frac{1}{g_\text{co}} + \frac{1}{2} \int_{k'} \mathcal{G}_{0,1} (k') \mathcal{G}_{0,2} (k' + k) = \alpha (\beta, \mu) + \bar{\gamma} | \nu_n | + \nu_n^2 + \bar{f}_i (\mathbf{k}) ,
\end{align}
where the charge propagator $\chi^{-1}_{i} (k)$ is actually peaked at the wave vector $Q_i'$, i.e., the ordering vector that corresponds to the order parameter $\Delta_i$. In the previous expression, $\nu_n = 2 \pi T n$ is the bosonic Matsubara frequency, coherent with the bosonic nature of the excitations of the charge ordered state, $\bar{\gamma}$ is the Landau damping coefficient and $\bar{f}_i (\mathbf{k})$ is a function of $\mathbf{k}$ of the kind:
\begin{align}
	& \bar{f}_1 (\mathbf{k}) = \Big( 1 - \frac{\eta}{2} \Big) k_x^2 - \sqrt{3} \eta k_x k_y + \Big( 1 + \frac{\eta}{2} \Big) k_y^2 , \nonumber \\
	& \bar{f}_2 (\mathbf{k}) = (1 + \eta) k_x^2 + (1 - \eta) k_y^2 , \nonumber \\
	& \bar{f}_3 (\mathbf{k}) = \Big( 1 - \frac{\eta}{2} \Big) k_x^2 + \sqrt{3} \eta k_x k_y + \Big( 1 + \frac{\eta}{2} \Big) k_y^2 ,
\end{align}
where $\eta$ is the anisotropy parameter $\eta \in (-1,1)$. These expressions come from including a gradient term of the fields in the free energy. In the following, to simplify the calculations, we assume $\eta = 0$, which makes $\bar{f}_1 (\mathbf{k}) = \bar{f}_2 (\mathbf{k}) = \bar{f}_3 (\mathbf{k}) = \bar{f} (\mathbf{k}) = k_x^2 + k_y^2$, i.e., the bare charge susceptibilities for the three order parameters are equal $\chi^{-1}_{i} (k) = \chi^{-1} (k)$. The same assumption, in the context of nematicity driven by spin fluctuations in the iron-based superconductors, has been shown not to qualitatively alter the results \cite{Fernandes2012_PRB}. The free energy becomes:
\begin{align}
	& \mathcal{F}_\text{eff}^{Q_{4 a}} = \sum_i  \int_k \Delta_i (k)  \chi^{-1} (k) \Delta_i (-k) + \frac{\gamma}{3} \int_x \Delta_1 (x) \Delta_2 (x) \Delta_3 (x) + \frac{\xi_1}{4} \int_x \big( \sum_i \Delta_i^2 (x) \big)^2 \nonumber \\
	& + \frac{\xi_2 - 2 \xi_1}{4} \int_x \sum_{i < j} \Delta_i^2 (x) \Delta_j^2 (x), \nonumber
\end{align}
where $\int_x = \int_0^\beta d \tau \int d^d r$ and $x = (\tau, \mathbf{r})$. We can decouple the quartic part of the interaction by introducing three auxiliary Hubbard-Stratonovich fields $A (x) \rightarrow \sum_i \Delta_i^2 (x) / \sqrt{3}$, $O_1 (x) \rightarrow \sqrt{\frac{2}{3}} \big( \Delta_1^2 (x) - \frac{\Delta_2^2 (x) + \Delta_3^2 (x)}{2} \big)$ and $O_2 (x) \rightarrow \frac{1}{\sqrt{2}} (\Delta_2^2 (x) - \Delta_3^2 (x))$, where $A (x)$ transforms as an A$_1$ irrep while $\mathbf{O}^t (x) = (O_1(x), O_2(x))$ transforms as the two dimensional E$_2$ irrep of D$_{6 \text{h}}$. Particularly, we rewrite the second quartic contribution to the free energy as:
\begin{align}
    \sum_{i < j} \Delta_i^2 \Delta_j^2 & = \Delta_1^2 \Delta_2^2 + \Delta_1^2 \Delta_3^2 + \Delta_2^2 \Delta_3^2 \nonumber \\
    & = \Big( \frac{\sum_i \Delta_i^2}{\sqrt{3}} \Big)^2 - \frac{1}{2} \Big[ \Big( \sqrt{\frac{2}{3}} \big( \Delta_1^2 (x) - \frac{\Delta_2^2 (x) + \Delta_3^2 (x)}{2} \big) \Big)^2 + \Big( \frac{\Delta_2^2 (x) - \Delta_3^2 (x)}{\sqrt{2}} \Big)^2 \Big] . \nonumber
\end{align}
This way, we might rewrite the free energy as:
\begin{align}
	& \mathcal{F}_\text{eff}^{Q_{4 a}} = \sum_i  \int_k \Delta_i (k)  \chi^{-1} (k) \Delta_i (-k) + \frac{\gamma}{3} \int_x \Delta_1 (x) \Delta_2 (x) \Delta_3 (x) - \int_k \Big[ \frac{A^2}{\xi_1 + \xi_2} - \frac{2 (O_1^2 + O_2^2)}{\xi_2 - 2 \xi_1} \Big] \nonumber  \\
	& + \sqrt{\frac{2}{3}} \int_x O_1 \Big( \Delta_1^2 (x) - \frac{\Delta_2^2 (x) + \Delta_3^2 (x)}{2} \Big) + \frac{1}{\sqrt{2}} \int_x O_2 \big( \Delta_2^2 (x) - \Delta_3^2 (x) \big) + \frac{1}{\sqrt{3}} \sum_i \int_x A \Delta_i^2 (x) . \nonumber
\end{align}
From the expression above, it follows that a nonzero value of $O_1$ ($O_2$) implies a nonzero expectation value of $\frac{2 O_1}{\xi_2 - 2 \xi_1} = \sqrt{\frac{3}{2}} \langle \Delta_1^2 - \frac{\Delta_2^2 + \Delta_3^2}{2} \rangle \neq 0$ ($\frac{2 O_2}{\xi_2 - 2 \xi_1} = \frac{3}{2 \sqrt{2}} \langle \Delta_2^2 - \Delta_3^2 \rangle \neq 0$) and thus the nematicity of the state if, at the same time, $\Delta_1 = \Delta_2 = \Delta_3 = 0$. Even if the nematic order parameter should be represented as a traceless symmetric tensor \cite{Fradkin2010_AnnRev}, we will refer to $\mathbf{N} = (\sqrt{\frac{2}{3}} \big( \Delta_1^2 - \frac{\Delta_2^2 + \Delta_3^2}{2} \big), \frac{1}{\sqrt{2}} (\Delta_2^2 - \Delta_3^2))$ as the nematic order parameter, as it became customary in this context. By including the Hubbard-Stratonovich fields into the definition of $\chi^{-1} (k)$, we arrive at the expression for the free energy:
\begin{align}
	\mathcal{F}_\text{eff}^{Q_{4 a}} & = \sum_i  \int_k \Delta_i (k)  \chi^{-1}_i (k) \Delta_i (-k) + \frac{\gamma}{3} \int_x \Delta_1 (x) \Delta_2 (x) \Delta_3 (x) - \int_k \Big[ \frac{A^2}{\xi_1 + \xi_2} - \frac{2 (O_1^2 + O_2^2)}{\xi_2 - 2 \xi_1} \Big] , \nonumber
\end{align}
where:
\begin{align}
	& \chi^{-1}_1 (k) = \chi^{-1} (k) + \frac{A}{\sqrt{3}} + \sqrt{\frac{2}{3}} O_1 , \label{eq_supl:chi1} \\
	& \chi^{-1}_2 (k) = \chi^{-1} (k) + \frac{A}{\sqrt{3}} - \frac{O_1}{\sqrt{6}} + \frac{O_2}{\sqrt{2}} , \label{eq_supl:chi2} \\
	& \chi^{-1}_3 (k) = \chi^{-1} (k) + \frac{A}{\sqrt{3}} - \frac{O_1}{\sqrt{6}} - \frac{O_2}{\sqrt{2}} . \label{eq_supl:chi3}
\end{align}
It is also possible to add a field $\mathbf{h} = (h_1, h_2)$ coupled to $\mathbf{N}$ by substituting, in Eqs.~\eqref{eq_supl:chi1}-\eqref{eq_supl:chi3}, $O_1 \rightarrow O_1 + h_1$ and $O_2 \rightarrow O_2 + h_2$. In the following, we assume this substitution to be performed. Moreover, we include the gaussian fluctuations of the fields $\Delta_1 (x)$, $\Delta_2 (x)$ and $\Delta_3 (x)$, which we generally rewrite as $\Delta_i (x) = \Delta_i + \delta \Delta_i (x)$, i.e., $\Delta_i = \langle \Delta_i (x) \rangle$ is the spatial average $\langle \cdots \rangle$ of the field $\Delta_i (x)$ (implying $\langle \delta \Delta_i (x) \rangle = 0$) \cite{Hecker2018_npj}. Formally, we treat them within the large-N expansion of the fields $\Delta_i$. This way, we arrive at the expression:
\begin{align}
	\mathcal{F}_\text{eff}^{Q_{4 a}} & \approx \int_k \delta \boldsymbol{\Delta}^t (k) \hat{\Lambda}^{-1} (k) \delta \boldsymbol{\Delta} (k) + \sum_i \Delta_i (0)  \chi^{-1}_i (0) \Delta_i (0) + \frac{\gamma}{3} \prod_i \Delta_i (0) - \int_k \Big[ \frac{A^2}{\xi_1 + \xi_2} - \frac{2 (O_1^2 + O_2^2)}{\xi_2 - 2 \xi_1} \Big] , \nonumber
\end{align}
having defined the vector $\delta \boldsymbol{\Delta}^t = (\delta \Delta_1, \delta \Delta_2, \delta \Delta_3)$ and the matrix:
\begin{align}
	\hat{\Lambda}^{-1} (k) & =  \frac{\chi^{-1}_1 (k)}{3} \big(  \lambda_0 + \frac{3 \lambda_3 + \sqrt{3} \lambda_8}{2}  \big) + \frac{\chi^{-1}_2 (k)}{3} \big( \lambda_0 - \frac{3 \lambda_3 - \sqrt{3} \lambda_8}{2}  \big) \nonumber \\
	& + \frac{\chi^{-1}_3 (k)}{3} \big( \lambda_0 - \sqrt{3} \lambda_8  \big) + \frac{\gamma}{3} ( \Delta_3 (0) \lambda_1 + \Delta_2 (0) \lambda_4 + \Delta_1 (0) \lambda_6) , \nonumber
\end{align}
where $\lambda_i$ are the Gell-Mann matrices:
\begin{align}
	& \lambda_0 = \begin{pmatrix}
	1 & 0 & 0 \\
	0 & 1 & 0 \\
	0 & 0 & 1
	\end{pmatrix} , \ \ \ \lambda_1 = \begin{pmatrix}
	0 & 1 & 0 \\
	1 & 0 & 0 \\
	0 & 0 & 0
	\end{pmatrix} , \ \ \ \lambda_3 = \begin{pmatrix}
	1 & 0 & 0 \\
	0 & -1 & 0 \\
	0 & 0 & 0
	\end{pmatrix} \nonumber
	 \\
	& \lambda_4 = \begin{pmatrix}
	0 & 0 & 1 \\
	0 & 0 & 0 \\
	1 & 0 & 0
	\end{pmatrix} , \ \ \ \lambda_6 = \begin{pmatrix}
	0 & 0 & 0 \\
	0 & 0 & 1 \\
	0 & 1 & 0
	\end{pmatrix} , \ \ \ \lambda_8 = \frac{1}{\sqrt{3}} \begin{pmatrix}
	1 & 0 & 0 \\
	0 & 1 & 0 \\
	0 & 0 & -2
	\end{pmatrix} . \nonumber
\end{align}
By integrating out the fluctuations of the order parameters, we remain with the effective free energy:
\begin{align} \label{eq_supl:free_en_eff_finQ4}
	\mathcal{F}_\text{eff}^{Q_{4 a}} & = \sum_i \Delta_i (0)  \chi^{-1}_i (0) \Delta_i (0) + \frac{\gamma}{3} \prod_i \Delta_i (0) - \int_k \Big[ \frac{A^2}{\xi_1 + \xi_2} - \frac{2 (O_1^2 + O_2^2)}{\xi_2 - 2 \xi_1} \Big] + \frac{1}{2}  \int_k  \text{Tr} \big[ \ln \big( \hat{\Lambda}^{-1} (k)  \big) \big] .
\end{align}

\subsubsection{Expression of the coefficients of the Landau theory}
Following \cite{Park2021_PRB}, we write:
\begin{align}
	& \alpha (\beta, \mu) = \frac{1}{g_\text{co}} + \frac{1}{2} \int_k \mathcal{G}_{0,1} (k) \mathcal{G}_{0,2} (k) = \frac{1}{g_\text{co}} + \frac{1}{2} \int_{-\Lambda}^\Lambda \frac{d k_x \ d k_y}{(2 \pi \Lambda)^2} \frac{f (\varepsilon_{M'_1, \mathbf{k}} - \mu) - f (\varepsilon_{M'_2, \mathbf{k}} - \mu)}{\varepsilon_{M'_1, \mathbf{k}} - \varepsilon_{M'_2, \mathbf{k}}} = \nonumber \\
	& = \frac{1}{g_\text{co}} + \frac{1}{8 \pi^2 \Lambda^2 t} \int_{-\Lambda \sqrt{t \beta}}^{\Lambda \sqrt{t \beta}} dx \ dy \frac{F (\tilde{\varepsilon}_1 (x,y) - \beta \mu) - F (\tilde{\varepsilon}_2 (x,y) - \beta \mu)}{\tilde{\varepsilon}_1 (x,y) - \tilde{\varepsilon}_2 (x,y)} \nonumber \\
	& \approx \frac{1}{g_\text{co}} + \frac{1}{8 \pi^2 \Lambda^2 t} \int_{-\infty}^{\infty} dx \ dy \frac{F (\tilde{\varepsilon}_1 (x,y) - \beta \mu) - F (\tilde{\varepsilon}_2 (x,y) - \beta \mu)}{\tilde{\varepsilon}_1 (x,y) - \tilde{\varepsilon}_2 (x,y)} \nonumber \\
	& = \frac{1}{g_\text{co}} + \frac{3}{2 \pi^2 \Lambda^2 t} \int_{0}^{\infty} dx \int_{0}^{x/\sqrt{3}} dy \frac{F (\tilde{\varepsilon}_1 (x,y) - \beta \mu) - F (\tilde{\varepsilon}_2 (x,y) - \beta \mu)}{\tilde{\varepsilon}_1 (x,y) - \tilde{\varepsilon}_2 (x,y)} , \nonumber
\end{align}
having defined $\tilde{\varepsilon}_1 (x,y) = - \frac{1}{2 (1 - \delta^2)} \big[ (1 + \frac{\delta}{2}) x^2 + \sqrt{3} \delta \ x y + (1 - \frac{\delta}{2}) y^2 \big]$, $\tilde{\varepsilon}_2 (x,y) = - \frac{1}{2 (1 - \delta^2)} \big[ (1 - \delta) x^2 + (1 + \delta) y^2 \big]$ and $\tilde{\varepsilon}_3 (x,y) = - \frac{1}{2 (1 - \delta^2)} \big[ (1 + \frac{\delta}{2}) x^2 - \sqrt{3} \delta \ x y + (1 - \frac{\delta}{2}) y^2 \big]$, the Fermi distribution function $f (\varepsilon) = 1/(1 + e^{\beta \varepsilon})$ and the rescaled Fermi distribution $F (\tilde{\varepsilon}) = 1/(1 + e^{\tilde{\varepsilon}})$. In the fist step of the expression above, we explicitly computed the summation over the Matsubara frequencies, then we performed a change of variable and the approximation ($\approx$) is valid as soon as $\mu, 1/\beta \ll t \Lambda^2$. The last step follows from the symmetry of the integrand over the integration domain. Similarly, we write:
\begin{align}
	& \gamma (\beta, \mu) = \frac{3}{2} \int_k \mathcal{G}_{0,1} (k) \mathcal{G}_{0,2} (k)  \mathcal{G}_{0,3} (k) \nonumber \\
	& = - \frac{3}{2} \int_{-\Lambda}^\Lambda \frac{d k_x \ d k_y}{(2 \pi \Lambda)^2} \Big[ \frac{f (\varepsilon_{M'_1, \mathbf{k}} - \mu)}{(\varepsilon_{M'_1, \mathbf{k}} - \varepsilon_{M'_2, \mathbf{k}}) (\varepsilon_{M'_1, \mathbf{k}} - \varepsilon_{M'_3, \mathbf{k}})} + \frac{f (\varepsilon_{M'_2, \mathbf{k}} - \mu)}{(\varepsilon_{M'_2, \mathbf{k}} - \varepsilon_{M'_1, \mathbf{k}}) (\varepsilon_{M'_2, \mathbf{k}} - \varepsilon_{M'_3, \mathbf{k}})} \nonumber \\
	& + \frac{f (\varepsilon_{M'_3, \mathbf{k}} - \mu)}{(\varepsilon_{M'_3, \mathbf{k}} - \varepsilon_{M'_1, \mathbf{k}}) (\varepsilon_{M'_3, \mathbf{k}} - \varepsilon_{M'_2, \mathbf{k}})} \Big]  \nonumber \\
	& \approx - \frac{9 \beta}{2 \pi^2 \Lambda^2 t} \int_{0}^{\infty} dx \int_{0}^{x/\sqrt{3}} dy \Big[ \frac{F (\tilde{\varepsilon}_1 (x,y) - \beta \mu)}{(\tilde{\varepsilon}_1 (x,y) - \tilde{\varepsilon}_2 (x,y)) (\tilde{\varepsilon}_1 (x,y) - \tilde{\varepsilon}_3 (x,y))} + \frac{F (\tilde{\varepsilon}_2 (x,y) - \beta \mu)}{(\tilde{\varepsilon}_2 (x,y) - \tilde{\varepsilon}_1 (x,y)) (\tilde{\varepsilon}_2 (x,y) - \tilde{\varepsilon}_3 (x,y))} \nonumber \\
	& + \frac{F (\tilde{\varepsilon}_3 (x,y) - \mu)}{(\tilde{\varepsilon}_3 (x,y) - \tilde{\varepsilon}_1 (x,y)) (\tilde{\varepsilon}_3 (x,y) - \tilde{\varepsilon}_2 (x,y))} \Big] , \nonumber
\end{align}

\begin{align}
	& \xi_2 (\beta, \mu) = \frac{1}{4} \int_k \mathcal{G}_{0,1} (k) \mathcal{G}_{0,2} (k) \mathcal{G}_{0,3}^2 (k) \nonumber \\
	& = \frac{1}{12} \int_k \big[ \mathcal{G}_{0,1} (k) \mathcal{G}_{0,2} (k) \mathcal{G}_{0,3}^2 (k) + \mathcal{G}_{0,1}^2 (k) \mathcal{G}_{0,2} (k) \mathcal{G}_{0,3} (k) + \mathcal{G}_{0,1} (k) \mathcal{G}_{0,2}^2 (k) \mathcal{G}_{0,3} (k) \big] \nonumber \\
	& = \frac{1}{12} \int_{-\Lambda}^\Lambda \frac{d k_x \ d k_y}{(2 \pi \Lambda)^2} \Big[ \frac{f' (\varepsilon_{M'_1, \mathbf{k}} - \mu)}{(\varepsilon_{M'_1, \mathbf{k}} - \varepsilon_{M'_2, \mathbf{k}}) (\varepsilon_{M'_1, \mathbf{k}} - \varepsilon_{M'_3, \mathbf{k}})} + \frac{f' (\varepsilon_{M'_2, \mathbf{k}} - \mu)}{(\varepsilon_{M'_2, \mathbf{k}} - \varepsilon_{M'_1, \mathbf{k}}) (\varepsilon_{M'_2, \mathbf{k}} - \varepsilon_{M'_3, \mathbf{k}})} \nonumber, \\
	& + \frac{f' (\varepsilon_{M'_3, \mathbf{k}} - \mu)}{(\varepsilon_{M'_3, \mathbf{k}} - \varepsilon_{M'_1, \mathbf{k}}) (\varepsilon_{M'_3, \mathbf{k}} - \varepsilon_{M'_2, \mathbf{k}})} \Big] \nonumber \\
	& \approx  \frac{\beta^2}{4 \pi^2 \Lambda^2 t}  \int_{0}^{\infty} dx \int_{0}^{x/\sqrt{3}} dy \Big[ \frac{F' (\tilde{\varepsilon}_1 (x,y) - \beta \mu)}{(\tilde{\varepsilon}_1 (x,y) - \tilde{\varepsilon}_2 (x,y)) (\tilde{\varepsilon}_1 (x,y) - \tilde{\varepsilon}_3 (x,y))} + \frac{F' (\tilde{\varepsilon}_2 - \beta \mu)}{(\tilde{\varepsilon}_2 (x,y) - \tilde{\varepsilon}_1 (x,y)) (\tilde{\varepsilon}_2 (x,y) - \tilde{\varepsilon}_3 (x,y))} \nonumber, \\
	& + \frac{F' (\tilde{\varepsilon}_3 - \beta \mu)}{(\tilde{\varepsilon}_3 (x,y) - \tilde{\varepsilon}_1 (x,y)) (\tilde{\varepsilon}_3 (x,y) - \tilde{\varepsilon}_2 (x,y))} \Big] \nonumber
\end{align}

\begin{align}
	& \xi_1 (\beta, \mu) = \frac{1}{8} \int_k \mathcal{G}_{0,1}^2 (k) \mathcal{G}_{0,2}^2 (k) \nonumber \\
	& = - \frac{1}{4} \int_{-\Lambda}^\Lambda \frac{d k_x \ d k_y}{(2 \pi \Lambda)^2} \Big[ \frac{f (\varepsilon_{M'_2, \mathbf{k}} - \mu) - f (\varepsilon_{M'_1, \mathbf{k}} - \mu)}{(\varepsilon_{M'_2, \mathbf{k}} - \varepsilon_{M'_1, \mathbf{k}})^3} - \frac{f' (\varepsilon_{M'_1, \mathbf{k}} - \mu) + f' (\varepsilon_{M'_2, \mathbf{k}} - \mu)}{2 (\varepsilon_{M'_1, \mathbf{k}} - \varepsilon_{M'_2, \mathbf{k}})^2} \Big]  \nonumber ,\\
	& \approx - \frac{\beta^2}{16 \pi^2 \Lambda^2 t} \int_{-\infty}^{\infty} dx \ dy \Big[ \frac{F (\tilde{\varepsilon}_2 (x,y) - \beta \mu) - F (\tilde{\varepsilon}_1 (x,y) - \beta \mu)}{(\tilde{\varepsilon}_2 (x,y) - \tilde{\varepsilon}_1 (x,y) )^3} - \frac{F' (\tilde{\varepsilon}_1 (x,y) - \beta \mu) + F' (\tilde{\varepsilon}_2 (x,y) - \beta \mu)}{2 (\tilde{\varepsilon}_1 (x,y) - \tilde{\varepsilon}_2 (x,y))^2} \Big] \nonumber \\
	& \approx \frac{\beta^2}{32 \Lambda^2 t} \int_{0}^{\infty} d \varepsilon \rho (\varepsilon) \Big[ \frac{F (- \varepsilon - \beta \mu) - F (\varepsilon - \beta \mu)}{ \varepsilon^3} + \frac{F' (- \varepsilon  - \beta \mu) + F' ( \varepsilon  - \beta \mu)}{\varepsilon^2} \Big] \nonumber \\
	& = \frac{\sqrt{1 - \delta^2} \beta^2}{128 \pi \Lambda^2 t} \int_{0}^{\infty} d \varepsilon \Big[ \frac{F (- \varepsilon - \beta \mu) - F (\varepsilon - \beta \mu)}{ \varepsilon^3} + \frac{F' (- \varepsilon  - \beta \mu) + F' ( \varepsilon  - \beta \mu)}{\varepsilon^2} \Big] , \nonumber
\end{align}
having assumed that only the contributions of the kind: $\tilde{\varepsilon}_1 (x,y) \rightarrow \varepsilon$, $\tilde{\varepsilon}_2 (x,y) \rightarrow - \varepsilon$ are relevant and having taken into account that the density of states for a two dimensional electron gas is $\rho (\varepsilon) = \sqrt{1 - \delta^2}/(4 \pi)$.

\subsubsection{Saddle point equations for the $4 \times 4$ charge order}
In the effective action Eq.~\eqref{eq_supl:free_en_eff_finQ4}, one might first bring $\boldsymbol{\hat{\Lambda}}^{-1} (k)$ to its diagonal form $\boldsymbol{\hat{\Lambda}}^{-1}_\text{d} (k)$ and then take the logarithm since $\text{Tr} [ \text{ln} (\boldsymbol{\hat{\Lambda}}^{-1} (k)) ] = \text{Tr} [ \text{ln} (\boldsymbol{\hat{\Lambda}}^{-1}_\text{d} (k)) ]$. The eigenvalues of $\boldsymbol{\hat{\Lambda}}^{-1} (k)$ read:
\begin{align} \label{eq_sup:eigenval_HSQ4}
	E_m (k, \Delta_0, A, O_1, O_2, h_1, h_2) = \chi^{-1} (k) + \frac{A}{\sqrt{3}} + \sqrt{- \frac{4 P}{3}} \cos \Big[ \frac{1}{3} \arccos \Big( \frac{3 Q}{2 P} \sqrt{ -\frac{3}{P}} \Big) - \frac{2 m \pi}{3} \Big] ,
\end{align}
with $m = 0,1,2$ and
\begin{align}
	& P (\Delta_0, O_1, O_2, h_1, h_2) = - \frac{1}{2} \Big[ (O_1 + h_1 )^2 + ( O_2 + h_2 )^2 + \frac{2 \gamma^2}{3} \Delta_0^2 \Big] \nonumber , \\
	& Q (\Delta_0, O_1, O_2, h_1, h_2) = - \frac{2 \gamma^3 \Delta_0^3}{27} - \frac{2}{3} \sqrt{\frac{2}{3}} (O_1 + h_1) \ \frac{(O_1 + h_1) - \sqrt{3} (O_2 + h_2)}{2} \ \frac{(O_1 + h_1) + \sqrt{3} (O_2 + h_2)}{2} , \nonumber
\end{align}
having made explicit the functional dependence of $P$ and $Q$. Eq.~\eqref{eq_sup:eigenval_HSQ4} provides three real independent solutions when $4 P^3 + 27 Q^2 < 0$, however it becomes meaningless if $P = 0$ since one would get a division by zero in the argument of the arc cosine. The effective free energy Eq.~\eqref{eq_supl:free_en_eff_finQ4} can be rewritten as:
\begin{align} \label{eq_sup:free_en_eff_finQ4}
	\mathcal{F}_\text{eff}^{Q_{4 a}} & = \frac{3 \Delta_0^2}{2} \Big( \chi^{-1} (0) + \frac{A}{\sqrt{3}} \Big)  + \frac{\gamma \Delta_0^3}{3}  - \int_k \Big[ \frac{A^2}{\xi_1 + \xi_2} - \frac{2 (O_1^2 + O_2^2)}{\xi_2 - 2 \xi_1} \Big] \nonumber \\
	& + \frac{1}{2} \int_k  \ln \big[ E_0 ( k, \Delta_0, A, O_1, O_2, h_1, h_2 ) \ E_1 ( k, \Delta_0, A, O_1, O_2, h_1, h_2 ) \ E_2 ( k, \Delta_0, A, O_1, O_2, h_1, h_2 ) \big] .
\end{align}
The expression of the derivative of the m-th eigenvalue with respect to the parameter $\theta = \Delta_0, A, O_1, O_2$ is:
\begin{align}
	& \frac{\partial E_m}{\partial \theta} = \frac{1}{\sqrt{3}} \frac{\partial A}{\partial \theta}  + \frac{1}{\sqrt{-3 P}} \frac{\partial P}{\partial \theta} \cos \Big[ \frac{1}{3} \arccos \Big( \frac{3 Q}{2 P} \sqrt{ -\frac{3}{P}} \Big) - \frac{2 m \pi}{3} \Big] \nonumber \\
	& - \sqrt{\frac{4P}{4 P^3 + 27 Q^2}} \Big[ \frac{\partial Q}{\partial \theta} - \frac{3}{2} \frac{Q}{P} \frac{\partial P}{\partial \theta} \Big] \sin \Big[ \frac{1}{3} \arccos \Big( \frac{3 Q }{2 P} \sqrt{ -\frac{3}{P}} \Big) - \frac{2 m \pi}{3} \Big] , \nonumber
\end{align}
with:
\begin{align}
	& \frac{\partial P}{\partial \Delta_0} = - \frac{2 \gamma^2}{3} \Delta_0 \nonumber , \\
	& \frac{\partial P}{\partial O_1} = - ( O_1 + h_1 ) \nonumber , \\
	& \frac{\partial P}{\partial O_2} = - ( O_2 + h_2 ) \nonumber , \\
	& \frac{\partial P}{\partial A} = 0 , \nonumber
\end{align}
and:
\begin{align}
	& \frac{\partial Q}{\partial \Delta_0} = - \frac{2 \gamma^3 \Delta_0^2}{9} \nonumber , \\
	& \frac{\partial Q}{\partial O_1} = - \frac{1}{\sqrt{6}} \big[ (O_1 + h_1)^2 - (O_2 + h_2)^2 \big]  \nonumber , \\
	& \frac{\partial Q}{\partial O_2} = \sqrt{\frac{2}{3}} (O_1 + h_1) (O_2 + h_2) \nonumber , \\
	& \frac{\partial Q}{\partial A} = 0 . \nonumber
\end{align}
The saddle-point solution of Eq.~\eqref{eq_sup:free_en_eff_finQ4} leads to the equations:
\begin{align}
	& \frac{\partial \mathcal{F}_\text{eff}^{Q_{4 a}}}{\partial \Delta_0} = 3 \Delta_0 (\chi^{-1} (0) + \frac{A}{\sqrt{3}}) + \gamma \Delta_0^2 + \sum_m  \int_k \frac{1}{2 E_m} \frac{\partial E_m}{\partial \Delta_0} = 0 \nonumber , \\
	& \frac{\partial \mathcal{F}_\text{eff}^{Q_{4 a}}}{\partial A} =  \sqrt{3} \Delta_0^2 - \frac{2 A}{\xi_1 + \xi_2} + \sum_m \int_k \frac{1}{\sqrt{3} E_m} = 0 \nonumber , \\
	& \frac{\partial \mathcal{F}_\text{eff}^{Q_{4 a}}}{\partial O_1} = \frac{4 O_1}{\xi_2 - 2 \xi_1}  + \sum_m \int_k \frac{1}{2 E_m} \frac{\partial E_m}{\partial O_1} = 0 \nonumber , \\
	& \frac{\partial \mathcal{F}_\text{eff}^{Q_{4 a}}}{\partial O_2} = \frac{4 O_2}{\xi_2 - 2 \xi_1} + \sum_m \int_k \frac{1}{2 E_m} \frac{\partial E_m}{\partial O_2} = 0 . \nonumber
\end{align}

\begin{figure}	\centerline{\includegraphics[width=1.0\textwidth]{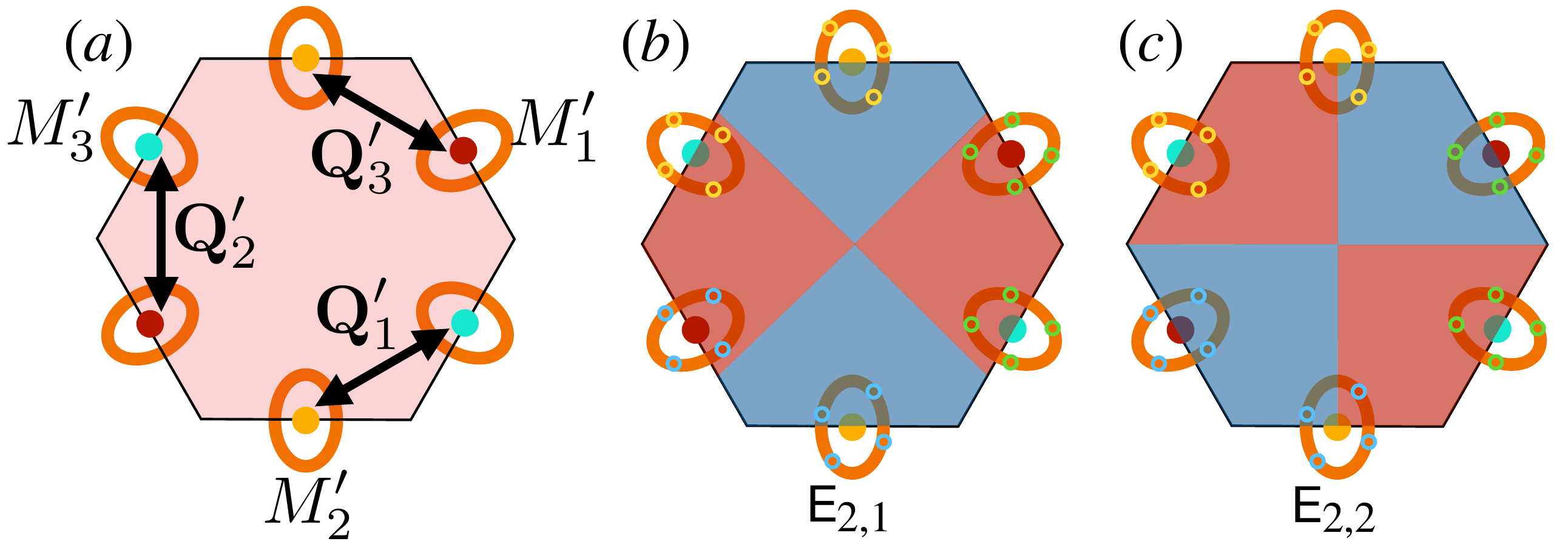}}
	\caption{ \textbf{Signs and nodes of the first order E$_2$ Brillouin zone harmonics -} (a) Representation of the first Brillouin zone of the $2 \times 2$ charge ordered state and of the Fermi pockets at the reconstructed M-points. (b-c) Blue (red) areas represent regions of the first Brillouin zone of negative (positive) values of the first order E$_2$ harmonics. The small coloured circles indicate the location of the hot-spots for different nesting vectors, clearly indicating that the first order harmonics do not change sign for all the hot-spots, i.e., one expects both the components of the E$_2$ irrep to be different from zero \cite{Kretzschmar2016_NatPhys}.} \label{fig_sup:E2_sym_BZ}
\end{figure}

\subsubsection{Rise of nematicity}
In the following, we analyze the condition $\Delta_i (0) = \Delta_0 = 0$ for each $i$, i.e., the three order parameters are equal and null. Indeed, we want to explore the possibility of having a nematic transition in the regime where no order parameter acquires a nonzero value. Using the saddle point approximation, one arrives at four coupled equations of state which have solutions characterized by $\Delta_0 = 0$, see the previous section of this Supplemental Material. We might write the nematic susceptibility $\hat{\chi}_\text{nem}^{Q_{4 a}}$, which is a $2 \times 2$ matrix \cite{Fernandes2012_SuScTe}:
\begin{align}
	& (\hat{\chi}_\text{nem}^{Q_{4 a}})_{l m} (k) = \langle N_l (k) N_m (-k) \rangle .
\end{align}
In the saddle point approximation, one arrives at the expressions at zero exchanged momentum \cite{Fernandes2010_PRL,Karahasanovic2015_PRB}:
\begin{align} \label{eq_supl:chi_nem}
	& (\hat{\chi}_\text{nem}^{Q_{4 a}})_{l m} = \frac{\partial O_l}{\partial h_m}\bigg|_{\mathbf{h}=\mathbf{0}} = - \delta_{l m} \frac{g_\text{nem}^{Q_{4 a}} \int_k \tilde{\chi}^2 (k)}{1 + g_\text{nem}^{Q_{4 a}} \int_k \tilde{\chi}^2 (k)} ,
\end{align}
with $l, m = 1,2$ and having introduced the propagator $\tilde{\chi}^{-1} (k) = \chi^{-1} (k) + A/\sqrt{3}$, where $A$ simply renormalizes the mass term due to fluctuations \cite{Dolgirev2020_PRB,Grandi2021_PRB}, and the nematic interaction is $g_\text{nem}^{Q_{4 a}} = - \frac{1}{4} (\xi_2 - 2 \xi_1)$. In the absence of a direct symmetry breaking in the E$_2$ sector, i.e., $\mathbf{O} = \mathbf{0}$, the nematic susceptibility has no off-diagonal elements and the diagonal components are identical, signaling the degeneracy and the lack of coupling of the two fields $O_1$ and $O_2$ \textit{at} the instability level, see Fig.~\ref{fig_sup:E2_sym_BZ}.

\newpage
\subsection{Zero-momentum ($Q_{1a}$) Pomeranchuk instability within the $2 \times 2$ charge ordered state}
Starting from the band dispersion defined in Eqs.~\eqref{eq_sup:dispersionsQ4}, i.e., from the effective model for the $2 \times 2$ charge ordered state, we can consider the interaction that drives a zero momentum instability in the charge sector, which reads:
\begin{align} 
	\mathcal{H}_\text{int}^{Q_{1a}} \approx - \frac{N}{2}  \sum_{i,j} g_{i j} n_i n_j , \nonumber
\end{align}
where $n_i = \sum_{|\mathbf{k}| < \Lambda, \sigma} c^\dagger_{i, \mathbf{k}, \sigma} c_{i, \mathbf{k}, \sigma}$, $i = 1,2,3$ is the index of the hole pockets centered at $M'_i$ and $g_{i j} = \delta_{i j} g_\text{d} + (1 - \delta_{i j}) g_\text{o}$, with $g_\text{d} = - \Gamma_\text{d}$ and $g_\text{o} = - 2 \Gamma_\text{o,b} + \Gamma_\text{o,c}$. We assume the system to be in the regime $g_\text{d} - g_\text{o} > 0$, which ensures the instability of the system towards the $d$-wave charge PI. The operators $n_i$ do not exchange any momentum differently from the other $\rho_i$ operators defined when we were analyzing the $4 \times 4$ charge order. Following \cite{Nandkishore2012_NatPhys}, we arrive at the expression:
\begin{align}
	\mathcal{F}^{Q_{1a}} = \sum_i \int_k \mathcal{G}_{0,i}^{-1} (i \omega_n, \mathbf{k}) n_{j, \mathbf{k}} - \frac{1}{2} \Phi^\dagger \tilde{\Lambda} \Phi , \nonumber
\end{align}
with $\mathcal{G}_{0,i}^{-1} (i \omega_n, \mathbf{k}) = -i \omega_n + \varepsilon_i (\mathbf{k}) - \mu$, $\Phi^t = (n_1, n_2, n_3)$ and
\begin{align}
	\tilde{\Lambda} = \begin{pmatrix}
	g_\text{d} & g_\text{o} & g_\text{o} \\
	g_\text{o} & g_\text{d} & g_\text{o} \\
	g_\text{o} & g_\text{o} & g_\text{d}
	\end{pmatrix} , \nonumber
\end{align}
having rescaled $g_\text{d} \rightarrow g_\text{d}/N$ and $g_\text{o} \rightarrow g_\text{o}/N$. The matrix $\tilde{\Lambda}$ has two degenerate eigenvalues $E_{1,2} = g_\text{d} - g_\text{o}$ and $E_3 = g_\text{d} + 2 g_\text{o}$. One can rewrite the free energy as:
\begin{align}
	\mathcal{F}^{Q_{1a}} = \sum_i \int_k \mathcal{G}_{0,i}^{-1} (i \omega_n, \mathbf{k}) n_{i, \mathbf{k}} - \frac{1}{2} \sum_i E_i V_i^2 , \nonumber
\end{align}
where $V_1 = \sqrt{\frac{2}{3}} (n_1 - \frac{n_2 + n_3}{2})$, $V_2 = \frac{1}{\sqrt{2}} (n_2 - n_3)$ and $V_3 = \frac{1}{\sqrt{3}} (n_1 + n_2 + n_3)$. Since $n_1$, $n_2$ and $n_3$ are actually constrained by the conservation of the total electronic charge ($n_1 + n_2 + n_3 = N$), the three eigenvectors $V_1$, $V_2$ and $V_3$ are not independent. For this reason, we perform only two Hubbard-Stratonovich transformations, mapping $V_1 \rightarrow O_1$ and $V_2 \rightarrow O_2$. We arrive writing:
\begin{align}
	\mathcal{F}_\text{eff}^{Q_{1a}} = \sum_i \int_k \mathcal{G}_{0,i}^{-1} (i \omega_n, \mathbf{k}) n_{i, \mathbf{k}} - \frac{O_1^2 + O_2^2}{2 (g_\text{o} - g_\text{d})} + \sqrt{\frac{2}{3}} O_1 \Big( n_1 - \frac{n_2 + n_3}{2} \Big) + O_2 \frac{n_2 - n_3}{\sqrt{2}} - \frac{g_\text{d} + 2 g_\text{o}}{6} N^2 . \nonumber
\end{align}
By defining:
\begin{align} \label{eq_sup:chi_zero_mom}
	\hat{\chi}^{-1} (i \omega_n, \mathbf{k}) & = \begin{pmatrix}
	\mathcal{G}_{0,1}^{-1} (i \omega_n, \mathbf{k})  + \sqrt{\frac{2}{3}} O_1 & 0 & 0 \\
	0 & \mathcal{G}_{0,2}^{-1} (i \omega_n, \mathbf{k})  - \frac{O_1}{\sqrt{6}} + \frac{O_2}{\sqrt{2}} & 0 \\
	0 & 0 & \mathcal{G}_{0,3}^{-1} (i \omega_n, \mathbf{k}) - \frac{O_1}{\sqrt{6}} - \frac{O_2}{\sqrt{2}}
	\end{pmatrix} \nonumber \\
	& = \begin{pmatrix}
	\chi^{-1}_1 (i \omega_n, \mathbf{k}) & 0 & 0 \\
	0 & \chi^{-1}_2 (i \omega_n, \mathbf{k}) & 0 \\
	0 & 0 & \chi^{-1}_3 (i \omega_n, \mathbf{k})
	\end{pmatrix} , 
\end{align}
the free energy becomes:
\begin{align}
	\mathcal{F}_\text{eff}^{Q_{1a}} = \sum_i \int_k \chi^{-1}_i (i \omega_n, \mathbf{k}) n_{i, \mathbf{k}} - \frac{O_1^2 + O_2^2}{2 (g_\text{o} - g_\text{d})} - \frac{g_\text{d} + 2 g_\text{o}}{6} N^2 , \nonumber
\end{align}
and, by integrating out the fermionic fields, we end up with the expression:
\begin{align} \label{eq_sup:free_en_beforeexp1Q}
	\mathcal{F}_\text{eff}^{Q_{1a}} = - \sum_i \int_k \ln \big( \chi^{-1}_i (i \omega_n, \mathbf{k}) \big) - \frac{O_1^2 + O_2^2}{2 (g_\text{o} - g_\text{d})} - \frac{g_\text{d} + 2 g_\text{o}}{6} N^2 . 
\end{align}
From the effective free energy above, we get the nematic susceptibility:
\begin{align} \label{eq_sup:chi_nem1Q}
	& (\hat{\chi}^{Q_{1a}}_\text{nem})_{l m} = \frac{\partial O_l}{\partial h_m}\bigg|_{\mathbf{h}=\mathbf{0}} = - \delta_{l m} \frac{g_\text{nem}^{Q_{1a}} \int_k \mathcal{G}_{0,i}^{2} (k)}{1 + g_\text{nem}^{Q_{1a}} \int_k \mathcal{G}_{0,i}^{2} (k)} ,
\end{align}
where $g_\text{nem}^{Q_{1a}} = g_\text{d} - g_\text{o}$, $i=1,2$ or $3$ and having substituted $O_1 \rightarrow O_1 + h_1$ and $O_2 \rightarrow O_2 + h_2$ into Eq.~\eqref{eq_sup:chi_zero_mom}. We might explicitly compute:
\begin{align} \label{eq_sup:chisq_zeromom}
	\int_k \mathcal{G}_{0,i}^{2} (k) & = \frac{\sqrt{1 - \delta^2}}{16 \pi^3 \Lambda^2 t} \int_{\tilde{\varepsilon} (\Lambda \sqrt{t \beta}, \Lambda \sqrt{t \beta})}^0 d \varepsilon F' (\varepsilon - \beta \mu) \nonumber \\
	& = - \frac{\sqrt{1 - \delta^2}}{16 \pi^3 \Lambda^2 t} \Big[ F (\tilde{\varepsilon} (\Lambda \sqrt{t \beta}, \Lambda \sqrt{t \beta}) - \beta \mu) - F ( - \beta \mu) \Big] .
\end{align}
The argument of the integral in Eq.~\eqref{eq_sup:chisq_zeromom} is negative since $F'$ is always smaller than zero ($\tilde{\varepsilon} (\Lambda \sqrt{t \beta}, \Lambda \sqrt{t \beta}) \le 0$). At high temperatures ($\beta \rightarrow 0^+$), the integral above goes to zero (indeed, $\tilde{\varepsilon} (0,0) = 0$), while at zero temperature ($\beta \rightarrow + \infty$) the overall result of the integral is $-1$. Indeed, at zero temperature the chemical potential of the system is negative and small $\mu = - | \mu |$, leading to $F (- \beta \mu) = F (\beta | \mu |) \xrightarrow[\beta \rightarrow + \infty]{} 0^+$ and to $F (\tilde{\varepsilon} (\Lambda \sqrt{t \beta}, \Lambda \sqrt{t \beta})- \beta \mu) = F (- | \tilde{\varepsilon} (\Lambda \sqrt{t \beta}, \Lambda \sqrt{t \beta}) | + \beta | \mu |) \xrightarrow[\beta \rightarrow + \infty]{} +1$, since the functional dependence of $\tilde{\varepsilon} (\Lambda \sqrt{t \beta}, \Lambda \sqrt{t \beta})$ by $\beta$ is still linear, but the overall prefactor is larger than $\mu$. Thus, we can say that, if:
\begin{align}
	1 - \frac{g_\text{d} - g_\text{o}}{2} \frac{\sqrt{1 - \delta^2}}{16 \pi^3 \Lambda^2 t} < 0 , \nonumber
\end{align}
a transition to a nematic state must have occurred in the system at some finite temperature, leading to the condition for the hopping strength:
\begin{align}
	\Lambda^2 t < \frac{\sqrt{1 - \delta^2}}{32 \pi^3} (g_\text{d} - g_\text{o}) . \nonumber
\end{align}
Generally speaking, we have a divergence in the nematic response function when:
\begin{align}
	\Lambda^2 t = \frac{\sqrt{1 - \delta^2}}{32 \pi^3} (g_\text{d} - g_\text{o}) \Big[ F (\tilde{\varepsilon} (\Lambda \sqrt{t \beta}, \Lambda \sqrt{t \beta}) - \beta \mu) - F ( - \beta \mu) \Big]. \nonumber
\end{align}
By rewriting $O_1 = O \cos (\theta)$ and $O_2 = O \sin (\theta)$ and by expanding the effective free energy Eq.~\eqref{eq_sup:free_en_beforeexp1Q} for small values of the nematic order parameter up to the fourth order in $O$ and taking into account that the integral over the three Green's functions are actually equivalent one to the other, we obtain:
\begin{align} \label{eq_sup:freeenZ3potts}
	& \mathcal{F}_\text{eff}^{Q_{1a}} \approx - 3 \int_k \ln \Big( \mathcal{G}_{0,1}^{-1} (i \omega_n, \mathbf{k}) \Big) + \frac{g_\text{d} + 2 g_\text{o}}{6} N^2 \nonumber \\
	& + \frac{O^2}{2 g_\text{nem}^{Q_{1a}}} \Big[ 1 + g_\text{nem}^{Q_{1a}} \int_k \mathcal{G}_{0,1}^{2} (i \omega_n, \mathbf{k}) \Big] - \frac{O^3}{3 \sqrt{6}} \cos (3 \theta) \int_k \mathcal{G}_{0,1}^{3} (i \omega_n, \mathbf{k}) + \frac{O^4}{8} \int_k \mathcal{G}_{0,1}^{4} (i \omega_n, \mathbf{k}) = \nonumber \\
	& = - 3 \int_k \ln \Big( \mathcal{G}_{0,1}^{-1} (i \omega_n, \mathbf{k}) \Big) + \frac{g_\text{d} + 2 g_\text{o}}{6} N^2 + \frac{\alpha}{2} \sum_{i=1}^2 O_i^2 - \frac{\gamma}{3} O_1 (O_1^2 - 3 O_2^2) + \frac{\xi}{4} \big( \sum_{i=1}^2 O_i^2 \big)^2 , 
\end{align}
which is the free energy for a three states Potts model \cite{Straley1973_JPA,Hecker2018_npj,Little2020_NatMat,Cho2020_NatComm}. The lack of broken $\mathbb{Z}_3$ symmetry \textit{at} the phase transition, which is reflected in the diagonal structure of the nematic response function Eq.~\eqref{eq_sup:chi_nem1Q}, is expected for symmetry reasons. However, as soon as the nematic parameter $O$ becomes finite, this symmetry breaks (\textit{below} the instability level). The divergence of the nematic response function is expected when the quadratic coefficient of Eq.~\eqref{eq_sup:freeenZ3potts} goes to zero. Moreover, the state that preserves the mirror symmetry is preferred, reducing the point group of the state to C$_{2 \text{v}}$. A similar expression to Eq.~\eqref{eq_sup:freeenZ3potts} is expected also by expanding $\mathcal{F}_\text{eff}^{Q_{4 a}}$, Eq.~\eqref{eq_supl:free_en_eff_finQ4}, for small values of the nematic fields $O_1$ and $O_2$ in the case of $\Delta_0 = 0$.

\noindent
In the last line of Eq.~\eqref{eq_sup:freeenZ3potts}, we have introduced the coefficients:
\begin{align}
	& \alpha (\beta, \mu) = \frac{1}{g_\text{nem}^{Q_{1a}}} + \int_k \mathcal{G}_{0,1}^2 (k) , \nonumber \\
	& \gamma (\beta, \mu) = \frac{3}{2} \int_k \mathcal{G}_{0,1}^3 (k) , \nonumber \\
	& \xi (\beta, \mu) = \frac{1}{8} \int_k \mathcal{G}_{0,1}^4 (k) , \nonumber
\end{align}
that can be easily computed considering the relation:
\begin{align}
    \int_k \mathcal{G}_{0,1}^n (k) = \frac{1}{(n-1)!} \int_{-\Lambda}^\Lambda \frac{dk_x \ dk_y}{(2 \pi \Lambda)^2} \ \partial^{n-1}_{\varepsilon} f (\varepsilon) \vert_{\varepsilon = \varepsilon_{M_1', \mathbf{k}} - \mu} . \nonumber
\end{align}
Similarly to the approach we took in the previous section, we might take into account the spatial fluctuations of $O_1$ and $O_2$ substituting \cite{Caprara2015_PRB,Paul2017_PRL}
\begin{align}
	\alpha \rightarrow 2 \chi^{-1}_{i} (k) & = \frac{1}{g_\text{nem}^{Q_{1a}}} + \frac{1}{2} \int_{k'} \mathcal{G}_{0,i} (k') \mathcal{G}_{0,i} (k' + k) = \alpha (\beta, \mu) + \bar{\gamma} \frac{| \nu_n |}{|\mathbf{k}|} + \bar{f} (\mathbf{k}) , \nonumber
\end{align}
where $\chi^{-1}_{i} (k)$ represents the propagator for the nematic fluctuations. Differently from the isotropic Fermi liquid, in which one has to distinguish between longitudinal and transverse fluctuations because the latter is a gapless Goldstone mode, in our anisotropic case we only have gapped (longitudinal) fluctuations \cite{Oganesyan2001_PRB,Zacharias2009_PRB,Garst2010_PRB}. In the expression above, $\nu_n = 2 \pi T n$ is the bosonic Matsubara frequency, coherent with the bosonic nature of the excitations of the charge ordered state, $\bar{\gamma}$ is the Landau damping coefficient and $\bar{f} (\mathbf{k}) = k_x^2 + k_y^2$, i.e., the bare susceptibilities are equal $\chi^{-1}_{i} (k) = \chi^{-1} (k)$. The same assumption, in the context of nematicity driven by spin fluctuations in the iron-based superconductors, has been shown not to qualitatively alter the results \cite{Fernandes2012_PRB}. The free energy becomes (neglecting the first two irrelevant contributions):
\begin{align}
	& \mathcal{F}_\text{eff}^{Q_{1a}}  = \sum_{i=1}^2 \int_k O_i (k) \chi^{-1} (k) O_i (k) - \frac{\gamma}{3} \int_x O_1 (x) (O_1^2 (x) - 3 O_2^2 (x)) + \frac{\xi}{4} \int_x \big( \sum_{i=1}^2 O_i^2 (x) \big)^2 . \nonumber
\end{align}

\newpage
\subsection{Finite stress}
The action of a finite stress (st) field $\boldsymbol{\sigma}_{\text{E}_2} = ( \sigma_{\text{E}_{2,1}}, \sigma_{\text{E}_{2,2}} )$ can be included by adding a contribution to the free energy of the kind:
\begin{align}
	\mathcal{F}_\text{st} & = g_a \int_x \sigma_{\text{E}_{2,1}} (x) \epsilon_{\text{E}_{2,1}} (x) + g_b \int_x \sigma_{\text{E}_{2,2}} (x) \epsilon_{\text{E}_{2,2}} (x),
\end{align}
where $g_a$ can be different from $g_b$ since $\boldsymbol{\sigma}_{\text{E}_2}$ is an external field and does not have to satisfy the symmetries of the lattice. By integrating out the elastic deformations, one arrives at the expression:
\begin{align}
	\tilde{\mathcal{F}}_\text{eff} & = \mathcal{F}_\text{eff} - \frac{\tilde{g}^2}{4 c_{\text{E}_2}} \int_x \mathbf{N}^2 (x) - \frac{\tilde{g}}{2 c_{\text{E}_2}} \int_x \Big( g_a \sigma_{\text{E}_{2,1}} (x) N_1 (x) + g_b \sigma_{\text{E}_{2,2}} (x) N_2 (x) \Big) \nonumber \\
	& - \frac{1}{4 c_{\text{E}_2}} \int_x \Big( g_a^2 \sigma_{\text{E}_{2,1}}^2 (x) + g_b^2 \sigma_{\text{E}_{2,2}}^2 (x) \Big) ,
\end{align}
where $\mathcal{F}_\text{eff}$ is one of the effective free energies described above and $\tilde{g}$ represents the interaction strength of the nemato-elastic coupling. The stresses have the effect of fields coupled to the nematic order parameters $N_1$ and $N_2$ so their presence can induce a non-zero expectation value of $N_1$ or $N_2$ irrespective of purely electronic effects or of the coupling to the elastic deformations. The relevant role played by externally applied stresses in stabilizing the nematic state is not in disagreement with recent experimental findings \cite{Guo2024_NatPhys}.

\end{document}